\documentclass[journal]{IEEEtran}

\usepackage{Sections/S0/Packages}
\usepackage{Sections/S0/Commands}

\newacronym{MC}{MC}{molecular communication}
\newacronym{ISI}{ISI}{inter-symbol interference}
\newacronym{DNA}{DNA}{deoxyribonucleic acid}
\newacronym{RV}{RV}{random variable}
\newacronym{PDE}{PDE}{partial differential equation}
\newacronym{CIR}{CIR}{channel impulse response}
\newacronym{AWGN}{AWGN}{additive white Gaussian noise}
\newacronym{PDF}{PDF}{probability density function}
\newacronym{PMF}{PMF}{probability mass function}
\newacronym{CDF}{CDF}{cumulative density function}
\newacronym{CLT}{CLT}{central limit theorem}
\newacronym{RMSE}{RMSE}{root mean square error}
\newacronym{SNR}{SNR}{signal-to-noise ratio}
\newacronym{LRE}{LRE}{law of rare events}
\newacronym{CSK}{CSK}{concentration shift keying}
\newacronym{OOK}{OOK}{on-off keying}
\newacronym{NCR}{NCR}{network of chemical reaction}
\newacronym{MSD}{MSD}{mean squared displacement}

\EndPreamble

\begin{document}


\title{Channel Modeling for Diffusive Molecular Communication -- A Tutorial Review }

\author{Vahid Jamali$\dag^\star$, Arman Ahmadzadeh$\dag^\star$, Wayan Wicke$\dag$, \\ Adam Noel$\ddag$, and Robert Schober$\dag$ \\
$\dag$ Friedrich-Alexander University of Erlangen-N\"urnberg, Germany \\
$\ddag$ University of Warwick, UK
\thanks{$^\star$ Co-first authors.} \vspace{-0.3cm}}
\maketitle

\begin{abstract}
Molecular communication (MC) is a new communication engineering paradigm where molecules are employed as information carriers. MC systems are expected to enable new revolutionary applications such as sensing of target substances in biotechnology, smart drug delivery in medicine, and monitoring of oil pipelines or chemical reactors in industrial settings. As for any other kind of communication, simple yet sufficiently accurate channel models are needed for the design, analysis, and efficient operation of MC systems. In this paper, we provide a tutorial review on mathematical channel modeling for diffusive MC systems. The considered end-to-end MC channel models incorporate the effects of the release mechanism, the MC environment, and the reception mechanism on the observed information molecules. Thereby, the various existing models for the different components of an MC system are presented under a common framework and the underlying biological, chemical, and physical phenomena are discussed. Deterministic models characterizing the expected number of molecules observed at the receiver and statistical models characterizing the actual number of observed molecules are developed. In addition, we provide channel models for time-varying MC systems with moving transmitters and receivers, which are relevant for advanced applications such as smart drug delivery with mobile nanomachines. For complex scenarios, where simple MC channel models cannot be obtained from first principles, we investigate simulation-driven and experimentally-driven channel models. Finally, we provide a detailed discussion of potential challenges, open research problems, and future directions in channel modeling for diffusive MC systems.
\end{abstract}


\section{Introduction}
\label{Sect:Intro}

Wireless communication networks have permeated throughout modern society, but existing systems are constrained by where conventional radio frequency technologies can be deployed. There are emerging applications where wireless communication could be a vital component, but where conventional implementations would be unsafe or impractical. An alternative approach that has received increasing attention within the communications research community over the last decade is \gls{MC}, where molecules are employed as the information carriers\footnote{We note that, in this paper, we use the terms ``molecule" and ``particle" interchangeably.}. \Gls{MC} was first proposed for the design of synthetic communication networks in \cite{Hiyama2005}. The topic has received steady growth since the seminal survey on nanonetworks in \cite{Survey_Mol_Nono}, which are networks of devices with nanoscale functional components. An attractive feature of \gls{MC} is its ubiquitous deployment in natural biochemical and biophysical systems, which lends credibility to its potential for biological applications such as targeting substances, smart drug delivery, and designing lab-on-a-chip systems \cite{Nariman_Survey}. Furthermore, \gls{MC} could be deployed in industrial settings, including the monitoring of chemical reactors and nanoscale manufacturing, or for larger activities such as monitoring the emission of pollutants or the transport of oil \cite{PARCERISAGINE20092753}.

Motivated by natural \gls{MC} systems, several different mechanisms have been considered for \gls{MC} in the literature including free diffusion \cite{PierobonJ3,EckfordDrift}, gap junctions \cite{NakanoGapJunc}, molecular motors \cite{Moore_Mmotor}, and bacterial motors \cite{Akyl_Bmototr}; see Fig.~\ref{Fig:MC_Mechanisms}. In particular, diffusion is referred to as the random movement of small particles suspended in a fluid medium as a result of their collisions with other particles in the fluid. Diffusion is one of the dominant propagation mechanisms in nature including communication inside cells and between cells, e.g., in quorum sensing among bacteria and in the synaptic cleft between neurons. Gap junctions enable another form of communication between cells where the molecules pass through small channels connecting the cytosol of neighboring cells. Calcium signaling is an example of this form of \gls{MC} that is used by adjacent cells to regulate a large number of cellular processes including fertilization,  proliferation, and death of mammalian cells \cite{NakanoGapJunc,AlbertsBook}. Molecular motors enable a form of active transportation of large signaling molecules via a special rail-like infrastructure, e.g., actin or microtubule filaments \cite{mallik2004molecular}. The motor moves along the rail by using repeated cycles of coordinated binding and unbinding of its legs to the rail.  This type of \gls{MC} is used for intracellular communication among organelles inside a cell \cite{Moore_Mmotor}. Finally, bacterial motors enable another kind of  active transport where the bacteria can pick up large signaling molecules, e.g., \gls{DNA}, and move in a specific direction, e.g., due to a food concentration gradient,  using their tiny propellers (known as flagella) \cite{Akyl_Bmototr}. 

\iftoggle{OneColumn}{%
\begin{figure}
	\centering
	\scalebox{0.8}{
		\pstool[width=1.2\linewidth]{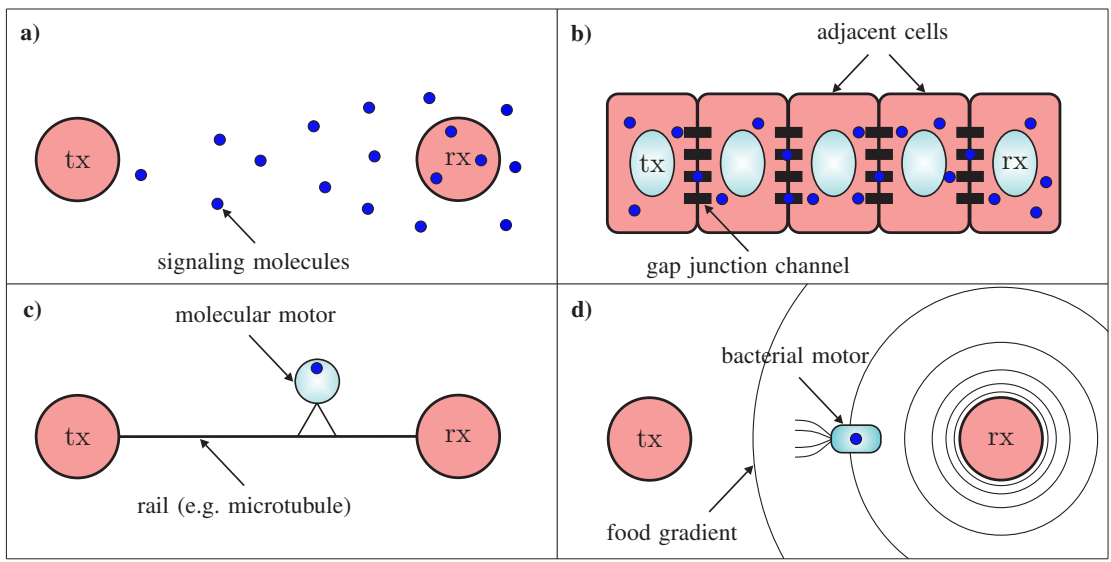}{
			\psfrag{A}[c][c][1]{\textbf{a)}}
			\psfrag{B}[c][c][1]{\textbf{b)}}
			\psfrag{C}[c][c][1]{\textbf{c)}}
			\psfrag{D}[c][c][1]{\textbf{d)}}
			\psfrag{T}[c][c][1.2]{$\mathrm{tx}$}
			\psfrag{R}[c][c][1.2]{$\mathrm{rx}$}
			\psfrag{E}[c][c][1]{signaling molecules}
			\psfrag{F}[c][c][1]{gap junction channel}
			\psfrag{G}[c][c][1]{rail (e.g. microtubule)}
			\psfrag{H}[c][c][1]{molecular motor}
			\psfrag{I}[c][c][1]{food gradient}
			\psfrag{J}[c][c][1]{bacterial motor}
			\psfrag{K}[c][c][1]{adjacent cells}
	} }
	\caption{Bio-inspired mechanisms for MC between a transmitter (tx) and a receiver (rx); a) Free diffusion, b) gap junctions, c) molecular motors, and d) bacterial motors.} 
	\label{Fig:MC_Mechanisms}
\end{figure}
}{%
\begin{figure*}
	\centering
	\scalebox{0.9}{
		\pstool[width=0.9\linewidth]{Sections/S1/Fig/MC/MC}{
			\psfrag{A}[c][c][1]{\textbf{a)}}
			\psfrag{B}[c][c][1]{\textbf{b)}}
			\psfrag{C}[c][c][1]{\textbf{c)}}
			\psfrag{D}[c][c][1]{\textbf{d)}}
			\psfrag{T}[c][c][1.2]{$\mathrm{tx}$}
			\psfrag{R}[c][c][1.2]{$\mathrm{rx}$}
			\psfrag{E}[c][c][1]{signaling molecules}
			\psfrag{F}[c][c][1]{gap junction channel}
			\psfrag{G}[c][c][1]{rail (e.g. microtubule)}
			\psfrag{H}[c][c][1]{molecular motor}
			\psfrag{I}[c][c][1]{food gradient}
			\psfrag{J}[c][c][1]{bacterial motor}
			\psfrag{K}[c][c][1]{adjacent cells}
	} }
	\caption{Bio-inspired mechanisms for MC between a transmitter (tx) and a receiver (rx); a) Free diffusion, b) gap junctions, c) molecular motors, and d) bacterial motors.} 
	\label{Fig:MC_Mechanisms}
\end{figure*}
}

Diffusion-based \gls{MC}, sometimes in combination with advection and chemical reaction networks (CRNs), has been the prevalent approach considered in the literature thus  far; see \cite[Table~4]{Nariman_Survey}. The main advantages of diffusion-based \gls{MC} include that, unlike gap junction-based MC, special infrastructure is not needed, and unlike motor-based \gls{MC}, external energy for propagation of the signaling molecules is not required. Moreover, the simplicity of diffusion makes it an attractive propagation scheme, especially for \emph{ad hoc} networks of devices with limited computational resources. Hence, in this tutorial, we focus on diffusion-based \gls{MC}, where we also consider environments with advection and CRNs.

\subsection{Scope}

A fundamental aspect in the analysis of any communication system is characterizing and understanding the physical layer, i.e., the channel between the transmitter and the receiver. Generally, we rely on a channel model that is simple and yet sufficiently accurate for us to design, analyze, and operate a system that communicates over the channel. A complete view of a diffusion-based channel includes the release of molecules from a transmitter, their propagation in the fluid environment, and the reception mechanism at the receiver. While there is rich historical literature on the physics of diffusion and characterizing \emph{expected} diffusion in environments of different shapes, cf. e.g., \cite{Crank1979,Carslaw}, the communications research community has expanded these models to account for the behavior of the \emph{end-to-end} system, for the inclusion of non-diffusive phenomena that play important roles in biophysical systems, and for the \emph{statistics} of molecular behavior.

Recent surveys, in particular \cite{Survey_Mol_Net,Nariman_Survey}, have provided excellent qualitative summaries of diffusive \gls{MC} and included some of the most common channel models available at the time of writing. A more complete mathematical treatment of diffusion-based modeling of \gls{MC} can be found in \cite{MC_Book}. However, there have been significant advances in channel modeling in the years since the publication of \cite{MC_Book}, and also since the most recent major survey of models in \cite{Nariman_Survey}. In particular, non-diffusive effects that can be coupled with diffusion, such as advective flow and chemical reaction kinetics, have been integrated in many channel models to make them more practical and more accurate.

Due to the rapid growth in channel models, it has become difficult for an interested researcher to enter the \gls{MC} field and become familiar with the state-of-the-art in diffusion-based channel modeling. It has also become more challenging for practitioners in this field to stay up to date. The aim of this tutorial review is to satisfy both audiences. We present a detailed and rigorous mathematical development of diffusive \gls{MC} channel models. We seek to provide a useful comprehensive reference on channel models that is both approachable for an audience that is new to the field and also convenient for active practitioners to assess and select a model. To do so, we begin with a review of the underlying fundamental laws that govern diffusive \gls{MC} channels and show how they are used in the literature to derive the \glspl{CIR} of different \gls{MC} systems. In addition, we present different deterministic and statistical models developed for the observation signal at the receiver. We also discuss the complementary roles of simulations and physical experiments to both support analytical modeling and to provide data-driven models when simple analytical models cannot be readily obtained.

\subsection{Contributions}


In this tutorial review, we make the following contributions:
\begin{enumerate}
	\item By taking a mathematically rigorous approach, we first provide a tutorial on the underlying phenomena from biology, chemistry, and physics, and their effect on the components of \gls{MC} systems. Specifically, we start with Fick's laws of diffusion and build towards the general advection-reaction-diffusion equation. We discuss the common assumptions and special cases that enable the general equation to be solved for the \gls{CIR} in closed form.
	\item We review the major end-to-end channel models in the diffusive \gls{MC} literature including the effects of release mechanisms, the physical channel, and reception mechanisms. In particular, we include the relevant classical models from the physical sciences literature, as well as a comprehensive presentation of the models that have been developed and the equations that have been derived within the communications engineering community over the last few years.
	\item We present a unified definition for the observed signal at a receiver. The unified definition encompasses both timing and counting receivers and helps to better understand the basic assumptions that have been made to arrive at the well-known signal models used in the \gls{MC} literature and how they relate to each other. Then, we focus on counting receivers and derive signal models relevant for different time scales. We further generalize these models to account for interfering noise molecules and \gls{ISI}. Finally, we study the correlation between the received signals observed at different time scales.
	\item  We discuss the integral role of simulations and experiments, in particular to gain insight from a data-driven model when closed-form solutions for the \gls{CIR} are not readily available. We also describe how to implement simple stochastic simulations as well as how to derive an example data-driven model based on experimental data.
\end{enumerate}

For clarity of presentation, the focus of the channel models presented in this work is on a \emph{single} communication link between one transmitter and one receiver. Many of the envisioned applications of diffusive \gls{MC} systems will depend on many links within a network of devices. While there have been a number of relevant contributions that consider the propagation of signals over multiple links, such as via relaying and cooperative detection (cf. e.g., \cite{Atakan2010,Einolghozati2013a,Nakano2013a,Ahmadzadeh2015a,Fang2017a}), these models can often be decomposed into a superposition of individual links. However, it is important to note that we cannot always directly apply single link analysis to multiple-link systems. In particular, we must exercise caution when there are multiple non-transparent devices (such as reactive receivers) in the system that molecules can collide or react with. The presence of one such device will impact the signal received at \emph{any} receiver. Some special cases can still be treated in closed form, such as having two absorbing receivers placed on either side of a transmitter in \cite{Lu2016}, but otherwise we need to resort to data-driven models, such as the simulation of multiple absorbing receivers in \cite{Arifler2017}.


\subsection{Organization}


The rest of this tutorial review is organized as follows, and also summarized in Table I to show how the content of Sections II-V is connected. We review the fundamental physical principles that govern diffusion-based \gls{MC} systems in Section~\ref{Sect:Prelim}. In particular, we model diffusion, advection, and chemical reactions, which lead to a general advection-reaction-diffusion \gls{PDE} to describe the spatio-temporal variation in molecule concentrations.

In Section~\ref{Sect:TxRxCh}, we discuss the components of \gls{MC} systems and their effect on the end-to-end \gls{CIR}. Our definition of the end-to-end channel includes the physical and chemical properties of the transmitter and receiver, as well as the fluid medium in which they are located. A table to summarize the reviewed \glspl{CIR} is also provided. 

In Section~\ref{Sect:RecSig}, we present a unified definition for the diffusive signal observed at the receiver. We focus on counting receivers to derive deterministic and statistical signal models that are valid for different time scales. We also consider the impact of interfering noise, including the interference caused by repeated transmissions at the transmitter, and the correlation among received signals sampled at different time~scales.

We discuss simulation- and experimental-driven models in Section~\ref{Sect:SimExp}. We describe the different physical scales for simulating diffusion-based systems, including existing simulation platforms for each scale, and discuss how to implement simple stochastic simulations. We review a selection of experimental platforms and discuss features that cannot be readily captured via modeling or simulation. Thereby, we propose to employ physically-motivated parametric models and neural networks whose parameters are found using experimental data. One experimental system is presented as a case study for data-driven modeling.

We end this tutorial review with a discussion of future work and open challenges in Section~\ref{Sect:ChalFuture} before presenting our conclusions in Section~\ref{Sect:Concl}. 

\iftoggle{OneColumn}{%
\begin{table}
		\caption{Organization and Content of Sections~II-V and Their Connections.} 
	\label{Fig:Organization}
	\centering
	\scalebox{0.65}{
		\pstool[width=1.6\linewidth]{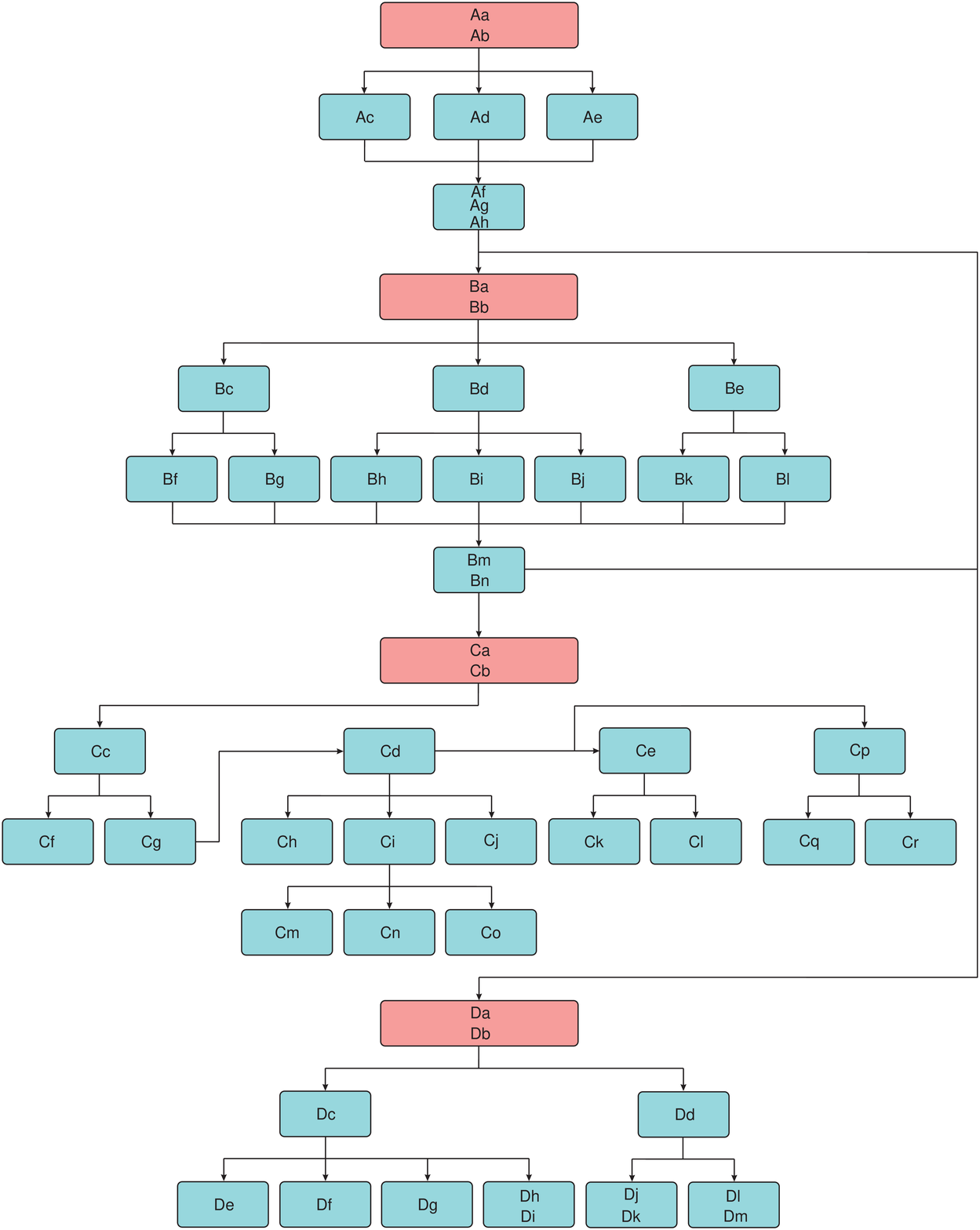}{
			\psfrag{Aa}[c][c][1]{Section~II}
			\psfrag{Ab}[c][c][1]{Governing Physical Principles}
			\psfrag{Ac}[c][c][1]{Diffusion}
			\psfrag{Ad}[c][c][1]{Advection}
			\psfrag{Ae}[c][c][1]{Reaction}
			\psfrag{Af}[c][c][1]{Reaction-}	
			\psfrag{Ag}[c][c][1]{Advection-}	
			\psfrag{Ah}[c][c][1]{Diffusion Eq.}	
			\psfrag{Ba}[c][c][1]{Section~III}
			\psfrag{Bb}[c][c][1]{End-to-End Channel Modeling}
			\psfrag{Bc}[c][c][1]{Transmitter}
			\psfrag{Bd}[c][c][1]{Channel}
			\psfrag{Be}[c][c][1]{Receiver}
			\psfrag{Bf}[c][c][1]{Point}
			\psfrag{Bg}[c][c][1]{Volume}
			\psfrag{Bh}[c][c][1]{Diffusive}
			\psfrag{Bi}[c][c][1]{Advective}
			\psfrag{Bj}[c][c][1]{Degradative}
			\psfrag{Bk}[c][c][1]{Passive}
			\psfrag{Bl}[c][c][1]{Active}
			\psfrag{Bm}[c][c][1]{End-to-End}
			\psfrag{Bn}[c][c][1]{CIR}
			\psfrag{Ca}[c][c][1]{Section~IV}
			\psfrag{Cb}[c][c][1]{Signal Modeling}
			\psfrag{Cc}[c][c][1]{Signal Type}
			\psfrag{Cd}[c][c][1]{Three Scales}
			\psfrag{Ce}[c][c][1]{Interference}
			\psfrag{Cf}[c][c][1]{Timing}
			\psfrag{Cg}[c][c][1]{Counting}
			\psfrag{Ch}[c][c][1]{Deterministic}
			\psfrag{Ci}[c][c][1]{Statistical}
			\psfrag{Cj}[c][c][1]{Time-Varying}
			\psfrag{Ck}[c][c][1]{External}
			\psfrag{Cl}[c][c][1]{ISI}
			\psfrag{Cm}[c][c][1]{Binomial}
			\psfrag{Cn}[c][c][1]{Gaussian}
			\psfrag{Co}[c][c][1]{Poisson}
			\psfrag{Cp}[c][c][1]{Correlation}
			\psfrag{Cq}[c][c][1]{Sample}
			\psfrag{Cr}[c][c][1]{CIR}
			\psfrag{Da}[c][c][1]{Section~V}
			\psfrag{Db}[c][c][1]{Data-Driven Modeling}
			\psfrag{Dc}[c][c][1]{Simulation}
			\psfrag{Dd}[c][c][1]{Experiment}
			\psfrag{De}[c][c][1]{Continuum}
			\psfrag{Df}[c][c][1]{Mesoscopic}
			\psfrag{Dg}[c][c][1]{Microscopic}
			\psfrag{Dh}[c][c][1]{Molecular}
			\psfrag{Di}[c][c][1]{Dynamics}
			\psfrag{Dj}[c][c][1]{Parametric}
			\psfrag{Dk}[c][c][1]{Model}
			\psfrag{Dl}[c][c][1]{Neural}
			\psfrag{Dm}[c][c][1]{Network}
	} }
\end{table}
}{%
\begin{table*}
	\caption{Organization and Content of Sections~II-V and Their Connections.} 
	\label{Fig:Organization}
	\centering
	\scalebox{0.75}{
		\pstool[width=1.2\linewidth]{Sections/S1/Fig/Organization/Organization}{
			\psfrag{Aa}[c][c][1]{Section~II}
			\psfrag{Ab}[c][c][1]{Governing Physical Principles}
			\psfrag{Ac}[c][c][1]{Diffusion}
			\psfrag{Ad}[c][c][1]{Advection}
			\psfrag{Ae}[c][c][1]{Reaction}
			\psfrag{Af}[c][c][1]{Reaction-}	
			\psfrag{Ag}[c][c][1]{Advection-}	
			\psfrag{Ah}[c][c][1]{Diffusion Eq.}	
			\psfrag{Ba}[c][c][1]{Section~III}
			\psfrag{Bb}[c][c][1]{End-to-End Channel Modeling}
			\psfrag{Bc}[c][c][1]{Transmitter}
			\psfrag{Bd}[c][c][1]{Channel}
			\psfrag{Be}[c][c][1]{Receiver}
			\psfrag{Bf}[c][c][1]{Point}
			\psfrag{Bg}[c][c][1]{Volume}
			\psfrag{Bh}[c][c][1]{Diffusive}
			\psfrag{Bi}[c][c][1]{Advective}
			\psfrag{Bj}[c][c][1]{Degradative}
			\psfrag{Bk}[c][c][1]{Passive}
			\psfrag{Bl}[c][c][1]{Active}
			\psfrag{Bm}[c][c][1]{End-to-End}
			\psfrag{Bn}[c][c][1]{CIR}
			\psfrag{Ca}[c][c][1]{Section~IV}
			\psfrag{Cb}[c][c][1]{Signal Modeling}
			\psfrag{Cc}[c][c][1]{Signal Type}
			\psfrag{Cd}[c][c][1]{Three Scales}
			\psfrag{Ce}[c][c][1]{Interference}
			\psfrag{Cf}[c][c][1]{Timing}
			\psfrag{Cg}[c][c][1]{Counting}
			\psfrag{Ch}[c][c][1]{Deterministic}
			\psfrag{Ci}[c][c][1]{Statistical}
			\psfrag{Cj}[c][c][1]{Time-Varying}
			\psfrag{Ck}[c][c][1]{External}
			\psfrag{Cl}[c][c][1]{ISI}
			\psfrag{Cm}[c][c][1]{Binomial}
			\psfrag{Cn}[c][c][1]{Gaussian}
			\psfrag{Co}[c][c][1]{Poisson}
			\psfrag{Cp}[c][c][1]{Correlation}
			\psfrag{Cq}[c][c][1]{Sample}
			\psfrag{Cr}[c][c][1]{CIR}
			\psfrag{Da}[c][c][1]{Section~V}
			\psfrag{Db}[c][c][1]{Data-Driven Modeling}
			\psfrag{Dc}[c][c][1]{Simulation}
			\psfrag{Dd}[c][c][1]{Experiment}
			\psfrag{De}[c][c][1]{Continuum}
			\psfrag{Df}[c][c][1]{Mesoscopic}
			\psfrag{Dg}[c][c][1]{Microscopic}
			\psfrag{Dh}[c][c][1]{Molecular}
			\psfrag{Di}[c][c][1]{Dynamics}
			\psfrag{Dj}[c][c][1]{Parametric}
			\psfrag{Dk}[c][c][1]{Model}
			\psfrag{Dl}[c][c][1]{Neural}
			\psfrag{Dm}[c][c][1]{Network}
	} }
\end{table*}
}

\section{Fundamental Governing Physical Principles in MC Systems}
\label{Sect:Prelim}
In this section, we review the fundamental laws that govern the propagation of molecules. In particular, we mathematically model the impact of diffusion, advection, and reaction on the spatio-temporal distribution of molecules. This modeling is essential for the development of channel models. A solid understanding of these phenomena is needed to develop intuition for molecule propagation in diffusive \gls{MC} systems. Furthermore, in Section~\ref{Sect:TxRxCh}, we will use the mathematical tools introduced in this section for the derivation of the \gls{CIR} for several different diffusive \gls{MC} systems. 

\subsection{Free Diffusion}
Molecules in a fluid environment, such as a liquid or a gas, are affected by thermal vibrations and
collisions with other molecules. The resulting movement of the molecules is a purely random without any preferred direction and is referred to as a random walk or Brownian motion. Let $\mathbf{d}_i(t)=[x,y,z]$ denote a vector specifying the position of the $i$-th molecule in three-dimensional (3D) Cartesian coordinates at time $t$. Thereby, the random walk is modeled by \cite[Eqs. (1.3) and (1.21)]{Berg}
\begin{IEEEeqnarray}{lll} \label{Eq:RandomWalk}
	\mathbf{d}_i(t+\Delta t) = \mathbf{d}_i(t) + \Normal{\mathbf{0}}{2D\Delta t \,\mathbf{I}},
\end{IEEEeqnarray}
where $\Delta t$ is the time step size and $D$ in [m$^2$s$^{-1}$] is the diffusion coefficient of the $i$-th molecule. Moreover, $\Normal{\boldsymbol{\mu}}{\boldsymbol{\Sigma}}$ denotes a multivariate Gaussian \gls{RV} with mean vector $\boldsymbol{\mu}$ and covariance matrix~$\boldsymbol{\Sigma}$, $\mathbf{0}$ represents a vector whose elements are all zeros, and $\mathbf{I}$ is the identity matrix. The diffusion coefficient determines how fast the molecule moves. The larger the diffusion coefficient, the larger the average displacement of the molecule in a given time interval. The value of the diffusion coefficient depends on the environment as well as the shape and the size of the particle. For large spherical particles, the diffusion coefficient can be determined based on the so-called Einstein relation \cite[Eq. (4.15)]{BioPhysic}
\begin{IEEEeqnarray}{lll} \label{Eq:Einstein}
	D=\frac{k_\textsc{b}T}{6\pi\eta R},
\end{IEEEeqnarray}
where $k_\textsc{b}=1.38\times 10^{-23}\,\,\text{J}\text{K}^{-1}$ is the Boltzmann constant, $T$ is the temperature in kelvin, $\eta$ is the (dynamic) viscosity of the fluid ($\eta=10^{-3}\,\,\text{kg}\,\text{m}^{-1}\text{s}^{-1}$ for water at $20\,^\circ\text{C}$), and $R$ is the radius of the particle. Note that larger particles have a smaller diffusion coefficient and are hence less affected by diffusion.

\begin{remk}
Besides the ideal free diffusion with constant diffusion coefficient discussed above, there are also other types of diffusion. For instance, in contrast to the typical free diffusion where the \gls{MSD} is linearly proportional to time, i.e., $\text{MSD}\propto D \Delta t$, in anomalous diffusion, the \gls{MSD} follows a nonlinear relation, i.e., $\text{MSD}\propto D \Delta t^\gamma$ where $\gamma\neq 1$. Sub-diffusion occurs when $\gamma<1$ and can be used to model diffusion inside biological cells where the presence of the organelles does not allow ideal free diffusion to take place \cite{Subdiffusion}. Super-diffusion occurs when $\gamma>1$ and can be used to model \emph{active} cellular transport processes \cite{desposito2011active}. Moreover, in (\ref{Eq:RandomWalk}), we assumed the diffusion coefficient to be \textit{constant}. However, the diffusion coefficient may depend on the local concentration of the molecules \cite{Diffusion_book}. For the constant diffusion coefficient assumption to hold, the temperature and viscosity of the environment are assumed to be uniform and constant and all solute molecules (dissolved molecules) are assumed to be locally dilute everywhere, i.e., the number of solute molecules is sufficiently small everywhere. These assumptions allow us to ignore potential collisions between solute molecules such that the diffusion coefficient does not vary with the local concentration \cite{Diffusion_book,Adam_Thesis}. We refer the readers to \cite{zoppou1999analytical} for the study of diffusion with non-constant diffusion coefficients. Another example of a complex diffusion process is the diffusion of protons in water. Here, the movement of the protons  is a combination of ideal free diffusion and the so-called structural diffusion where protons hop from one water molecule to the next. Nevertheless, it has been shown in \cite{kornyshev2003kinetics} that proton transport can be well approximated by free diffusion with an effective diffusion coefficient. \QEDwhite
\end{remk}

We let $c(\mathbf{d},t)$ denote the concentration of the solute molecules, i.e., the \emph{average number} of solute molecules per unit volume, at coordinate $\mathbf{d}$ and time $t$. The random movement of molecules due to diffusion, described by \Eqref{Eq:RandomWalk}, leads to variation in $c(\mathbf{d},t)$ across time and space that obeys Fick's second law of diffusion\footnote{Fick's first law of diffusion relates the diffusive flux, denoted by $\mathbf{J}(\mathbf{d},t)$, to the concentration as  $\mathbf{J}(\mathbf{d},t)= -D\nabla  c(\mathbf{d},t)$.}
 \begin{IEEEeqnarray}{lll} \label{Eq:PDE_Diff}
 	\frac{\partial c(\mathbf{d},t)}{\partial t} = D \nabla^2 c(\mathbf{d},t),
 \end{IEEEeqnarray}
where $\nabla^2$ is the Laplace operator, e.g., $\nabla^2= \frac{\partial^2}{\partial x^2}+\frac{\partial^2}{\partial y^2}+\frac{\partial^2}{\partial z^2}$ in Cartesian coordinates. The \gls{PDE} in \Eqref{Eq:PDE_Diff} can be solved for simple initial conditions (ICs) and simple boundary conditions (BCs). In the following, we consider a simple example, namely diffusion in an unbounded 3D environment with an impulsive point release, which has been the most widely studied case in the \gls{MC} literature due to its simplicity \cite{Akyl_Receiver_MC,ConsCIR,Nariman_Survey,NoelPro1,Akyildiz_MC_E2E,Equ_MC,PhY_MC,NoelPro3,TCOM_MC_CSI,DistanceEstLett}. 
In the remainder of this paper, we denote the solutions of the considered \glspl{PDE} by $c^{\ast}(\mathbf{d}, t)$.
 
\begin{examp}[Diffusion in an Unbounded 3D Environment with Impulsive Point Release]
\label{Ex:3D_Diff_unbounded}
	Consider a 3D diffusion process with instantaneous release of $N$ solute molecules from $\mathbf{d}_0$ at time $t_0$. To obtain $c^{\ast}(\mathbf{d},t)$, we have to solve \eqref{Eq:PDE_Diff} with the following initial and boundary conditions
\begin{IEEEeqnarray}{rCl}
	\label{Eq:3D_Diff_example_IC} 
	\mathrm{IC_1:} &\,\,& c(\mathbf{d}_0,t \to t_0)  =  N \delta \left( \mathbf{d} - \mathbf{d}_0 \right) \\
	\label{Eq:3D_Diff_example_BC}  
	\mathrm{BC_1:} &\,\,& c(\|\mathbf{d}\| \to \infty ,t) =  0,
\end{IEEEeqnarray} 
where $\delta(\mathbf{d})=\delta(x)\delta(y)\delta(z)$, and $\delta(\cdot)$ is the Dirac delta function. Solving \eqref{Eq:PDE_Diff} with $\mathrm{IC_1}$ and $\mathrm{BC_1}$ yields \cite[Eq. (2.8)]{Berg}
\begin{IEEEeqnarray}{C}
	\label{Eq:Solution_3D_Diff_example}
	c^{\ast}(\mathbf{d},t) =  \frac{N}{(4\pi D (t-t_0))^{3/2}} \exp \left( -\frac{\|\mathbf{d}-\mathbf{d}_0\|^2}{4D(t-t_0)} \right).
\end{IEEEeqnarray}
\QEDwhite
\end{examp}

In Fig.~\ref{Fig:Diffusion}, the molecule concentration $c^{\ast}(\mathbf{d},t)$ \big[molecules/m$^3$\big] in (\ref{Eq:Solution_3D_Diff_example}) is plotted versus time [$\mu$s] at distance $\mathbf{d}=[d,0,0]$ with $d\in\{300,400,500\}$ nm for an initial release of $N=10^4$ molecules with $D=4.5\times 10^{-10}$ m$^2$/s from the origin $\mathbf{d}_0=[0,0,0]$ at time $t_0=0$. From Fig.~\ref{Fig:Diffusion}, we observe that first $c^{\ast}(\mathbf{d},t)$  increases with time, which is due to the non-zero propagation time that the molecules need to reach $\mathbf{d}$, before it  decreases since the molecules diffuse away. Moreover, as distance increases, the peak of the concentration decreases since the molecules are spread over a larger volume. Furthermore, the time when the concentration peak occurs, denoted by $t^{\mathrm{p}}$, increases with distance.

\iftoggle{OneColumn}{%
\begin{figure}
	\centering   
		\resizebox{0.7\linewidth}{!}{
			\psfragfig{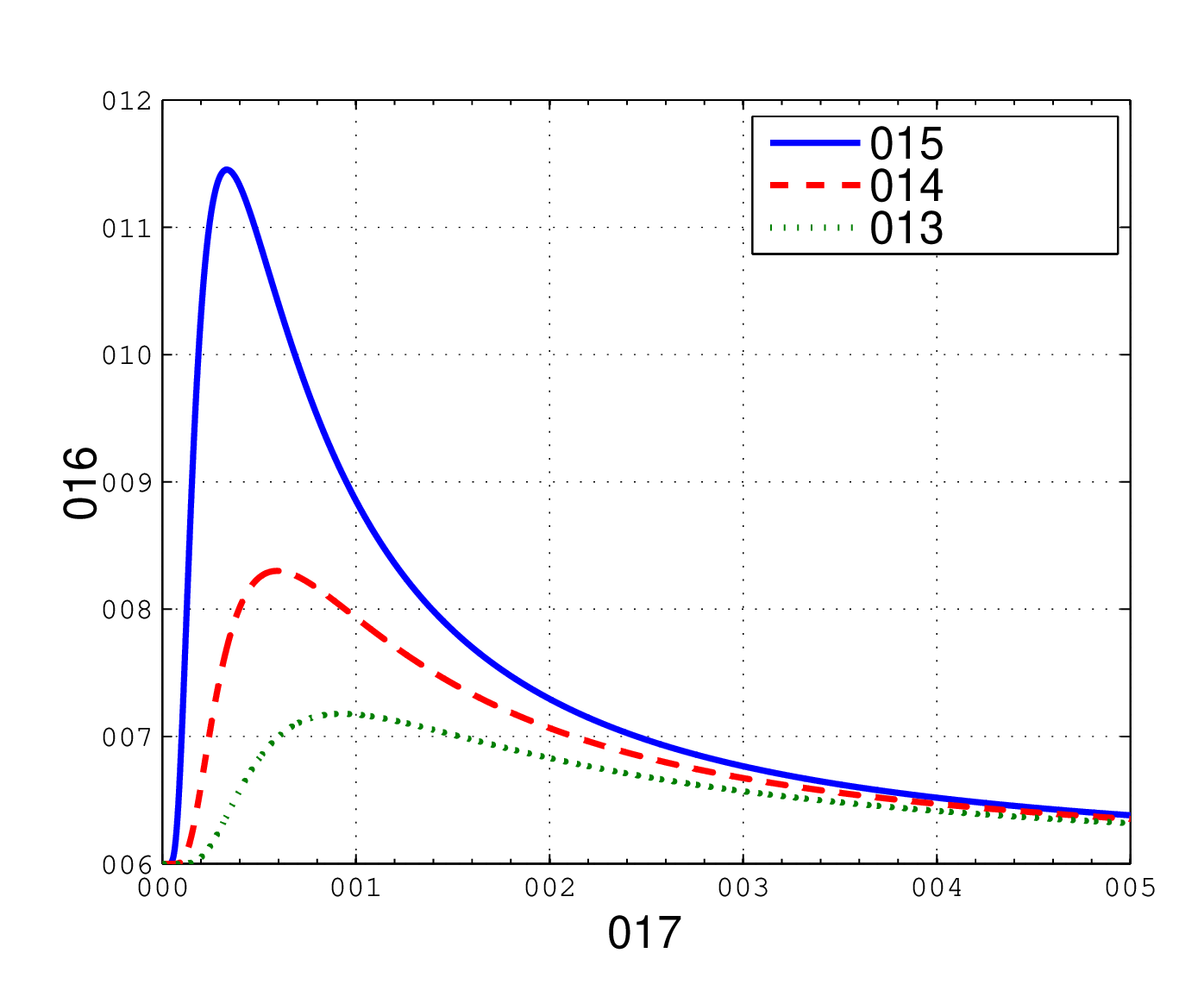}} 
	\caption{Molecule concentration $c^{\ast}(\mathbf{d},t)$ \big[molecules/m$^3$\big] versus  time [$\mu$s] at distance $\mathbf{d}=[d,0,0]$ with $d\in\{300,400,500\}$ nm for initial release of $N=10^4$ molecules with $D=4.5\times 10^{-10}$ m$^2$/s from the origin $\mathbf{d}_0=[0,0,0]$ at time $t_0=0$.}
	\label{Fig:Diffusion}
\end{figure}
}{%
\begin{figure}
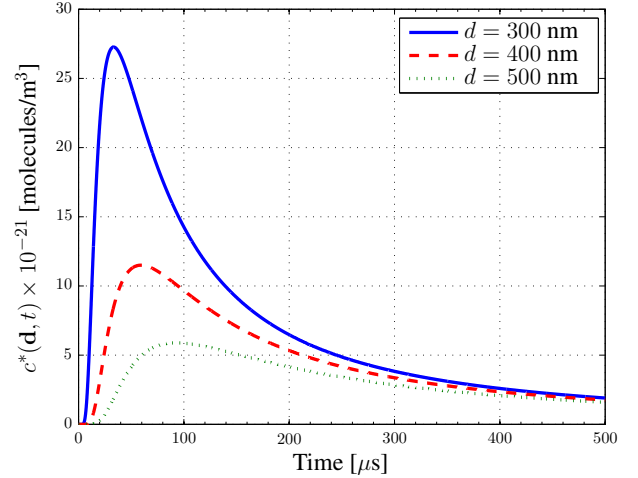

	\centering   
	\resizebox{1\linewidth}{!}{
		\psfragfig{Sections/S2/Fig/Diffusion/CIR_Diff}} \vspace{-0.5cm}
	\caption{Molecule concentration $c^{\ast}(\mathbf{d},t)$ \big[molecules/m$^3$\big] versus  time [$\mu$s] at distance $\mathbf{d}=[d,0,0]$ with $d\in\{300,400,500\}$ nm for initial release of $N=10^4$ molecules with $D=4.5\times 10^{-10}$ m$^2$/s from the origin $\mathbf{d}_0=[0,0,0]$ at time $t_0=0$.}
	\label{Fig:Diffusion}
\end{figure}
}

The assumption of an unbounded environment is accurate when the actual boundaries of the system are far away from the region of interest (i.e., from transmitter and receiver), such that the impact of the boundaries on the diffusing molecules can be neglected. In the following, we present an example where the effect of the boundaries cannot be neglected.

\begin{examp}[Diffusion in an Unbounded Straight Duct with Impulsive Release from Cross-Section]\label{Ex:DiffusionDuct}
	We assume a straight duct\footnote{A duct is a pipe, tube, or channel which carries a liquid or gas.} channel with circular cross-section and for convenience, we employ cylindrical coordinates, i.e., $\mathbf{d}=[\rho, \varphi, z]$ with $ 0\leq \rho \leq \acirc,\, 0\leq\varphi\leq 2\pi,\,-\infty < z < +\infty$,  where $\acirc$ denotes the radius of the circular cross-section of the duct. We assume that, at the time of release, $t_0$, the molecules are uniformly distributed across the cross-section at $z=z_0$. Therefore, we have the following initial and boundary conditions 
		\begin{IEEEeqnarray}{rCl}
				\mathrm{IC_2:} &\,\,& c(\mathbf{d}_0=[\rho,\varphi,z],t\to t_0)  =  \frac{N}{\pi \acirc^2} \delta(z-z_0) \\
			\label{Eq:CirDuct_BCs} 
			\mathrm{BC_2:} &\,\,& \frac{\partial c(\mathbf{d},t)}{\partial \rho}\Big|_{\rho = a_{\mathrm{c}}} =  0 \\ 
			\mathrm{BC_3:} &\,\,& c( \mathbf{d} = [\rho, \varphi, z\to \pm\infty] ,t)= 0,  
		\end{IEEEeqnarray} 
		where $\mathrm{BC}_2$ enforces the reflection of the molecules at the wall, i.e., a fully reflective wall is assumed. Solving \eqref{Eq:PDE_Diff} with $\mathrm{IC_2}$, $\mathrm{BC_2}$, and $\mathrm{BC_3}$ yields \cite{Robert_MCnote}    
		\iftoggle{OneColumn}{%
			\begin{IEEEeqnarray}{rCl}
				\label{Eq:Bounded_Vessel_1D} 
				c^{\ast}(\mathbf{d},t) = \frac{N}{\pi \acirc^2 \sqrt{4\pi D (t-t_0))}} \exp\left( -\frac{(z-z_0)^2}{4D(t-t_0)} \right),\quad \rho<\acirc.
			\end{IEEEeqnarray} 
		}{%
			\begin{IEEEeqnarray}{lCl}
				\label{Eq:Bounded_Vessel_1D} 
				c^{\ast}(\mathbf{d},t) \nonumber \\
				= \frac{N}{\pi \acirc^2 \sqrt{4\pi D (t-t_0))}} \exp\left( -\frac{(z-z_0)^2}{4D(t-t_0)} \right),\quad \rho<\acirc. \quad
			\end{IEEEeqnarray} 
		}		 
		\QEDwhite
\end{examp}

As can be seen from (\ref{Eq:Bounded_Vessel_1D}), $c^{\ast}(\mathbf{d},t)$ does not depend on variables $\rho$ and  $\varphi$ due to the symmetry of  the initial condition and the environment with respect to $\rho$ and  $\varphi$.  This model can be used to characterize the propagation of molecules in blood vessels as is necessary for drug delivery applications of \gls{MC} in the cardiovascular system  \cite{FELICETTI201627,chahibi2015molecular,femminella2015molecular,chude2017molecular,mosayebi2018early}.

\subsection{Advection} 
\label{subsection;advection}
Besides diffusion, advection is another fundamental mechanism for solute particle transport in a fluid environment.
In the following, we first specify how mass transport by advection affects a single solute particle.
Subsequently, we distinguish between two types of advection, namely drift and fluid flow, and give the particle velocity vector for some example cases.
Moreover, we present the advection equation which describes the change in molecule concentration due to advection.
Finally, we introduce the advection-diffusion equation, which captures the joint impact of diffusion and advection, and characterize the relative importance of diffusion and advection.

In general, transport by advection can be described by a velocity vector $\mathbf{v}(\mathbf{d},t)$ which generally may depend on position $\mathbf{d}$ and time $t$.
When considering the movement of the $i$-th particle at position $\mathbf{d}_i$ due to advection, its position at time $t+\Delta t$ can be modeled by
\begin{IEEEeqnarray}{lll} \label{Eq:Flow}
	\mathbf{d}_i(t+\Delta t) = \mathbf{d}_i(t) + \mathbf{v}(\mathbf{d}_i(t),t)\Delta t,
\end{IEEEeqnarray}
where $\Delta t$ should be small enough such that the velocity vector is constant between $\mathbf{d}_i(t)$ and $\mathbf{d}_i(t+\Delta t)$. Next, we discuss what may cause the velocity vector $\mathbf{v}(\mathbf{d},t)$ and what form it may take.

\subsubsection{Velocity Vector Field} Transport by advection can be mediated by different physical mechanisms which we categorize as \emph{force-induced drift} and \emph{bulk flow} \cite{Adam_OptReciever,WayanPro2}.   

\textbf{Force-Induced Drift:} Advection can be caused by external forces acting on the particles but not on the fluid containing the particles.
An external force can be modeled by force vector $\mathbf{F}(\mathbf{d},t)$ which describes the force on a particle at position $\mathbf{d}$ at time $t$.
These external forces can be electrical, e.g., if the particles are ions, or magnetic, e.g., if the particles are magnetic nanoparticles, or gravitational, e.g., if the particles have sufficient mass, or a combination of forces \cite{WayanPro2,tehrani2015novel}. When the force is not too large, the velocity vector can be determined from the corresponding force by Stokes' law via \cite[Eq. (2.65)]{bruus_theoretical_2007}
\begin{equation}
    \label{Eq:StokesLaw}
    \mathbf{v}(\mathbf{d},t) = \frac{\mathbf{F}(\mathbf{d},t)}{\zeta},
\end{equation}
where $\zeta$ is a proportionality constant referred to as the friction coefficient.
The friction coefficient can be related to the diffusion coefficient via $\zeta D=k_\textsc{b}T$.
In other words, using \eqref{Eq:Einstein}, we obtain $\zeta=6\pi\eta R$. Force $\mathbf{F}(\mathbf{d},t)$ may vary with time (e.g., for ions if the electric field changes over time) and space (e.g., for magnetic nanoparticles, the magnetic force generally decreases rapidly with increasing distance from the magnet) \cite{WayanPro2,tehrani2015novel}.

\textbf{Bulk Flow:} If the particle movement is induced by the movement of the fluid, then the resulting transport by advection is referred to as flow.
Flow can be encountered in many \gls{MC} environments such as blood vessels and microfluidic channels \cite{probstein_physicochemical_2005}.
In \gls{MC}, we typically have dilute particle suspensions, where the flow velocity $\mathbf{v}(\mathbf{d},t)$ is independent from the particle concentration.
Thereby, the velocity vector will depend on space if there are boundaries or obstacles in the environment, e.g., in a duct, the flow velocity is typically largest in the center and smallest at the boundary where the fluid is subject to friction.
The flow may also depend on time, e.g., in a blood vessel the flow is generated by the periodic contractions of the heart. 

\begin{remk}
	Although both flow and external force cause the particles to drift, which can be modeled by \eqref{Eq:Flow}, they may require quite different considerations. For instance, any object in the environment influences the velocity vector caused by bulk flow since the flow may not be able to penetrate the object and has to go around the object. On the other hand, the drift velocity vector caused by an external force is not necessarily influenced by objects in the environment. \QEDwhite
\end{remk}

Flow can be also categorized into two classes, namely \emph{turbulent} and \emph{laminar} flow. In particular, when the variations of the flow velocity, over space and/or time, are stochastic, e.g., due to rough surfaces and high flow velocities \cite{white_fluid_2016}, we refer to the flow as turbulent. If the flow is not turbulent, it is referred to as laminar. For flow in a bounded environment of effective length $\deff$ and with an effective velocity of $\veff$, the Reynolds number can be used as a criterion for predicting laminar or turbulent flow and is given by \cite[Eq. (1.24)]{white_fluid_2016}
\begin{equation}
    \mathrm{Re} = \frac{\deff\cdot \veff}{\nu},
\end{equation}
where $\nu$ is the kinematic viscosity [$\text{m}^2/\text{s}$] of the fluid\footnote{Kinematic viscosity $\nu$ is related to (dynamic) viscosity $\eta$ according to $\nu=\eta/\rho_{d}$ where $\rho_d$ [$\text{kg}\,\text{m}^{-3}$] is the fluid density.}.
For example, for flow in a straight pipe with circular cross-section of radius $\acirc$, the flow can be assumed to be laminar and turbulent for $\mathrm{Re}\ll2100$ and $\mathrm{Re}\gg2100$, respectively, where $\deff=\acirc$  \cite{white_fluid_2016}.
For microfluidic settings, typically $\mathrm{Re}\ll10$ and hence laminar flow can be assumed \cite{bruus_theoretical_2007}. For most blood vessels, $\mathrm{Re}<500$ holds and hence the blood flow is typically laminar \cite{back1986measurement,Arendt_Flow}. Only in large arteries such as the aorta (the largest artery in the human body), the Reynolds number can be in the range $[3400, 4500]$ and thereby blood flow exhibits turbulent behavior \cite{Arendt_Flow}.

Generally, for a given environment, the flow velocity vector $\mathbf{v}(\mathbf{d},t)$ as a function of space and time can be determined by solving the so-called Navier-Stokes equation with appropriate boundary conditions, see e.g. \cite[Eq.~(5.22)]{bruus_theoretical_2007}. Let us review two special cases of $\mathbf{v}(\mathbf{d},t)$, which have been widely studied in the \gls{MC} literature \cite{NoelPro3,Nariman_Survey,Adam_OptReciever,WayanPro1,koike2017molecular} and are also considered in Section~\ref{Sect:TxRxCh}.

\begin{examp}[Uniform and/or Constant Advection]\label{Ex:Uniform}
    For uniform advection, the velocity vector is constant across space but can be time-dependent, i.e., $\mathbf{v}(\mathbf{d},t)=\mathbf{v}(t)$  \cite{koike2017molecular}. For advection by flow in an unbounded environment, uniform flow solves the Navier-Stokes equation and hence can be physically plausible. Moreover, for advection by drift, uniform drift is applicable when the corresponding force vector does not depend on space, see \eqref{Eq:StokesLaw}. As a special case, the velocity vector may be constant across both space and time, i.e., $\mathbf{v}(\mathbf{d},t)=\mathbf{v}$. Due to its simplicity, advection with constant velocity is the most widely-studied advection model in the \gls{MC} literature \cite{NoelPro3,Nariman_Survey,Adam_OptReciever}.   \QEDwhite
\end{examp}

\begin{examp}[Steady Flow in an Infinite Straight Duct with Circular Cross-Section]\label{Ex:Poiseuille}
    For this example, we concentrate on advection by fluid flow because force-induced drift is completely specified by \eqref{Eq:StokesLaw}. In particular, in this case, the  flow velocity vector in cylindrical coordinates $[\rho,\varphi,z]$ can be obtained as \cite[Eq. (4.134)]{white_fluid_2016}
    \begin{equation}
        \label{Eq:Poiseuille}
        \mathbf{v}(\rho) = \left[0,0,v_0\left(1-\frac{\rho^2}{\acirc^2}\right)\right],\qquad 0\leq \rho \leq \acirc,
    \end{equation}
    where $v_0$ is the center velocity. The flow described in \eqref{Eq:Poiseuille} is laminar and can be interpreted as follows.
    For a given $\rho$, the flow velocity in \eqref{Eq:Poiseuille} is constant but increases from the boundary where $\mathbf{v}(\acirc)=[0,0,0]$ towards the center where $\mathbf{v}(0)=[0,0,v_0]$, i.e., for each $\rho$, we can think of a circular layer within the duct that slides along its neighboring layers with a constant velocity. The velocity vector in (\ref{Eq:Poiseuille}) is known as the Poiseuille flow profile and is a common model for the flow in blood capillaries \cite{WayanPro1}. \QEDwhite
\end{examp}

While for other environments and boundary conditions the velocity vector can still in principle be obtained from the Navier-Stokes equation, it is often not possible to do so analytically. In these cases, the Navier-Stokes equation can be solved by numerical algorithms that are well-established in computational fluid dynamics \cite{white_fluid_2016}.

\subsubsection{Advection Equation} Given $\mathbf{v}(\mathbf{d},t)$, the change in concentration with respect to time due to advective transport is modeled by the following \gls{PDE}, which is referred to as the advection equation or continuity equation \cite[Eq. (4.14)]{bruus_theoretical_2007}
\begin{IEEEeqnarray}{lll} \label{Eq:PDE_Flow}
	\frac{\partial c(\mathbf{d},t)}{\partial t} = - \nabla\cdot \left(\mathbf{v}(\mathbf{d},t)c(\mathbf{d},t)\right),
\end{IEEEeqnarray}
where $\nabla=[\frac{\partial}{\partial x},\frac{\partial}{\partial y},\frac{\partial}{\partial z}]$ denotes the gradient operator and $\mathbf{x}\cdot\mathbf{y}$ denotes the inner product of two vectors $\mathbf{x}$ and $\mathbf{y}$. In general, \eqref{Eq:PDE_Flow} cannot be readily solved for a given velocity vector and numerical methods have to be employed \cite{chung_computational_2010}. Nevertheless, for the velocity vectors in Examples~\ref{Ex:Uniform} and \ref{Ex:Poiseuille}, (\ref{Eq:PDE_Flow}) can be solved as shown in the following.

\begin{examp}
	Assuming initial condition $c(\mathbf{d},0)$ at $t=0$, the advection equation \eqref{Eq:PDE_Flow} has the following solution for $t>0$
	\iftoggle{OneColumn}{%
		\begin{IEEEeqnarray}{lll} \label{Eq:Sol_AdvecUniformLaminar}		
		c^{\ast}(\mathbf{d},t)  = \begin{cases}
			c\left(\mathbf{d}-\displaystyle\int_0^t\mathbf{v}(\tau)\de\tau,0\right),\quad &\text{Uniform Flow} \\
			c\left(\mathbf{d}-\mathbf{v}t,0\right),\quad &\text{Constant Uniform Flow} \\ 
			c\left(\mathbf{d}-\mathbf{v}(\rho)t,0\right),\quad &\text{Poiseuille Flow}.
		\end{cases}
	\end{IEEEeqnarray}	
	}{%
			\begin{IEEEeqnarray}{lll} \label{Eq:Sol_AdvecUniformLaminar}		
			c^{\ast}(\mathbf{d},t)  \nonumber \\ 
			= \begin{cases}
				c\left(\mathbf{d}-\displaystyle\int_0^t\mathbf{v}(\tau)\de\tau,0\right), &\text{Uniform Flow} \\
				c\left(\mathbf{d}-\mathbf{v}t,0\right), &\text{Constant Uniform Flow} \\ 
				c\left(\mathbf{d}-\mathbf{v}(\rho)t,0\right), &\text{Poiseuille Flow}.
			\end{cases}\quad
		\end{IEEEeqnarray}
	} 
	\QEDwhite
\end{examp}

We note that while the solutions in \eqref{Eq:Sol_AdvecUniformLaminar} appear similar, they are actually fundamentally different.
In particular, for constant uniform flow and uniform flow (space-independent flow profiles), the initial concentration is simply translated to a different position without changing its shape. However, for Poiseuille flow (space-dependent flow profile), the concentration generally spreads in space over time depending on the initial concentration.

\subsubsection{Advection-Diffusion Equation}
In many application scenarios, such as drug delivery via the capillary networks \cite{FELICETTI201627,chahibi2015molecular,femminella2015molecular,chude2017molecular,mosayebi2018early}, advection and diffusion are both present in the \gls{MC} environment. Thereby, the combined effect of both advection and diffusion is characterized by the following \gls{PDE} known as the advection-diffusion equation 
\begin{IEEEeqnarray}{lll} \label{Eq:PDE_FlowDiff}
	\frac{\partial c(\mathbf{d},t)}{\partial t} = D \nabla^2 c(\mathbf{d},t) - \nabla\cdot \left(\mathbf{v}(\mathbf{d},t)c(\mathbf{d},t)\right).
\end{IEEEeqnarray}
Similar to diffusion equation (\ref{Eq:PDE_Diff}), (\ref{Eq:PDE_FlowDiff}) cannot be solved analytically for general velocity vectors $\mathbf{v}(\mathbf{d},t)$ and general boundary and initial conditions. In the following, we first provide the solution of (\ref{Eq:PDE_FlowDiff}) for constant uniform flow in an unbounded environment. Subsequently, we quantify the relative impact of diffusion  over advection by introducing the notions of P\'eclet number and dispersion factor. 
 
\begin{examp}
 	Consider an unbounded 3D environment with instantaneous release of $N$ solute molecules at $\mathbf{d}_0$ at time $t_0$. Solving (\ref{Eq:PDE_FlowDiff}) with initial condition $\mathrm{IC_1}$ in \eqref{Eq:3D_Diff_example_IC}, boundary condition $\mathrm{BC_1}$ in \eqref{Eq:3D_Diff_example_BC}, and constant uniform velocity vector $\mathbf{v}$ yields \cite[Eq. (18)]{NoelPro3} 
 	\iftoggle{OneColumn}{%
 	 \begin{IEEEeqnarray}{C}
 		\label{Eq:Solution_DiffAdvec}
 		c^{\ast}(\mathbf{d},t) =  \frac{N}{(4\pi D (t-t_0))^{3/2}} \exp \left( -\frac{\|\mathbf{d}-(t-t_0)\mathbf{v}-\mathbf{d}_0\|^2}{4D(t-t_0)} \right),\quad t>t_0.
 	\end{IEEEeqnarray}		
 	}{%
 		 \begin{IEEEeqnarray}{ll}
 			\label{Eq:Solution_DiffAdvec}
 			c^{\ast}(\mathbf{d},t) =  &\frac{N}{(4\pi D (t-t_0))^{3/2}} \nonumber \\
 			&\times \exp \left( -\frac{\|\mathbf{d}-(t-t_0)\mathbf{v}-\mathbf{d}_0\|^2}{4D(t-t_0)} \right),\quad t>t_0.\quad\,\,
 		\end{IEEEeqnarray}	
 	}
 \QEDwhite
\end{examp}

In Fig.~\ref{Fig:Advection}, we show molecule concentration $c^{\ast}(\mathbf{d},t)$ [molecules/m$^3$] in \eqref{Eq:Solution_DiffAdvec} versus  time [$\mu$s] at $\mathbf{d}=[400,0,0]$ nm for initial release of $N=10^4$ molecules with $D=4.5\times 10^{-10}$ m$^2$/s from $\mathbf{d}_0=[0,0,0]$ at $t_0=0$,  and flow velocity vector $\mathbf{v}=[v,0,0]$ with $v\in\{0,2,5\}\times 10^{-3}$ m/s. From Fig.~\ref{Fig:Advection}, we observe that as the flow velocity increases, the concentration peak increases and $t^{\mathrm{p}}$ decreases. This is mainly due to the fact that the flow is in the same direction as the point where the concentration is measured, i.e.,  parallel flow is considered. Parallel flow can considerably enhance the coverage of a diffusion-based \gls{MC} system, e.g., in blood vessels. Moreover, by increasing $v$, the tail of $c^{\ast}(\mathbf{d},t)$ over time is decreased, which is useful for \gls{ISI} reduction in \gls{MC} systems \cite{farsad_tabletop_2013,NoelPro3}.

\iftoggle{OneColumn}{%
\begin{figure}
	\centering  
	\resizebox{0.7\linewidth}{!}{
		\psfragfig{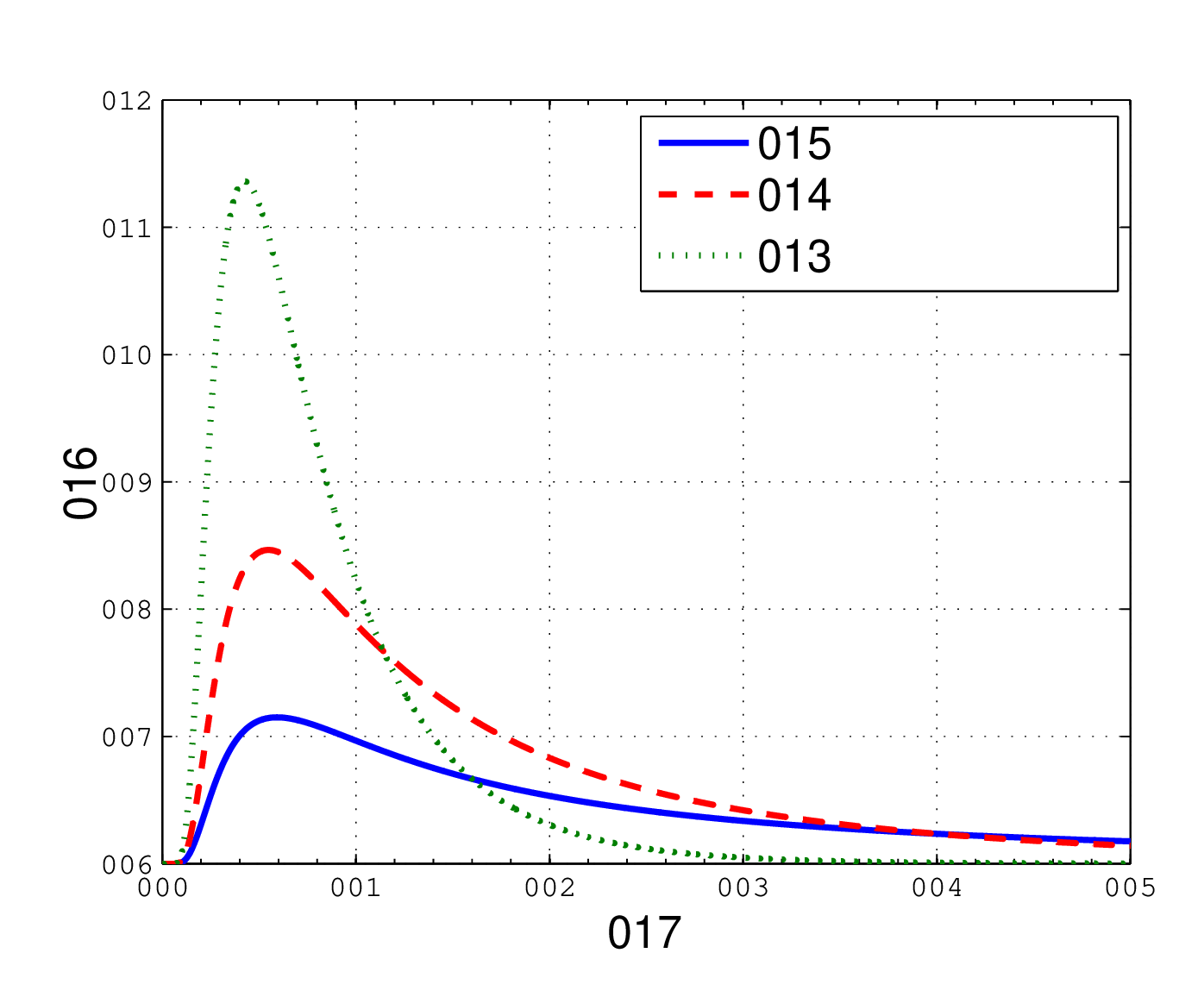}} 
	\caption{Molecule concentration $c^{\ast}(\mathbf{d},t)$ [molecules/m$^3$] versus  time [$\mu$s] at $\mathbf{d}=[400,0,0]$ nm for initial release of $N=10^4$ molecules with $D=4.5\times 10^{-10}$ m$^2$/s from $\mathbf{d}_0=[0,0,0]$ at $t_0=0$, and flow velocity $\mathbf{v}=[v,0,0]$ with $v\in\{0,2,5\}\times 10^{-3}$ m/s.}
	\label{Fig:Advection}
\end{figure}
}{%
	\begin{figure}
		\centering  
		\resizebox{1\linewidth}{!}{
			\psfragfig{Sections/S2/Fig/Advection/CIR_Flow}} \vspace{-0.5cm}
		\caption{Molecule concentration $c^{\ast}(\mathbf{d},t)$ [molecules/m$^3$] versus  time [$\mu$s] at $\mathbf{d}=[400,0,0]$ nm for initial release of $N=10^4$ molecules with $D=4.5\times 10^{-10}$ m$^2$/s from $\mathbf{d}_0=[0,0,0]$ at $t_0=0$, and flow velocity $\mathbf{v}=[v,0,0]$ with $v\in\{0,2,5\}\times 10^{-3}$ m/s.}
		\label{Fig:Advection}
	\end{figure}
}

\textbf{Relative Importance of Advection over Diffusion for Molecule Transport:} Advection and diffusion can both displace and transport molecules, albeit in different ways. An important question is under what conditions is one more effective than the other. The P\'eclet number, denoted by $\mathrm{Pe}$, can be used to answer this question.  Let us assume a velocity vector with strength $v$ and transport over a distance $d_{c}$ which is referred to as the characteristic length. The P\'eclet number quantifies the ratio of time required for particles to be transported by diffusion over distance $d_c$ (which is proportional to $d_c^2/D$) with the time required for particles to be transported by advection over distance $d_c$ (given by $d_c/v$). This ratio is given by    \cite[Eq. (4.44)]{bruus_theoretical_2007}
\begin{equation}
\mathrm{Pe}=\frac{\frac{d_c^2}{D}}{\frac{d_c}{v}}=\frac{v\cdot d_{c}}{D}. 
\end{equation}
Note that $\mathrm{Pe}$ is a dimensionless number. If $\mathrm{Pe}\ll 1$ holds, diffusion dominates advection and the spreading of molecules is almost isotropic despite a weak biased transport in the direction of the flow. In this case, the solution of the diffusion equation (\ref{Eq:PDE_Diff}) provides an accurate estimate of the molecule concentration. On the other hand, if $\mathrm{Pe}\gg 1$ holds, advection dominates diffusion and is the main cause for molecule transport. In this case, the advection equation (\ref{Eq:PDE_Flow}) can be solved to obtain an accurate estimate of the molecule concentration. Finally, for $\mathrm{Pe} \approx 1$, molecule transport is sensitive to both diffusion and advection and the advection-diffusion equation in (\ref{Eq:PDE_FlowDiff}) should be solved.

\textbf{Relative Importance of Advection over Diffusion for Dispersion:} Let us consider a straight duct with a circular cross-section, see Examples~\ref{Ex:DiffusionDuct} and \ref{Ex:Poiseuille}, where advection is the main transport mechanism along the duct. In other words, $\mathrm{Pe}_z\triangleq \frac{v_{\mathrm{eff}}d_z}{D}\gg 1$ holds where $\mathrm{Pe}_z$ denotes the P\'eclet number for transport along the $z$-axis, $v_{\mathrm{eff}}=v_0/2$ is the effective flow velocity in the duct (see (\ref{Eq:Poiseuille})), and $d_z$ is the desired transport length along the $z$-axis. In this case, we are interested in studying the dispersion (spatial spreading) of individual particles across the cross-section over the time when transport along the $z$-axis occurs. In particular, one may distinguish between the following two extreme regimes, namely the non-dispersive and dispersive regimes: 

\textit{i) Non-dispersive regime:} Here, particles do not considerably diffuse across the cross-section while being transported by advection. Therefore, each particle is simply transported along the $z$-axis by advection with a velocity strength that depends on the radial position of the particle, $\rho$, according to \eqref{Eq:Poiseuille}. We note that although the dispersion of individual particles is negligible in this regime, the shape of the concentration profile varies over time since the flow has a different effect at different radial positions, i.e., particles closer to the center of the duct travel faster. 

\textit{ii) Dispersive regime:} In the dispersive regime, particles fully diffuse across the cross-section while also being transported along the $z$-axis by advection. 
In addition to the dispersion across the cross-section, there is also dispersion along the $z$-axis, due to the combined impact of diffusion and advection with  space-dependent flow profile (\ref{Eq:Poiseuille}).

In the following, we mathematically quantify the dispersive and non-dispersive regimes in terms of system parameters, i.e., $v_{\mathrm{eff}}$, $D$, $d_z$, and $\acirc$. We choose the characteristic length $d_{c}$ as the distance over which the velocity vector changes (usually a fraction of $\acirc$). Moreover, we define $\bar{d}_z\triangleq d_z/d_c$ as the corresponding dimensionless normalized distance with respect to characteristic distance $d_c$. Then, we can compare the characteristic time required for particles to be transported by advection over distance $d_z$  (given by $d_z/v_{\mathrm{eff}}$) with the time required for diffusion over distance $d_c$ (which is proportional to $d_c^2/D$). To compare these two time scales, we can define a dispersion factor $\alpha_d$ as
\begin{equation}
    \label{Eq:Timescales}
    \alpha_d = \frac{\frac{d_z}{v_{\mathrm{eff}}}}{\frac{d_c^2}{D}}=\frac{Dd_z}{v_{\mathrm{eff}} d_c^2}=\frac{\bar{d}_z^2}{\mathrm{Pe}_{z}}.
\end{equation}
Here, $\alpha_d\ll 1$ signifies that there is not enough time for particles to diffuse across the cross-section while being transported by advection over distance $d_z$, i.e., we are in the non-dispersive regime. On the other hand, for $\alpha_d\gg 1$, diffusion causes considerable dispersion across the cross-section, which in turn causes significant dispersion along the $z$-axis due to space-dependent flow velocity (\ref{Eq:Poiseuille}), i.e., we are in the dispersive regime. In other words, in terms of the P\'eclet number $\mathrm{Pe}_{z}$, we have non-dispersive and dispersive regimes if $\mathrm{Pe}_{z}\gg\bar{d}_z^2$ and $\mathrm{Pe}_{z}\ll\bar{d}_z^2$ hold, respectively. 

Fig.~\ref{Fig:Dispersion} illustrates different dispersion regimes for a 3D straight duct. For clarity of presentation, we only show those particles where the $x$-component of their position lies in interval $[-0.1\acirc,0.1\acirc]$.  As can be seen from Fig.~\ref{Fig:Dispersion}, for $\alpha_d=0.1$, the positions of the particles simply follow the velocity profile in (\ref{Eq:Poiseuille}) whereas for $\alpha_d=10$, particles are significantly dispersed in the environment. 

\iftoggle{OneColumn}{%
\begin{figure}
	\centering  
	\resizebox{0.7\linewidth}{!}{
		\psfragfig{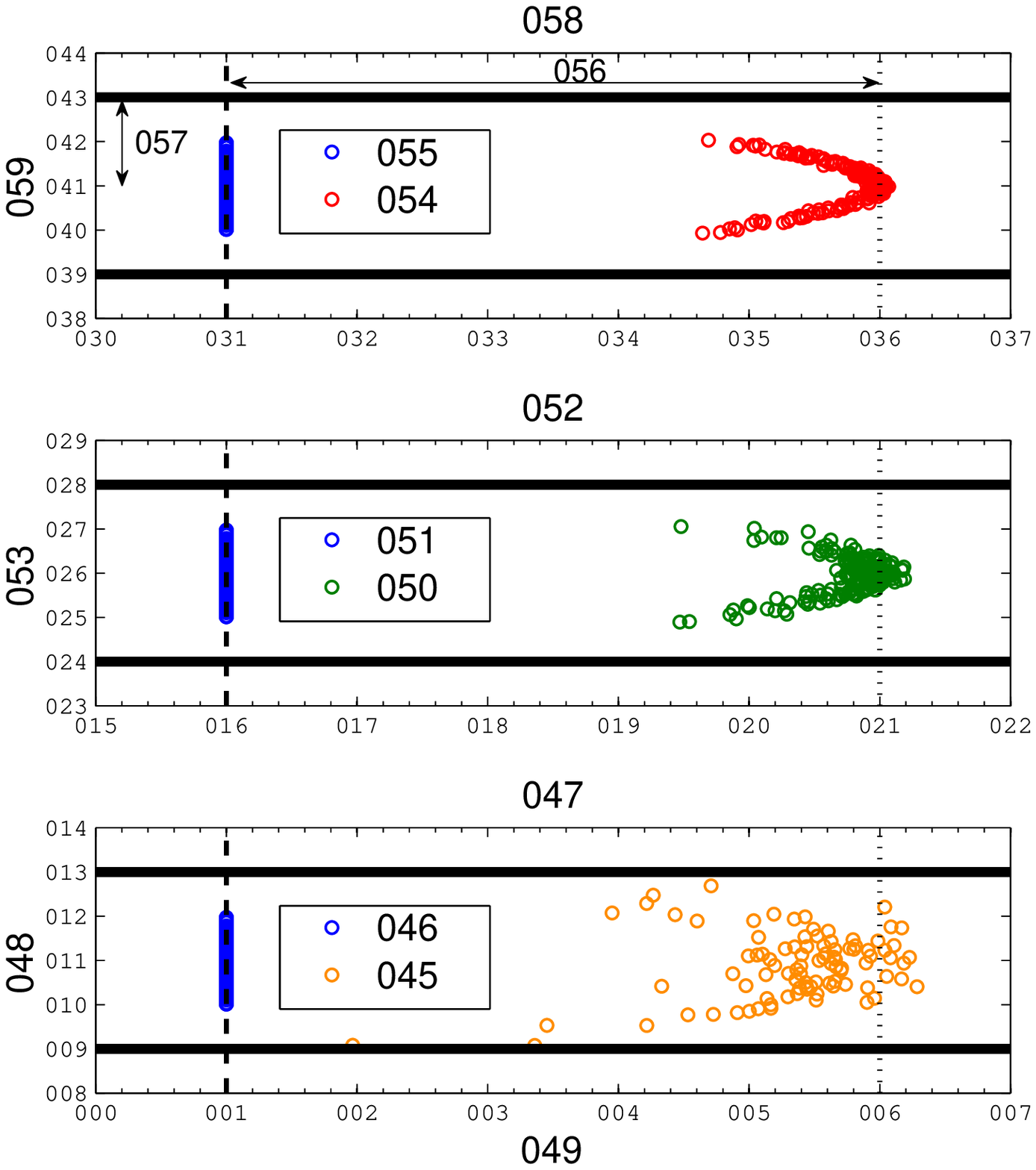}} \vspace{-1cm}
	\caption{Illustration of different dispersion regimes in a 3D straight duct with reflective walls, $D=10^{-11}\text{m}^2$/s, $\acirc=10\mu$m, $d_c=0.1\acirc$, $d_z=50\mu$m, flow velocity profile in (\ref{Eq:Poiseuille}), and $v_0= 10^{-2},10^{-3},10^{-4}$m/s which leads to $\alpha_d=0.1,1,10$, respectively. For clarity of presentation, we only show those particles where the $x$-component of their position lies in interval $[-0.1\acirc,0.1\acirc]$. The particles are initially placed at $z=0$ and uniformly distributed in a disk with radius $5~\mu$m centered at $(x,y)=(0,0)$. The solid horizontal lines represent the duct walls, the dashed vertical lines denote the initial positions of the particles on the $z$-axis, and the dotted vertical lines denote the distance of interest on the $z$-axis, i.e., $d_z$. }
	\label{Fig:Dispersion}
\end{figure}
}{%
\begin{figure}
	\centering  
	\resizebox{1\linewidth}{!}{
		\psfragfig{Sections/S2/Fig/Dispersion/Dispersion}} \vspace{-1cm}
	\caption{Illustration of different dispersion regimes in a 3D straight duct with reflective walls, $D=10^{-11}\text{m}^2$/s, $\acirc=10\mu$m, $d_c=0.1\acirc$, $d_z=50\mu$m, flow velocity profile in (\ref{Eq:Poiseuille}), and $v_0= 10^{-2},10^{-3},10^{-4}$m/s which leads to $\alpha_d=0.1,1,10$, respectively. For clarity of presentation, we only show those particles where the $x$-component of their position lies in interval $[-0.1\acirc,0.1\acirc]$. The particles are initially placed at $z=0$ and uniformly distributed in a disk with radius $5~\mu$m centered at $(x,y)=(0,0)$. The solid horizontal lines represent the duct walls, the dashed vertical lines denote the initial positions of the particles on the $z$-axis, and the dotted vertical lines denote the distance of interest on the $z$-axis, i.e., $d_z$. }
	\label{Fig:Dispersion}
\end{figure}
}

\subsection{Chemical Reactions}\label{Sec:Fund_reaction}

Another important phenomenon affecting the propagation of signaling molecules in diffusive \gls{MC} systems is chemical reactions. On the one hand, chemical reactions may occur naturally in \gls{MC} environments and their impact must be taken into account for communication design. On the other hand, chemical reactions have been exploited in the  \gls{MC} literature to achieve certain objectives, such as \gls{ISI} reduction \cite{Adam_Enzyme,Nariman_AcidBasePlatform,PhY_MC,NoelPro3} and ligand-based reception modeling \cite{ArmanJ2,Deng2015}. Therefore, in the following, we first review  general chemical reactions, the corresponding reaction equations, and  examples of reactions widely considered in the \gls{MC} literature. Subsequently, we study the joint impact of all three phenomena discussed in this section, namely diffusion, advection, and reaction, on the propagation of the molecules and solve the corresponding advection-reaction-diffusion equation for a simple example.

\subsubsection{Reaction Equation}
 Consider a general reaction of the form \cite[Eq. (13)]{CoxNatureCommun}
\begin{IEEEeqnarray}{lll} \label{Eq:Reaction}
	\sum_{I\in\mathcal{I}}n_I I \overset{\kappa}{\rightarrow} \sum_{J\in\mathcal{J}}n_J J,
\end{IEEEeqnarray}
where $I\in\mathcal{I}$ are reactant molecules, $\mathcal{I}$ is the set of reactant molecules, $J\in\mathcal{J}$ are product molecules, $\mathcal{J}$ is the set of product molecules, $n_I$ and $n_J$ are non-negative integers, and $\kappa$ is the reaction rate constant. Let $c_{I}(\mathbf{d},t)$ and $c_{J}(\mathbf{d},t)$ denote the concentration of type-$I$ and type-$J$ molecules at coordinate $\mathbf{d}$ and time $t$, respectively. Reactions locally change the concentration of particles over time which is described by the following \glspl{PDE},  known as reaction equations
\begin{IEEEeqnarray}{lll} \label{Eq:ReactionRate}
  \frac{\partial c_{I}(\mathbf{d},t)}{\partial t} 	&= -n_I f(\kappa,c_{I},\forall I\in\mathcal{I}), \quad \forall I\in\mathcal{I}\IEEEyesnumber\IEEEyessubnumber \\
  \frac{\partial c_{J}(\mathbf{d},t)}{\partial t} 	&=  n_J f(\kappa,c_{I},\forall I\in\mathcal{I}), \quad \forall J\in\mathcal{J},\IEEEyessubnumber
\end{IEEEeqnarray}
where $f(\kappa,c_{I},\forall I\in\mathcal{I})$ denotes the reaction rate function, which depends on the reaction rate constant and the concentrations of the reactant molecules. The reaction rate function has the following general form, known as the rate law \cite[Eq. (9.2)]{PhysicChemistry}
\begin{IEEEeqnarray}{lll} \label{Eq:ReactionLaw}
	f(\kappa,c_{I},\forall I\in\mathcal{I}) = \kappa \prod_{I\in\mathcal{I}} c_{I}^{\varepsilon_I}(\mathbf{d},t),
\end{IEEEeqnarray}
where $\varepsilon_I$ is the order of the reaction with respect to type-$I$ reactant molecules and typically takes an integer value (but in principle may also assume real values). The overall reaction order is defined as $\sum_{I\in\mathcal{I}}\varepsilon_I$ \cite{PhysicChemistry,Robert_MCnote}. Note that the units of reaction rate function $f(\kappa,c_{A},c_{B})$ and reaction rate constant $\kappa$ are $\frac{\text{molecule}}{\text{s}\cdot\text{m}^3}$ and $\frac{1}{\text{s}}\big(\frac{\text{molecule}}{\text{m}^3}\big)^{1-\sum_{I\in\mathcal{I}}\varepsilon_I}$, respectively.

In the following, we present three important classes of reactions, namely unimolecular degradation, bimolecular reactions, and enzymatic reactions, which can all play important roles in \gls{MC} systems \cite{NoelPro3,Adam_Enzyme,Heren,TCOM_NonCoherent,cho2017effective}. In particular, degradation is a natural characteristic of some types of  molecules and its effect has to be accounted for in communication design, see Section~\ref{Sub;ChaMod} and \cite{NoelPro3,Heren}. Bimolecular reactions can be used to analayze ligand-receptor binding \cite{ArmanJ2, Deng2015} and reactive signaling \cite{Nariman_AcidBasePlatform,ICC_Reactive}. In addition, enzymatic reactions have been studied in the \gls{MC} literature for the purpose of \gls{ISI} reduction \cite{Adam_Enzyme,cho2017effective}.

\begin{examp}[Unimolecular Degradation]
This reaction is used to describe the degradation of a desired type of molecule, e.g., type $A$, into a new type of molecule, denoted by $\phi$, which is of no interest for the considered communications. In fact, unimolecular degradation is often used as a first-order approximation of more complex reactions such as bimolecular and enzymatic reactions, see Examples~\ref{Exmp:Bimolecular} and \ref{Exmp:Enzyme}. Unimolecular degradation is modeled by \cite[Ch. 9]{PhysicChemistry}
\begin{IEEEeqnarray}{lll} \label{Eq:Deg}
	A \overset{\kappa}{\rightarrow} \phi,
\end{IEEEeqnarray}
where $\kappa$ [$\frac{1}{\text{s}}\big(\frac{\text{molecule}}{\text{m}^3}\big)^{1-\varepsilon_A}$] is the reaction rate constant, $f(\kappa,c_{A})=\kappa c_{A}^{\varepsilon_A}(\mathbf{d},t)$ is the reaction rate function, and $\varepsilon_A$ is the reaction order. In the \gls{MC} literature, first-order reactions are used to model degradation, i.e., $\varepsilon_A=1$ \cite{NoelPro3,Heren}. However, depending on the speed of reaction, higher and lower order reactions may be relevant, e.g., zero-order ($\varepsilon_A=0$) or second-order (Type-I) ($\varepsilon_A=2$) reactions \cite[Ch. 9]{PhysicChemistry}.  Assuming an initial condition $c_A(\mathbf{d},t_0)$ at $t_0$, \Eqref{Eq:ReactionRate} has the following solution for $t>t_0$
\begin{IEEEeqnarray}{lll} \label{Eq:Sol_Degradation}
	c^{\ast}_A(\mathbf{d},t)  = \begin{cases}
		[c_A(\mathbf{d},t_0)-\kappa (t-t_0)]^{+},\quad &\mathrm{if}\,\,\varepsilon_A=0 \\
		c_A(\mathbf{d},t_0)\exp(-\kappa (t-t_0)),\quad &\mathrm{if}\,\,\varepsilon_A=1 \\
		1/\left(\kappa (t-t_0)+1/c_A(\mathbf{d},t_0)\right),\quad &\mathrm{if}\,\,\varepsilon_A=2, 		
	\end{cases} \quad
\end{IEEEeqnarray}
where $[x]^{+}=\max\{0,x\}$. Note that the speed of  molecule concentration decay is hyperbolic for second-order degradations, which is faster than the exponential decay for first-order degradations, which in turn is faster than the linear decay for  zero-order degradations. Nevertheless, for sufficiently large $t$, $c^{\ast}_A(\mathbf{d},t)$ for second-order degradations is larger than that for first-order degradations, whereas $c^{\ast}_A(\mathbf{d},t)=0,\,\,t\geq t_0+\frac{c_A(\mathbf{d},t_0)}{\kappa}$, holds for zero-order degradations.  
\QEDwhite
\end{examp}

\begin{examp}[Bimolecular Reactions]\label{Exmp:Bimolecular} Some reactions may involve the interaction of two reactant chemical species, e.g., $A$ and $B$, to produce product molecule(s), e.g., $C$. For instance, in \cite{ArmanJ2}, the activation of ligand receptors via signaling molecules was modeled by a second-order bimolecular reaction. Moreover, in \cite{Nariman_AcidBasePlatform} and \cite{ICC_Reactive}, acids and bases were used as reactive signaling molecules to reduce \gls{ISI}. Acids and bases cancel each other out to produce salt and water. This process is modeled by a second-order bimolecular reaction. In particular, the second-order (Type-II) bimolecular reaction is given by \cite{Tro2015Chemistry}
\begin{IEEEeqnarray}{lll} \label{Eq:BiDeg} 
	A + B \underset{\kappa_b}{\overset{\kappa_f}{\rightleftharpoons}} C,
\end{IEEEeqnarray}
where $\kappa_f$ is the forward reaction rate constant \big[$\frac{\text{m}^3}{\text{s}\cdot\text{molecule}}$\big], $\kappa_b$ \big[$\frac{1}{\text{s}}$\big] is the backward reaction rate constant, and $f(\kappa,c_{A},c_{B})=\kappa_f c_{A}(\mathbf{d},t)c_{B}(\mathbf{d},t)$ is the reaction rate function. The  \glspl{PDE} corresponding to (\ref{Eq:BiDeg}) are \textit{nonlinear} and challenging to solve. However, after introducing some approximations, in Section~\ref{Sect:TxRxCh}, we use  (\ref{Eq:BiDeg}) to derive the \glspl{CIR} of \gls{MC} systems affected by bimolecular reactions. Moreover, let us assume $\kappa_b\to 0$ and that the concentration of type-$B$ molecules is sufficiently large such that the reaction in (\ref{Eq:BiDeg}) does not considerably change $c_{B}(\mathbf{d},t)$ over time, i.e., $c_{B}(\mathbf{d},t)\approx c_{B}(\mathbf{d},t=0) \triangleq c_{B}(\mathbf{d})$. In this case, the bimolecular reaction in (\ref{Eq:BiDeg}) can be approximated by the first-order unimolecular reaction in (\ref{Eq:Deg}) with $\kappa=\kappa_fc_{B}(\mathbf{d})$ \cite{ArmanJ2}. \QEDwhite
\end{examp}

\begin{examp}[Enzymatic Reactions]\label{Exmp:Enzyme}
For typical scenarios, the speed of natural degradation might be too slow compared to the desired time scale of communication. In this case, enzymes can be used to accelerate the reaction process. Enzymes, denoted by $E$, are specific proteins that bind to the desired molecule $A$ (also referred to as the substrate), and lower the activation energy needed for a reaction to occur. Enzymatic degradations are modeled by the following reactions \cite[Eq. (1)]{Adam_Enzyme}
\begin{IEEEeqnarray}{lll} \label{Eq:EnzamiticDeg}
	A + E \underset{\kappa_b}{\overset{\kappa_f}{\rightleftharpoons}} AE \overset{\kappa_d}{\rightarrow} E + \phi,
\end{IEEEeqnarray}
where $AE$ is an intermediate chemical species and $\phi$ is the product molecule. Moreover, $\kappa_f$ \big[$\frac{\text{m}^3}{\text{s}\cdot\text{molecule}}$\big], $\kappa_b$ \big[$\frac{1}{\text{s}}$\big], and $\kappa_d$ \big[$\frac{1}{\text{s}}$\big] denote the reaction rate constants of the forward, backward\footnote{The forward and backward reaction rate constants are also referred to as binding and unbinding reaction rate constants, respectively.}, and degradation reactions, respectively. As can be seen from (\ref{Eq:EnzamiticDeg}), the enzyme molecules are \emph{not} consumed in the reaction process. The following set of \glspl{PDE}, known as Michaelis-Menten kinetics, describe the evolution of the concentrations of the participating molecules 
\begin{IEEEeqnarray}{lll} \label{Eq:EnzamiticPDE}
	\frac{\partial c_A(\mathbf{d},t)}{\partial t} =  -\kappa_f c_A(\mathbf{d},t)c_E(\mathbf{d},t) + \kappa_b c_{AE}(\mathbf{d},t) \IEEEyesnumber \IEEEyessubnumber\\
	\frac{\partial c_E(\mathbf{d},t)}{\partial t} =  -\kappa_f c_A(\mathbf{d},t)c_E(\mathbf{d},t) + (\kappa_b+\kappa_d) c_{AE}(\mathbf{d},t) \quad \IEEEyessubnumber\\	
    \frac{\partial c_{AE}(\mathbf{d},t)}{\partial t} =  \kappa_f c_A(\mathbf{d},t)c_E(\mathbf{d},t) - (\kappa_b+\kappa_d) c_{AE}(\mathbf{d},t). \quad\quad\,\, \IEEEyessubnumber	
\end{IEEEeqnarray}
Solving the above system of \textit{coupled} and \textit{nonlinear} \glspl{PDE} is challenging. Let us consider very fast degradation reactions, i.e., $\kappa_d\to\infty$, slow backward reactions, i.e., $\kappa_b\to 0$, and that the concentration of enzyme molecules is much larger than the concentration of type-$A$ molecules. In this case, the formation of intermediate $AE$ molecules does not last long and hence, we obtain $c_E(\mathbf{d},t)\approx c_E(\mathbf{d},t=0) \triangleq c_E(\mathbf{d})$. In \cite{Adam_Enzyme}, it was shown that under the aforementioned assumptions, the enzymatic reaction in \eqref{Eq:EnzamiticDeg} can be approximated by the first-order unimolecular reaction in \eqref{Eq:Deg} with reaction rate constant $\kappa=\frac{\kappa_f\kappa_d}{\kappa_b+\kappa_d}c_E(\mathbf{d}) \approx \kappa_f c_E(\mathbf{d})$.  \QEDwhite
\end{examp}

\subsubsection{Advection-Reaction-Diffusion Equation}

Next, we consider the joint effects of diffusion, drift, and reactions. For simplicity, we focus on a single molecule type and drop the corresponding subscript. In this case, the general advection-reaction-diffusion equation is given by the following \gls{PDE} \cite{Berg,Andrews2004}
\iftoggle{OneColumn}{%
\begin{IEEEeqnarray}{lll} \label{Eq:Fick_general}
	\frac{\partial c(\mathbf{d},t)}{\partial t} = D \nabla^2 c(\mathbf{d},t)   - \nabla\cdot \left(\mathbf{v}(\mathbf{d},t)c(\mathbf{d},t)\right) + q f\left(\kappa,c(\mathbf{d},t)\right),
\end{IEEEeqnarray}
}{%
	\begin{IEEEeqnarray}{lll} \label{Eq:Fick_general}
		\frac{\partial c(\mathbf{d},t)}{\partial t} = \,& D \nabla^2 c(\mathbf{d},t) \nonumber \\
		&  - \nabla\cdot \left(\mathbf{v}(\mathbf{d},t)c(\mathbf{d},t)\right) + q f\left(\kappa,c(\mathbf{d},t)\right),
	\end{IEEEeqnarray}
}
where $q=1$ and $q=-1$ hold if the considered molecule is the product and the reactant of the reaction, respectively. Solving (\ref{Eq:Fick_general}) for general initial and boundary conditions is again difficult for most practical \gls{MC} environments. Hence, in the following, we make some simplifying assumptions that enable us to solve (\ref{Eq:Fick_general}) in closed form for one example scenario \cite{Adam_Enzyme}.

\begin{examp}
Let us assume the impulsive release of $N$ molecules at time $t_0$  by a point source located at $\mathbf{d}_0$ into an unbounded 3D environment, i.e., initial condition $\mathrm{IC_1}$ in \eqref{Eq:3D_Diff_example_IC} and boundary condition $\mathrm{BC_1}$ in \eqref{Eq:3D_Diff_example_BC} hold. Moreover, we assume uniform flow $\mathbf{v}(\mathbf{d},t)=\mathbf{v}$ and the first-order degradation reaction in (\ref{Eq:Deg}), i.e., $q=-1$ and $f\left(\kappa,c(\mathbf{d},t)\right)=\kappa c(\mathbf{d},t)$. Based on these assumptions, (\ref{Eq:Fick_general}) has the following closed-form solution \cite{NonlinearPDE_Debnath,Adam_Universal_Noise}
\iftoggle{OneColumn}{%
\begin{IEEEeqnarray}{lll} \label{Eq:Fick_sol}
	c^{\ast}(\mathbf{d},t) = \frac{N}{\left(4\pi D (t-t_0)\right)^{3/2}} \exp\left(-\kappa (t-t_0)-\frac{\|\mathbf{d}-(t-t_0)\mathbf{v}-\mathbf{d}_0\|^2}{4D(t-t_0)}\right), \qquad t>t_0.
\end{IEEEeqnarray}
}{%
	\begin{IEEEeqnarray}{lll} \label{Eq:Fick_sol}
		c^{\ast}(\mathbf{d},t) = \,\frac{N}{\left(4\pi D (t-t_0)\right)^{3/2}} \nonumber \\
		\times \exp\left(-\kappa (t-t_0)-\frac{\|\mathbf{d}-(t-t_0)\mathbf{v}-\mathbf{d}_0\|^2}{4D(t-t_0)}\right), \,\,\, t>t_0.\quad\,\,\,
	\end{IEEEeqnarray}
}
\QEDwhite
\end{examp}

In Fig.~\ref{Fig:Reaction}, the molecule concentration $c^{\ast}(\mathbf{d},t)$ [molecules/m$^3$] is shown versus time [$\mu$s] at $\mathbf{d}=[400,0,0]$~nm for an initial release of $N=10^4$ molecules from $\mathbf{d}_0=[0,0,0]$ and at $t_0=0$, $D=4.5\times 10^{-10}$ m$^2$/s, flow velocity $\mathbf{v}=[10^{-3},0,0]$~m/s, and $\kappa\in\{0,1,2\}\times 10^{4}$ 1/s. This figure shows that as the degradation rate constant increases, the concentration peak decreases, which is not desirable for an \gls{MC} system, in general. However, the tail of the concentration for large $t$ fades away much faster for larger degradation rates, which was exploited for \gls{ISI} reduction in  \cite{Adam_Enzyme}.

\iftoggle{OneColumn}{%
\begin{figure}[!t]
	\centering 
	\resizebox{0.7\linewidth}{!}{
		\psfragfig{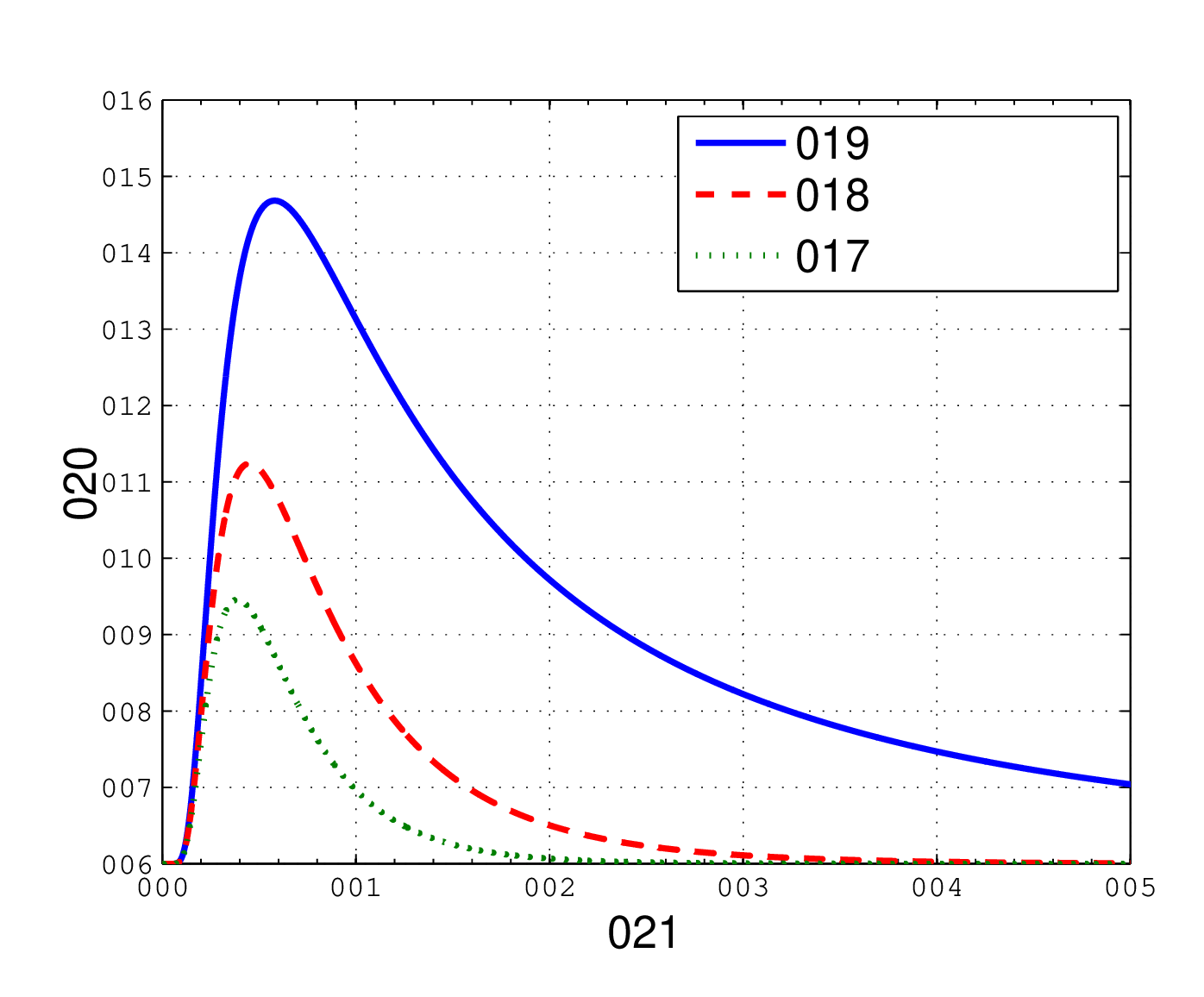}} 
	\caption{Molecule concentration $c^{\ast}(\mathbf{d},t)$ [molecules/m$^3$] versus time [$\mu$s] at $\mathbf{d}=[400,0,0]$~nm for an initial release of $N=10^4$ molecules from $\mathbf{d}_0=[0,0,0]$ and at $t_0=0$, $D=4.5\times 10^{-10}$ m$^2$/s, flow velocity $\mathbf{v}=[10^{-3},0,0]$ m/s, and $\kappa\in\{0,1,2\}\times 10^{4}$~1/s.}
	\label{Fig:Reaction}
\end{figure}
}{%
	\begin{figure}[!t]
		\centering 
		\resizebox{1\linewidth}{!}{
			\psfragfig{Sections/S2/Fig/Reaction/CIR_Reaction}} \vspace{-0.5cm}
		\caption{Molecule concentration $c^{\ast}(\mathbf{d},t)$ [molecules/m$^3$] versus time [$\mu$s] at $\mathbf{d}=[400,0,0]$~nm for an initial release of $N=10^4$ molecules from $\mathbf{d}_0=[0,0,0]$ and at $t_0=0$, $D=4.5\times 10^{-10}$ m$^2$/s, flow velocity $\mathbf{v}=[10^{-3},0,0]$ m/s, and $\kappa\in\{0,1,2\}\times 10^{4}$~1/s.}
		\label{Fig:Reaction}
	\end{figure}
}

\section{Component Modeling}
\label{Sect:TxRxCh}
In this section, we review the existing component models for the transmitter, receiver, and physical channel of diffusive \gls{MC} systems. To this end, in Section~\ref{Sub;CIR}, we first define the end-to-end \gls{CIR} of single-link diffusive \gls{MC} systems, and discuss the relevant mechanisms of each component and their impact on the end-to-end \gls{CIR}. We use the \gls{CIR} to characterize the components of MC systems, since the impulse response fully characterizes the behaviour of linear systems, and linearity is commonly assumed in the \gls{MC} literature. Subsequently, in Sections~\ref{Sub;RecMod}, \ref{Sub;TraMod}, and \ref{Sub;ChaMod}, we review the existing models that have been developed by taking into account the impact of the receiver, transmitter, and physical channel on the end-to-end \gls{CIR}, respectively. Finally, in Section~\ref{Sub;SumCIR}, we provide a summary table of all reviewed end-to-end \gls{CIR} models.

\subsection{Channel Impulse Response}
\label{Sub;CIR}

\iftoggle{OneColumn}{%
\begin{figure}[!t]	
	\centering
	\includegraphics[scale=1]{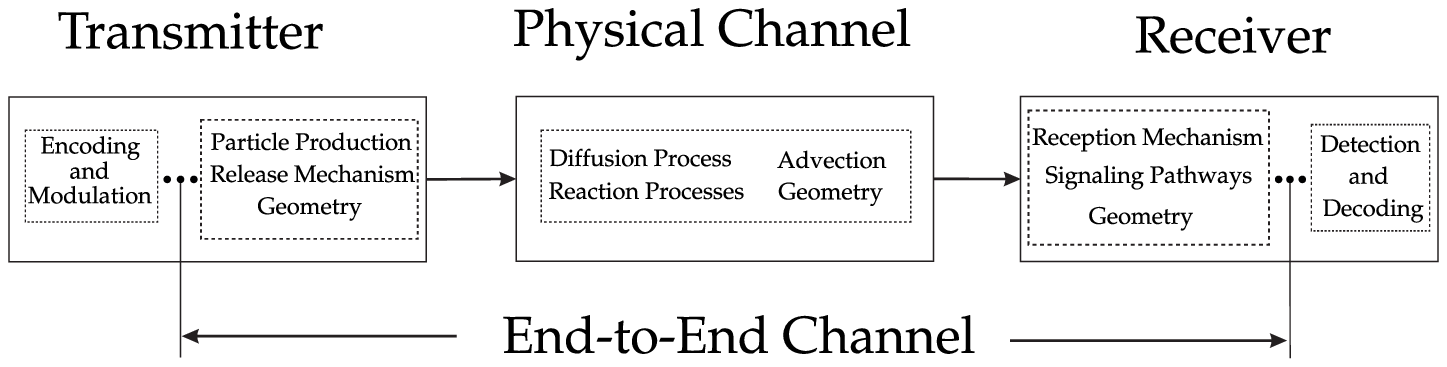} 
	\caption{Schematic presentation of the end-to-end chain of communication in typical diffusive \gls{MC} systems.}
	\label{Fig:General_System_Model} 
\end{figure}
}{%
  \begin{figure*}[!t]	
	\centering
	\includegraphics[scale=1]{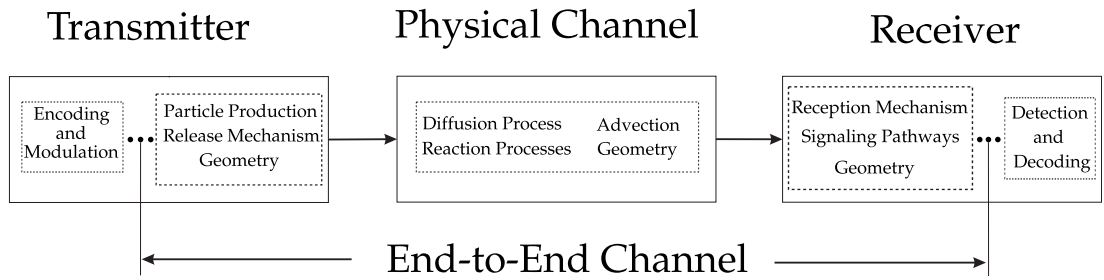} 
	\caption{Schematic presentation of the end-to-end chain of communication in typical diffusive \gls{MC} systems.}
	\label{Fig:General_System_Model} 
\end{figure*}
}

In this subsection, we first briefly discuss the relevant mechanisms that characterize the functionalities of the transmitter and receiver, and the phenomena and impairments that occur in the physical channel of diffusive \gls{MC}  systems. Then, we provide a formal definition of what we refer to as the end-to-end channel of diffusive \gls{MC} systems and we show how the \gls{CIR} corresponding to the end-to-end channel can be obtained using the tools introduced in Section~\ref{Sect:Prelim}. 

Similar to traditional communication systems, the end-to-end chain of diffusive \gls{MC} systems consists of three components, namely the transmitter, the physical channel, and the receiver; see Fig.~\ref{Fig:General_System_Model}. Each of these components has unique features and responsibilities, which are outlined below; see also Fig.~\ref{Fig:Physical_System_Model}.

\begin{itemize}
\item \textbf{Transmitter:} The transmitter is responsible for the encoding and modulation of information bits. In \gls{MC}, the information is typically encoded in the number, type, or time of release of signaling molecules. Furthermore, the transmitter has to generate the signaling molecules, (e.g. by CRNs inside the transmitter), store the signaling molecules, (e.g. in vesicles), and control their release into the physical channel.
 
\item \textbf{Physical Channel:} The physical channel is the environment in which the signaling molecules move and propagate once they leave the transmitter. In diffusive \gls{MC} systems, the movement of signaling molecules, at its most basic level, is described by the diffusion process. However, during the course of diffusion, the random walk of signaling molecules may be affected by several other factors and noise sources such as advection, CRNs degrading the signaling molecules, environment geometry, and obstacles inside the physical channel, see Section~\ref{Sect:Prelim}. 

\iftoggle{OneColumn}{%
\begin{figure}[!t]	
	\centering
	\includegraphics[scale=0.8]{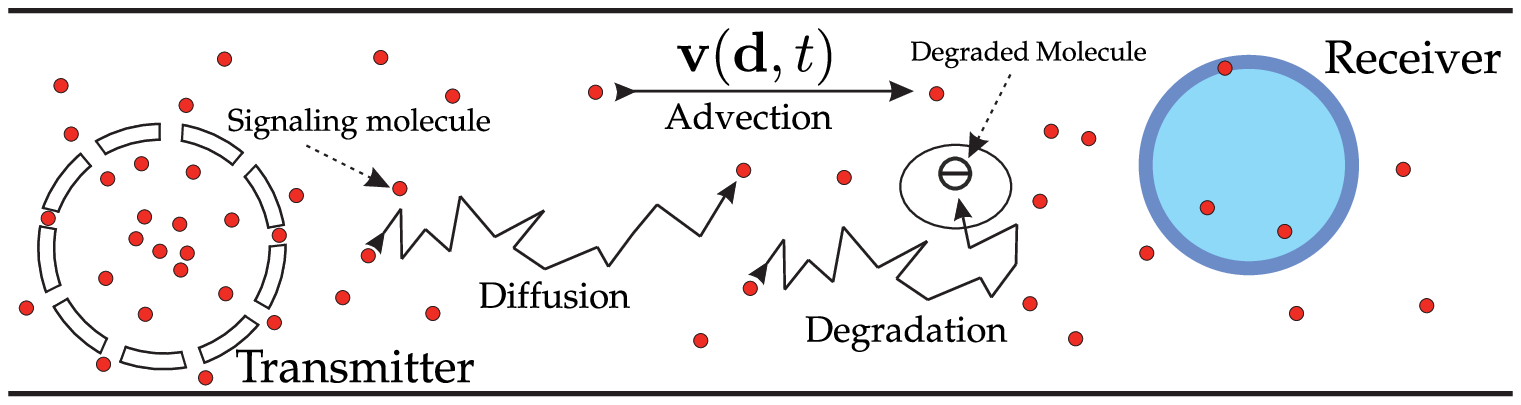} 
	\caption{Example of a physical system model including a transmitter, physical channel, and receiver.}
	\label{Fig:Physical_System_Model} 
\end{figure}
}{%
\begin{figure*}[!t]	
	\centering
	\includegraphics[scale=0.8]{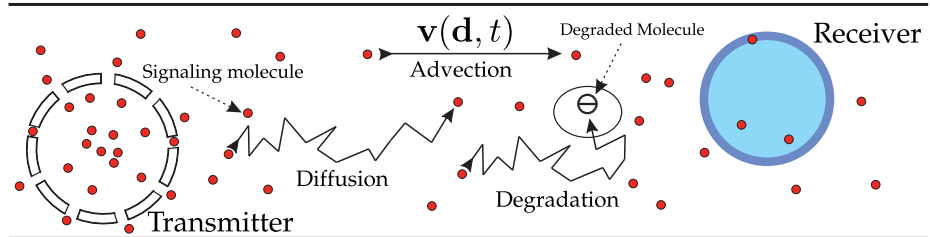} 
	\caption{Example of a physical system model including a transmitter, physical channel, and receiver.}
	\label{Fig:Physical_System_Model} 
\end{figure*}
}
 
\item \textbf{Receiver:} Signaling particles that reach the vicinity of the receiver can be observed and processed by the receiver to extract the information that is necessary for performing detection and decoding. The reception mechanism of the receiver \emph{may} include the following functionalities, depending on its structure: \textit{i)} external sensory units for detecting the presence of signaling molecules, membrane receptors of cells in nature, or sensing component(s) of macro-scale receivers such as the alcohol sensor in \cite{farsad_tabletop_2013} and the magnetic coils of the susceptometer in \cite{unterweger_experimental_2018}; \textit{ii)} internal relaying and interface components to convey and convert the measurements of the sensory unit into quantitites suitable for detection and decoding of the information bits. For instance, in nature, this task is performed by the CRNs \emph{inside} cells, which are referred to as \emph{downstream signaling pathways} \cite{AlbertsBook}. Downstream signaling pathways may be driven by activated receptors or directly by signaling molecules that passively enter the cells. 
\end{itemize}


In the following, we formally define the end-to-end channel to study the reviewed CIR models in a unified manner. 

\begin{defin}[\textbf{End-to-end Channel}]
 We define the \emph{end-to-end channel} as the effective channel that not only includes the physical channel but also the impact of the physical and chemical properties of the transmitter and receiver, including the effects of signaling molecule generation, release mechanisms, sensory units, and internal receiver components. \QEDwhite
\end{defin}
 
  Note that our definition of the end-to-end channel does \emph{not} include the coding, modulation, detection, and decoding operations that the transmitter and receiver may perform; see also Fig.~\ref{Fig:General_System_Model}. This definition of the end-to-end channel is analogous to that in traditional wireless communication systems, where the antennas, power amplifiers, and filters of the transmitter and receiver are also included in the model for the wireless end-to-end channel. The input to the end-to-end channel is the signal representing the modulated information symbol, which we also refer to as the \emph{stimulation signal}. The stimulation signal can be an electrical (voltage or current), magnetic, mechanical, optical, chemical, or temperature signal. The output of the end-to-end channel is referred to as the observed signal and should be in a form that is suitable for the subsequent detection and decoding operations. Depending on the structure of the receiver, the observed signal can be either a number of \emph{output molecules} or any secondary signal derived from the output molecules. In particular, output molecules may represent: \textit{i)} signaling molecules that can passively enter the receiver; \textit{ii)} absorbed molecules that hit the receiver surface; or \textit{iii)} activated receptors. Furthermore, the secondary signal derived from output molecules may be an electrical signal, e.g., the output voltage or output current of the alcohol sensor in \cite{farsad_tabletop_2013}.    
In the following, for the definition of the CIR of the end-to-end channel, we emphasize that we consider the number of the output molecules as the observed signal, as it is commonly assumed in the \gls{MC} literature, although our definition can be easily extended to other forms of the observed signal.    

\begin{defin}[\textbf{Channel Impulse Response}]
We define the \gls{CIR} of the \emph{end-to-end channel}, denoted by $h(t)$, as the probability of \emph{observation} of one output molecule at time $t$ at the receiver when the transmitter is stimulated in an impulsive manner at time $t_0 = 0$. \QEDwhite 
\end{defin}

We note that defining the \gls{CIR} as a probability has several advantages. In particular, it facilitates the definition of the received signal in Section~\ref{Sect:RecSig}. There, we propose a general received signal model that takes into account both the \emph{arrival time} and the \emph{numbers} of observed output molecules. As will be shown in Section~\ref{Sect:RecSig}, both of these quantities can be readily obtained from the probability of observation of one output molecule.
  
In our definition of the CIR, the quantitative meaning of the term \emph{observation} depends on the type of receiver and is defined for each considered receiver model in detail in the next subsection, e.g., for passive receivers the observed signal is defined as the number of signaling molecules inside the receiver, while for reactive receivers it is defined as the number of activated receptor molecules. Furthermore, we assume that the transmitter stimulation is an impulsive input that either controls the opening and closing of the signaling molecule reservoir or drives the CRNs inside the transmitter responsible for the generation of the signaling molecules.
 
In this section, we assume that the parameters of the considered MC system are constant, i.e., the end-to-end CIR $h(t)$ is time-invariant. In the following, we refer to the signaling molecules as $A$ molecules. The following phenomena may affect the propagation of the $A$ molecules, and as a result, $h(t)$: 
\begin{enumerate} 
	\item \textbf{Particle generation:} Generation of the $A$ molecules is performed, e.g., by the CRNs inside the transmitter. 
	\item \textbf{Release mechanism:} The release mechanism can be chemical, electrical, or mechanical and controls the release of the $A$ molecules into the physical channel.
	\item \textbf{Diffusion:} Diffusion refers to the propagation of molecules by Brownian motion.
	\item \textbf{Degradation and production:} CRNs may degrade or produce $A$ molecules in the physical channel. 
	\item \textbf{Advection:} Advection may affect the transportation of the $A$ molecules in the physical channel.
	\item \textbf{Geometry:} Potentially, the geometry of the individual components of the end-to-end channel can influence the propagation of signaling molecules.
	\item \textbf{Receptor kinetics:} Receptor kinetics affect the interaction of the $A$ molecules with the receptors of the sensory unit at the receiver. 
	\item \textbf{Signaling pathways:} The signaling pathways transducing the observed $A$ molecules into secondary signal affect the received signal.            
\end{enumerate}

In order to obtain $h(t)$ for a specific \gls{MC} system, one has to solve the advection-reaction-diffusion equation \eqref{Eq:Fick_general} or a simplified version thereof, depending on the \gls{MC} system under consideration, with the appropriate initial and boundary conditions. 
The initial conditions of the system capture the initial states of the CRNs, the time of production of the $A$ molecules, and the location of the produced $A$ molecules. The boundary conditions capture the physical and chemical properties of the components of the end-to-end channel. As discussed in the previous section, the solution to this system of PDEs does not exist in closed-form for many environments. However, as we will see in the remainder of this section, in the \gls{MC} literature, different approximations have been developed to arrive at approximate yet meaningful solutions for $h(t)$ that can still capture the main effects and phenomena of the end-to-end channel. These approximate models focus on one of the components of the MC system and make simplifying assumptions about the other two. Accordingly, we will consider such receiver, transmitter, and channel centric models in the following three subsections.

\subsection{Receiver Models} 
\label{SubSec;Receiver}
\label{Sub;RecMod}
\iftoggle{OneColumn}{%
\begin{figure}[!t]
\centering
\includegraphics[scale=.75]{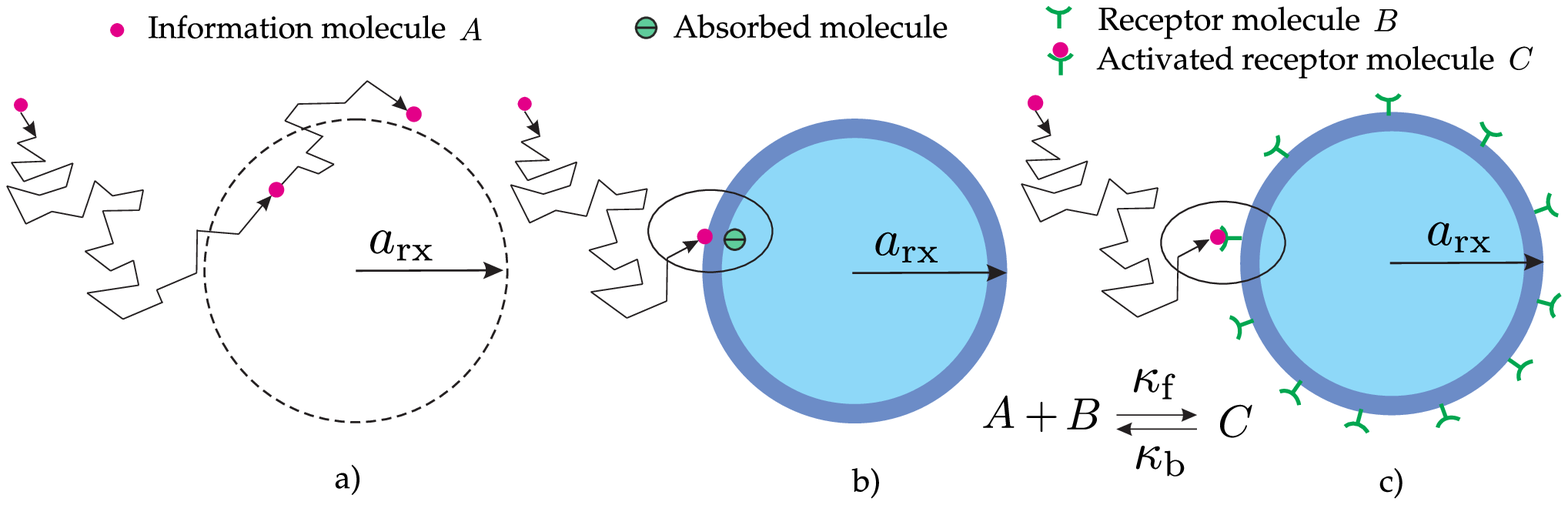}
\caption{Schematic depiction of three common receiver models; a) passive receiver, b) fully absorbing receiver, and c) reactive receiver.}
\label{Fig1}       
\end{figure}
}{%
\begin{figure*}[!t]
\centering
\includegraphics[scale=.65]{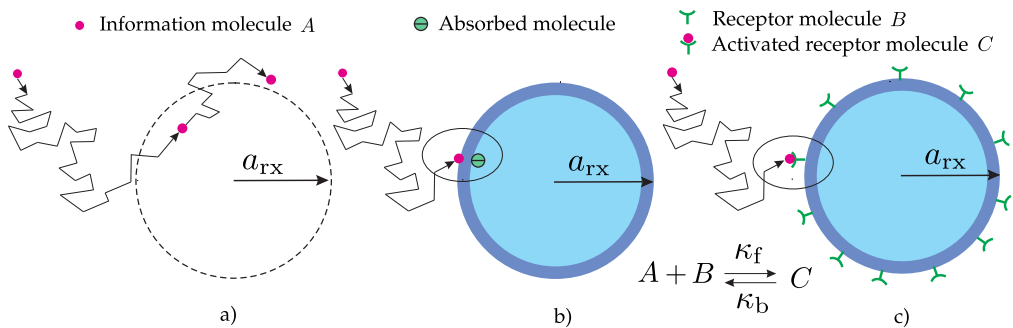}
\caption{Schematic depiction of three common receiver models; a) passive receiver, b) fully absorbing receiver, and c) reactive receiver.}
\label{Fig1}       
\end{figure*}
}
In this section, we review some of the existing end-to-end \gls{CIR} models that focus particularly on the properties of the receiver, while simplifying assumptions for the transmitter and MC environment are made. The reception mechanism of the receiver can be categorized into two classes: \textit{i}) passive reception, where the receiver does not impede the movement of signaling molecules; and \textit{ii}) active reception, where the receiver may affect the movement of signaling molecules either by their absorption on its surface, or by chemically reacting with them via receptors (and thereby forming ligand-receptor complexes) embedded in the receiver surface. For active reception, both mechanisms can be described by a form of chemical reaction. Moreover, the received signaling molecules may be converted via signaling pathways into secondary molecules, which can later be used for detection or decoding of the information. In nature, cells have diverse types of signaling pathways, each of which is responsible for relaying a particular type of measurement taken in the extracellular space to the organelles in the cytosol, which ultimately causes a response by the cell. For more information on the signaling pathways in natural cells, we refer the interested reader to \cite{AlbertsBook}. 

For the \gls{CIR} models considered in the following, we adopt rather simple models for the transmitter and the physical channel. Specifically, we assume that the transmitter is a \emph{point} that releases one $A$ molecule \emph{instantaneously} upon stimulation at time $t_0 = 0$ at location $\mathbf{d}_{\tx}$, where $\mathbf{d}_{\tx}$ denotes the location of the center of the transmitter; see Section~\ref{Sub;TraMod} for more details on the point transmitter model. In other words, a point transmitter implicitly implies that 
upon stimulation, the $A$ signaling molecule is \emph{immediately} produced and enters the physical channel. We denote the location of the center of the receiver by $\mathbf{d}_{\rx}$, and the distance between the center of the transmitter and the center of the receiver by $d_0 = \| \mathbf{d}_{\tx} - \mathbf{d}_{\rx} \|$.
Furthermore, for the physical channel, we consider an unbounded environment affected only by diffusion noise; see Section~\ref{Sub;ChaMod} for more complex \gls{MC} environments. 

\textbf{Passive receiver:} Passive receivers (also referred to as transparent receivers or perfect monitoring receivers) employ passive reception mechanisms and are commonly considered in the \gls{MC} literature, see e.g. \cite{Akyl_Receiver_MC,ConsCIR,Nariman_Survey,NoelPro1,Akyildiz_MC_E2E,Equ_MC,PhY_MC,NoelPro3,TCOM_MC_CSI,DistanceEstLett}. In particular, signaling $A$ molecules in the vicinity of the receiver can enter and leave the receiver via free diffusion; see e.g. Fig.~\ref{Fig1}a). The passive receiver model is a good approximation for the diffusion of small uncharged molecules such as ethanol, urea, and oxygen. These molecules can enter and leave a cell by passive diffusion across the plasma membrane \cite{AlbertsBook}. A passive receiver model is also valid for the experimental system in \cite{unterweger_experimental_2018}, where the susceptometer that serves as the receiver does not impede the movement of the magnetic nanoparticles passing through it. For passive receivers, the set of all points $\mathbf{d}$ inside the volume of the receiver, $\mathcal{V}_{\rx}$, constitutes the sensing area, and the number of $A$ molecules in $\mathcal{V}_{\rx}$ constitutes the observed signal. Let $N_{\tx}$ denotes the number of molecules that the transmitter releases. Since we are interested in computing CIR $h(t)$, i.e., the probability that a molecule released by the transmitter at $t_0=0$ is observed at the receiver at time $t$, we set $N_{\tx} = 1$. Moreover, we use the notation $p(\mathbf{d},t) = c(\mathbf{d},t)|_{N_{\tx}=1}$ which can be interpreted as the PDF of a molecule released by the transmitter at $t_0=0$ with respect to $\mathbf{d}$ at time $t$. In other words, $p^{\ast}(\mathbf{d},t) \mathrm{d}x \mathrm{d}y \mathrm{d}z$ is the probability that the molecule is observed in volume $\mathrm{d}x \mathrm{d}y \mathrm{d}z$ at time $t$. Since we focus on linear systems, solving $c^{\ast}(\mathbf{d},t)$ with $N_{\tx} \neq 1$ and solving $p^{\ast}(\mathbf{d},t)$ for $N_{\tx} =1$ are related as $p^{\ast}(\mathbf{d},t) = c^{\ast}(\mathbf{d},t)/ N_{\tx}$. For the considered MC system with a point transmitter and unbounded environment, the CIR of a passive receiver can be obtained by first finding $p(\mathbf{d},t)$ from \eqref{Eq:PDE_Diff} with 
the following initial and boundary conditions 
\begin{IEEEeqnarray}{rCl}
	\label{Eq;CIR_passive_IC2} 
	\mathrm{IC_3:} &\,\,& p(\mathbf{d},t_0)  =  \delta \left( \mathbf{d} - \mathbf{d}_{\tx} \right) \\
	\label{Eq;CIR_passive_BC3} 
	\mathrm{BC_3:} &\,\,& p(\| \mathbf{d} \| \to \infty ,t) =  0.
\end{IEEEeqnarray}
Given the solution of \eqref{Eq:PDE_Diff}, $p^{\ast}(\mathbf{d},t)$, $h(t)$ can be written as 
\begin{IEEEeqnarray}{C}
	\label{Eq;CIR_passive_ProbObs} 
	h(t) = \int_{\mathbf{d} \in \mathcal{V}_{\rx}} p^{\ast}(\mathbf{d},t) \mathrm{d}\mathbf{d}. 
\end{IEEEeqnarray} 

The solution of the integral in \eqref{Eq;CIR_passive_ProbObs} can be readily obtained when the receiver is sufficiently far away from the transmitter, i.e., $d_0$ is very large relative to the largest dimension of the receiver. In this case, a common approach, which is referred to as the \emph{uniform concentration assumption} (UCA), is to approximate $p^{\ast}(\mathbf{d},t)$ everywhere inside the volume of the receiver by its value at the center of the receiver, i.e., $p^{\ast}(\mathbf{d},t) \simeq p^{\ast}(\mathbf{d}_{\rx},t), \forall \mathbf{d} \in \mathcal{V}_{\rx}$. This leads to the following simple expression for $h(t)$, \cite{Akyl_Receiver_MC,ConsCIR,Nariman_Survey,NoelPro1,Akyildiz_MC_E2E,Equ_MC,PhY_MC,NoelPro3,TCOM_MC_CSI,DistanceEstLett} 

\begin{equation}
	\label{Eq.Passive_mean2} 
	h(t) = \frac{V_{\rx}}{(4\pi D t)^{3/2}} \exp \left( - \frac{d_0^2}{4Dt}\right),
\end{equation}
where $V_{\rx}$ is a constant denoting the volume of the receiver. We note that \eqref{Eq.Passive_mean2} is valid independent of the geometry of the receiver. Specifically, the UCA is one of the most useful approximation methods in the \gls{MC} literature, since it directly relates the solution of \eqref{Eq:PDE_Diff}, \eqref{Eq:PDE_FlowDiff}, and \eqref{Eq:Fick_general} to the CIR of the corresponding system. Thus, many results in the rich literature on solving PDEs, see \cite{Carslaw}, can be used to obtain the CIR in \gls{MC} systems with passive receivers under the UCA.    
   
The problem of solving \eqref{Eq;CIR_passive_ProbObs} may become cumbersome when the receiver is close to the transmitter. In this case, the solution of the integral depends on the geometry of the receiver and the UCA cannot be hold. It has been shown in \cite[Eq. (27)]{NoelPro1} that for a \emph{spherical} passive receiver with radius $a_{\rx}$, $h(t)$ is given by
\iftoggle{OneColumn}{%
\begin{IEEEeqnarray}{rCl} 
	\label{Eq;CIR_Passive_woUCA}
	h(t) & = & \frac{1}{2} \left( \mathrm{erf} \left( \frac{a_{\rx}-d_0}{\sqrt{4D t}} \right) + \mathrm{erf} \left( \frac{a_{\rx} + d_0}{\sqrt{4D t}} \right) \right) \nonumber \\ 
	 && \> + \frac{\sqrt{D t}}{a_{\rx}\sqrt{\pi}} \left( \mathrm{exp} \left( -\frac{(a_{\rx}-d_0)^2}{4D t} \right) + \mathrm{exp} \left( -\frac{(a_{\rx}+d_0)^2}{4D t} \right) \right),  
\end{IEEEeqnarray} 
}{%
\begin{IEEEeqnarray}{rCl} 
	\label{Eq;CIR_Passive_woUCA}
	h(t) & = & \frac{1}{2} \left( \mathrm{erf} \left( \frac{a_{\rx}-d_0}{\sqrt{4D t}} \right) + \mathrm{erf} \left( \frac{a_{\rx} + d_0}{\sqrt{4D t}} \right) \right) \nonumber \\ 
	 && \> + \frac{\sqrt{D t}}{a_{\rx}\sqrt{\pi}} \left( \mathrm{exp} \left( -\frac{(a_{\rx}-d_0)^2}{4D t} \right) \right. \nonumber \\
	 && \>+ \left. \mathrm{exp} \left( -\frac{(a_{\rx}+d_0)^2}{4D t} \right) \right),  
\end{IEEEeqnarray} 
}  
where $\mathrm{erf}(\cdot)$ denotes the error function. Eq.~\eqref{Eq.Passive_mean2} provides an accurate approximation for (\ref{Eq;CIR_Passive_woUCA}) if $a_{\rx} < 0.15\, d_0$ \cite{NoelPro1}. 

\begin{remk}
	We refer the interested reader to \cite{NoelPro1} for an analytical expression for $h(t)$ for a passive receiver with rectangular geometry. \QEDwhite
\end{remk}

\textbf{Fully-absorbing Receiver:}
For fully-absorbing receivers \cite{YilmazL1,Akkaya,Heren,Deng2017,Arifler2017,Zabini2018,Dinc2017} (also referred to as perfect sinks), unlike the passive receiver model, the physical and chemical properties of the receiver geometry are taken into account. In particular, the signaling $A$ molecules that reach the receiver via diffusion are absorbed as soon as they hit the receiver surface, see Fig.~\ref{Fig1}b). The sensing area of a fully-absorbing receiver is defined as all points $\mathbf{d}$ on the surface of the receiver, $\mathcal{S}_{\rx}$, and the observed signal is the number of \emph{absorbed} molecules during an infinitesimally small time $\mathrm{d}t$. 
Here, a useful quantity that facilitates the derivation of $h(t)$ is the rate of absorption of the $A$ molecule, which we denote by $k(t)$. Given $k(t)$, we have $h(t) = k(t)\mathrm{d}t$. Now, to derive $h(t)$, we first have to solve \eqref{Eq:PDE_Diff} with $\mathrm{IC}_3$ \eqref{Eq;CIR_passive_IC2}, $\mathrm{BC}_3$ \eqref{Eq;CIR_passive_BC3}, and the following boundary condition that models the absorption of the $A$ molecule on the surface of the receiver
\begin{IEEEeqnarray}{C}
	\label{Eq;CIR_FullAbsorbing_BC4} 
	\mathrm{BC}_4 : p(\mathbf{d} \in \mathcal{S}_{\rx},t) = 0,
\end{IEEEeqnarray}
where in a spherical coordinate system, $\mathbf{d} = [\rho, \varphi, \theta]$, for a spherical receiver with radius $a_{\rx}$ located at the origin of the coordinate system, i.e., $\mathbf{d}_{\rx}=[0,0,0]$, we have $\mathcal{S}_{\rx} = \{ \mathbf{d}| \rho = a_{\rx} \}$. Given $p^{\ast}(\mathbf{d},t)$, i.e., the solution of \eqref{Eq:PDE_Diff} with $\mathrm{IC}_3$, $\mathrm{BC}_3$, and $\mathrm{BC}_4$, $k(t)$ is given by \cite[Eq.~(3.106)]{SchultenL1} 
\begin{IEEEeqnarray}{C}
	\label{Eq;CIR_FullAbsorbing_Rate}
	k(t) = 4\pi a_{\rx}^2 D  \frac{\partial p^{\ast}(\mathbf{d},t)}{\partial \rho}\bigg|_{\rho = a_{\rx}}.  
\end{IEEEeqnarray} 
In \cite{YilmazL1}, $p^{\ast}(\mathbf{d},t)$ for a spherical absorbing receiver is introduced to the MC community and $h(t)$ is calculated as \cite[Eq.~(22)]{YilmazL1} 
\begin{IEEEeqnarray}{C}
	\label{Eq.CIR_FullAbsorbing} 
	h(t)  = \frac{ a_{\rx}(d_0-a_{\rx})}{td_0\sqrt{4\pi D t}} \exp \left( -\frac{(d_0-a_{\rx})^2}{4D t} \right) \mathrm{d}t. 
\end{IEEEeqnarray}     

Another quantity of interest is the probability that a given $A$ molecule is absorbed by time $t$, $\tilde{g}(t)$, which can be obtained as 
\begin{equation}
	\label{Eq.Absorbing_number} 
	\tilde{g}(t) = \int_{t^{\prime} = 0}^{t} k(t^{\prime}) \mathrm{d}t^{\prime} =  \frac{ a_{\rx}}{d_0} \mathrm{erfc} \left( \frac{d_0-a_{\rx}}{\sqrt{4D t}} \right),
\end{equation}
where $\mathrm{erfc}(\cdot)$ is the complementary error function.
 
\begin{remk} Alternatively, when the receiver counts the number of absorbed molecules during observation window $[t_u,\, t_l]$, $h(t)$ can be defined as 
\iftoggle{OneColumn}{%
\begin{IEEEeqnarray}{C}
	\label{Eq.CIR_FullAbsorbing_window} 
	h(t) = \tilde{g}(t_u) - \tilde{g}(t_l) = \frac{ a_{\rx}}{d_0} \left[ \mathrm{erfc} \left( \frac{d_0-a_{\rx}}{\sqrt{4D t_u}} \right) - \mathrm{erfc} \left( \frac{d_0-a_{\rx}}{\sqrt{4D t_l}} \right) \right].  
\end{IEEEeqnarray} 
}{%
\begin{IEEEeqnarray}{rCl}
	\label{Eq.CIR_FullAbsorbing_window} 
	h(t) & = & \tilde{g}(t_u) - \tilde{g}(t_l) \nonumber \\ 
	& = & \frac{ a_{\rx}}{d_0} \left[ \mathrm{erfc} \left( \frac{d_0-a_{\rx}}{\sqrt{4D t_u}} \right) - \mathrm{erfc} \left( \frac{d_0-a_{\rx}}{\sqrt{4D t_l}} \right) \right].   
\end{IEEEeqnarray}
}
\QEDwhite
\end{remk} 

\begin{remk}
For a \emph{fully-absorbing receiver}, it is implicitly assumed that the \emph{whole} surface of the receiver is fully-absorbing. The extension of this model to the case where the receiver surface is \emph{partially} covered by \emph{fully} absorbing receptor patches is considered in \cite{Akkaya}. Moreover, the extension of the fully-absorbing receiver to take the impact of degradation and production noise into account, is considered in \cite{Heren}.
\QEDwhite
\end{remk} 

\begin{remk}
We note that one of earliest CIR models taking the absorption of particles in a \emph{1D} diffusion channel with uniform drift into account is proposed in \cite{Eckford2012}. There, a closed-form expression is derived for the probability of the time of absorption of the signaling molecules.  
\QEDwhite
\end{remk}

\textbf{Reactive Receiver:}
Large or polar signaling molecules cannot passively diffuse through the membrane of cells and are detected by external receptors embedded in the cell membrane. In particular, the diffusive signaling $A$ molecules that reach the cell \emph{may} participate in a \emph{reversible} bimolecular second-order reaction with receptor protein $B$ molecules on the cell surface and form a ligand-receptor complex $C$ molecules; see e.g. Fig.~\ref{Fig1}c). The ligand-receptor interaction can be modelled as shown in \eqref{Eq:BiDeg}  
with binding reaction rate constant $\kappa_f$ in [$\mathrm{molecule^{-1}\cdot m^3\cdot s^{-1}}$] and unbinding reaction rate constant $\kappa_b$ in [$\mathrm{s}^{-1}$]. For such reactive receivers, the sensing area is that part of the receiver surface that is covered by receptors, denoted by $\tilde{\mathcal{S}}_{\rx}$, and the number of activated receptors $C$ constitute the received signal. We refer the interested reader to \cite{ArmanJ2} for a closed-form CIR expression for reactive receivers.

\begin{remk}
In \cite{Deng2015}, a reactive receiver with an infinite number of receptor $B$ molecules covering the whole surface of the recevier, $\mathcal{S}_{\rx}$ (i.e., a homogenous receiver surface, which is a special case of \cite{ArmanJ2}), was considered and the corresponding \gls{CIR} was numerically evaluated. Furthermore, in the \gls{MC} literature, first steps to take the impact of ligand-receptor interaction on the \gls{CIR} into account are made in \cite{PierobonJ3} and \cite{ShahMohammadianProc1}. There, for the evaluation of $h(t)$, the  diffusion equation and the reaction equation are solved separately, unlike \cite{ArmanJ2, Deng2015} where a \emph{coupled} diffusion-reaction equation is considered. \QEDwhite 
\end{remk}  

\begin{remk}
The fully-absorbing receiver is a special case of the reactive receiver when the whole surface of the receiver is covered with infinitely many $B$ molecules, $\kappa_b = 0$, and $\kappa_f \to \infty$. In this case, reaction equation \eqref{Eq:BiDeg} becomes a pseudo first-order reaction of the form $A \mathop{\longrightarrow} C$, with binding reaction rate constant $\kappa_f \to \infty$, where now $C$ corresponds to the number of absorbed molecules. However, $\kappa_f \to \infty$ implies that any collision of a signaling $A$ molecule with the receiver surface leads to the formation of a $C$ molecule, i.e., the reaction is deterministic. We refer the interested reader to \cite{ArmanJ2} where it is shown how the CIR of the reactive receiver, under the above assumptions, is simplified to the CIR of the fully-absorbing receiver. 
\QEDwhite
\end{remk} 

\begin{remk}
A receiver model that, unlike the \gls{CIR} models reviewed in this section so far, also accounts for the impact of the \emph{signaling pathways}, is proposed in \cite{Chou_Rx}. In that model, two simple approximate signaling pathways, modeled via first-order and second-order CRNs, are considered. The \gls{CIR} model in \cite{Chou_Rx} is derived based on a mesoscopic modeling approach; see Section~\ref{Sect:SimExp} for more details on mesoscopic modeling.
\QEDwhite
\end{remk} 

\subsection{Transmitter Models}
\label{Sub;TraMod}
\iftoggle{OneColumn}{%
\begin{figure}[!t]
\centering
\includegraphics[scale=.75]{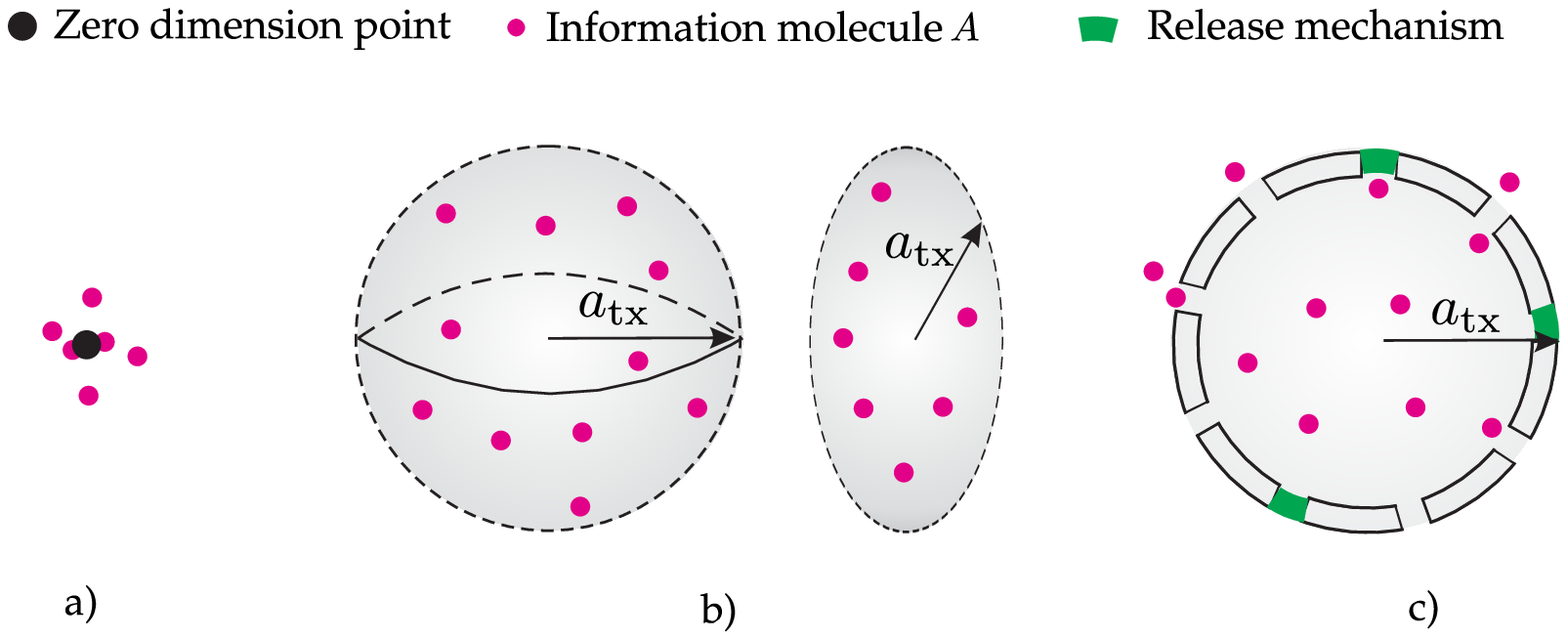}
\caption{Schematic depiction of transmitter models; a) point transmitter, b) volume transmitter, and c) ion-channel based transmitter.}
\label{Fig;TransmitterModels}       
\end{figure}
}{%
\begin{figure}[!t]
\centering
\includegraphics[scale=.5]{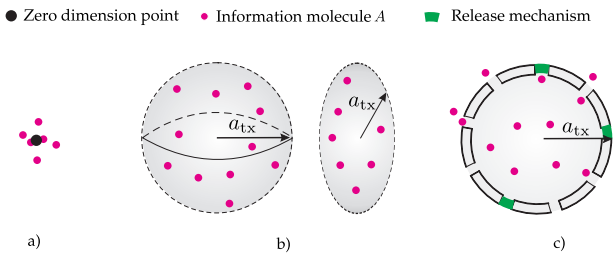}
\caption{Schematic depiction of transmitter models; a) point transmitter, b) volume transmitter, and c) ion-channel based transmitter.}
\label{Fig;TransmitterModels}       
\end{figure}
}
In this section, we review some of the existing end-to-end \gls{CIR} models developed in the \gls{MC} literature that mainly focus on the properties of the transmitter. The main features of the transmitter that can potentially affect the end-to-end \gls{CIR} include: \textit{i}) the geometry of the transmitter, i.e., the volume, boundaries, and shape of the transmitter \cite{Survey_Mol_Nono,Survey_Mol_Net}; \textit{ii}) the particle generation via chemical reactions, which can take different forms ranging from a simple zero-order production reaction to more complex CRNs that take several aspects of $A$ molecule generation into account including, e.g., energy consumption via hydrolization of adenosine triphosphate (ATP) molecules \cite{AlbertsBook}; and \textit{iii}) the release mechanism controlling the release of the $A$ molecules into the physical channel. In particular, after production, the $A$ molecules can leave the transmitter either \emph{passively}, for instance by passive diffusion through channels or gates embedded in the hull of the transmitter, or \emph{actively}, for example via pumps integrated in the hull of the transmitter. In nature, passive and active transportation occur in cells via ion channels and transporters, respectively, see \cite{AlbertsBook}. In the following, we study transmitter models that partially take the effects of the geometry, release mechanisms, and particle generation into account. 

\textbf{Point Transmitter:} The point transmitter is the most widely used transmitter model in the \gls{MC} literature mainly due to its simplicity, see \cite{Nariman_Survey}. However, this model takes none of the above mentioned features into account. In particular, the point transmitter, as the name suggests, is modelled as a zero-dimensional point, i.e., the impact of the geometry of a physical transmitter is not included in the model; see Fig.~\ref{Fig;TransmitterModels}a). Furthermore, it is commonly assumed that the $A$ molecules are produced \emph{instantaneously} and enter the physical channel \emph{immediately}. These assumptions imply that the effects of the particle generation and the release mechanism on $h(t)$ are neglected. 

\textbf{Volume Transmitter:} Unlike point transmitters, where all $A$ molecules are generated at the same location, volume transmitter models take the transmitter geometry into account by assuming that the $A$ molecules are initially distributed over the transmitter volume\footnote{We note that, here, the term ``volume'' is generic and may refer to a volume or a surface in a 3D space, a surface or a line in a 2D space, and a line in a 1D space.} \cite{Noel_CT}; see Fig.~\ref{Fig;TransmitterModels}b). This leads to more realistic models since, in reality, signaling molecules are physical quantities that occupy space. However, volume transmitter models assume that the $A$ molecules are generated \emph{instantaneously}, and that the surface of the transmitter is transparent and does not impede the diffusion of the $A$ molecules. With these two assumptions, volume transmitters neglect the effect of the particle generation and the impact of the release mechanisms. Let us, for the moment, denote the \gls{CIR} models obtained for a point transmitter model, e.g., \eqref{Eq;CIR_Passive_woUCA}, \eqref{Eq.Passive_mean2}, \eqref{Eq.CIR_FullAbsorbing}, by $h^{\bullet}(t,d_0)$. Then, employing the principle of superposition and assuming a uniform particle distribution over the volume of the transmitter, $\mathcal{V}_{\tx}$, the \gls{CIR} of the corresponding volume transmitter can be written as \cite[Eq.~(12)]{Noel_CT} 
\begin{IEEEeqnarray}{rCl} 
	\label{Eq;CIR_VolTX} 
	h(t) = \frac{1}{V_{\tx}} \int_{\mathbf{d} \in \mathcal{V}_{\tx}} h^{\bullet}\left(t, \| \mathbf{d} - \mathbf{d}_{\rx} \| \right) \mathrm{d}\mathbf{d},
\end{IEEEeqnarray}
where $V_{\tx}$ denotes the volume of the transmitter. 

\begin{remk} 
In \cite{Noel_CT}, \eqref{Eq;CIR_VolTX} is solved numerically for a 3D spherical transmitter and both passive and fully-absorbing receivers. Furthermore, in \cite{Noel_CT}, closed-form expressions are given for corresponding \emph{one-dimensional} scenarios. \QEDwhite  
\end{remk}

One useful approximation of \eqref{Eq;CIR_VolTX} can be obtained when the transmitter is sufficiently far away from the receiver. Then, the distance of any point inside the transmitter to the receiver can be approximated by $d_0$ and \eqref{Eq;CIR_VolTX} simplifies to
\iftoggle{OneColumn}{%
\begin{IEEEeqnarray}{rCl} 
	\label{Eq;CIR_VolTX_app} 
	h(t) \approx \frac{h^{\bullet}\left(t, d_0 \right)}{V_{\tx}} \int_{\mathbf{d} \in \mathcal{V}_{\tx}} \mathrm{d}\mathbf{d} = \frac{h^{\bullet}\left(t, d_0 \right)}{V_{\tx}} \times V_{\tx} = h^{\bullet}(t).   
\end{IEEEeqnarray} 
}{%
\begin{IEEEeqnarray}{rCl} 
	\label{Eq;CIR_VolTX_app} 
	h(t) \approx \frac{h^{\bullet}\left(t, d_0 \right)}{V_{\tx}} \int_{\mathbf{d} \in \mathcal{V}_{\tx}} \mathrm{d}\mathbf{d} = \frac{h^{\bullet}\left(t, d_0 \right)}{V_{\tx}} \times V_{\tx} = h^{\bullet}(t). \nonumber \\*   
\end{IEEEeqnarray}
} 
We refer the interested reader to \cite{Noel_CT}, where the accuracy of the above approximation has been investigated for several environments. 

\begin{remk}
	The analytical transmitter models in \cite{Noel_CT} assume that the $A$ molecules are generated throughout $\mathcal{V}_{\tx}$. \cite{Noel_CT} and \cite{Yilmaz2017} simulated a volume transmitter model where the $A$ molecules are generated on the surface of a reflective spherical transmitter. In \cite{Yilmaz2017}, a parametric model is proposed for the CIR of an MC system employing the considered transmitter and a fully-absorbing receiver. A machine learning approach is used to obtain the parameters of the parametric model. \QEDwhite    
\end{remk}
 
\textbf{Ion-Channel Based Transmitter:} Ion-channel based (IC) transmitters are considered in \cite{Arjmandi2016} to model the effect of the release of the signaling molecules into the physical channel. IC transmitters take both the transmitter geometry and the release mechanism into account. In particular, IC transmitters are modelled as spherical objects with \emph{ion-channels} embedded in their membrane; see Fig.~\ref{Fig;TransmitterModels}c). The opening and closing of the ion-channels is controlled via a so-called gating parameter such as a voltage or a ligand. When the gating parameter is applied, e.g., in the form of a voltage across the transmitter membrane, the ion-channels open and the $A$ molecules can leave the transmitter via passive diffusion. The impact of the particle generation is neglected in \cite{Arjmandi2016}. In particular, it is assumed that the $A$ molecules diffuse with different diffusion coefficients inside and outside the transmitter. In \cite{Arjmandi2016} an expression is derived for the average rate of signaling molecules entering the physical channel upon transmitter stimulation.  
Furthermore, an approximate solution for the CIR of the corresponding end-to-end channel is provided under the conditions that the entire surface of the transmitter is covered by a large number of open ion-channels and that the signaling molecules diffuse with the same diffusion coefficient inside and outside the transmitter. There, a passive receiver under the UCA and an unbounded environment is considered. Then, the CIR is approximated as \cite[Eq.~(42)]{Arjmandi2016} 
\begin{IEEEeqnarray}{C}
	\label{Eq;CIR_IBM}
\hspace{-2mm}	h(t) = \frac{a_{\tx}}{d_0 \sqrt{2Dt}} \exp \left( \frac{-\left(d_0^2 + a_{\tx}^2 \right)}{4Dt} \right) \mathrm{sinh} \left( \frac{d_0 a_{\tx}}{2Dt} \right)\!,
\end{IEEEeqnarray}
where $\mathrm{sinh}(\cdot)$ denotes the hyperbolic sine function.    
In fact, \eqref{Eq;CIR_IBM} is actually the CIR of a volume transmitter, since the assumption of having many open ion-channels is equivalent to assuming that the entire surface of the transmitter is a transparent membrane.

  
\begin{remk}
None of the transmitter models reviewed so far consider the impact of the particle generation via CRNs inside the transmitter. This is mainly due to the fact that to take the particle generation into account, a coupled reaction-diffusion equation has to be solved, which is a challenging task. Neverdeless, the effect of particle generation is studied in \cite{Chou2015,Awan2017a,Awan2017b}. There, a common methodology for solving the corresponding reaction-diffusion equation is to adopt mesoscopic models and \emph{numerically} solve the problem. \QEDwhite 
\end{remk}
   
\subsection{Physical Channel Models}
\label{Sub;ChaMod}
In this section, we review some of the existing end-to-end \gls{CIR} models that emphasize the phenomena or impairments of the physical channel. In diffusive \gls{MC} systems, the signaling molecules that enter the physical channel may be affected by several factors and noise sources besides diffusion, including: \textit{i}) advection that can be constructive or destructive depending on the direction and strength of the velocity vector; \textit{ii}) the geometry of the physical channel, e.g., bounded or unbounded environments, constraining the dispersion of the particles; and \textit{iii}) degradation and production of $A$ molecules. For the \gls{CIR} models reviewed in this section, in order to be able to focus on how $h(t)$ is affected by the phenomena in the physical channel, we adopt the point or volume transmitter model and the passive receiver model. 

\iftoggle{OneColumn}{%
\begin{figure}[!t]
\centering
\includegraphics[scale=.75]{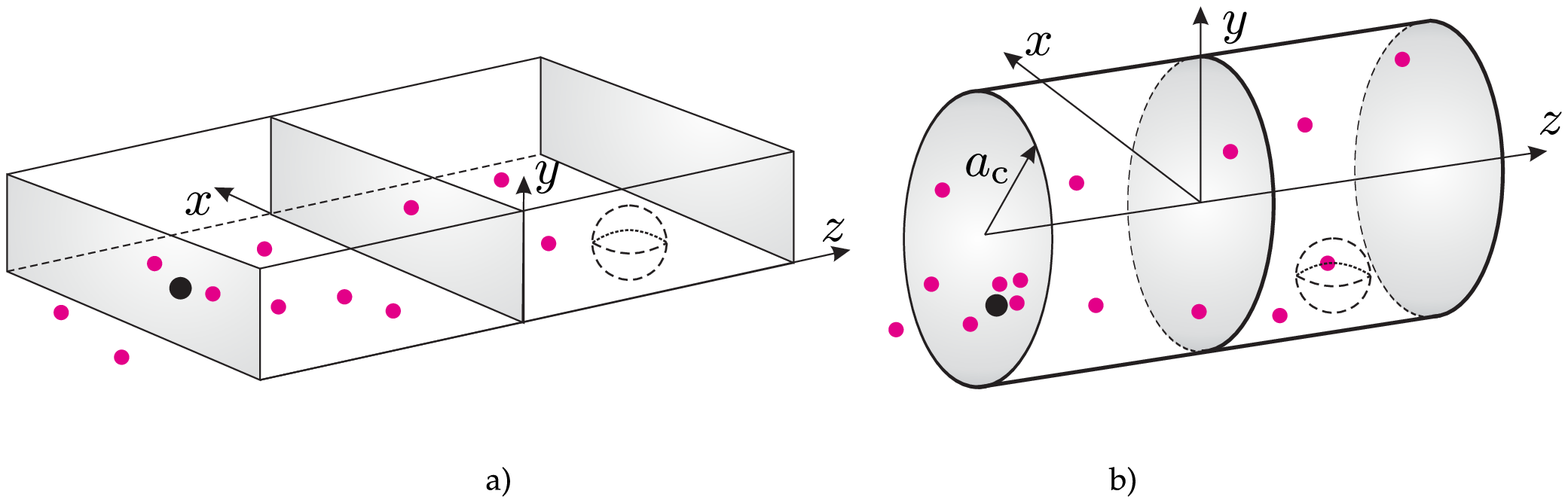}
\caption{Schematic presentation of two duct channels with a) rectangular and b) circular cross sections.}
\label{Fig;BoundedChannels}       
\end{figure} 
}{%
\begin{figure}[!t]
\centering
\includegraphics[scale=.40]{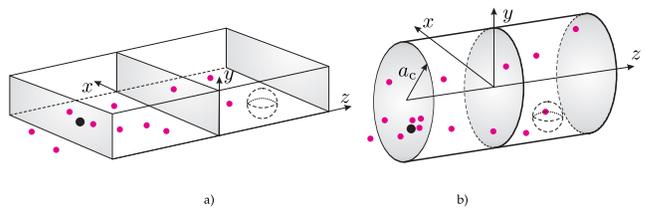}
\caption{Schematic presentation of two duct channels with a) rectangular and b) circular cross sections.}
\label{Fig;BoundedChannels}       
\end{figure} 
}  
\textbf{Bounded Diffusion Channels:} The \gls{CIR} models reviewed previously were obtained under the assumption of an unbounded physical channel. Now, we focus on \gls{CIR} models that assume a more elaborate physical channel geometry. To determine $h(t)$ for \emph{bounded} physical channels, generally, one has to solve diffusion equation \eqref{Eq:PDE_Diff} with appropriate boundary conditions reflecting the physical and chemical properties of the geometry of the  channel. Unfortunately, for many practical geometries, simple and insightful solutions of \eqref{Eq:PDE_Diff} do not exist. Thus, approximations are needed to model practical geometries. In the following, we focus on a class of bounded physical channels that are referred to as \emph{duct channels}. In particular, we consider duct channels with \emph{rectangular} and \emph{circular} cross sections; see Fig.~\ref{Fig;BoundedChannels}. These two duct channels are of particular importance since they approximate the geometry of microfluidic channels and blood vessels, respectively.
  
\begin{itemize}
	\item \textit{Rectangular Duct Channel:} For a rectangular duct channel with dimensions $-\infty < z < + \infty,\, 0<x<l_x,\, 0<y<l_y$, fully reflective walls, a point transmitter at $\mathbf{d}_{\tx} = [x_{\tx}, y_{\tx}, z_{\tx}]$, and a receiver at $\mathbf{d}_{\rx} = [x_{\rx}, y_{\rx}, z_{\rx}]$, the \gls{CIR} can be obtained by solving \eqref{Eq:PDE_Diff} for $p(\mathbf{d},t)$ with $\mathrm{IC}_3$ and the following boundary conditions
\begin{IEEEeqnarray}{rCl}
	\label{Eq;CIR_RecDuct_BCs} 
	\mathrm{BC_5:} &\,\,& \frac{\partial p(\mathbf{d},t)}{\partial x}\bigg|_{x = \{0,l_x\}} =  0, \\ 
	\mathrm{BC_6:} &\,\,& \frac{\partial p(\mathbf{d},t)}{\partial y}\bigg|_{y = \{0,l_y\}} =  0, \\ 
	\mathrm{BC_7:} &\,\,& p( \mathbf{d} = [x,y,z\to \pm\infty] ,t)= 0,  
\end{IEEEeqnarray}
where $\mathrm{BC}_5$ and $\mathrm{BC}_6$ capture the reflection of the $A$ molecule on the duct walls. Since, for the considered geometry, the diffusion of the $A$ molecule in one Cartesian coordinate does not influence its diffusion in the other coordinates, we can write $p(\mathbf{d},t) = p(x,t)\times p(y,t) \times p(z,t)$. Now, solving \eqref{Eq:PDE_Diff} for $p(x,t)$, $p(y,t)$, and $p(z,t)$ with $\mathrm{BC}_5$, $\mathrm{BC}_6$, and $\mathrm{BC}_7$, respectively, and considering a passive receiver under the UCA, $h(t)$ can be obtained as follows \cite[Eq.~(14.4.4)]{Carslaw}
\iftoggle{OneColumn}{%
	\begin{IEEEeqnarray}{rCl} 
		\label{Eq;CIR_Bounded_Duct} 
		h(t) & = & \frac{ V_{\rx}}{l_x l_y} \left[ 1 + 2\sum_{n=1}^{\infty} e^{-D n^2 \pi^2 t/l_x^2} \cos\left( \frac{n\pi x_{\rx}}{l_x} \right) \cos\left( \frac{n\pi x_{\tx}}{l_x} \right) \right] \times \left[ 1 + 2\sum_{n=1}^{\infty} e^{-D n^2 \pi^2 t/l_y^2} \right. \nonumber \\ 
		&& \>\times \left.  \cos\left( \frac{n\pi y_{\rx}}{l_y} \right) \cos\left( \frac{n\pi y_{\tx}}{l_y} \right) \right] \times \left[ \frac{1}{\sqrt{4D \pi t}} \exp \left( \frac{-(z_{\rx} - z_{\tx})^2}{4Dt} \right)\right].  
	\end{IEEEeqnarray}
	}{%
	\begin{IEEEeqnarray}{rCl} 
		\label{Eq;CIR_Bounded_Duct} 
		h(t) & = & \frac{ V_{\rx}}{l_x l_y} \Bigg[ 1 + 2\sum_{n=1}^{\infty} e^{-D n^2 \pi^2 t/l_x^2} \cos\left( \frac{n\pi x_{\rx}}{l_x} \right)  \nonumber \\ 
		&&\>\times  \cos\left( \frac{n\pi x_{\tx}}{l_x} \right) \Bigg] \times \Bigg[ 1 + 2\sum_{n=1}^{\infty} e^{-D n^2 \pi^2 t/l_y^2} \nonumber \\ 
		&& \>\times \cos\left( \frac{n\pi y_{\rx}}{l_y} \right) \cos\left( \frac{n\pi y_{\tx}}{l_y} \right) \Bigg] \nonumber \\ 
		&& \>\times \Bigg[ \frac{1}{\sqrt{4D \pi t}} \exp \left( \frac{-(z_{\rx} - z_{\tx})^2}{4Dt} \right)\Bigg].  
	\end{IEEEeqnarray}
	}
	\item \textit{Circular Duct Channel:} For a circular duct channel with dimensions $0<\rho<a_{\mathrm{c}},\, 0<\theta<2\pi\, -\infty < z < +\infty$ in cylindrical coordinates, fully reflective walls, a point transmitter at $\mathbf{d}_{\tx} = [\rho_{\tx}, \varphi_{\tx}, z_{\tx}]$, and a receiver at $\mathbf{d}_{\rx} = [\rho_{\rx}, \varphi_{\rx}, z_{\rx}]$, the \gls{CIR} can be derived by solving \eqref{Eq:PDE_Diff} with $\mathrm{IC}_3$ \eqref{Eq;CIR_passive_IC2} and the following boundary conditions 
\begin{IEEEeqnarray}{rCl}
	\label{Eq;CIR_CirDuct_BCs} 
	\mathrm{BC_8:} &\,\,& \frac{\partial p(\mathbf{d},t)}{\partial \rho}\bigg|_{\rho = a_{\mathrm{c}}} =  0, \\ 
	\mathrm{BC_9:} &\,\,& p( \mathbf{d} = [\rho, \varphi, z\to \pm\infty] ,t)= 0.  
\end{IEEEeqnarray} 
Here, $\mathrm{BC}_8$ models the reflection of the $A$ molecule at the wall of the duct with radius $a_{\mathrm{c}}$. 
Employing the same technique as for rectangular duct channels, using \cite[Eq.~(14.13.7)]{Carslaw} and considering a passive receiver under the UCA, $h(t)$ can be obtained as follows
\iftoggle{OneColumn}{%
	 \begin{IEEEeqnarray}{rCl}
	 	\label{Eq;CIR_Bounded_Vessel} 
	 	h(t) & = & \frac{V_{\rx}\exp\left( -(z_{\rx}-z_{\tx})^2/4D t \right)}{2\pi a_{\mathrm{c}}^2\sqrt{\pi D t}} \nonumber \\ 
	 	&&\>\times \left[ 1 + \sum_{n = -\infty }^{+\infty} \cos \left( n \left(\varphi_{\rx} - \varphi_{\tx} \right)\right) \sum_{\alpha} \exp \left( -D \alpha^2 t\right) 
	 \frac{\alpha^2 J_n(\alpha \rho_{\rx})J_n(\alpha \rho_{\tx})}{\left( \alpha^2 - n^2/a_{\mathrm{c}}^2 \right)J_n^2(a_{\mathrm{c}}\alpha)}\right],
	 \end{IEEEeqnarray} 
}{%
\begin{IEEEeqnarray}{lll}
	 	\label{Eq;CIR_Bounded_Vessel} 
	 	h(t)  = & \frac{V_{\rx}\exp\left( -(z_{\rx}-z_{\tx})^2/4D t \right)}{2\pi a_{\mathrm{c}}^2\sqrt{\pi D t}} \nonumber \\ 
	 	&\times \left[ 1 + \sum_{n = -\infty }^{+\infty} \cos \left( n \left(\varphi_{\rx} - \varphi_{\tx} \right)\right) \right. \nonumber \\ 
	 	&\times \left. \sum_{\alpha} \exp \left( -D \alpha^2 t\right) 
	 \frac{\alpha^2 J_n(\alpha \rho_{\rx})J_n(\alpha \rho_{\tx})}{\left( \alpha^2 - n^2/a_{\mathrm{c}}^2 \right)J_n^2(a_{\mathrm{c}}\alpha)}\right], \quad\,\,\,\,
	 \end{IEEEeqnarray} 
}
	 where the summation in $\alpha$ is over the positive roots of $J_n^{\prime}(\alpha a_{\mathrm{c}}) = 0$. Here, $J_n(\cdot)$ denotes the $n$-th order Bessel function of the first kind and $J_n^{\prime}(\cdot)$ denotes its derivative.    
\end{itemize}
 
The CIR expressions \eqref{Eq;CIR_Bounded_Duct} and \eqref{Eq;CIR_Bounded_Vessel} indicate that even for simple bounded environments, the solution to $h(t)$ may be quite complicated and difficult to interpret. To gain more insight, in Fig.~\ref{Fig;BoundedChannels_Analysis}, we compare the CIR of an unbounded physical channel to that of a  circular duct channel for system parameters $\mathbf{d}_{\tx} = [0,0, -1.15]$ $\mathrm{\mu m}$, $\mathbf{d}_{\rx} = [0,0,0]$ $\mathrm{\mu m}$, receiver radius $a_{\rx} = 0.15$ $\mathrm{\mu m}$, and $a_{\mathrm{c}} \in \{5,6,9,12\}\times a_{\rx}$. Fig.~\ref{Fig;BoundedChannels_Analysis} shows that when duct radius $a_{\mathrm{c}}$ is small, the CIR of the duct channel is much larger than the CIR of the unbounded channel, i.e., for a given time $t$ it is more likely to observe the signaling molecule. This is because when $a_{\mathrm{c}}$ is small, the signaling molecule is reflected more frequently on the duct walls which increases its chance of being observed at the receiver compared to the unbounded case where the $A$ molecules can diffuse away. However, for large $a_{\mathrm{c}}$, the CIR of the duct channel approaches the CIR of the unbounded environment, i.e., the CIR of the unbounded channel provides a good approximation for the CIR of a large bounded circular duct channel.       

\iftoggle{OneColumn}{%
\begin{figure}[!t]
\centering
\includegraphics[scale=.75]{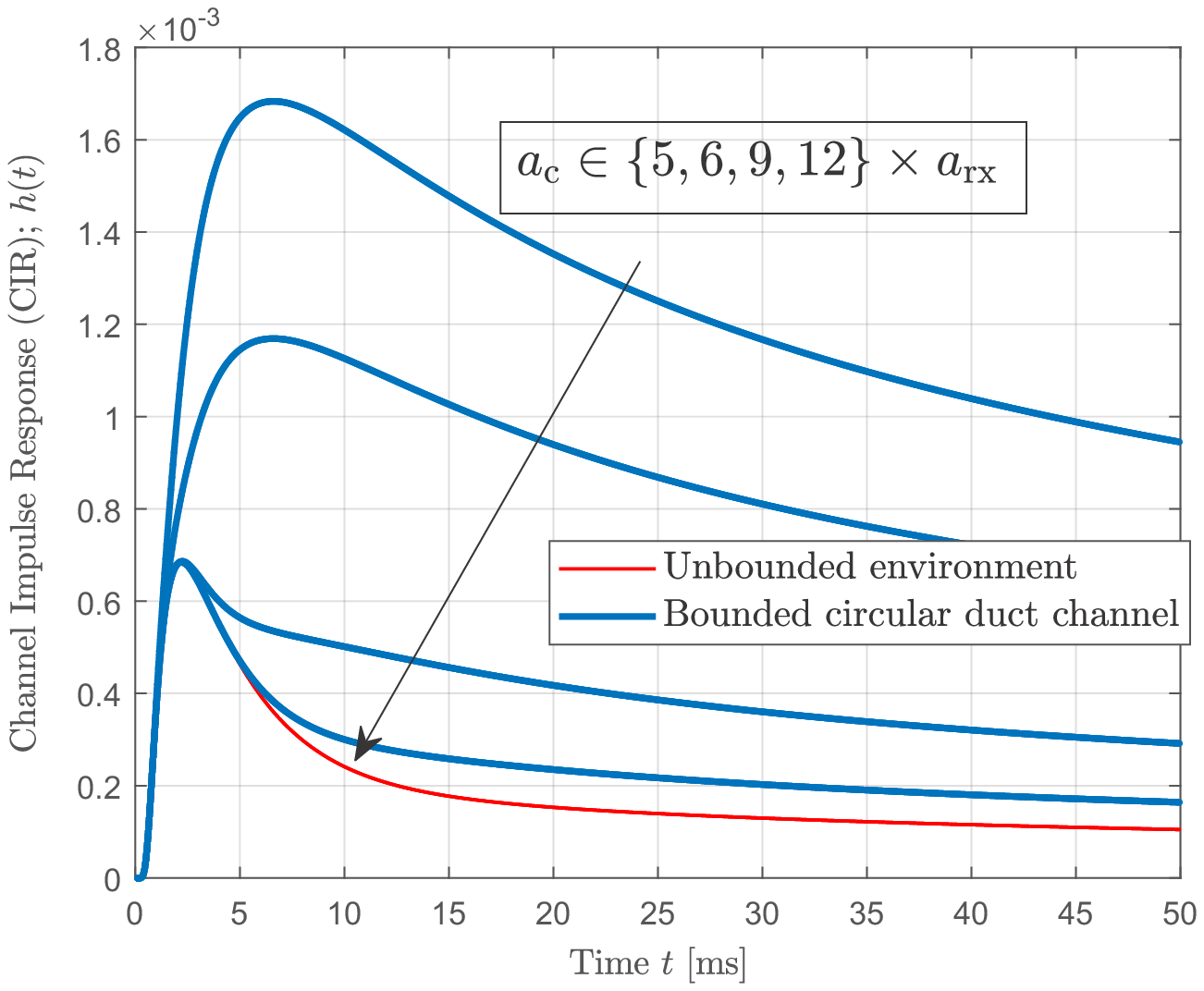}
\caption{Channel impulse response, $h(t)$, as a function of time $t$, for an unbounded environment and a bounded circular duct channel. The duct radius increases in the direction of the arrow.}
\label{Fig;BoundedChannels_Analysis}       
\end{figure}
}{%
\begin{figure}[!t]
\centering
\includegraphics[scale=.6]{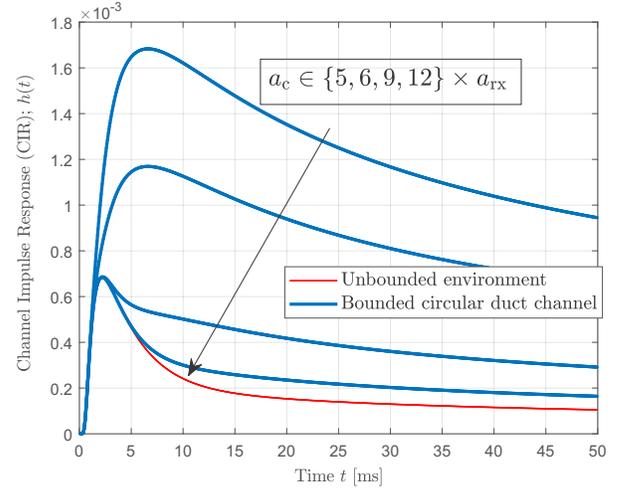}
\caption{Channel impulse response, $h(t)$, as a function of time $t$, for an unbounded environment and a bounded circular duct channel. The duct radius increases in the direction of the arrow.}
\label{Fig;BoundedChannels_Analysis}       
\end{figure}
}  

\begin{remk}
We note that the necessary condition for the validity of the UCA developed for passive receivers in unbounded channels, i.e., $a_{\rx} < 0.15 d_0$, is not applicable for bounded physical channels. However, we expect that as $a_{\rx}/d_0 \to 0$, the accuracy of \eqref{Eq;CIR_Bounded_Duct} and \eqref{Eq;CIR_Bounded_Vessel} improves. \QEDwhite 
\end{remk} 

\textbf{Advection Channels:} Next, we consider physical channels in which the signaling molecules experience advection in addition to their random walk. In particular, for the \gls{CIR} models reviewed in this section, for mathematical tractability, we consider advection processes with a  time-invariant velocity field, i.e., we assume $\mathbf{v}(\mathbf{d},t) = \mathbf{v}(\mathbf{d}),\,\, \forall t > t_0$.  
\begin{itemize}
	\item \textbf{Uniform constant advection:} In this case, the magnitude and the direction of the velocity field are identical at any point $\mathbf{d}$ in space, i.e., $\mathbf{v}(\mathbf{d}) = \mathbf{v} = [v_x, v_y, v_z], \,\, \forall \mathbf{d} \in \mathcal{R}^3$, where $\mathcal{R}^3$ is the set of all points in the 3D Cartesian system, see Example~3 in Section~\ref{Sect:Prelim}. Vector $\mathbf{v}$ can be effectively decomposed into two components, a parallel component $v_{\|}$ and an orthogonal component $v_{\bot}$ with respect to $\mathbf{d}_0 = \mathbf{d}_{\tx} - \mathbf{d}_{\rx}$. Let us assume, without loss of generality, a point transmitter at $\mathbf{d}_{\tx} = [0,0,-z_{\tx}]$ and a passive receiver located at $\mathbf{d}_{\rx} = [0,0,0]$, such that $\mathbf{d}_0 = [0,0,-z_{\tx}]$. Then, $v_{\|} = v_z$, and we can write $v_{\bot} = \sqrt{v_x^2 + v_y^2}$.
	
	\begin{itemize}	 
	\item \textit{Unbounded Channel with UCA:} For an unbounded channel and a passive receiver under the UCA, $h(t)$ can be obtained by solving advection-diffusion equation \eqref{Eq:PDE_FlowDiff}. Using the method of \emph{moving reference frame}, i.e., assuming that the reference of the coordinate system is moving with $\mathbf{v}$, it can be readily verified that $h(t)$ can be obtained from \eqref{Eq.Passive_mean2} as \cite[Eq.~(18)]{NoelPro3}
\iftoggle{OneColumn}{%
\begin{IEEEeqnarray}{rCl} 
	\label{Eq;CIR_UniDRF_Unbouonded_PassiveRX} 
	h(t) = \frac{ V_{\rx}}{\left(4\pi D t \right)^{3/2}} \exp \left( -\frac{(v_{\bot}t)^2 + (z+z_{\tx}-v_{\|}t)^2}{4Dt} \right). 
\end{IEEEeqnarray}
}{%
\begin{IEEEeqnarray}{ll} 
	\label{Eq;CIR_UniDRF_Unbouonded_PassiveRX} 
	h(t) = &\frac{ V_{\rx}}{\left(4\pi D t \right)^{3/2}} \nonumber \\ 
	& \times \exp \left( -\frac{(v_{\bot}t)^2 + (z+z_{\tx}-v_{\|}t)^2}{4Dt} \right). \quad
\end{IEEEeqnarray}
}
Eq.~\eqref{Eq;CIR_UniDRF_Unbouonded_PassiveRX} can be also directly obtained from \eqref{Eq:Solution_DiffAdvec} after setting $N=1$, multiplying $c^{\ast}(\mathbf{d},t)$ with $V_{\rx}$, and using the $\mathbf{v}$, $\mathbf{d}_{\tx}$ and $\mathbf{d}_{\rx}$ mentioned above. 
	\item \textit{Unbounded channel without UCA:} For the case when the UCA does not hold, $v_{\|} \neq 0$, and $v_{\bot} \neq 0$, $h(t)$ can be solved numerically. However, it is shown in \cite{NoelPro3} that for the special case of $v_{\bot} = 0$, $h(t)$ can be obtained from \eqref{Eq;CIR_Passive_woUCA} after substituting $d_0$ with $-(z_{\tx} - v_{\|}t)$.
	\item \textit{Bounded channel with UCA:} In this case, i.e., when we have bounded channels such as duct channels, and for the general case where $v_{\|} \neq 0 $, $v_{\bot} \neq 0$, we cannot apply the technique of \emph{moving reference frame} in the dimensions of the coordinate system where the physical channel is bounded. Thus, $h(t)$ has to be directly evaluated via the corresponding advection-diffusion equation. However, for the special case of $v_{\bot} = 0$, after substituting $z_{\tx}$ with $z_{\tx} - v_{\|}t$, the corresponding \gls{CIR}s of the rectangular and circular duct channels can be obtained from \eqref{Eq;CIR_Bounded_Duct} and \eqref{Eq;CIR_Bounded_Vessel}, respectively. 
	\item \textit{Bounded channel without UCA:} In this case, the general form of $h(t)$ depends on the geometries of the bounded physical channel and the passive receiver. However, for a rectangular duct channel, a rectangular passive receiver, and $v_{\|} \neq 0$, $v_{\bot} \neq 0$, an analytical expression for $h(t)$ is derived in \cite{WayanPro2}. We note that in \cite{WayanPro2}, it is assumed that $v_{\|}$ and $v_{\bot}$ are a fluid velocity field and a drift velocity caused by a magnetic field, respectively. However, the derived expression for $h(t)$ is valid independent of the origin of $v_{\|}$ and $v_{\bot}$. Furthermore, in \cite{WayanPro2}, the case of partially absorbing duct channel walls is also considered.      
	\end{itemize}
	
\item \textbf{Laminar flow:} In this case, we only focus on bounded channels, and in particular on circular duct channels, since laminar flow arises in bounded environments. Thus, we consider $\mathbf{v}(\mathbf{d})$ given in \eqref{Eq:Poiseuille}. For the \gls{CIR} models reviewed here, we distinguish between point and volume transmitter models with axial position $z_{\tx} = 0$, and consider the passive receiver model with the following dimensions in cylindrical coordinates $  a_{\mathrm{c}} - l_\rho \leq \rho_{\rx} \leq a_{\mathrm{c}}, |\varphi_{\rx}|\leq l_{\varphi}/2 ,|z_{\rx}-d_z|\leq l_z/2 $; see Fig.~\ref{Fig;Circular_Duct_Laminar}. In particular, we distinguish between two cases, namely the dispersion regime ($\alpha_d \ll 1$) and the flow dominant regime ($\alpha_d \gg 1$), see \eqref{Eq:Timescales}.
\iftoggle{OneColumn}{%
\begin{figure}[!t]
\centering
\includegraphics[scale=1.75]{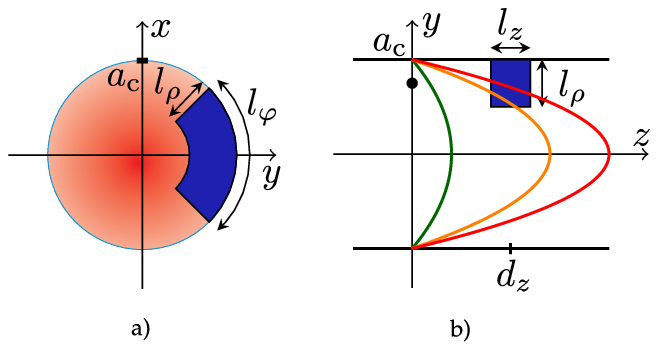}
\caption{Schematic presentation of a circular duct channel with radius $a_{\mathrm{c}}$ and laminar flow; a) cross section and b) along the $z$ axis. The receiver is depicted in blue color.}
\label{Fig;Circular_Duct_Laminar}       
\end{figure}
}{%
\begin{figure}[!t]
\centering
\includegraphics[scale=1]{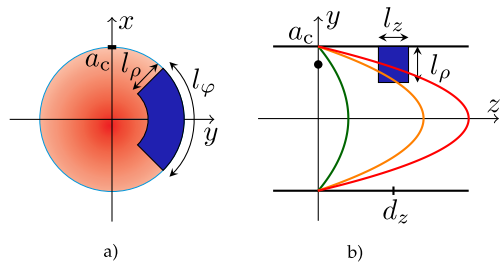}
\caption{Schematic presentation of a circular duct channel with radius $a_{\mathrm{c}}$ and laminar flow; a) cross-section and b) along the $z$ axis. The receiver is depicted in blue color.}
\label{Fig;Circular_Duct_Laminar}       
\end{figure}
}  
	\begin{itemize}
		\item \textit{Dispersion regime with UCA:} In the dispersion regime, $\alpha_d \gg 1$ holds in \eqref{Eq:Timescales}. As a result, the time required for transportation of the $A$ molecule in the $z$ direction via flow, $d_z/v_{\mathrm{eff}}$, is much larger than $a_{\mathrm{c}}^2/D$, which is the characteristic time for diffusion of the $A$ molecule over distance $a_{\mathrm{c}}$. This fact has two immediate consequences: \textit{i)} by the time that the released $A$ molecule reaches the receiver, it experiences the average flow velocity, i.e., $v_{\mathrm{eff}}$, due to its fast diffusion across the cross section; \textit{ii)} there is no difference between point and uniform release and $h(t)$ only depends on $z$. Thus, the corresponding advection-diffusion equation in three dimensional space can be effectively \emph{approximated} by its one dimensional equation with effective velocity $v_{\mathrm{eff}}$ and effective diffusion coefficient $D_{\mathrm{eff}}$ as follows 
		\begin{IEEEeqnarray}{C} 
		\label{Eq;Effective_1D_Diffusion} 
			\partial_t p(z,t) = D_{\mathrm{eff}} \partial_z^2 p(z,t) - v_{\mathrm{eff}}t, 
		\end{IEEEeqnarray}
where $D_{\mathrm{eff}}$ is the Aris-Taylor effective diffusion coefficient and can be obtained as \cite[Eq.~(4.6.35)]{probstein_physicochemical_2005} 
		\begin{IEEEeqnarray}{C}
			\label{Eq;Aris-Tylor} 
			D_{\mathrm{eff}} = \left( 1 + \frac{1}{48} \left( \frac{(v_{\mathrm{eff}}a_{\mathrm{c}})^2}{D} \right) \right).
		\end{IEEEeqnarray} 
Solving \eqref{Eq;Effective_1D_Diffusion} with the UCA approximation, $\mathrm{BC}_9$,  and the following initial condition for uniform release across the cross section
\begin{IEEEeqnarray}{rCl}
	\label{Eq;IC2} 
	\mathrm{IC_4:} &\,\,& p(z, t_0) = \frac{1}{4\pi a_{\mathrm{c}}^2} \delta(z) 
\end{IEEEeqnarray}
leads to \cite[Eq.~(11)]{WayanPro1}
\iftoggle{OneColumn}{%
		\begin{IEEEeqnarray}{rCl} 
			\label{Eq;CIR_Laminar_DisReg_UCA} 
			h(t) = \frac{ V_{\rx}}{\pi a_{\mathrm{c}}^2} \times \frac{1}{\sqrt{4\pi D_{\mathrm{eff}}t}} \exp \left( -\frac{\left(d_z - v_{\mathrm{eff}}t \right)^2}{4D_{\mathrm{eff}}t} \right).
		\end{IEEEeqnarray}
		}{%
		\begin{IEEEeqnarray}{rCl} 
			\label{Eq;CIR_Laminar_DisReg_UCA} 
			h(t) = \frac{ V_{\rx}}{\pi a_{\mathrm{c}}^2} \times \frac{1}{\sqrt{4\pi D_{\mathrm{eff}}t}} \exp \left( -\frac{\left(d_z - v_{\mathrm{eff}}t \right)^2}{4D_{\mathrm{eff}}t} \right). \nonumber \\*
		\end{IEEEeqnarray}
		}
		
		\item \textit{Dispersion regime without UCA:} In this case, $h(t)$ can be obtained by taking the integral of the solution of \eqref{Eq;Effective_1D_Diffusion} over the volume of the receiver, which leads to \cite[Eq.~(13)]{WayanPro1}
		\iftoggle{OneColumn}{%
		\begin{IEEEeqnarray}{rCl}
			\label{Eq;CIR_Laminar_DisReg_woUCA} 
			 h(t) = \frac{l_{\varphi}(2 a_{\mathrm{c}} l_{\rho} -l_{\rho}^2)}{2\pi a_{\mathrm{c}}^2} \times \left[ \mathrm{Q} \left( \frac{d_z - l_z/2 - v_{\mathrm{eff}}t}{2D_{\mathrm{eff}}t} \right) +  \mathrm{Q} \left( \frac{d_z + l_z/2 - v_{\mathrm{eff}}t}{2D_{\mathrm{eff}}t} \right) \right],
		\end{IEEEeqnarray}
		}{%
		\begin{IEEEeqnarray}{rCl}
			\label{Eq;CIR_Laminar_DisReg_woUCA} 
			 h(t) & = & \frac{l_{\varphi}(2 a_{\mathrm{c}} l_{\rho} -l_{\rho}^2)}{2\pi a_{\mathrm{c}}^2} \times \left[ \mathrm{Q} \left( \frac{d_z - l_z/2 - v_{\mathrm{eff}}t}{2D_{\mathrm{eff}}t} \right) \right. \nonumber \\
			 && \>+  \left. \mathrm{Q} \left( \frac{d_z + l_z/2 - v_{\mathrm{eff}}t}{2D_{\mathrm{eff}}t} \right) \right],
		\end{IEEEeqnarray}
		}
where $Q(\cdot)$ denotes the Gaussian Q-function. 
\begin{remk}
We note that the accuracy of both \eqref{Eq;CIR_Laminar_DisReg_UCA} and \eqref{Eq;CIR_Laminar_DisReg_woUCA} depends on the value of $\alpha_d$ in \eqref{Eq:Timescales}. For example, by increasing $D$ and $d_z$, $\alpha_d$ decreases and the accuracy of the dispersion regime improves; see \cite{WayanPro1}. \QEDwhite    
\end{remk}
 
		\item \textit{Flow dominant regime with volume transmitter:} In this case, i.e., $\alpha_d \ll 1$, the impact of diffusion is negligible. Thus, the signaling molecules do not have sufficient time to disperse across the cross section of the duct channel before they arrive at the receiver. As a result, a particle released at radial position $\rho$ is observed approximately at the same radial position at the receiver. Thus, we have to distinguish between the volume and point transmitter models. For the case of uniform release, $h(t)$ can be approximated as \cite[Eq.~(16)]{WayanPro1}
		\iftoggle{OneColumn}{%
		 \begin{IEEEeqnarray}{C} 
		 \label{Eq;CIR_Laminar_FlowReg_UnifromRel}
		 h(t) = \begin{cases} 
		 0, & t \leq t_1 \\ 
		  \frac{c_{\varphi}(2 a_{\mathrm{c}} l_{\rho} -l_{\rho}^2)}{2\pi a_{\mathrm{c}}^2} - \frac{l_{\varphi}}{2 \pi} \frac{d_z - l_z/2}{2v_{\mathrm{eff}}t}, & t_1 < t < t_2 \\
		 \frac{l_{\varphi}}{2\pi} \frac{l_z}{2v_{\mathrm{eff}}t}, & t \geq t_2.
		 \end{cases} 
		 \end{IEEEeqnarray} 
		 }{%
		 \begin{IEEEeqnarray}{C} 
		 \label{Eq;CIR_Laminar_FlowReg_UnifromRel}
		 h(t) = \begin{cases} 
		 0, & t \leq t_1 \\ 
		  \frac{c_{\varphi}(2 a_{\mathrm{c}} l_{\rho} -l_{\rho}^2)}{2\pi a_{\mathrm{c}}^2} - \frac{l_{\varphi}}{2 \pi} \frac{d_z - l_z/2}{2v_{\mathrm{eff}}t}, & t_1 < t < t_2 \\
		 \frac{l_{\varphi}}{2\pi} \frac{l_z}{2v_{\mathrm{eff}}t}, & t \geq t_2. 
		 \end{cases} \quad\,\,
		 \end{IEEEeqnarray} 
		 }
		 where 
		 \begin{IEEEeqnarray}{C} 
		 	\label{Eq;t1t2_Laminar}
		 	t_{1,2} = \frac{d_z \mp l_z/2}{2v_{\mathrm{eff}}(1- (1-l_{\rho}/a_{\mathrm{c}})^2)}.
		 \end{IEEEeqnarray}
In \eqref{Eq;t1t2_Laminar}, $t_1$ and $t_2$ are the times when the parabolic velocity field first hits and leaves the receiver volume, respectively.
		 
		 \item \textit{Flow dominant regime with point transmitter:} For the case of a point transmitter, when the $A$ molecule is released $d_z$ away from the receiver but within the $\rho$ and $\varphi$ coordinates defining the dimensions of the receiver, i.e., at $\rho_{\tx} \in [a_{\mathrm{c}}-l_{\rho}, a_{\mathrm{c}}]$ and $\varphi_{\tx} \in [-l_{\varphi}/2, l_{\varphi}/2]$, we observe the $A$ molecule with certainty if $d_z - l_z/2 \leq v(\rho_{\tx})t \leq d+l_z/2$ 
		 \begin{IEEEeqnarray}{C}
		 \label{Eq;CIR_Laminar_FlowReg_PoifromRel}
		 	h(t) = \mathrm{rect} \left( \frac{v(\rho_{\tx})t - d_z}{l_z} \right),
		 \end{IEEEeqnarray}
		 where $\mathrm{rect}(x) = 1$ if $ -1/2 \leq x \leq 1/2$. 
	\end{itemize}	  
\end{itemize}

\begin{remk}
In the \gls{MC} literature, first steps towards the extension of some of the CIR models reveiwed for the advection channel to more complex networks of interconnected bounded duct channels are provided in \cite{Chahibi2013, chahibi2015molecular, mosayebi2018early, Iwasaki2018}. For example, in \cite{Chahibi2013}, the Aris-Taylor effective diffusion coefficient approximation is employed to calculate the end-to-end CIR of multiple interconnected blood vessels for drug delivery applications. Furthermore, in \cite{mosayebi2018early}, the uniform advection model is adopted to model blood vessel networks for abnormality detection applications in biological systems. \QEDwhite 
\end{remk}
   
\textbf{Degradation Channels:} In degradation channels, the arrival of the signaling molecules is affected by their possible degradation and production. In this case, $h(t)$ can be obtained by solving the diffusion-reaction equation \eqref{Eq:Fick_general} (with $\mathbf{v}(\mathbf{d},t)$) given appropriate initial and boundary conditions. However, the solution to \eqref{Eq:Fick_general} depends very much on the structure of the corresponding CRN described by reaction rate function $f(\cdot)$. Often, the reaction terms in $f(\cdot)$ are highly nonlinear and coupled, which makes the problem of solving  \eqref{Eq:Fick_general} very challenging. Here, in order to arrive at mathematically tractable and insightful results describing the \emph{general} behaviour of degradation channels, we focus on two particular forms of degradation and production noise, namely \emph{first-order degradation} and \emph{enzymatic degradation}; see Examples 7 and 9, and \cite{Adam_Enzyme,Nakano2013a,ArmanJ2,Heren}. 
\begin{itemize}
	\item \textit{First-order degradation:} Let us for the moment denote the CIR expressions developed in the previous sections by $\tilde{h}(t)$. It can be shown that $h(t)$, for a physical channel with first-order degradation reaction of the form \eqref{Eq:Deg} and reaction rate constant $\kappa$ can be readily obtained from $\tilde{h}(t)$ when the following assumptions hold. \text{A1}) The signaling $A$ molecule is affected by the degradation reaction uniformly and equally throughout the entire \emph{end-to-end channel}, and \text{A2}) it is not involved in any other CRN from stimulation time $t_0$ until observation time $t$. In this case, we can write 
	\begin{IEEEeqnarray}{C}
		\label{Eq;CIR_Degradation_1storder}
		h(t) = \tilde{h}(t) \times \exp \left( - \kappa t\right).
	\end{IEEEeqnarray}
In \eqref{Eq;CIR_Degradation_1storder}, the term $\exp \left( - \kappa t\right)$ captures the \emph{surviving probability} of the signaling molecule, which is a monotonically decreasing function of time, i.e., as $t$ increases, it becomes more likely that the signaling molecule $A$ is degraded in the channel. As a result, for degradation channels, at any instant, $h(t)$ is smaller than the corresponding CIR without degradation.
 	 
Assumption \text{A2}) holds for all \gls{CIR} models presented so far except the \gls{CIR} model for the reactive receiver in Section~\ref{SubSec;Receiver}. This is because for reactive receivers, the signaling molecule is involved in a ligand-receptor kinetic reaction \eqref{Eq:BiDeg}, in addition to the degradation reaction, and may experience several binding/unbinding events before reception time $t$. We refer the interested reader to \cite{ArmanJ2}, where a closed-form expression is derived for the CIR of an MC system with a reactive receiver and first-order degradation channel. 
 
\item \textit{Enzymatic degradation:} The impact of enzymatic degradation reactions in the channel is studied for passive receiver and point transmitter models in \cite{Adam_Enzyme}. Enzymatic reactions include a second-order reaction, and as a result, in order to obtain $h(t)$, we have to solve \eqref{Eq:Fick_general} with $\mathrm{IC}_3$, $\mathrm{BC}_3$, and $f(\cdot)$ driven by \eqref{Eq:EnzamiticPDE}. However, this system of nonlinear equations does not facilitate a closed-form solution for $h(t)$. As a result, in \cite{Adam_Enzyme}, several approximate solutions were proposed for the CIR of an \gls{MC} system with point transmitter, passive receiver, and unbounded environment as follows: 
\begin{itemize}
	\item If the concentration of signaling molecules, $c_A(\mathbf{d},t)$, and the concentration of the intermediate species $AE$, $c_{AE}(\mathbf{d},t)$, remain constant over time and space, then $f(\cdot)$ is described only by (28a) and $h(t)$ has the following solution
	\iftoggle{OneColumn}{%
	\begin{IEEEeqnarray}{C}
		\label{Eq;CIR_Passive_Enzymatic_app1}
		h(t) \approx \frac{V_{\rx}}{\left( 4\pi D t\right)^{3/2}} \exp \left(-\kappa_f c_E t - \frac{d_0}{4D t}\right) + \kappa_b c_{AE}t.
	\end{IEEEeqnarray}
	}{%
	\begin{IEEEeqnarray}{C}
		\label{Eq;CIR_Passive_Enzymatic_app1}
		h(t) \approx \frac{V_{\rx}}{\left( 4\pi D t\right)^{3/2}} \exp \left(-\kappa_f c_E t - \frac{d_0}{4D t}\right) + \kappa_b c_{AE}t. \nonumber \\*
	\end{IEEEeqnarray}
	}
	\item If $\kappa_d \to \infty$ and $\kappa_b \to 0$, then the total concentration of the enzyme $E$ molecules remains constant and $c_{AE} = 0$. Then, a lower bound for $h(t)$ is obtained via \eqref{Eq;CIR_Passive_Enzymatic_app1} after setting $c_{AE} = 0$. 
	\item Another useful approximation is obtained by assuming that $c_{AE}$ is constant \cite{Adam_Enzyme}. This is a valid assumption when $\kappa_f \to \infty$ and $\kappa_b \to 0$. Then, as explained in Section~\ref{Sect:Prelim}, the enzymatic reaction in \eqref{Eq:EnzamiticDeg} can be approximated by the first-order unimolecular reaction in \eqref{Eq:Deg} and the corresponding $h(t)$ can be written as
	\iftoggle{OneColumn}{%
	\begin{IEEEeqnarray}{C}
	\label{Eq;CIR_Passive_Enzymatic_app3}
	h(t) \approx \frac{V_{\rx}}{\left( 4\pi D t\right)^{3/2}} \exp \left(- \frac{\kappa_f \kappa_d}{\kappa_b + \kappa_d} c_{E_t} t - \frac{d_0}{4D t}\right),
	\end{IEEEeqnarray} 
	}{%
	\begin{IEEEeqnarray}{C}
	\label{Eq;CIR_Passive_Enzymatic_app3}
	h(t) \approx \frac{V_{\rx}}{\left( 4\pi D t\right)^{3/2}} \exp \left(- \frac{\kappa_f \kappa_d}{\kappa_b + \kappa_d} c_{E_t} t - \frac{d_0}{4D t}\right), \nonumber \\*
	\end{IEEEeqnarray} 
	}
	where $c_{E_t}$ denotes the total concentration of the enzyme $E$ molecules, including both bound and unbound enzyme. We refer the interested reader to \cite{Adam_Enzyme} for verification of the accuracy of the proposed approximate expressions for $h(t)$.
 \end{itemize}

\begin{remk}
Signaling molecules of different types may also degrade each other. For instance, the \gls{MC} testbed in \cite{Nariman_AcidBasePlatform} uses acids and bases as signaling molecules that can participate in a bimolecular reaction and cancel each other out, cf. (\ref{Eq:BiDeg}). Unfortunately, the underlying \glspl{PDE} that describe the bimolecular reaction are coupled and nonlinear and closed-form expressions for the CIR are not available. In  \cite{ICC_Reactive}, a numerical method was developed which decouples reaction and diffusion in each time slot and computes the channel response in an iterative manner. \QEDwhite
\end{remk}

\end{itemize}   

\subsection{Summary of End-to-End CIR Models}
\label{Sub;SumCIR}
To conclude this section, we provide a summary of the reviewed CIR models in Table~\ref{Table;ReviewCIR}. For conciseness of presentation, we use the following abbreviations. For the transmitter, ``$\Omega_{\tx}$" indicates whether a point transmitter is assumed or a volume transmitter releasing the molecules from its volume ($\mathcal{V}_{\tx}$) or surface ($\mathcal{S}_{\tx}$). In the later case, we also specify whether the surface is transparent (``trans") or reflective (``refl"). Furthermore, ``PG" and ``REM" specify whether particle generation and the release mechanism are taken into account, respectively. For the physical channel, ``Diff" and ``Adv" denote whether diffusion and advection processes are taken into account, respectively. In the case of advection, ``uni" and ``lam" specify uniform and laminar advections in the physical channel. ``Geo" specifies whether a bounded (``$\bigcirc$") or unbounded (``$\bigotimes$") geometry is considered. ``D\&P" indicates degradation and production reactions of signaling molecules in the physical channel. For the receiver, ``$\Omega_{\rx}$" indicates whether the volume of the receiver, $\mathcal{V}_{\rx}$, constitutes the sensing area of the receiver or a surface, $\mathcal{S}_{\rx}$, or a partial surface, $\tilde{\mathcal{S}}_{\rx}$. Furthermore, ``passive" and ``active" refer to the reception mechanism of the receiver. In the later case, ``Det" and ``Sto" specify whether the corresponding reaction in active reception is considered deterministically or stochastically, respectively. ``SP" indicates whether signaling pathways are considered. Moreover, ``Dim" denotes the dimension of the considered end-to-end channel. ``Num" indicates that the CIR $h(t)$ is obtained numerically. Whenever possible, we also refer to the equation number of the corresponding CIR $h(t)$. Finally, whenever the reaction-diffusion equation was involved for obtaining $h(t)$, we emphasize whether the reaction and diffusion processes are solved separately (``Sep") or jointly (``Joi"). Thus, the reaction process can potentially describe an active reception mechanism, D\&P reaction, or both.   

\iftoggle{OneColumn}{%
\begin{sidewaystable}
\renewcommand{\arraystretch}{1.4}
\caption{Summary of the reviewed CIR models}
\label{Table;ReviewCIR}
\centering
\begin{scriptsize}
\scalebox{0.9}{ 
\begin{tabular}{|c|c|c|c|c|c|c|c|c|c|c|c|c|c|c|c|c|c|c|c|c|} 
\hline 
    & \multicolumn{4}{|c|}{Transmitter} & \multicolumn{4}{|c|}{Physical Channel} & \multicolumn{6}{|c|}{Receiver} & \multirow{3}{*}{Dim} & \multicolumn{2}{|c|}{\eqref{Eq:Fick_general}}  & \multicolumn{2}{|c|}{$h(t)$}  \\ \cline{2-15} \cline{17-20}  
  Ref.  & \multirow{2}{*}{Surface} & \multirow{2}{*}{$\Omega_{\tx}$} & \multirow{2}{*}{PG} & \multirow{2}{*}{REM} & \multirow{2}{*}{Diff} & \multirow{2}{*}{Adv} & \multirow{2}{*}{Geo} & \multirow{2}{*}{D\&P} & \multirow{2}{*}{$\Omega_{\rx}$} & \multicolumn{2}{|c|}{Passive} & \multicolumn{2}{|c|}{Active} & \multirow{2}{*}{SP} & & \multirow{2}{*}{Joi} & \multirow{2}{*}{Sep} &\multirow{2}{*}{Ana} & \multirow{2}{*}{Num} \\ \cline{11-14} 
  &&&&&&&&&& w/o UCA & w/ UCA & Sto & Det &&&&&& \\ \hline 
  \cite{NoelPro1}&&point&&&\checkmark && $\bigotimes$ & & $\mathcal{V}_{\rx}$ & \checkmark & \checkmark &&&&3D&&& \eqref{Eq.Passive_mean2}, \eqref{Eq;CIR_Passive_woUCA}  & \\ \hline
\cite{YilmazL1} &&point&&& \checkmark && $\bigotimes$ & & $\mathcal{S}_{\rx}$ & & & & 1st-order && 3D & \checkmark & & \eqref{Eq.CIR_FullAbsorbing} & \\ \hline
\cite{Akkaya} &&point&&& \checkmark && $\bigotimes$ & & $\mathcal{ \tilde{S}}_{\rx}$ & & & & 1st-order && 3D & \checkmark & & \cite[Eq.~(13)]{Akkaya} & \\ \hline 
\cite{Heren} &&point&&& \checkmark && $\bigotimes$ & \eqref{Eq:Deg} & $\mathcal{S}_{\rx}$ & & & & 1st-order && 3D & \checkmark & & \cite[Eq.~(12)]{Heren} & \\ \hline
\cite{Deng2015} &&point&&& \checkmark && $\bigotimes$ & & $\mathcal{S}_{\rx}$ & & & 2nd-order & && 3D & \checkmark & & & \checkmark \\ \hline
\cite{ArmanJ2} &&point&&& \checkmark && $\bigotimes$ & \eqref{Eq:Deg} & $\mathcal{\tilde{S}}_{\rx}$ & & & 2nd-order & && 3D & \checkmark & & \cite[Eq.~(29)]{ArmanJ2} & \\ \hline
\cite{Chou_Rx} &trans.&$\mathcal{V}_{\tx}$&&& \checkmark && $\bigotimes$ & & $\mathcal{V}_{\rx}$ & & \checkmark &&& \checkmark & 3D & \checkmark & & & \checkmark \\ \hline 
\cite{Noel_CT} &trans.& $\mathcal{V}_{\tx}  / \mathcal{S}_{\tx}$ &&& \checkmark && $\bigotimes$ & & $\mathcal{V}_{\rx} / \mathcal{S}_{\rx}$ & & \checkmark & & 1st-order && 1D / 3D & \checkmark &  & \checkmark \, (1D) & \eqref{Eq;CIR_VolTX}\, 3D \\ \hline
\cite{Yilmaz2017} &refl.& $ \mathcal{S}_{\tx}$ &&& \checkmark && $\bigotimes$ & & $\mathcal{S}_{\rx}$ & & & & 1st-order && 3D & \checkmark &  &  & \checkmark \\ \hline
\cite{Arjmandi2016} &refl.& $ \mathcal{V}_{\tx}$ & & \checkmark & \checkmark && $\bigotimes$ & & $\mathcal{V}_{\rx}$ & & \checkmark & &  && 3D & \checkmark &  &  & \checkmark \\ \hline 
\cite{Chou2015,Awan2017a,Awan2017b} &trans.& $ \mathcal{V}_{\tx}$ & \checkmark & & \checkmark && $\bigotimes$ & & $\mathcal{V}_{\rx}$ & & \checkmark & &  & \checkmark & 3D & \checkmark &  &  & \checkmark \\ \hline
This article &&point&&& \checkmark & uni. & $\bigcirc$ & & $\mathcal{V}_{\rx}$ & & \checkmark & & && 3D & & & \eqref{Eq;CIR_Bounded_Duct}, \eqref{Eq;CIR_Bounded_Vessel} & \\ \hline
\cite{NoelPro3} &&point&&& \checkmark & uni. & $\bigotimes$ & & $\mathcal{V}_{\rx}$ & \checkmark & \checkmark & & && 3D & & & \eqref{Eq;CIR_UniDRF_Unbouonded_PassiveRX} & \\ \hline 
\cite{WayanPro2} &&point&&& \checkmark & uni. & $\bigcirc$ & \eqref{Eq:Deg} & $\mathcal{V}_{\rx}$ & \checkmark & & & && 3D & \checkmark & & \cite[Eq.~(15)]{WayanPro2} & \\ \hline
\cite{WayanPro1} & trans. &point / $\mathcal{S}_{\tx}$&&& \checkmark & lam. & $\bigcirc$ &  & $\mathcal{V}_{\rx}$ & \checkmark & & & && 3D & & & \eqref{Eq;CIR_Laminar_DisReg_UCA}, \eqref{Eq;CIR_Laminar_DisReg_woUCA}, \eqref{Eq;CIR_Laminar_FlowReg_UnifromRel}, \eqref{Eq;CIR_Laminar_FlowReg_PoifromRel} & \\ \hline 
\cite{PierobonJ3} & trans. & $\mathcal{V}_{\tx}$&&& \checkmark & & $\bigotimes$ &  & $\mathcal{V}_{\rx}$&& \checkmark & 2nd-order & && 3D &  & \checkmark  & \checkmark & \\ \hline  
 \cite{ShahMohammadianProc1} &&point&&& \checkmark & & $\bigotimes$ &  & $\mathcal{V}_{\rx}$ & & \checkmark & & 2nd-order & & 3D & & \checkmark & \cite[Eq.~(8)]{ShahMohammadianProc1} & \\ \hline 
 \cite{Adam_Enzyme} &&point&&& \checkmark &  & $\bigotimes$ & \eqref{Eq:EnzamiticPDE} & $\mathcal{V}_{\rx}$ & & \checkmark & & && 3D & \checkmark & & \eqref{Eq;CIR_Passive_Enzymatic_app1}, \eqref{Eq;CIR_Passive_Enzymatic_app3} & \\ \hline 
\cite{Eckford2012} &&point&&& \checkmark & uni.  & $\bigotimes$ & & & & \checkmark & & 1st-order && 1D & \checkmark & & \cite[Eq.~(12)]{Eckford2012} & \\ \hline               
\end{tabular}}
\end{scriptsize}
\end{sidewaystable}
}{%

\begin{table*}[!h]
\renewcommand{\arraystretch}{2.4}
\caption{Summary of CIR Models Reviewed in Section~\ref{Sect:TxRxCh}}
\label{Table;ReviewCIR}
\centering
\rotatebox{90}{%
\scalebox{0.9}{ 
\begin{tabular}{|c|c|c|c|c|c|c|c|c|c|c|c|c|c|c|c|c|c|c|} 
\hline 
  \multirow{3}{*}{$h(t)$} & \multirow{3}{*}{Ref} & \multicolumn{4}{|c|}{Transmitter} & \multicolumn{4}{|c|}{Physical Channel} & \multicolumn{6}{|c|}{Receiver} & \multirow{3}{*}{Dim} & \multicolumn{2}{|c|}{\eqref{Eq:Fick_general}}  \\ \cline{3-16} \cline{18-19}  
 && \multirow{2}{*}{Surface} & \multirow{2}{*}{$\Omega_{\tx}$} & \multirow{2}{*}{PG} & \multirow{2}{*}{REM} & \multirow{2}{*}{Diff} & \multirow{2}{*}{Adv} & \multirow{2}{*}{Geo} & \multirow{2}{*}{D\&P} & \multirow{2}{*}{$\Omega_{\rx}$} & \multicolumn{2}{|c|}{Passive} & \multicolumn{2}{|c|}{Active} & \multirow{2}{*}{SP} & & \multirow{2}{*}{Joi} & \multirow{2}{*}{Sep} \\ \cline{12-15} 
  &&&&&&&&&&& w/o UCA & w/ UCA & Sto & Det &&&& \\ \hline 
 \eqref{Eq.Passive_mean2}, \eqref{Eq;CIR_Passive_woUCA}  & \cite{NoelPro1}&&point&&&\checkmark && $\bigotimes$ & & $\mathcal{V}_{\rx}$ & \checkmark & \checkmark &&&&3D&& \\ \hline
\eqref{Eq.CIR_FullAbsorbing} & \cite{YilmazL1} &&point&&& \checkmark && $\bigotimes$ & & $\mathcal{S}_{\rx}$ & & & & 1st-order && 3D & \checkmark &  \\ \hline
\cite[Eq.~(13)]{Akkaya} & \cite{Akkaya} &&point&&& \checkmark && $\bigotimes$ & & $\mathcal{ \tilde{S}}_{\rx}$ & & & & 1st-order && 3D & \checkmark & \\ \hline 
\cite[Eq.~(12)]{Heren} & \cite{Heren} &&point&&& \checkmark && $\bigotimes$ & \eqref{Eq:Deg} & $\mathcal{S}_{\rx}$ & & & & 1st-order && 3D & \checkmark & \\ \hline
Num & \cite{Deng2015} &&point&&& \checkmark && $\bigotimes$ & & $\mathcal{S}_{\rx}$ & & & 2nd-order & && 3D & \checkmark & \\ \hline
 \cite[Eq.~(29)]{ArmanJ2} &\cite{ArmanJ2} &&point&&& \checkmark && $\bigotimes$ & \eqref{Eq:Deg} & $\mathcal{\tilde{S}}_{\rx}$ & & & 2nd-order & && 3D & \checkmark &  \\ \hline
Num & \cite{Chou_Rx} &trans&$\mathcal{V}_{\tx}$&&& \checkmark && $\bigotimes$ & & $\mathcal{V}_{\rx}$ & & \checkmark &&& \checkmark & 3D & \checkmark & \\ \hline 
\eqref{Eq;CIR_VolTX} & \cite{Noel_CT} &trans/refl& $\mathcal{V}_{\tx}  / \mathcal{S}_{\tx}$ &&& \checkmark && $\bigotimes$ & & $\mathcal{V}_{\rx} / \mathcal{S}_{\rx}$ & & \checkmark & & 1st-order && 1D / 3D & \checkmark & \\ \hline
Num & \cite{Yilmaz2017} &refl& $ \mathcal{S}_{\tx}$ &&& \checkmark && $\bigotimes$ & & $\mathcal{S}_{\rx}$ & & & & 1st-order && 3D & \checkmark &  \\ \hline
Num & \cite{Arjmandi2016} &refl& $ \mathcal{V}_{\tx}$ & & \checkmark & \checkmark && $\bigotimes$ & & $\mathcal{V}_{\rx}$ & & \checkmark & &  && 3D & \checkmark &  \\ \hline 
Num & \cite{Chou2015,Awan2017a,Awan2017b} &trans& $ \mathcal{V}_{\tx}$ & \checkmark & & \checkmark && $\bigotimes$ & & $\mathcal{V}_{\rx}$ & & \checkmark & &  & \checkmark & 3D & \checkmark &  \\ \hline
\eqref{Eq;CIR_Bounded_Duct}, \eqref{Eq;CIR_Bounded_Vessel} & This article &&point&&& \checkmark & uni & $\bigcirc$ & & $\mathcal{V}_{\rx}$ & & \checkmark & & && 3D & &  \\ \hline
\eqref{Eq;CIR_UniDRF_Unbouonded_PassiveRX} & \cite{NoelPro3} &&point&&& \checkmark & uni & $\bigotimes$ & & $\mathcal{V}_{\rx}$ & \checkmark & \checkmark & & && 3D & &  \\ \hline 
\cite[Eq.~(15)]{WayanPro2} & \cite{WayanPro2} &&point&&& \checkmark & uni & $\bigcirc$ & \eqref{Eq:Deg} & $\mathcal{V}_{\rx}$ & \checkmark & & & && 3D & \checkmark &  \\ \hline
\eqref{Eq;CIR_Laminar_DisReg_UCA}, \eqref{Eq;CIR_Laminar_DisReg_woUCA}, \eqref{Eq;CIR_Laminar_FlowReg_UnifromRel}, \eqref{Eq;CIR_Laminar_FlowReg_PoifromRel} &\cite{WayanPro1} & trans &point / $\mathcal{S}_{\tx}$&&& \checkmark & lam & $\bigcirc$ &  & $\mathcal{V}_{\rx}$ & \checkmark & & & && 3D & &  \\ \hline 
 & \cite{PierobonJ3} & trans & $\mathcal{V}_{\tx}$&&& \checkmark & & $\bigotimes$ &  & $\mathcal{V}_{\rx}$&& \checkmark & 2nd-order & && 3D &  & \checkmark  \\ \hline  
\cite[Eq.~(8)]{ShahMohammadianProc1} & \cite{ShahMohammadianProc1} &&point&&& \checkmark & & $\bigotimes$ &  & $\mathcal{V}_{\rx}$ & & \checkmark & & 2nd-order & & 3D & & \checkmark  \\ \hline 
\eqref{Eq;CIR_Passive_Enzymatic_app1}, \eqref{Eq;CIR_Passive_Enzymatic_app3} & \cite{Adam_Enzyme} &&point&&& \checkmark &  & $\bigotimes$ & \eqref{Eq:EnzamiticPDE} & $\mathcal{V}_{\rx}$ & & \checkmark & & && 3D & \checkmark &  \\ \hline 
\cite[Eq.~(12)]{Eckford2012} &\cite{Eckford2012} &&point&&& \checkmark & uni & $\bigotimes$ & & & & & & 1st-order && 1D & \checkmark &  \\ \hline               
\end{tabular}
}
}
\end{table*}
}

\section{Received Signal Modeling}
\label{Sect:RecSig}
In this section, we provide mathematical models for the signals used for estimation of the system parameters and detection of the transmitted data by \gls{MC} receivers. To this end, we first present a unified signal representation for \gls{MC} systems. Next, we introduce three time scales for the  signal observed at the receiver, and subsequently, we provide signal models for each of these time scales. In addition, we generalize these models to account for the interfering noise molecules in the environment. Subsequently, time-slotted communication is considered and a corresponding signal model is developed which accounts for the impact of \gls{ISI}. Finally, the correlation of the signals received at different time instants is discussed for the considered time scales. 

The models that we present in this section are general in the sense that they apply to all \gls{MC} systems discussed in Section~\ref{Sect:TxRxCh}. More specifically, these models only depend on the \gls{CIR} $h(t)$  within the considered observation window or at the considered sampling times. We note that for most \gls{MC} environments, derivation of the \gls{CIR} in closed form, as was done for specific cases in Section~\ref{Sect:TxRxCh}, is challenging. In Section~\ref{Sect:SimExp}, we present numerical and simulation methods to obtain the \gls{CIR} of more complex \gls{MC} systems. In addition, in practical \gls{MC} systems, the transmitter may send known pilot symbols that enable the receiver to estimate the \gls{CIR} from its observations (e.g., see \cite{TCOM_MC_CSI} and \cite{Adam_Channel} for  channel parameter estimators for \gls{MC} systems). The models developed in this section are applicable for analytically derived, simulated, and estimated \glspl{CIR}.

\subsection{Unified Signal Definition}\label{Sec:UnifSig}
In the \gls{MC} literature, different physical quantities have been modeled as the received signal. Important examples include \textit{i)} the number of molecules observed at a given time within the volume of a transparent receiver \cite{Akyl_Receiver_MC,ConsCIR,NoelPro1,Akyildiz_MC_E2E,Equ_MC,PhY_MC,NoelPro3,TCOM_MC_CSI,DistanceEstLett}, \textit{ii)} the number of molecules bound at a given time to the receptors of a reactive receiver \cite{ArmanJ2,Deng2015,Akkaya}, \textit{iii)} the accumulated number of molecules observed by a fully-absorbing receiver within a given observation time window \cite{Yilmaz_Poiss,YilmazL1,Yilmaz2014a,damrath2017equivalent,cao2018diffusive}, and \textit{iv)} the arrival times of the molecules at a fully-absorbing receiver \cite{InvGaussian,Yonathan_Timimg,Nariman_Timing,Rose_T1,Rose_T2,li2014capacity}. In the following, we first provide a unified definition of the received signal of general \gls{MC} receivers including the aforementioned special cases. Since the presented general signal model is difficult to analyze, subsequently,  we introduce the concept of counting receivers, which are widely considered in the literature and allow for simple mathematical modeling. The main purpose for introducing a general representation of the received signal is to highlight the basic assumptions that have been made to arrive at specific signal models used in the literature and to unveil the connections between different signal models. 

\subsubsection{Generalized Receivers} Since different molecules of the same type are indistinguishable for the receiver, the most detailed information that the receiver could access at a given time $t$ is the arrival (and departure, if relevant) times of the molecules at (or from) the receiver up to that time. We use this fact to introduce a unified representation of the received signal of general \gls{MC} receivers. For mathematical rigor, let us first formally distinguish between two types of receivers, namely recurrent and non-recurrent receivers.

\begin{defin} 
If a given molecule can be observed by the receiver at most once, then the receiver is referred to as \textit{non-recurrent}; otherwise, it is referred to as \textit{recurrent}.
\end{defin}

Transparent and reactive receivers with unbinding rate $\kappa_b\neq 0$ are recurrent since a given molecule can be observed multiple times by the receiver. On the other hand, fully-absorbing receivers and reactive receivers with  $\kappa_b= 0$ are non-recurrent since after a molecule has been observed at the receiver, it cannot be observed again. For non-recurrent receivers, the time instants at which the molecules are observed constitute the most general signal representation. Let us define  
\begin{IEEEeqnarray}{lll}\label{Eq:Tin}
	\vec{\mathbf{T}}^{\mathrm{arv}}(t) = \left[t_1,t_2,\dots,t_{n^{\mathrm{arv}}(t)}\right],
\end{IEEEeqnarray}
as the vector containing the arrival times $t_n,\,\,n=1,2,\dots,n^{\mathrm{arv}}(t)$, of all $n^{\mathrm{arv}}(t)$ molecules \textit{observed} by time $t$ in an ascending order. We note that both the number of molecules observed by time $t$, i.e., $n^{\mathrm{arv}}(t)$, and their arrival times $t_n, \,\,n=1,\dots,n^{\mathrm{arv}}(t)$, are \glspl{RV}. On the other hand, for recurrent receivers, in addition to $\vec{\mathbf{T}}^{\mathrm{arv}}(t)$, we need to keep track of the molecules that have been \textit{un-observed}, i.e., have left the receiver. To this end, let us define 
\begin{IEEEeqnarray}{lll}\label{Eq:Tout}
\vec{\mathbf{T}}^{\mathrm{dpr}}(t) = \left[t'_1,t'_2,\dots,t'_{n^{\mathrm{dpr}}(t)}\right],
\end{IEEEeqnarray}
as the vector containing the departure times $t'_n,\,\,n=1,2,\dots,n^{\mathrm{dpr}}(t)$, of all $n^{\mathrm{dpr}}(t)$ molecules that have left the receiver by time $t$ in an ascending order. We note that by the above formulation, non-recurrent receivers can be seen as a special case of recurrent receivers where $n^{\mathrm{dpr}}(t)=0,\,\,\forall t$. In summary, $\vec{\mathbf{T}}^{\mathrm{arv}}(t)$ and $\vec{\mathbf{T}}^{\mathrm{dpr}}(t)$ constitute a complete and unified representation of the received signal of \gls{MC} receivers. As will be shown in the following, different notions of received signal used in the \gls{MC} literature can be seen as special cases of $\vec{\mathbf{T}}^{\mathrm{arv}}(t)$ and $\vec{\mathbf{T}}^{\mathrm{dpr}}(t)$.


\subsubsection{Timing-based Receivers}
In the \gls{MC} literature, timing channels have been used as a model for non-recurrent receivers \cite{InvGaussian,Rose_T1,Rose_T2,Yonathan_Timimg,Nariman_Timing}. Let $\mathbf{T}^{\mathrm{rls}}$ denote the vector containing the release times of the molecules by the transmitter and let  $\mathbf{T}^{\mathrm{arv}}$ be the vector containing the respective arrival times of the molecules at the receiver. Thus, $\mathbf{T}^{\mathrm{arv}}$ is related to $\mathbf{T}^{\mathrm{rls}}$ according to \cite{InvGaussian,Yonathan_Timimg,Nariman_Timing,Rose_T1,Rose_T2,li2014capacity}
\begin{IEEEeqnarray}{lll}\label{Eq:Tmol}
	\mathbf{T}^{\mathrm{arv}} = \mathbf{T}^{\mathrm{rls}} + \mathbf{T}^{\mathrm{dly}},
\end{IEEEeqnarray}
where $\mathbf{T}^{\mathrm{dly}}$ is a vector containing the random delays between the release of the molecules by the transmitter and their observation at the receiver. Moreover, it is typically assumed that the release, propagation, and reception of molecules are independent from each other, which we refer to as the \textit{independent molecule behavior} assumption \cite{Rose_T1,Rose_T2}. Based on this assumption, the elements in $\mathbf{T}^{\mathrm{dly}}$ are independent and identically distributed and assume only non-negative real values. For an unbounded 1D environment, the random observation delay follows a Levy distribution if no flow is present \cite{Nariman_Timing} and the inverse Gaussian distribution if flow in the direction of the receiver is present \cite{InvGaussian}. 

We note that, in practice, $\mathbf{T}^{\mathrm{arv}}$ is not available at the receiver since \textit{i)} different molecules of the same type are indistinguishable by the receiver and \textit{ii)} out of the total number of released molecules, only $n^{\mathrm{arv}}(t)$ molecules are observed by time $t$. In fact, $\vec{\mathbf{T}}^{\mathrm{arv}}(t)$ is the actual observation signal available to the receiver. To arrive at a model for $\vec{\mathbf{T}}^{\mathrm{arv}}(t)$, we introduce the following definitions and assumptions. Let us assume that $\Ntx$ molecules are released by the transmitter within interval $[0,t]$ and their release times are collected in $\mathbf{T}^{\mathrm{rls}}$. Since the $n^{\mathrm{arv}}(t)$ molecules observed at the receiver are indistinguishable, we do not know which $n^{\mathrm{arv}}(t)$ molecules out of the total $\Ntx$ released molecules have been observed. In general, there are at most $\frac{\Ntx!}{(n^{\mathrm{arv}}(t)-1)!}$ possibilities for selecting $n^{\mathrm{arv}}(t)$ (observed) molecules from the $\Ntx$ (released) molecules. Therefore, we define $\mathbf{p}_p,\,\,p=1,\dots,\frac{\Ntx!}{(n^{\mathrm{arv}}(t)-1)!}$, as a vector which contains the $p$-th possible order index of the observed molecules. Moreover, let $f_{X}(x)$ and $F_X(x)$ denote the \gls{PDF} and \gls{CDF} of \gls{RV} $X$ at $X=x$, respectively.  We note that due to causality, $f_{T^{\mathrm{dly}}}(t)=0,\,\,t<0$, has to hold where \gls{RV} $T^{\mathrm{dly}}$ denotes the random delay of a given molecule. Following a similar framework as developed in \cite{Rose_T1,Rose_T2}, the \gls{PDF} of the observation vector $\vec{\mathbf{T}}^{\mathrm{arv}}(t)=\vec{\mathbf{t}}^{\mathrm{arv}}$ conditioned on the molecule release time vector $\mathbf{T}^{\mathrm{rls}}$, denoted by $f_{\vec{\mathbf{T}}^{\mathrm{arv}}(t)|\mathbf{T}^{\mathrm{rls}}}\left(\vec{\mathbf{t}}^{\mathrm{arv}}|\mathbf{T}^{\mathrm{rls}}\right)$, is obtained as
\iftoggle{OneColumn}{%
\begin{IEEEeqnarray}{lll}\label{Eq:TimeModel}
	f_{\vec{\mathbf{T}}^{\mathrm{arv}}(t)|\mathbf{T}^{\mathrm{rls}}}\left(\vec{\mathbf{t}}^{\mathrm{arv}}|\mathbf{T}^{\mathrm{rls}}\right) = \sum_{p=1}^{\frac{\Ntx!}{(n^{\mathrm{arv}}(t)-1)!}} f_{\vec{\mathbf{T}}^{\mathrm{arv}}(t)|\mathbf{T}^{\mathrm{rls}}}\left(\vec{\mathbf{t}}^{\mathrm{arv}}|\mathbf{T}^{\mathrm{rls}},\mathbf{p}_p\right) \nonumber \\
	=  \sum_{p=1}^{\frac{\Ntx!}{(n^{\mathrm{arv}}(t)-1)!}} \left[\prod_{m=1}^{n^{\mathrm{arv}}(t)}f_{T^{\mathrm{dly}}}\left(t_m-\mathbf{T}^{\mathrm{rls}}[{\mathbf{p}_p}[m]]\right) 
	\prod_{m=n^{\mathrm{arv}}(t)+1}^{\Ntx}\Big[1-F_{T^{\mathrm{dly}}}\left(t-\mathbf{T}^{\mathrm{rls}}[{\mathbf{p}_p}[m]] \right)\Big] \right]  \nonumber \\
	\overset{(a)}{=} \frac{\Ntx!}{(n^{\mathrm{arv}}(t)-1)!} \big(f_{T^{\mathrm{dly}}}(t_m)\big)^{n^{\mathrm{arv}}(t)} \big(1-F_{T^{\mathrm{dly}}}(t_m)\big)^{\Ntx-n^{\mathrm{arv}}(t)},
\end{IEEEeqnarray}
}{%
\begin{IEEEeqnarray}{lll}\label{Eq:TimeModel}
	f_{\vec{\mathbf{T}}^{\mathrm{arv}}(t)|\mathbf{T}^{\mathrm{rls}}}\left(\vec{\mathbf{t}}^{\mathrm{arv}}|\mathbf{T}^{\mathrm{rls}}\right) \nonumber \\
	= \sum_{p=1}^{\frac{\Ntx!}{(n^{\mathrm{arv}}(t)-1)!}} f_{\vec{\mathbf{T}}^{\mathrm{arv}}(t)|\mathbf{T}^{\mathrm{rls}}}\left(\vec{\mathbf{t}}^{\mathrm{arv}}|\mathbf{T}^{\mathrm{rls}},\mathbf{p}_p\right) \nonumber \\
	=  \sum_{p=1}^{\frac{\Ntx!}{(n^{\mathrm{arv}}(t)-1)!}} \Bigg[\prod_{m=1}^{n^{\mathrm{arv}}(t)}f_{T^{\mathrm{dly}}}\left(t_m-\mathbf{T}^{\mathrm{rls}}[{\mathbf{p}_p}[m]]\right) \nonumber \\
	\qquad\qquad\quad\times \prod_{m=n^{\mathrm{arv}}(t)+1}^{\Ntx}\Big[1-F_{T^{\mathrm{dly}}}\left(t-\mathbf{T}^{\mathrm{rls}}[{\mathbf{p}_p}[m]] \right)\Big] \Bigg]  \nonumber \\
	\overset{(a)}{=} \frac{\Ntx!}{(n^{\mathrm{arv}}(t)-1)!} \big(f_{T^{\mathrm{dly}}}(t_m)\big)^{n^{\mathrm{arv}}(t)} \nonumber\\ \qquad \qquad \qquad \qquad \times \big(1-F_{T^{\mathrm{dly}}}(t_m)\big)^{\Ntx-n^{\mathrm{arv}}(t)},
\end{IEEEeqnarray}
}
where equality $(a)$ holds when all $\Ntx$ molecules are released at time zero. The above formulation provides a general framework for modeling the arrival times of non-recurrent receivers. Unfortunately, (\ref{Eq:TimeModel}) cannot be easily simplified and its generalization to recurrent receivers or the cases when interfering noise molecules or \gls{ISI} are present is cumbersome. In fact, the results reported in \cite{InvGaussian,Yonathan_Timimg,Nariman_Timing,Rose_T1,Rose_T2,li2014capacity} are valid for non-recurrent receivers when \gls{ISI} and interfering noise molecules do not exist. In addition, perfect synchronization is a key underlying assumption for most timing channels considered in the literature \cite{InvGaussian,Yonathan_Timimg,Nariman_Timing,Rose_T1,Rose_T2,li2014capacity} and hence the performance of timing receivers is very sensitive to synchronization errors. Therefore, in the remainder of this section, we consider special receivers, namely \textit{molecule counting} receivers, whose signal is a function of $n^{\mathrm{arv}}(t)$ and $n^{\mathrm{dpr}}(t)$ only.  Molecule counting receivers are widely adopted in the literature and the corresponding received signal lends itself to more tractable models and analysis.

\subsubsection{Counting Receivers} These receivers consider the number of observed molecules as the received signal. In general, the receiver may count the number of observed molecules multiple times, which is referred to as a multi-sample detector \cite{Equ_MC,ConsCIR,Akyl_Receiver_MC,Adam_OptReciever,TNBC_Sync,Yilmaz2017}. Let $r(t_m)$ denote the received signal at sample time $t_m=m\Delta t,\,\,m=1,2,\dots$, where $\Delta t$ is the sample interval. To formally characterize $r(t_m)$, we distinguish two types of counting receivers, namely arrival-counting and observation-counting receivers.


\begin{defin} 
If a receiver counts the number of molecules that have \textit{arrived within the observation window $(t_m-\Delta t,t_m]$} at its reception site, i.e., $r(t_m)=n^{\mathrm{arv}}(t_m)-n^{\mathrm{arv}}(t_m-\Delta t)$, then it is referred to as an \textit{accumulative-molecule-counting} (AMC) receiver, whereas if the receiver counts the number of molecules that are \textit{observed at a given time $t$} at its reception site, i.e., $r(t_m)=n^{\mathrm{arv}}(t_m)-n^{\mathrm{dpr}}(t_m)$, then it is referred to as an \textit{instantaneous-molecule-counting} (IMC) receiver.
\end{defin}

In general, there are four types of receivers based on the recurrent/non-recurrent and AMC/IMC  classifications. In the following, we present the different counting receivers used in the \gls{MC} literature as special cases of these four categories:

\textbf{Non-Recurrent Accumulative-Molecule-Counting (nR-AMC) Receivers:} The signal in this case is  $r(t_m)=n^{\mathrm{arv}}(t_m)-n^{\mathrm{arv}}(t_m-\Delta t)$ where $n^{\mathrm{arv}}(t_m)\geq n^{\mathrm{arv}}(t_m-\Delta t) \geq 0$. For instance, for fully-absorbing receivers, $r(t_m)$ denotes the number of molecules that have arrived within interval $(t_m-\Delta t,t_m]$ \cite{Yilmaz_Poiss,YilmazL1,Yilmaz2014a,damrath2017equivalent,cao2018diffusive}. 


\textbf{Recurrent Accumulative-Molecule-Counting (R-AMC) Receivers:} The signal in this case is  $r(t_m)=n^{\mathrm{arv}}(t_m)-n^{\mathrm{arv}}(t_m-\Delta t)$ where $n^{\mathrm{arv}}(t_m)\geq n^{\mathrm{arv}}(t_m-\Delta t) \geq 0$. Although the mathematical form looks identical to that for nR-AMC receivers, the modeling for R-AMC receivers is much more cumbersome since one molecule might be counted multiple times within the observation window $(t_m-\Delta t,t_m]$.  Furthermore, we note that the expected number of observed molecules for R-AMC receivers is larger than that for nR-AMC receivers because some molecules may be counted multiple times.


\textbf{Recurrent Instantaneous-Molecule-Counting (R-IMC) Receivers:} The signal in this case is  $r(t_m)=n^{\mathrm{arv}}(t_m)-n^{\mathrm{dpr}}(t_m)$ where $n^{\mathrm{arv}}(t_m)\geq n^{\mathrm{dpr}}(t_m)\geq 0$. For instance, for transparent receivers, $r(t_m)$ denotes the number of molecules within the receiver volume at time $t_m$ \cite{Akyl_Receiver_MC,ConsCIR,NoelPro1,Akyildiz_MC_E2E,Equ_MC,PhY_MC,NoelPro3,TCOM_MC_CSI,DistanceEstLett}, and for reactive receivers, $r(t_m)$ is the number of molecules bound to the receiver's receptors at time $t_m$ \cite{ArmanJ2,Deng2015,Akkaya}. 


\textbf{Non-Recurrent Instantaneous-Molecule-Counting (nR-IMC) Receivers:} The signal in this case is  $r(t_m)=n^{\mathrm{arv}}(t_m)-n^{\mathrm{dpr}}(t_m)=n^{\mathrm{arv}}(t_m)$ where $n^{\mathrm{arv}}(t_m)\geq 0$ and $n^{\mathrm{dpr}}(t_m)=0$. We note that since the received molecules do not leave the receiver, $r(t_m)$ is a non-decreasing function~of~time.



In the remainder of this section, we focus on the modeling of $r(t_m)$ for R-IMC receivers as a function of \gls{CIR} $h(t)$, i.e., the probability of a molecule being observed at time $t$ seconds after its release by the transmitter; see Section~\ref{Sect:TxRxCh}. This model is also valid for nR-AMC and nR-IMC receivers if $h(t_m)$ is substituted by the probability of observing a molecule within intervals $(t_m-\Delta t,t_m]$ and $(0,t_m]$, respectively, after its release by the transmitter at time $t=0$, cf. (\ref{Eq.CIR_FullAbsorbing_window}).  Modeling of $r(t_m)$ for R-AMC receivers is cumbersome due to the possibility of counting a given molecule multiple times within the observation window. This type of signal is relevant,  e.g., for ligand-based receivers when a ligand molecule can activate the receptors on the receiver surface multiple times.  However, this problem has not yet been studied in the \gls{MC} literature and is a potential topic for future research. Finally, in the following, we assume that the sampling interval $\Delta t$ is sufficiently large such that consecutive samples are statistically independent. Therefore, we drop index $m$ in Sections~IV-B, IV-C, and IV-D for simplicity. How large $\Delta t$ should be chosen to guarantee sample independence will be discussed in Section~IV-F.

\subsection{Three Time-Scale Signal Representation}\label{Sec:ThreeScale}

 Let us define $r(t,\tau)$ as the number of molecules observed at the receiver $t$ seconds after its release is stimulated by the transmitter at time $\tau$. Then, $r(t,\tau)$ can be modeled as
\begin{IEEEeqnarray}{lll}\label{Eq:Sig_3Scale} 
	r(t,\tau) = \bar{r}(t,\tau)+w(t,\tau),
\end{IEEEeqnarray}
where $\bar{r}(t,\tau)=\Ex{r(t,\tau)}$ denotes the mean of the signal for a fixed set of channel parameters, $w(t,\tau)$ denotes the random fluctuations around the mean (e.g., caused by diffusion), and $\Ex{\cdot}$ denotes expectation. We note that the channel parameters may also change over time; however, this is over a scale that is much slower than the signal variations. In other words, the mean of the signal, $\bar{r}(t,\tau)$, varies over time $t$ due to diffusion, advection, and reactions in the channel, but it also varies over the larger time scale $\tau$ due to variations of system parameters such as temperature, viscosity, and the distance between a mobile transmitter and receiver \cite{Arman_TimeVariant,Qiu2017,TCOM_NonCoherent}. In summary, we have variations on three time scales in $r(t,\tau)$: \textit{Time Scale 1)} variations of $r(t,\tau)$ around its mean $\bar{r}(t,\tau)$, i.e., noisy fluctuations  $w(t,\tau)$, \textit{Time Scale 2)} variations of the signal mean $\bar{r}(t,\tau)$ over observation time $t$, which are slower than the variations of $w(t,\tau)$, and \textit{Time Scale 3)} variations of $\bar{r}(t,\tau)$ over the release time $\tau$, which are slower than the variations of $\bar{r}(t,\tau)$ with respect to $t$. For instance, for typical \gls{MC} systems at microscale, e.g., cell-to-cell communication, the noisy fluctuations vary on the order of a few $\mu$s, the variations of the signal mean over time $t$ are on the order of tens or hundreds of $\mu$s, and the change in the parameters, e.g., due to the mobility of the nodes, can be on time scales much larger than ms \cite{ArmanJ2}. Fig.~\ref{Fig:CIR_3Tier} illustrates $r(t,\tau)$ versus $t$ for three values of $\tau$. The aforementioned three time scales are illustrated in this figure: \textit{1)} the actual received signals, $r(t,\tau)$, are denoted by colored solid lines;  \textit{2)} the black dashed lines denote the signal mean $\bar{r}(t,\tau)$; and \textit{3)} the variations of the signal due to changes in the system parameters over time scale $\tau$ are represented by different colors. 

\iftoggle{OneColumn}{%
\begin{figure}
	\centering  
	\resizebox{0.6\linewidth}{!}{
		\psfragfig{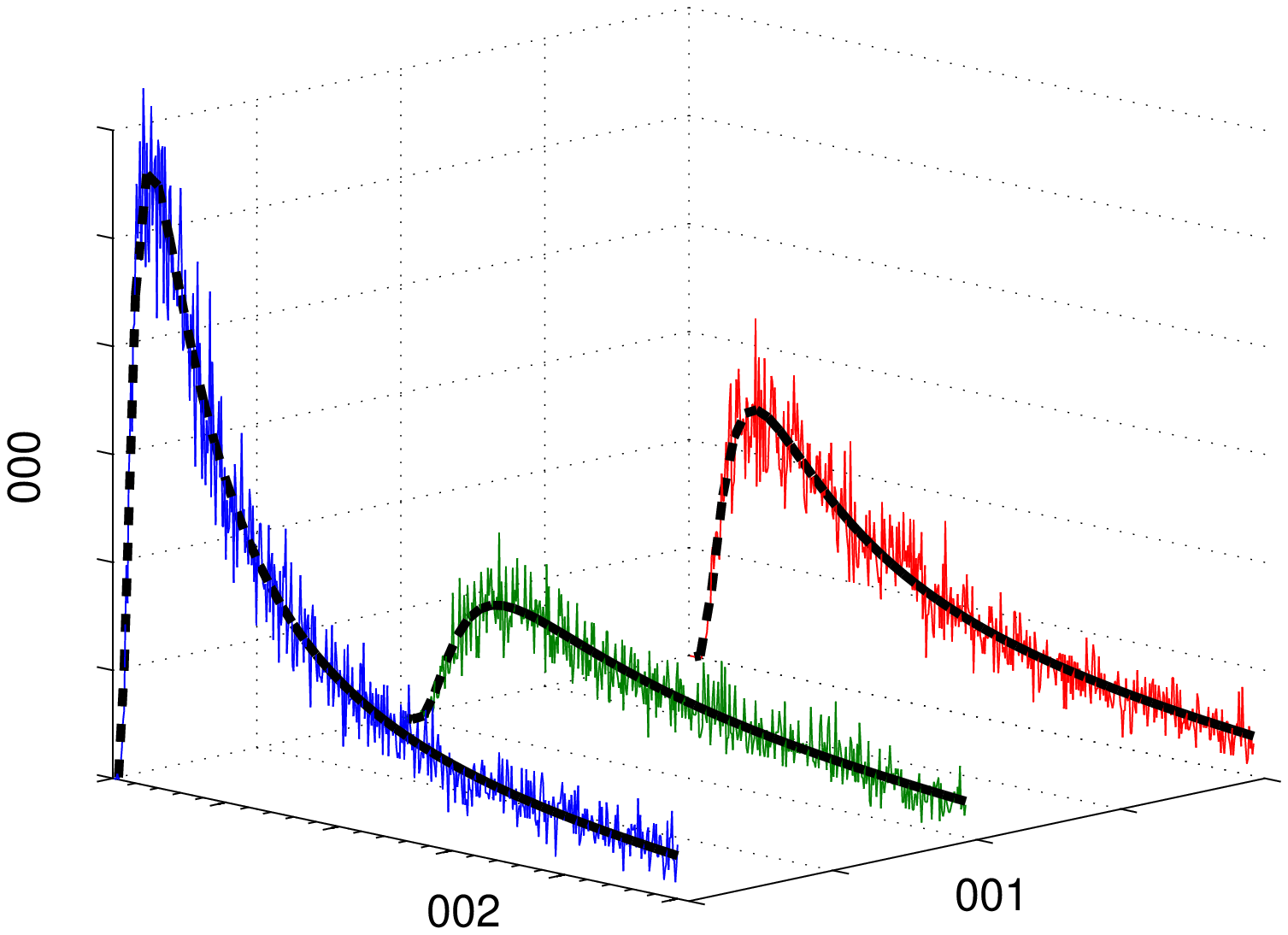}} \vspace{-0.5cm}
	\caption{Number of molecules observed at the receiver $t$ seconds after the release by the transmitter at time $\tau$. The three time scales are illustrated as follows: \textit{1)} the actual received signals, $r(t,\tau)$, are denoted by colored solid lines;  \textit{2)} the black dashed lines denote the signal mean $\bar{r}(t,\tau)$; and \textit{3)} the variations of the signal due to changes in the system parameters over time scale $\tau$ are represented by different colors. }
	\label{Fig:CIR_3Tier}
\end{figure}
}{%
\begin{figure}
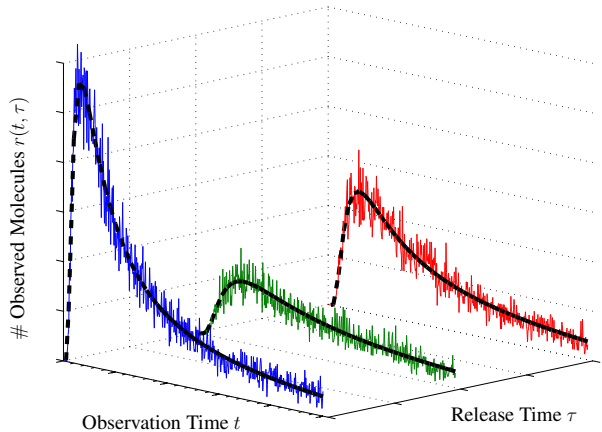

	\centering  
	\resizebox{1\linewidth}{!}{
		\psfragfig{Sections/S4/Fig/CIR/CIR_3Tier}} \vspace{-0.5cm}
	\caption{Number of molecules observed at the receiver $t$ seconds after the release by the transmitter at time $\tau$. The three time scales are illustrated as follows: \textit{1)} the actual received signals, $r(t,\tau)$, are denoted by colored solid lines;  \textit{2)} the black dashed lines denote the signal mean $\bar{r}(t,\tau)$; and \textit{3)} the variations of the signal due to changes in the system parameters over time scale $\tau$ are represented by different colors. }
	\label{Fig:CIR_3Tier}
\end{figure}
}

\begin{remk}
The three time-scale signal representation for \gls{MC} is analogous to a similar signal representation in wireless communications. In particular, in a wideband wireless communication system impaired by \gls{AWGN} and fading, the \gls{AWGN} is analogous to the random fluctuations of the signal in \gls{MC}, the \gls{CIR} of the wireless communication channel is analogous to the signal mean in \gls{MC}, and the variations of the \gls{CIR} over time (due to the movement of the transmitter and/or receiver) are analogous to the time-variant signal mean in \gls{MC} \cite{rappaport2002wireless}. \QEDwhite
\end{remk}

\subsection{Signal Models}
In the following, we first derive the expected number of molecules observed at the receiver, which we refer to as the deterministic model of the received signal. Subsequently, we derive statistical models of the received signal that capture the random fluctuations of the observed molecules. Finally, we study time-variant channels and derive stochastic models to capture the effect of the time variance.

\subsubsection{Deterministic Models}

In Section~\ref{Sect:TxRxCh}, we derived the \gls{CIR} $h(t)$ which can be interpreted as the probability of a  molecule released at time $t=0$ being observed at the receiver at time $t$. Let us define $h(t,\tau)$ as the probability of a molecule released by the transmitter at time $\tau$ being observed at the receiver at time $t$. In the following, we first assume a time-invariant \gls{MC} channel which leads to  $h(t,\tau)=h(t-\tau),\,\,t\geq \tau$. Then, in Section~\ref{Sec:TimeVarying}, we analyze the impact of  time variance of the channel. Following the independent molecule behavior assumption \cite{Adam_Thesis,Yilmaz_Poiss}, the \textit{expected} number of molecules observed at the receiver at time $t$ due to the release of $\Ntx$ molecules by the transmitter at time $\tau=0$ is readily obtained as
\begin{IEEEeqnarray}{lll} \label{Eq:Dtrm}
	\bar{r}(t,\tau) = \Ntx h(t,\tau).
\end{IEEEeqnarray}
For a given set of system parameters, the expected behavior is non-random and we have a deterministic signal model. Thus, each of the \gls{CIR} expressions derived in Section~\ref{Sect:TxRxCh} constitutes a deterministic representation of the respective \gls{MC} system.

\begin{remk}
The independent molecule behavior assumption has to hold for  (\ref{Eq:Dtrm}) to be valid. However, for some \gls{MC} systems, this assumption does not hold. For instance, if the number of receptors on the surface of a reactive receiver is finite, $\bar{r}(t,\tau)$ becomes a nonlinear function of the released molecules $\Ntx$ and cannot be described by the simple linear expression in (\ref{Eq:Dtrm}). This effect is known as \textit{receptor occupancy} \cite{ArmanJ2}. In these cases, $\bar{r}(t,\tau)$ has to be found for a given $\Ntx$ either numerically or via simulation, cf. Section~\ref{Sect:SimExp}-A for a detailed discussion on simulation methods. \QEDwhite
 \end{remk}

\begin{remk}\label{Remk:Pulse}
	The deterministic model in (\ref{Eq:Dtrm}) assumes an impulsive release of $\Ntx$ molecules at time $\tau=0$ by the transmitter. In general, the transmitter may release the molecules continuously over a finite time interval $[0,T^{\mathrm{rls}}]$ of length $T^{\mathrm{rls}}$. Let $g(t)$ denote the release function satisfying $\int_{t=0}^{T^{\mathrm{rls}}}g(t)\mathrm{d}t=\Ntx$ and $g(t)=0,\,\,t\notin [0,T^{\mathrm{rls}}]$. Then, the expected number of molecules observed at the receiver at time $t$ due to the release of molecules by the transmitter with release function $g(t)$ is given by
	\begin{IEEEeqnarray}{lll} \label{Eq:DtrmPulse}
		\bar{r}(t,\tau) = \int_{t'=0}^{t} g(t')h(t-t',\tau)\mathrm{d}t'.
	\end{IEEEeqnarray}
	We note that (\ref{Eq:DtrmPulse}) reduces to (\ref{Eq:Dtrm}) for $g(t)=\Ntx \delta(t)$. In the remainder of this section, we focus on impulsive release, as this is typically assumed in the \gls{MC}  literature. \QEDwhite
\end{remk}

\subsubsection{Statistical Models}\label{Sec:Statistical}

In the following, we develop statistical models for the number of molecules observed at the receiver as a function of $h(t,\tau)$ for time-invariant \gls{MC} channels. 

\textbf{Binomial Model:} Based on the independent molecule behavior assumption and since any given molecule released by the transmitter is either observed by the receiver or not, a binary state model applies and the number of observed molecules follows the Binomial distribution with $\Ntx$ trials and success probability $h(t,\tau)$, i.e., 
\begin{IEEEeqnarray}{lll} \label{Eq:Bin}
	r(t,\tau)\sim\Bin{\Ntx}{h(t,\tau)},
\end{IEEEeqnarray}
where $\Bin{N}{p}$ represents a Binomial distribution with parameters $N$ and $p$ denoting the number of trials and the success probability, respectively. Under the Binomial model, the \gls{PMF} of $r(t,\tau)$, denoted by $f_{r}^{\mathcal{B}}(n)$, is given by
\iftoggle{OneColumn}{%
\begin{IEEEeqnarray}{lll} \label{Eq:Bin_PMF}
	f_{r}^{\mathcal{B}}(n) = {\Ntx \choose n} \big(h(t,\tau)\big)^n
	\big(1-h(t,\tau)\big)^{\Ntx-n},\quad n\in\{0,1,\dots,\Ntx\}.
\end{IEEEeqnarray}
}{%
\begin{IEEEeqnarray}{lll} \label{Eq:Bin_PMF}
	f_{r}^{\mathcal{B}}(n) = {\Ntx \choose n} \big(h(t,\tau)\big)^n
	\big(1-h(t,\tau)\big)^{\Ntx-n},\quad 
\end{IEEEeqnarray}
for $n\in\{0,1,\dots,\Ntx\}$. 	
}
Unfortunately, the Binomial distribution considerably complicates the analysis of \gls{MC} systems. Therefore, in the following, we present two approximations of the Binomial model with better mathematical tractability.

\textbf{Gaussian Model:} If the expected number of molecules observed at the receiver, i.e., $\bar{r}(t,\tau)$, is sufficiently large, then we can apply the \gls{CLT} and approximate $r(t,\tau)$ by a Gaussian \gls{RV} with mean and variance identical to that of the Binomial \gls{RV}. This leads to 
\begin{IEEEeqnarray}{lll} \label{Eq:Gaus}
	r(t,\tau)\sim\Normal{\Ntx h(t,\tau)}{\Ntx h(t,\tau)(1-h(t,\tau))}.  
\end{IEEEeqnarray}
Under the Gaussian model, the \gls{PDF} of $r(t,\tau)$, denoted by $f_{r}^{\mathcal{N}}(n)$, is given by
\iftoggle{OneColumn}{%
\begin{IEEEeqnarray}{lll} \label{Eq:Gaus_PDF}
	f_{r}^{\mathcal{N}}(n) = \frac{1}{\sqrt{2\pi\Ntx h(t,\tau)(1-h(t,\tau))}}\mathrm{exp}\left(-\frac{\left(n-\Ntx h(t,\tau)\right)^2}{2\Ntx h(t,\tau)(1-h(t,\tau))}\right),\quad n\in \mathbb{R}. 
\end{IEEEeqnarray}
}{%
\begin{IEEEeqnarray}{lll} \label{Eq:Gaus_PDF}
	f_{r}^{\mathcal{N}}(n) = \, &\frac{1}{\sqrt{2\pi\Ntx h(t,\tau)(1-h(t,\tau))}} \nonumber \\ &\times \mathrm{exp}\left(-\frac{\left(n-\Ntx h(t,\tau)\right)^2}{2\Ntx h(t,\tau)(1-h(t,\tau))}\right),\,\,\, n\in \mathbb{R}. \quad\,\,
\end{IEEEeqnarray}
}
The Gaussian distribution is much more amenable to analysis than the Binomial distribution. However,  the basic assumption behind the applicability of the Gaussian distribution, namely large $\bar{r}(t,\tau)$, may not hold in \gls{MC} systems. In fact, although the number of released molecules $\Ntx$ can be quite large, the expected number of observed molecules $\bar{r}(t,\tau)$ can be very small. Moreover, Gaussian \glspl{RV} are continuous and can assume non-integer and negative values, which contradicts the true nature of \gls{RV} $r(t,\tau)$ as discrete and non-negative. 

\textbf{Poisson Model:} For the case when the number of trials is large and the mean of the Binomial \gls{RV} is small, the Binomial distribution can be well approximated by a Poisson distribution with the same mean $\bar{r}(t,\tau)=\Ntx h(t,\tau)$, i.e.,
\begin{IEEEeqnarray}{lll} \label{Eq:Poiss}
	r(t,\tau)\sim\Poisson{\Ntx h(t,\tau)},
\end{IEEEeqnarray}
where $\Poisson{\lambda}$ represents the Poisson distribution with parameter $\lambda$ denoting the mean of the \gls{RV}. Under the Poisson model, the \gls{PMF} of $r(t,\tau)$, denoted by $f_{r}^{\mathcal{P}}(n)$, is given by
\begin{IEEEeqnarray}{lll} \label{Eq:Poiss_PDF}
	f_{r}^{\mathcal{P}}(n) = \frac{\left(\Ntx h(t,\tau)\right)^n}{n!}\exp\left(-\Ntx h(t,\tau)\right),\quad n\in\mathbb{N}.\quad
\end{IEEEeqnarray}
In fact, assuming $\bar{r}(t,\tau)$ is fixed, the proof simply follows from \cite{Robert_MCnote}
\iftoggle{OneColumn}{%
\begin{IEEEeqnarray}{lll} \label{Eq:Poiss_PDF_proof}
	\underset{\Ntx\to\infty}{\lim}f_r^{\mathcal{B}}(n) &= 
	\underset{\Ntx\to\infty}{\lim} {\Ntx \choose n}  \left(\frac{\bar{r}(t,\tau)}{\Ntx}\right)^n\left(1-\frac{\bar{r}(t,\tau)}{\Ntx}\right)^{\Ntx-n} \nonumber \\
	&\overset{(a)}{=} \frac{\left(\bar{r}(t,\tau)\right)^n}{n!}\exp\left(-\bar{r}(t,\tau)\right) =f_r^{\mathcal{P}}(n),
\end{IEEEeqnarray}
}{%
\begin{IEEEeqnarray}{lll} \label{Eq:Poiss_PDF_proof}
	\underset{\Ntx\to\infty}{\lim} & f_r^{\mathcal{B}}(n) \nonumber \\
	&= \underset{\Ntx\to\infty}{\lim} {\Ntx \choose n}  \left(\frac{\bar{r}(t,\tau)}{\Ntx}\right)^n\left(1-\frac{\bar{r}(t,\tau)}{\Ntx}\right)^{\Ntx-n} \nonumber \\
	&\overset{(a)}{=} \frac{\left(\bar{r}(t,\tau)\right)^n}{n!}\exp\left(-\bar{r}(t,\tau)\right) =f_r^{\mathcal{P}}(n),
\end{IEEEeqnarray}
}
where for equality $(a)$ we used $\underset{x\to\infty}{\lim}{x \choose y}=\frac{x^y}{y!}$ and $\underset{x\to\infty}{\lim} \left(1-\frac{y}{x}\right)^x=\exp(-y)$ \cite{TableIntegSerie}.  

\textbf{Comparison:} In order to quantify the accuracy of the Gaussian and Poisson approximations, we define the \gls{RMSE} between the the approximated Gaussian and Poisson \glspl{CDF}, denoted by $F_r^{x}(n),\,\,x\in\{\mathcal{N},\mathcal{P}\}$, and the Binomial \gls{CDF}, denoted by $F_r^{\mathcal{B}}(n)$, as \cite{Yilmaz_Poiss,damrath2017equivalent,marcone2017gaussian}
\begin{IEEEeqnarray}{lll} \label{Eq:RMSE}
	\mathrm{RMSE}^x = \sqrt{\frac{1}{\Ntx+1}\sum_{n=0}^{\Ntx}|F_r^{x}(n)-F_r^{\mathcal{B}}(n)|^2}.
\end{IEEEeqnarray}
In Fig.~\ref{Fig:RMSE}, the \gls{RMSE} between the approximate Gaussian and Poisson \glspl{CDF} and the Binomial \gls{CDF} versus $h(t,\tau)$ is shown for $\Ntx\in\{10^{2},10^{3},10^{4},10^{5}\}$. We observe from this figure that by increasing $h(t,\tau)$, the accuracy of the Poisson model deteriorates  whereas the accuracy of the Gaussian model improves, which is consistent with the respective  assumptions that led to their derivation. Moreover, as $\Ntx$ increases, the Gaussian model becomes more accurate whereas this is not true for the Poisson model if $h(t,\tau)$ is very small. In fact, for small $h(t,\tau)$ and $\Ntx$, both the Binomial and Poisson distributions approach the binary distribution, i.e., either zero or one molecule is observed and the probability of observing more than one molecule becomes negligible. Since for typical \gls{MC} systems, the value of $h(t,\tau)$ is expected to be much smaller than $0.1$, the Poisson model is generally a more accurate model.  Nevertheless, the fact that the accuracy of the Gaussian model increases with increasing $\Ntx$ makes it a suitable model for macroscale applications when $\Ntx$ is potentially very large. Moreover, the Gaussian model is attractive for asymptotically high \gls{SNR} analysis. These observations are consistent with the results reported in  \cite{Yilmaz_Poiss}. 

\iftoggle{OneColumn}{%
\begin{figure}
	\centering  
	\resizebox{0.7\linewidth}{!}{
		\psfragfig{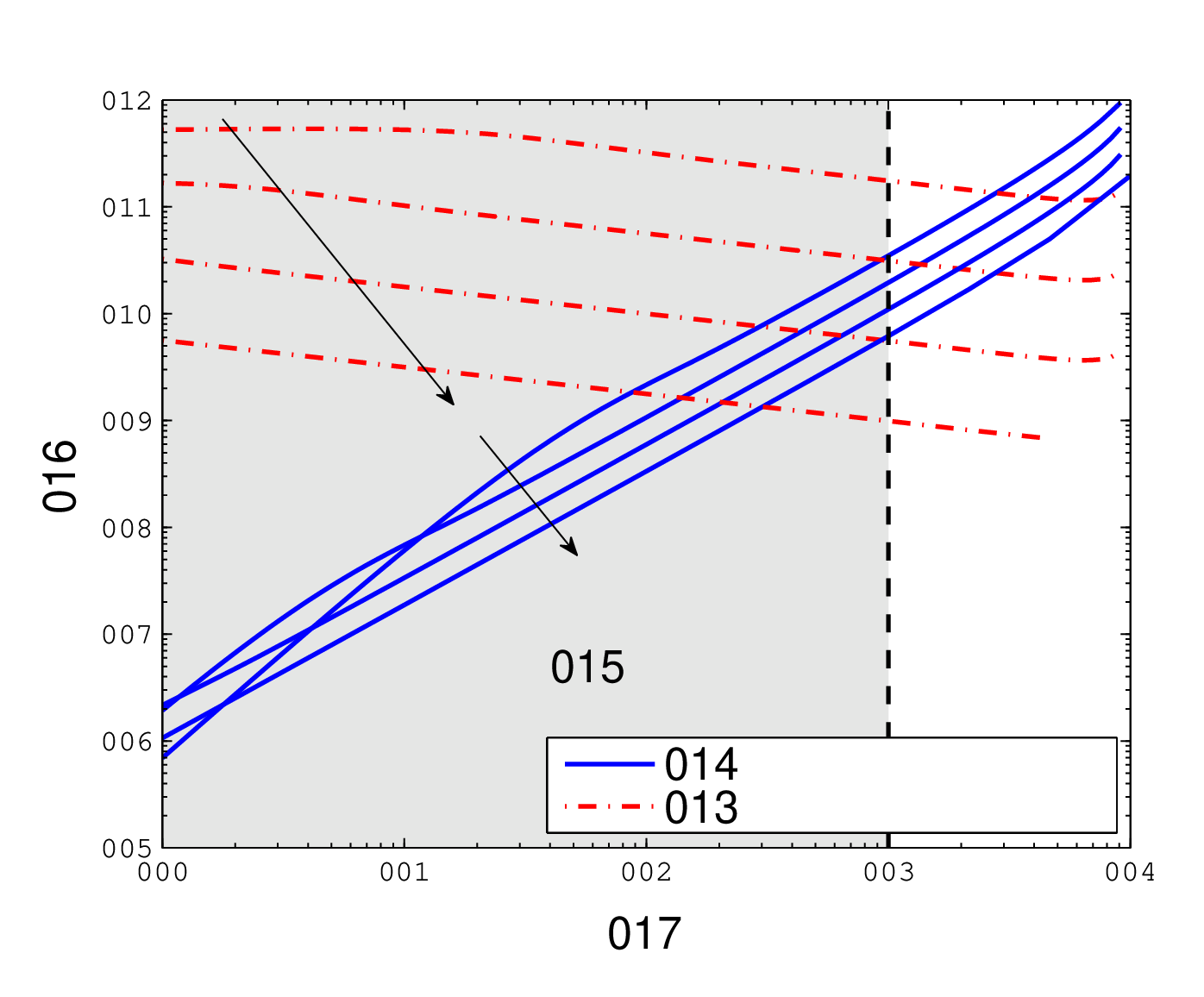}} \vspace{-0.5cm}
	\caption{RMSE between the approximate Gaussian/Poisson CDF and the Binomial CDF versus $h(t,\tau)$ for $\Ntx\in\{10^{2},10^{3},10^{4},10^{5}\}$. The number of released molecules increases in the direction of the arrows. For typical \gls{MC} systems, $h(t,\tau)<0.1$ holds, which is indicated by the shaded area and the dashed vertical line and denoted by \textit{practical regime}.}
	\label{Fig:RMSE}
\end{figure}
}{%
\begin{figure}
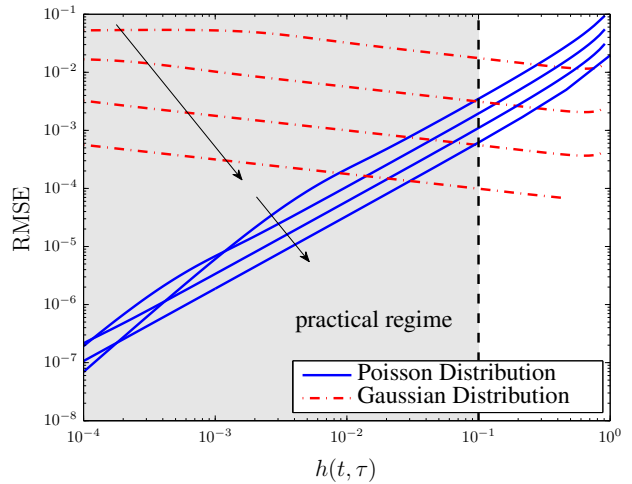

	\centering  
	\resizebox{1\linewidth}{!}{
		\psfragfig{Sections/S4/Fig/MSE/MSE}} \vspace{-0.5cm}
	\caption{RMSE between the approximate Gaussian/Poisson CDF and the Binomial CDF versus $h(t,\tau)$ for $\Ntx\in\{10^{2},10^{3},10^{4},10^{5}\}$. The number of released molecules increases in the direction of the arrows. For typical \gls{MC} systems, $h(t,\tau)<0.1$ holds, which is indicated by the shaded area and the dashed vertical line and denoted by \textit{practical regime}.}
	\label{Fig:RMSE}
\end{figure}
}

\subsubsection{Time-Variant Models}\label{Sec:TimeVarying}

Until now, we have assumed time-invariant \gls{MC} channels where the channel parameters are fixed. Hence, $h(t,\tau)$ and consequently $\bar{r}(t,\tau)$ were only  functions of $t-\tau$.  Now, we consider time-variant \gls{MC} channels where $h(t,\tau)$ and $\bar{r}(t,\tau)$ are in general functions of both $t$ and $\tau$. More specifically, we study the impact of system parameter variations on the mean received signal $\bar{r}(t,\tau)$. In principle, each of the system parameters such as $D$, $\mathbf{v}(\mathbf{d},t)$,   the physical and chemical properties of the boundaries of the end-to-end channel, the reaction rates of the involved CRNs, and $\mathbf{d}_{\tx}$ and $\mathbf{d}_{\rx}$ can potentially vary over time, which in turn leads to a variation of $\bar{r}(t,\tau)$. For instance, the diffusion coefficient $D$ appears in the expressions for $h(t,\tau)$ for all diffusive \gls{MC} systems, and consequently in $\bar{r}(t,\tau)$. As we can see from \eqref{Eq:Einstein}, changes in the fluid environment channel parameters, e.g., the viscosity or temperature, will result in a change in $D$. In fact, the impact of variations in $D$ on $\bar{r}(t,\tau)$ for a point transmitter and passive receiver in 1D is investigated in \cite{Qiu2017}. The authors in \cite{TCOM_NonCoherent} consider a point transmitter with  impulsive release, a passive receiver with the UCA, and an unbounded 3D environment with uniform flow and uniformly distributed enzymes, cf. \eqref{Eq;CIR_passive_ProbObs} and \eqref{Eq:Fick_sol}.  There, the impact of Gaussian variations in the diffusion coefficient, flow velocity, and enzyme concentration is modeled by a parametric model where the parameters of the model are obtained via curve fitting. The impact of the mobility of a point transmitter and a passive receiver on the \gls{CIR} $\bar{r}(t,\tau)$ is studied in \cite{Arman_TimeVariant} and a stochastic model for $\bar{r}(t,\tau)$ is derived. Similarly, a stochastic model  for mobile \gls{MC} systems with a point transmitter and fully-absorbing receiver is derived in \cite{cao2018diffusive}. We note that mobile transceivers are relevant for many envisioned applications of synthetic \gls{MC} systems such as targeted drug delivery and health monitoring \cite{chahibi2015molecular,femminella2015molecular,chude2017molecular,mosayebi2018early}. Therefore, in the following, we focus on diffusive mobile \gls{MC} systems and review some of the results reported in \cite{Arman_TimeVariant}.

We assume a point transmitter, a passive receiver with the UCA approximation, and an unbounded diffusive channel without advection. Furthermore, we model the mobility of transmitter and receiver via 3D diffusion, since diffusion is a common cause of mobility and can also be used to model more elaborate movements such as cell migrations and bacteria chemotaxis \cite{alt1980biased,hong2007chemotaxis}. In particular, we denote the diffusion coefficients of transmitter and receiver by $D_{\tx}$ and $D_{\rx}$, respectively, and their corresponding locations at time $\tau$ by $\mathbf{d}_{\tx}(\tau)$ and $\mathbf{d}_{\rx}(\tau)$, respectively. Then, it can be shown that $\mathbf{d}(\tau) = \mathbf{d}_{\tx}(\tau) - \mathbf{d}_{\rx}(\tau)$ follows a Gaussian distribution \cite[Eq.~(2)]{Arman_TimeVariant}  
\begin{IEEEeqnarray}{C}
	\label{Eq;PDFd(t)} 
	f_{\mathbf{d}(\tau)}(\mathbf{d}) = \frac{1}{(4\pi D_2 \tau)^{3/2}}\exp \left( \frac{-\|\mathbf{d} - \mathbf{d}(0)\|^2}{4 D_2 \tau} \right), 
\end{IEEEeqnarray}
where $D_2 = D_{\tx} + D_{\rx}$ is an effective diffusion coefficient capturing the relative motion of transmitter and receiver, see \cite[Eq.~(10)]{ArmanMobileMC}. Then, given \eqref{Eq.Passive_mean2}, the CIR of the end-to-end channel can be rewritten as  
\begin{IEEEeqnarray}{C}
	\label{Eq;CIRTimVar}
	h(t,\tau) = \frac{V_{\mathrm{rx}}}{(4\pi D_1 t)^{3/2}} \exp \left( \frac{- d^2(\tau)}{4 D_1 t} \right),
\end{IEEEeqnarray}
where $d(\tau) = \| \mathbf{d}(\tau) \|$ and $D_1 = D + D_{\rx}$ is the effective diffusion coefficient capturing the relative motion of the signaling molecules and the receiver, see \cite[Eq.~(8)]{ArmanMobileMC}. The movement of the receiver affects both \eqref{Eq;PDFd(t)} and \eqref{Eq;CIRTimVar} via $D_2$ and $D_1$, respectively, as long as its movement with respect to the transmitter and the signaling molecules is accounted for. For any given $t$, $h(t,\tau)$ is a stochastic process with \glspl{RV} $h(t, \tau_i), \, i \in \{1,2,3,\dots \}$. In the following, we analyze the mean, the variance, and an approximate expression for the \gls{PDF} of $\bar{r}(t,\tau)$.      

\textbf{Mean:} Let $d_0$ denote the distance between the transmitter and receiver at $\tau=0$, i.e., $\|\mathbf{d}(0)\|=d_0$. Given \eqref{Eq;PDFd(t)} and \eqref{Eq;CIRTimVar}, the mean of the time-variant channel, denoted by $m(t,\tau)$, can be evaluated as \cite[Eq.~(14)]{Arman_TimeVariant} 
\iftoggle{OneColumn}{%
\begin{IEEEeqnarray}{rCl}
	\label{Eq.mean_h_def} 
	m_{\bar{r}}(t,\tau) & = & \Ex{\bar{r}(t,\tau)} = \int \limits_{\mathbf{d} \in \mathbb{R}^3} \bar{r}(t,\tau|d_0)\,\, f_{\mathbf{d}(\tau)}(\mathbf{d}) \mathrm{d} \mathbf{d}, \nonumber \\ 
	& = & \frac{\Ntx V_{\rx}}{\left( 4\pi \left(D_1 t + D_2 \tau \right) \right)^{3/2}} \exp \left( \frac{-d_0^2}{4 \left(D_1 t + D_2 \tau \right) } \right), \quad	
\end{IEEEeqnarray}
}{%
\begin{IEEEeqnarray}{lll}
	\label{Eq.mean_h_def} 
	m_{\bar{r}}(t,\tau)  =  \Ex{\bar{r}(t,\tau)} \nonumber \\ 
	= \int \limits_{\mathbf{d} \in \mathbb{R}^3} \bar{r}(t,\tau|d_0)\,\, f_{\mathbf{d}(\tau)}(\mathbf{d}) \mathrm{d} \mathbf{d}, \nonumber \\ 
	=  \frac{\Ntx V_{\rx}}{\left( 4\pi \left(D_1 t + D_2 \tau \right) \right)^{3/2}} \exp \left( \frac{-d_0^2}{4 \left(D_1 t + D_2 \tau \right) } \right), \quad	
\end{IEEEeqnarray}
}
where in $\Ex{\bar{r}(t,\tau)}$, the expectation is taken with respect to the \gls{RV} $\mathbf{d}(\tau)$. As we expected, $\bar{r}(t,\tau)$ is a function of $\tau$, because of the mobility of transmitter and receiver. As a result, $\bar{r}(t,\tau)$ is a non-stationary stochastic process. Moreover, due to the assumption of an unbounded environment, on average the transmitter and receiver diffuse away from each other. Therefore, when at least one of the transceivers is mobile, i.e., $D_2\neq 0$, we obtain $\bar{r}(t,\tau)\to 0$ as $\tau\to\infty$.  

\textbf{Variance:} The variance of $\bar{r}(t,\tau)$, denoted by $\sigma_{\bar{r}}^2(t,\tau)$, is given by
\begin{IEEEeqnarray}{C}\label{Eq:Var_TVarying}
	\sigma_{\bar{r}}^2(t,\tau) = \Ex{\bar{r}^2(t,\tau)} - m_{\bar{r}}^2(t,\tau),
\end{IEEEeqnarray}
where the second order moment $\phi_{\bar{r}}(t,\tau)\triangleq\Ex{\bar{r}^2(t,\tau)}$ is obtained as \cite[Eq.~(21)]{Arman_TimeVariant} 
\begin{IEEEeqnarray}{C}
	\label{Eq.ACF_equaltime}
	\phi_{\bar{r}}(t,\tau) = \frac{(\Ntx)^2V_{\rx}^2 \exp \left( \frac{-d_0^2}{ 2 \left(D_1 t + 2 D_2 \tau \right)} \right)}{\left( 4 \pi D_1 t \right)^{3/2} \left( 4\pi \left(D_1 t + 2 D_2 \tau \right) \right)^{3/2}}. 
\end{IEEEeqnarray}
We note that $\sigma_{\bar{r}}^2(t,\tau)\to 0$ as $\tau\to\infty$, which is due to the fact that   $\bar{r}(t,\tau)\to 0$ as $\tau\to\infty$. On the other hand, it can be shown that the normalized variance $\frac{\sigma_{\bar{r}}^2(t,\tau)}{m_{\bar{r}}^2(t,\tau)}\to \infty$ as $\tau\to\infty$. In other words, the normalized variance increases as $\tau$ increases. This in turn implies that due to the random walk, the uncertainty that we have about $\bar{r}(t,\tau)$ increases as $\tau$ increases.

\textbf{Approximate \gls{PDF}:} In the following, we present the approximated \gls{PDF} of the considered time-variant channel with mobile transceivers and refer the interested reader to \cite{Arman_TimeVariant} for the exact expressions of the \gls{CDF} and \gls{PDF}. In particular, it is shown in \cite{Arman_TimeVariant} that when $D_2 \tau \leq d_0^2/200$ holds for any $\tau > 0$, then the \gls{PDF} of the \gls{CIR} can be accurately approximated via a log-normal distribution \cite[Eq.~(29)]{Arman_TimeVariant} 
\iftoggle{OneColumn}{%
\begin{IEEEeqnarray}{rCl}
	\label{Eq.PDFofCIRLognormal} 
	f_{h(t,\tau)}(h)  \sim  \text{Lognormal} \left( \tilde{\mu}, \tilde{\sigma}^2 \right)
	\quad\text{with}\quad
	\begin{cases}
		\tilde{\mu}  =  \ln \left( \frac{V_{\rx}}{(4\pi D_1 t)^{3/2}} \right) - \frac{D_2 \tau}{4 D_1 t} \left( 6+\frac{d_0^2}{D_2 \tau} \right) \\
		\tilde{\sigma}^2  = \left( \frac{D_2 \tau}{2D_1 t} \right)^2 \left( 6 + \frac{2d_0^2}{D_2 \tau} \right) 
	\end{cases}
\end{IEEEeqnarray}
}{%
\begin{IEEEeqnarray}{rCl}
	\label{Eq.PDFofCIRLognormal} 
	f_{h(t,\tau)}(h)  \sim  \text{Lognormal} \left( \tilde{\mu}, \tilde{\sigma}^2 \right)
\end{IEEEeqnarray}
with
\begin{IEEEeqnarray}{rCl}
	\begin{cases}
		\tilde{\mu}  =  \ln \left( \frac{V_{\rx}}{(4\pi D_1 t)^{3/2}} \right) - \frac{D_2 \tau}{4 D_1 t} \left( 6+\frac{d_0^2}{D_2 \tau} \right) \\
		\tilde{\sigma}^2  = \left( \frac{D_2 \tau}{2D_1 t} \right)^2 \left( 6 + \frac{2d_0^2}{D_2 \tau} \right) 
	\end{cases}\nonumber
\end{IEEEeqnarray}
}
where $\tilde{\mu}$ and $\tilde{\sigma}^2$ denote the mean and the variance of the log-normal distribution. Given \eqref{Eq.PDFofCIRLognormal}, the \gls{PDF} of $\bar{r}(t,\tau)$, denoted by $f_{\bar{r}(t,\tau)}(\bar{r})$, can be written as 
\begin{IEEEeqnarray}{C}
	\label{Eq;PDFofMeanSignal}
	f_{\bar{r}(t,\tau)}(\bar{r}) = \frac{1}{\Ntx} \times f_{h(t,\tau)}\left(\frac{\bar{r}}{\Ntx}\right). 
\end{IEEEeqnarray}
The above stochastic model can be used for the design and performance analysis of time-variant \gls{MC} systems. For instance, (\ref{Eq;PDFofMeanSignal}) was used in \cite{ArmanMobileMC} to compute the expected error probability of a mobile \gls{MC} system when the knowledge of \gls{CIR} $h(t,\tau)$ used for detection becomes gradually outdated due to the mobility of the transceivers. Moreover, in \cite{TCOM_NonCoherent}, a stochastic channel model was used to develop non-coherent detectors. In contrast to the stochastic model in (\ref{Eq;PDFofMeanSignal}) for mobile \gls{MC} systems, it was shown in \cite{TCOM_NonCoherent} that the Gamma distribution is a good fit for (Gaussian) variations in the diffusion coefficient, flow velocity, and enzyme concentration in a non-mobile \gls{MC} system.

\subsection{Interfering Noise Molecules}

In the previous section, we have considered statistical models for the number of molecules observed at the receiver due to the release of signaling molecules by the transmitter. However,  \gls{MC} systems may be impaired by noise molecules that are not released by the transmitter but originate from interfering natural or synthetic sources. In the following, we introduce statistical models to account for the number of noise molecules that are observed at the receiver. Since information and noise molecules are indistinguishable, the receiver treats the total numbers of observed signaling molecules, denoted by $r_{\mathrm{sig}}(t,\tau)$, and interfering noise molecules, denoted by $r_{\mathrm{int}}(t,\tau)$, as the received signal $r(t,\tau)$, i.e., 
\begin{IEEEeqnarray}{lll} \label{Eq:Signal_Noise}
	r(t,\tau) = r_{\mathrm{sig}}(t,\tau)+r_{\mathrm{int}}(t,\tau).
\end{IEEEeqnarray}

To derive a statistical model for $r_{\mathrm{int}}(t,\tau)$, we focus on a passive receiver. Similar arguments apply for other receiver types. We make the following assumptions. A1) Let $\bar{r}_{\mathrm{int}}(\tau)$ denote the expected number of noise molecules observed within the receiver volume $V_{\mathrm{rx}}$ at a given sample time $t$. We assume that the value of $\bar{r}_{\mathrm{int}}(\tau)$ is constant over observation time $t$. Nevertheless, $\bar{r}_{\mathrm{int}}(\tau)$ may change over larger time scale $\tau$ due to variations in the system parameters such as the temperature, cf. Section~\ref{Sec:ThreeScale} and Section~\ref{Sec:TimeVarying}. A2) It is further assumed that the observation of one noise molecule at the receiver is independent from the observations of other noise molecules. A3) Finally, we assume that the expected number of noise molecules observed within a given volume in space is proportional to the size of that volume.

Based on assumptions A1-A3, the statistics of the observed noise molecules is Poisson following the \gls{LRE} \cite{Falk2010LRE}. In particular, suppose the receiver volume is divided into $J$ subvolumes where $J\gg \bar{r}_{\mathrm{int}}(\tau)$. Thus, $\frac{\bar{r}_{\mathrm{int}}(\tau)}{J}$ can be interpreted as the probability that one noise molecule is observed in one of these subvolumes at the receiver. The probability that two noise molecules are simultaneously observed  in one subvolume becomes negligible for large $J$. Therefore, the number of noise molecules observed over the entire volume of the receiver follows a Binomial distribution $\Bin{J}{\frac{\bar{r}_{\mathrm{int}}(\tau)}{J}}$ with $J$ trials and success probability $\frac{\bar{r}_{\mathrm{int}}(\tau)}{J}$. Consequently, since $J$ is a free variable, one can assume $J\to\infty$ such that the Binomial distribution approaches the Poisson distribution $\Poisson{\bar{r}_{\mathrm{int}}(\tau)}$, cf. (\ref{Eq:Poiss_PDF}). In summary, under assumptions A1-A3, we obtain $r_{\mathrm{int}}(t,\tau)\sim\Poisson{\bar{r}_{\mathrm{int}}(\tau)}$.

\begin{remk}
	The choice of the Poisson distribution for the number of  environmental noise molecules observed at the receiver, $r_{\mathrm{int}}(t,\tau)$, can be further justified from an information-theoretic perspective \cite{Bialek2012biophysics}. Let us define  \gls{RV} $\mathbf{D}=[\mathbf{d}_1,\mathbf{d}_2,\dots,\mathbf{d}_{r_{\mathrm{int}}}]$ where $\mathbf{d}_i$ denotes the coordinates of the $i$-th noise molecule observed at the receiver and we drop argument $(t,\tau)$ of $r_{\mathrm{int}}(t,\tau)$ in $\mathbf{D}$ for notational simplicity. In particular, the maximum entropy distribution for $\mathbf{D}$ corresponds to a Poisson distribution for the number of observed noise molecules  $r_{\mathrm{int}}(t,\tau)$. Therefore, the most random noise under assumptions A1-A3 is Poisson noise, i.e., a worst-case scenario. To see this, let $f_{\mathbf{D}}(\mathbf{D})$ denote the distribution of \gls{RV} $\mathbf{D}$. Using the chain rule, we have  $f_{\mathbf{D}}(\mathbf{D})=f_{\mathbf{D}|r_{\mathrm{int}}}(\mathbf{D}|r_{\mathrm{int}})f_{r_{\mathrm{int}}}(r_{\mathrm{int}})$ where $f_{\mathbf{D}|r_{\mathrm{int}}}(\mathbf{D}|r_{\mathrm{int}})$ is the conditional distribution of $\mathbf{D}$ given $r_{\mathrm{int}}(t,\tau)$ and $f_{r_{\mathrm{int}}}(r_{\mathrm{int}})$ denotes the distribution of $r_{\mathrm{int}}(t,\tau)$. For maximum entropy, $f_{\mathbf{D}|r_{\mathrm{int}}}(\mathbf{D}|r_{\mathrm{int}})$ should be a uniform distribution across the receiver volume. Substituting this result in $f_{\mathbf{D}}(\mathbf{D})$, we obtain that $f_{r_{\mathrm{int}}}(r_{\mathrm{int}})$ has to be the Poisson distribution to maximize the entropy of $\mathbf{D}$ \cite[Appendix~8]{Bialek2012biophysics}. \QEDwhite
\end{remk}

We note that the Poisson distribution $\Poisson{\lambda}$ approaches a Gaussian distribution $\Normal{\lambda}{\lambda}$ for $\lambda\to\infty$. Therefore, for very noisy environments, the approximation $r_{\mathrm{int}}(t,\tau)\sim\Normal{\bar{r}_{\mathrm{int}}(\tau)}{\bar{r}_{\mathrm{int}}(\tau)}$ becomes valid.

\begin{remk}
More detailed models for the number of noise molecules observed at the receiver can be found in \cite{Adam_Universal_Noise} where it was assumed that the noise molecules originate from external sources that continuously release molecules or transmitters of other communication links which use the same type of molecule for signaling. There, the expected number of noise molecules observed at the receiver at time $t$ after the noise sources start releasing molecules, denoted by $\bar{r}_{\mathrm{int}}(t,\tau)$, was derived as a function of the system parameters such as the distances between the noise sources and the receiver. It was shown that asymptotically as $t\to\infty$, $\bar{r}_{\mathrm{int}}(t,\tau)$ converges to a  constant value, i.e., $\bar{r}_{\mathrm{int}}(\tau)$, which is consistent with assumption A1 made earlier in this section. \QEDwhite
\end{remk}

\subsection{Continuous Transmission}

The statistical models developed so far are appropriate for one-shot transmission. Nevertheless, in most communication systems, the transmitter may send multiple symbols consecutively to the receiver. To develop a model valid for continuous transmission, we consider a time-slotted communication system where one symbol is transmitted in each time slot, also referred to as a symbol interval, of length $T^{\mathrm{symb}}$. We focus on \gls{CSK} modulation  where the transmitter releases $s[k]\Ntx$ molecules at the beginning of the $k$-th symbol interval to convey information symbol $s[k]\in[0,1]$ \cite{Nariman_Survey}. We assume synchronous transmission and that the receiver counts the number of observed molecules multiple times in each symbol interval with sampling interval $\Delta t$ \cite{TNBC_Sync}. Because of the memory of the \gls{MC} channel, \gls{ISI} occurs. To take this into account, we assume  that the \gls{MC} channel has a memory of $L$ symbol intervals, i.e., the \gls{ISI} in symbol interval $k$ originates from the symbols transmitted in the $L-1$ previous symbol intervals.  We further take into account that communication may be impaired by noise molecules that originate from interfering natural or synthetic sources. Finally, we assume that the \gls{MC} channel parameters remain unchanged for the considered observation window, and hence, we drop argument $\tau$ in $r(t,\tau)$, $\bar{r}_{\mathrm{int}}(\tau)$, and $h(t,\tau)$ for notational simplicity. In the following, we first provide the signal model for a general case and subsequently simplify it for extreme \gls{SNR} regimes to obtain further insight.

\subsubsection{General Case} Let $r[k,m]$ denote the total number of  molecules observed at the receiver for sample $m$ in symbol interval $k$, i.e., $r[k,m]=r(t_{k,m})$ where $t_{k,m}=(k-1)T^{\mathrm{symb}}+m\Delta t$  and $r(t_{k,m})$ is given in \eqref{Eq:Sig_3Scale}. Then, following the discussion in Section~\ref{Sec:Statistical}, $r[k,m]$ can be accurately modeled as a Poisson \gls{RV}, i.e., 
\begin{IEEEeqnarray}{lll} \label{Eq:ChannelInOut_General_Poisson}
r[k,m]  \sim \Poisson{\sum_{l=1}^L \left(\bar{r}_{\mathrm{sig}}[l,m] s[k-l+1]\right)
	+\bar{r}_{\mathrm{int}}}, 
\end{IEEEeqnarray}
where $\bar{r}_{\mathrm{sig}}[l,m]=\Ntx h\left(t_{l,m}\right)$. Moreover,  we used the  superposition property of Poisson \gls{RV}s, i.e., if $X$ and $Y$ are two independent Poisson \gls{RV}s with means $\lambda_x$ and $\lambda_y$, respectively, then $X+Y$ is also a Poisson \gls{RV} with mean $\lambda_x+\lambda_y$ \cite{Poisson_Channel}. Alternatively, defining $\tilde{r}[k,m]=r[k,m]-\bar{r}_{\mathrm{int}}$, one can obtain the following more familiar additive signal model
\iftoggle{OneColumn}{%
\begin{IEEEeqnarray}{lll} \label{Eq:ChannelInOut_General_Anternative}
	\tilde{r}[k,m]  = \underset{\text{signal component}}{\underbrace{\sum_{l=1}^L \left(\bar{r}_{\mathrm{sig}}[l,m] s[k-l+1]\right)}}
	+ \underset{\text{diffusion noise}}{\underbrace{r_{\mathrm{dfn}}[k,m]}} 
	+ \underset{\text{interference noise}}{\underbrace{r_{\mathrm{int}}[k,m]}},
\end{IEEEeqnarray}
}{%
\begin{IEEEeqnarray}{lll} \label{Eq:ChannelInOut_General_Anternative}
	\tilde{r}[k,m]  = &\underset{\text{signal component}}{\underbrace{\sum_{l=1}^L \left(\bar{r}_{\mathrm{sig}}[l,m] s[k-l+1]\right)}}
	\nonumber \\
	&+ \underset{\text{diffusion noise}}{\underbrace{r_{\mathrm{dfn}}[k,m]}} 
	+ \underset{\text{interference noise}}{\underbrace{r_{\mathrm{int}}[k,m]}},
\end{IEEEeqnarray}
}
where $r_{\mathrm{dfn}}[k,m]\sim \mathcal{P}_0\left(\sum_{l=1}^L \bar{r}_{\mathrm{sig}}[l,m] s[k-l+1]\right)$ denotes the diffusion noise and $r_{\mathrm{int}}[k,m]\sim \mathcal{P}_0\left(\bar{r}_{\mathrm{int}}\right)$ denotes the interfering noise molecules. Here, we use the notation $X\sim\mathcal{P}_0(\lambda)$ when $X=Y-\lambda$ where $Y\sim\Poisson{\lambda}$, i.e., $X$ is a Poisson \gls{RV} whose mean has been subtracted. 

When the expected numbers of information and interfering noise molecules are large, one may use the Gaussian model for the number of observed molecules, i.e., $r[k,m]  \sim \Normal{\bar{r}[k,m]}{\bar{r}[k,m]}$ where $\bar{r}[k,m]={\sum_{l=1}^L \bar{r}_{\mathrm{sig}}[l,m] s[k-l+1]+\bar{r}_{\mathrm{int}}}$. One can also write  $\tilde{r}[k,m]$ in the form of (\ref{Eq:ChannelInOut_General_Anternative}) where
for the Gaussian model, we have $r_{\mathrm{dfn}}[k,m]\sim \Normal{0}{\sum_{l=1}^L \bar{r}_{\mathrm{sig}}[l,m] s[k-l+1]}$  and $r_{\mathrm{int}}[k,m]\sim \Normal{0}{\bar{r}_{\mathrm{int}}}$. We note that unlike for the \gls{AWGN} channel in conventional wireless communication, the Gaussian diffusion noise in \gls{MC} is signal dependent. 

\begin{remk}
In (\ref{Eq:ChannelInOut_General_Anternative}), we distinguish between two types of additive noise, namely $r_{\mathrm{dfn}}[k,m]$, which originates from signaling molecules, and $r_{\mathrm{int}}[k,m]$, which originates from external interfering noise molecules. We note that the randomness of $r_{\mathrm{dfn}}[k,m]$ and $r_{\mathrm{int}}[k,m]$ can be attributed to the random Brownian motion of the signaling and noise molecules, respectively. In addition to the aforementioned noises,  other types of noises may be present. For instance, in a reactive receiver, the noisy measurements of the activated receptors, caused by the randomness of diffusion and ligand-receptor interactions, may be relayed by  signaling pathways to the interior of the receiver (e.g. a cell), which may add extra noise \cite{PierobonJ3,marcone2017gaussian}. We refer to this noise as \textit{counting noise} to contrast it with the diffusion noise. \QEDwhite
\end{remk}

\subsubsection{Simplifications for Extreme SNR Regimes}
In the following, we further simplify the model in (\ref{Eq:ChannelInOut_General_Anternative}) for two asymptotic \gls{SNR} regimes, namely the diffusion-noise-limited and interference-limited regimes. To do so, we first formally define \gls{SNR} as \cite{CC_TCOM}
\iftoggle{OneColumn}{%
\begin{IEEEeqnarray}{lll} \label{Eq:SNR}
	\mathrm{SNR} = \frac{\text{Power of Signal}}{\text{Variance of Diffusion Noise}+\text{Variance of Interfering Noise}}=\frac{\bar{r}_{\mathrm{sig}}^2}{\bar{r}_{\mathrm{sig}}+\bar{r}_{\mathrm{int}}},
\end{IEEEeqnarray}
}{%
\begin{IEEEeqnarray}{lll} \label{Eq:SNR}
	\mathrm{SNR} \nonumber \\
	= \frac{\text{\small Power of Signal}}{\text{\small Variance of Diffusion Noise}+\text{\small Variance of Interfering Noise}} \nonumber\\
	=\frac{\bar{r}_{\mathrm{sig}}^2}{\bar{r}_{\mathrm{sig}}+\bar{r}_{\mathrm{int}}},
\end{IEEEeqnarray}
}
where $\bar{r}_{\mathrm{sig}}$ denotes the expected number of signaling molecules received at the sampling time. In the following, we focus on the \gls{ISI}-free channel, i.e., $L=1$, and a single-sample detector. Therefore, we drop indices $l$ and $m$ for notational simplicity. 

\iftoggle{OneColumn}{%
	\begin{figure}
		\centering  
		\resizebox{0.7\linewidth}{!}{
			\psfragfig{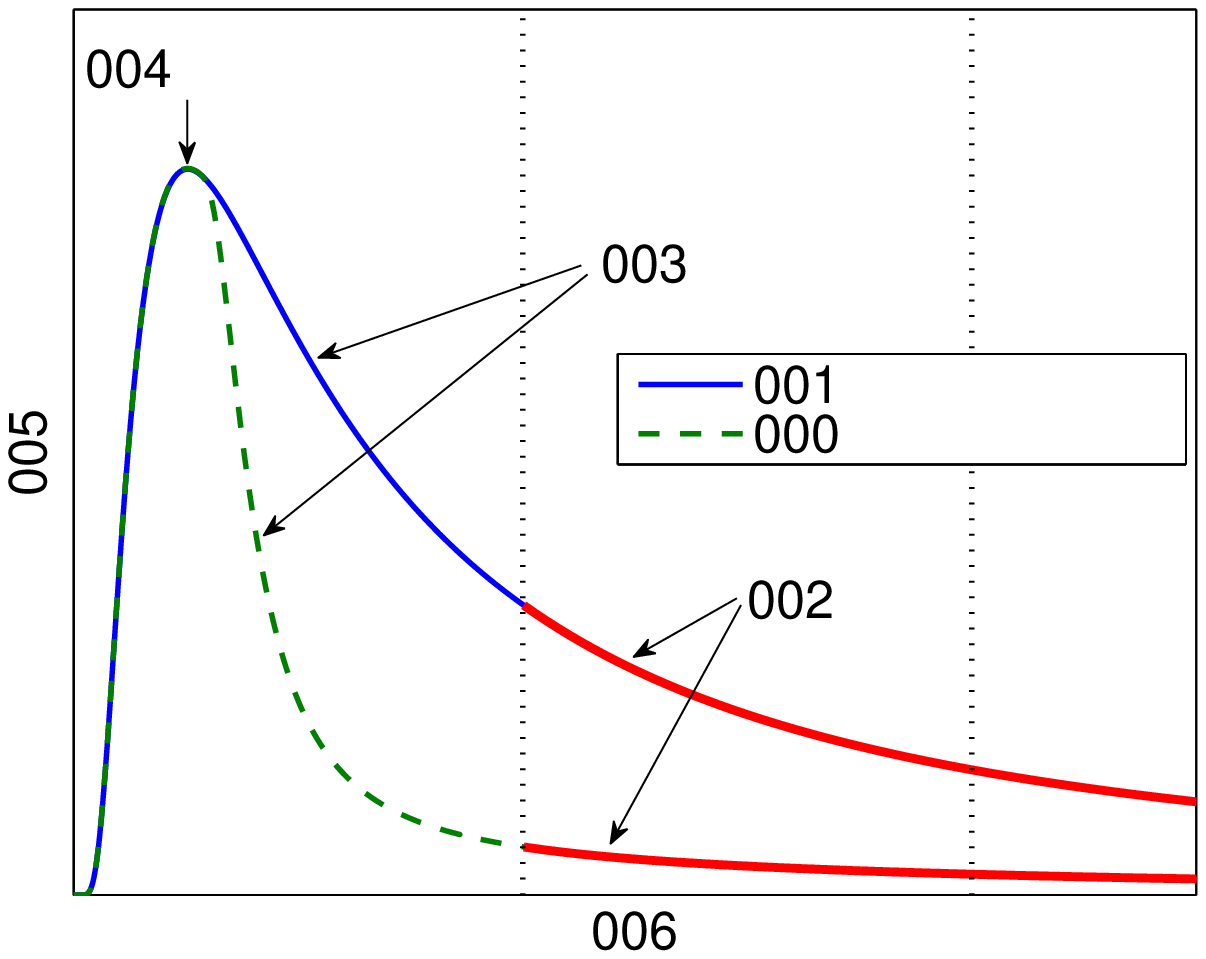}} \vspace{-0.5cm} 
		\caption{Expected number of molecules observed at the receiver versus time. The dotted vertical lines indicate the beginning of symbol intervals. The injection of the reactive cleaning signal helps to shorten the CIR.}
		\label{Fig:ISI_acid}
	\end{figure}
}{%
	\begin{figure}
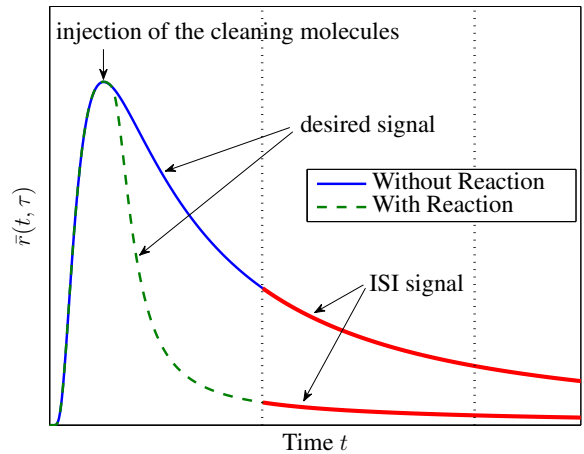

		\centering  
		\resizebox{1\linewidth}{!}{
			\psfragfig{Sections/S4/Fig/ISI/ISI}} \vspace{-0.5cm} 
		\caption{Expected number of molecules observed at the receiver versus time. The dotted vertical lines indicate the beginning of symbol intervals. The injection of the reactive cleaning signal helps to shorten the CIR.}
		\label{Fig:ISI_acid}
	\end{figure}
}

\begin{remk}
One approach to obtain an approximately \gls{ISI}-free channel is to choose a sufficiently large symbol interval such that the \gls{CIR} practically fully decays to zero within one symbol interval. In such a case, the transmission rate may be severely reduced which may lead to an inefficient system design.  Fortunately, it has been shown in the literature that reactions can be beneficial for \gls{ISI} mitigation  \cite{Adam_Enzyme,Nariman_AcidBasePlatform,ICC_Reactive}. In particular, enzymes  \cite{Adam_Enzyme} and reactive signaling molecules, such as acid and base molecules \cite{Nariman_AcidBasePlatform,ICC_Reactive}, may be used to speed up the decay of the \gls{CIR} as a function of time, which would increase the accuracy of the assumption of an \gls{ISI}-free channel, see Fig.~\ref{Fig:Reaction}. For instance, in \cite{ICC_Reactive}, a reactive signaling \gls{MC} system was assumed where the transmitter employs different molecules that react with each other, e.g., acids and bases. Then, after the release of the signaling molecules (e.g., an acid), the transmitter may release so-called cleaning molecules (e.g., a base). It is shown that the resulting \gls{CIR} is considerably shortened, which makes the \gls{ISI}-free channel an accurate model, see Fig.~\ref{Fig:ISI_acid}. Moreover, the peak of the received signal remains unchanged since the cleaning molecules are released after the peak is observed at the receiver. \QEDwhite
\end{remk}

\textbf{Diffusion-Noise-Limited SNR Regime:} In this case, we assume $\bar{r}_{\mathrm{sig}}\gg \bar{r}_{\mathrm{int}}$ holds. Thus, the model in (\ref{Eq:ChannelInOut_General_Anternative}) simplifies to 
\begin{IEEEeqnarray}{lll} \label{Eq:DiffusionLimited}
	\tilde{r}[k]  = \bar{r}_{\mathrm{sig}} s[k] + r_{\mathrm{dfn}}[k],
\end{IEEEeqnarray}
where $r_{\mathrm{dfn}}[k]\sim\mathcal{P}_0\left(\bar{r}_{\mathrm{sig}} s[k]\right)$ and $r_{\mathrm{dfn}}[k]\sim\Normal{0}{\bar{r}_{\mathrm{sig}} s[k]}$ hold for the Poisson and Gaussian models, respectively. The \gls{SNR} in this case is obtained as $\mathrm{SNR}=\bar{r}_{\mathrm{sig}}$.

\textbf{Interference-Limited SNR Regime:} In this case, we assume $\bar{r}_{\mathrm{sig}}\ll \bar{r}_{\mathrm{int}}$ holds. Thus, the model in (\ref{Eq:ChannelInOut_General_Anternative}) simplifies to 
\begin{IEEEeqnarray}{lll} \label{Eq:Gaussian_InterferenceLimited}
	\tilde{r}[k]  = \bar{r}_{\mathrm{sig}} s[k] + r_{\mathrm{int}}[k],
\end{IEEEeqnarray}
where $r_{\mathrm{int}}[k]\sim\mathcal{P}_0\left(\bar{r}_{\mathrm{int}} \right)$ and $r_{\mathrm{int}}[k]\sim\Normal{0}{\bar{r}_{\mathrm{int}}}$ hold for the Poisson and Gaussian models, respectively. The \gls{SNR} in this case is obtained as $\mathrm{SNR}=\bar{r}_{\mathrm{sig}}^2/\bar{r}_{\mathrm{int}}$. We note that this special case yields a signal-independent (Gaussian) model as it is widely adopted in conventional wireless communications.

Finally, we note that it may be necessary to use a combination of both of the above special cases for analysis of \gls{MC} systems. For instance, for a simple \gls{OOK} modulation, i.e., $s[k]\in\{0,1\}$, the interference noise molecules are dominant for bit $s[k]=0$ whereas the diffusion noise is dominant for bit $s[k]=1$. This is schematically illustrated in Fig.~\ref{Fig:Noise} where it can be observed that the noise power for symbol $s[k]=1$ (diffusion-noise-limited regime) is larger than that for symbol $s[k]=0$ (interference-limited regime). A similar observation has also been reported for photon-counting receivers in optical wireless communications where the shot noise at the receiver has two components, one generated by the laser transmitter (analogous to diffusion noise) and one generated by the ambient background light (analogous to interfering noise molecules) \cite{UysalFSOsurvey}.

\iftoggle{OneColumn}{%
\begin{figure}
	\centering  
	\resizebox{0.7\linewidth}{!}{
		\psfragfig{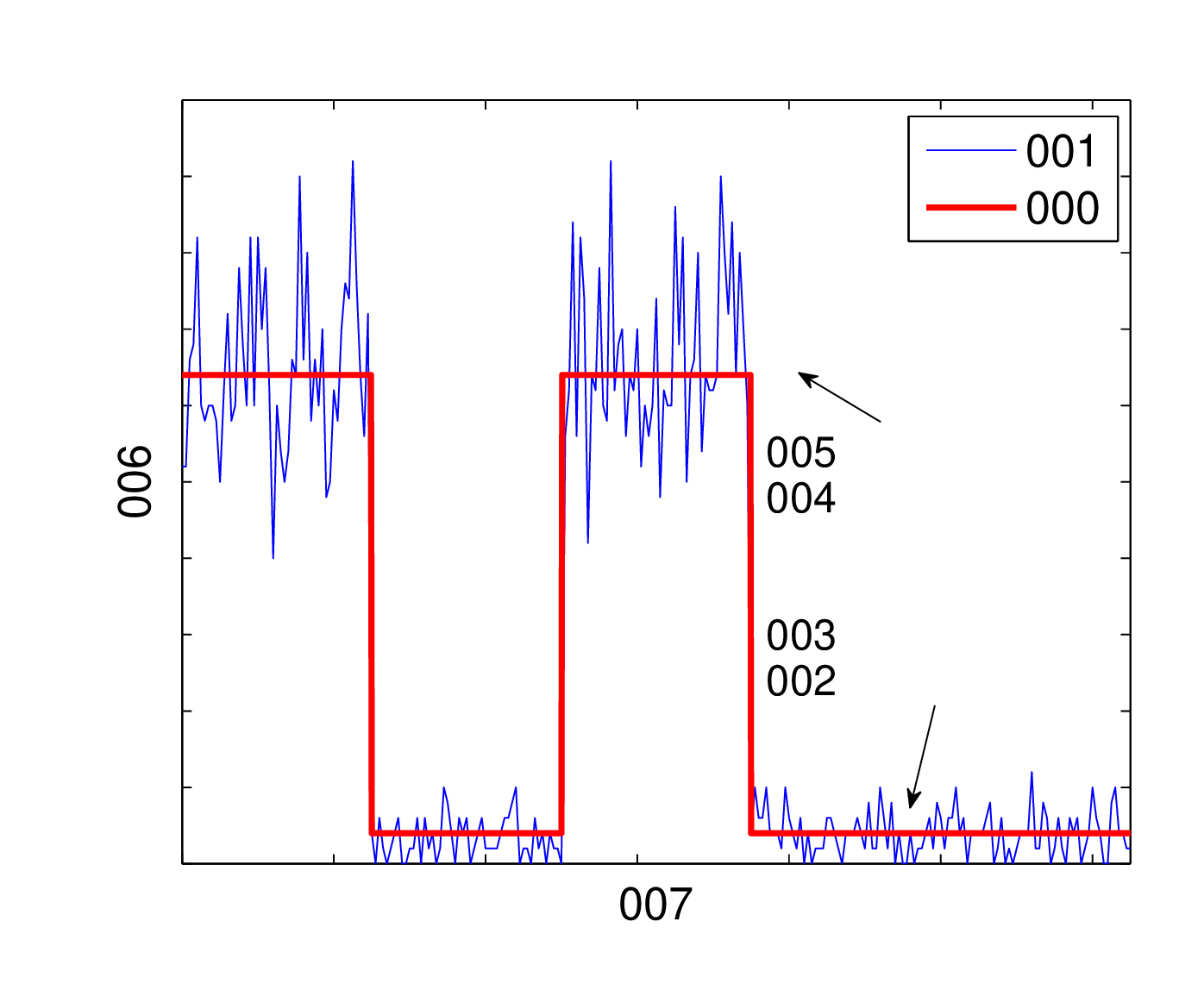}}\vspace{-0.5cm} 
	\caption{Received signal versus time. Illustration of diffusion-noise-limited and interference-limited regimes.}
	\label{Fig:Noise}
\end{figure}
}{%
\begin{figure}
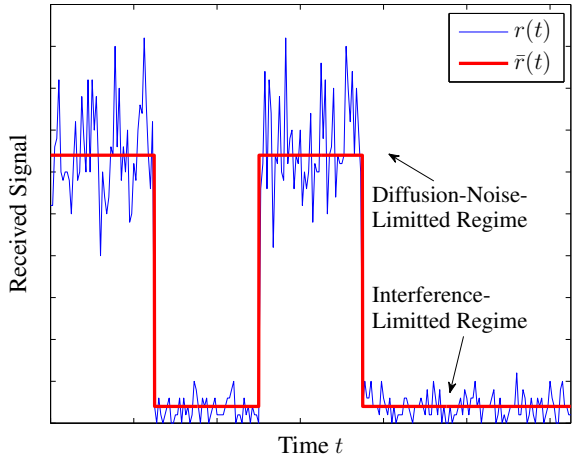

	\centering  
	\resizebox{1\linewidth}{!}{
		\psfragfig{Sections/S4/Fig/Noise/Noise}}\vspace{-0.5cm} 
	\caption{Received signal versus time. Illustration of diffusion-noise-limited and interference-limited regimes.}
	\label{Fig:Noise}
\end{figure}
}

\subsection{Time Correlation}

In the following, we discuss the signal correlation with respect to observation time scale $t$ and release time scale $\tau$.

\subsubsection{Sample Correlation}\label{Sect:SampCorrelation} 
In (\ref{Eq:ChannelInOut_General_Poisson}) and (\ref{Eq:ChannelInOut_General_Anternative}), we assume that the number of molecules counted at different time instants $t$ within one symbol interval or in different symbol intervals are independent from each other. However, this assumption holds only if the sampling interval is chosen large enough such that the independence of consecutive samples is guaranteed. In \cite{Adam_OptReciever}, the mutual information between two samples $r(t_1,\tau)$ and $r(t_2,\tau)$ was numerically computed and the minimum spacing needed to ensure independence between consecutive samples was found such that the corresponding mutual information is below some threshold. Since the mutual information between two samples is difficult to derive in closed form, one may consider the  Pearson correlation coefficient \cite{benesty2009pearson} among two consecutive samples instead, i.e.,
\iftoggle{OneColumn}{%
\begin{IEEEeqnarray}{lll}
	\rho_t(t_1,t_2) 
	&=\frac{\Ex{\big(r(t_1,\tau)-\bar{r}(t_1,\tau)\big)\big(r(t_2,\tau)-\bar{r}(t_2,\tau)\big)}}{\sqrt{\Ex{\big(r(t_1,\tau)-\bar{r}(t_1,\tau)\big)^2}\Ex{\big(r(t_2,\tau)-\bar{r}(t_2,\tau)\big)^2}}} \nonumber \\
	&\overset{(a)}{=}\frac{\Ex{r(t_1,\tau)r(t_2,\tau)}-\bar{r}(t_1,\tau)\bar{r}(t_2,\tau)}{\sqrt{\bar{r}(t_1,\tau)\bar{r}(t_2,\tau)}},
\end{IEEEeqnarray}
}{%
\begin{IEEEeqnarray}{lll}
	\rho_t(t_1,t_2) \nonumber \\
	=\frac{\Ex{\big(r(t_1,\tau)-\bar{r}(t_1,\tau)\big)\big(r(t_2,\tau)-\bar{r}(t_2,\tau)\big)}}{\sqrt{\Ex{\big(r(t_1,\tau)-\bar{r}(t_1,\tau)\big)^2}\Ex{\big(r(t_2,\tau)-\bar{r}(t_2,\tau)\big)^2}}} \nonumber \\
	\overset{(a)}{=}\frac{\Ex{r(t_1,\tau)r(t_2,\tau)}-\bar{r}(t_1,\tau)\bar{r}(t_2,\tau)}{\sqrt{\bar{r}(t_1,\tau)\bar{r}(t_2,\tau)}},
\end{IEEEeqnarray}
}
where equality $(a)$ follows from the fact that under both Poisson and Gaussian statistics, the variance of $r(t,\tau)$ is $\bar{r}(t,\tau)$. The cross-correlation term $\Ex{r(t_1,\tau)r(t_2,\tau)}$ depends on the specific adopted receiver type. Note that by definition, $-1\leq \rho_t(t_1,t_2) \leq 1$ holds. Typically, the sample times $t_1$ and $t_2$ should be separated such that $\rho_t(t_1,t_2)$ falls below a certain threshold, denoted by $\zeta_t$, i.e., $|t_2-t_1|$ should be large enough such that $\rho_t(t_1,t_2)<\zeta_t$ holds. In Fig.~\ref{Fig:Rho_t}, we show the absolute correlation $|\rho_t(t^{\mathrm{p}},t^{\mathrm{p}}+\Delta t)|$ versus $\Delta t$ where $t^{\mathrm{p}}$ denotes the peak of the expected received signal. As can be seen from this figure, the correlation decreases as $\Delta t$ increases. Moreover, as an example, we choose the value of the threshold as $\zeta_t=0.2$.  One can observe from Fig.~\ref{Fig:Rho_t} that as the diffusion coefficient of the molecules increases, the minimum sample spacing $\Delta t$ needed to ensure   $|\rho_t(t^{\mathrm{p}},t^{\mathrm{p}}+\Delta t)|<\zeta_t$ decreases.

\iftoggle{OneColumn}{%
\begin{figure}
	\centering  
	\resizebox{0.7\linewidth}{!}{
		\psfragfig{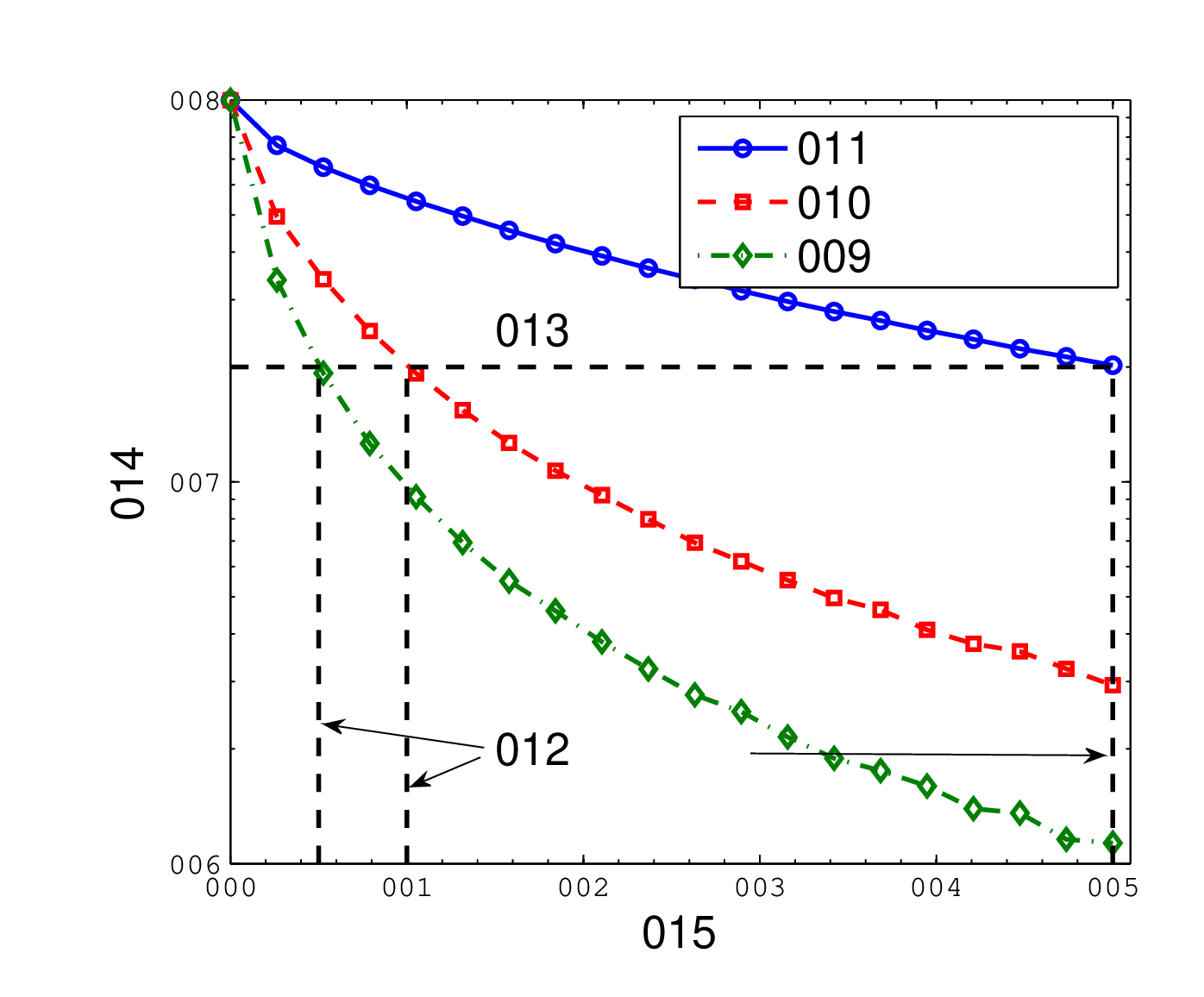}} \vspace{-0.5cm}
	\caption{Absolute correlation $|\rho_t(t^{\mathrm{p}},t^{\mathrm{p}}+\Delta t)|$ versus $\Delta t$ [$\mu$s] where $t^{\mathrm{p}}=\max_t \bar{r}(t,\tau)$ for a point transmitter, an unbounded environment, a passive receiver of radius $a_{\rx}=50$~nm, $d=200$~nm, $\Ntx=2000$, and $D=\{1,5,10\}\times 10^{-11}\,\,\text{m}^2/\text{s}$.}
	\label{Fig:Rho_t}
\end{figure}
}{%
\begin{figure}
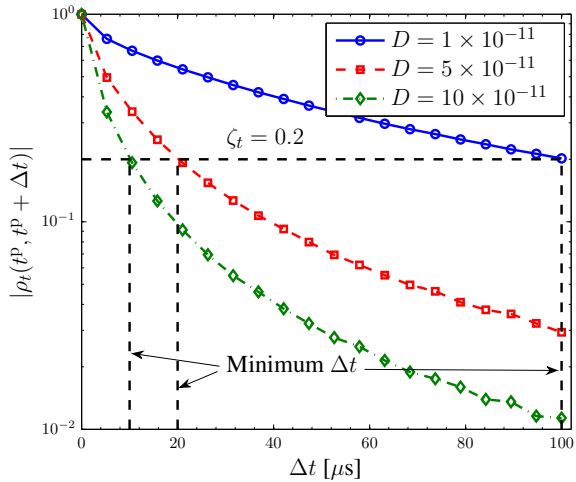

	\centering  
	\resizebox{1\linewidth}{!}{
		\psfragfig{Sections/S4/Fig/Rho_t/Rho_t}} \vspace{-0.5cm}
	\caption{Absolute correlation $|\rho_t(t^{\mathrm{p}},t^{\mathrm{p}}+\Delta t)|$ versus $\Delta t$ [$\mu$s] where $t^{\mathrm{p}}=\max_t \bar{r}(t,\tau)$ for a point transmitter, an unbounded environment, a passive receiver of radius $a_{\rx}=50$~nm, $d=200$~nm, $\Ntx=2000$, and $D=\{1,5,10\}\times 10^{-11}\,\,\text{m}^2/\text{s}$.}
	\label{Fig:Rho_t}
\end{figure}
}

\subsubsection{Mean Correlation}
Recall that if the system parameters change, the mean signal $\bar{r}(t,\tau)$ varies over time scale $\tau$. In a similar manner as for sample correlation, one can define a correlation factor $\rho_{\tau}(\tau_1,\tau_2)$ between the mean signals at time $\tau_1$ and $\tau_2$ as follows
\begin{IEEEeqnarray}{lll}\label{Eq:Rho_tau}
	\rho_{\tau}(\tau_1,\tau_2)=\frac{\Ex{\bar{r}(t,\tau_1)\bar{r}(t,\tau_2)}-m_{\bar{r}}(t,\tau_1)m_{\bar{r}}(t,\tau_2)}{\sigma_{\bar{r}}(t,\tau_1)\sigma_{\bar{r}}(t,\tau_2)}.\quad
\end{IEEEeqnarray}
For the case when transmitter and receiver mobility are the cause of the variations in $\bar{r}(t,\tau)$, cf. Section~\ref{Sec:TimeVarying}, $m_{\bar{r}}(t,\tau)$ and $\sigma_{\bar{r}}(t,\tau)$ are given by (\ref{Eq.mean_h_def}) and (\ref{Eq:Var_TVarying}), respectively. Moreover, the cross-correlation, denoted by $\phi_{\bar{r}}(t,\tau_1,\tau_2)\triangleq \Ex{\bar{r}(t,\tau_1) \bar{r}(t,\tau_2)}$, for two arbitrary times $\tau_1$ and $\tau_2 > \tau_1$ is derived in \cite[Eq.~(19)]{Arman_TimeVariant}
\iftoggle{OneColumn}{%
\begin{IEEEeqnarray}{ll}
	\label{Eq.ACF}
	\phi_{\bar{r}}(t,\tau_1,\tau_2)  = \iint \limits_{\mathbf{d}_1,\,\mathbf{d}_2 \in \mathbb{R}^3} \bar{r}(t,\tau_1|\mathbf{d}(\tau_1) = \mathbf{d}_1)
	\bar{r}(t,\tau_2|\mathbf{d}(\tau_2) = \mathbf{d}_2) f_{\mathbf{d}(\tau_1),\,\mathbf{d}(\tau_2)} \left(\mathbf{d}_1,\, \mathbf{d}_2 \right) \mathrm{d} \mathbf{d}_1 \mathrm{d} \mathbf{d}_2, \nonumber \\ 
	=  \frac{(\Ntx)^2(2\pi)^3 \phi^2 \lambda(\tau_1) \lambda(\tau_2 - \tau_1)}{\big(4\theta(\tau_1,\tau_2)\big)^{3/2}} 
	\exp \left( - \beta\left(\tau_1\right)d_0^2\left[1-\frac{\left(\alpha+\beta(\tau_2-\tau_1)\right)\beta(\tau_1)}{\theta(\tau_1,\tau_2)}\right]\right), 
\end{IEEEeqnarray} 
}{%
\begin{IEEEeqnarray}{ll}
	\label{Eq.ACF}
	\phi_{\bar{r}}(t,\tau_1,\tau_2) \nonumber \\
	 = \iint \limits_{\mathbf{d}_1,\,\mathbf{d}_2 \in \mathbb{R}^3} \bar{r}(t,\tau_1|\mathbf{d}(\tau_1) = \mathbf{d}_1)
	\bar{r}(t,\tau_2|\mathbf{d}(\tau_2) = \mathbf{d}_2) \nonumber \\ \,\,\times f_{\mathbf{d}(\tau_1),\,\mathbf{d}(\tau_2)} \left(\mathbf{d}_1,\, \mathbf{d}_2 \right) \mathrm{d} \mathbf{d}_1 \mathrm{d} \mathbf{d}_2, \nonumber \\ 
	=  \frac{(\Ntx)^2(2\pi)^3 \phi^2 \lambda(\tau_1) \lambda(\tau_2 - \tau_1)}{\big(4\theta(\tau_1,\tau_2)\big)^{3/2}} \nonumber \\
	\,\,\times \exp \left( - \beta\left(\tau_1\right)d_0^2\left[1-\frac{\left(\alpha+\beta(\tau_2-\tau_1)\right)\beta(\tau_1)}{\theta(\tau_1,\tau_2)}\right]\right), \quad
\end{IEEEeqnarray} 
} 
where for compactness $\phi$, $\lambda(\tau)$, $\alpha$, $\beta(\tau)$, and $\theta(\tau_1,\tau_2)$ are respectively defined as 
\iftoggle{OneColumn}{%
\begin{IEEEeqnarray}{C}
	\label{Eq.Notations}
	\phi = \frac{V_{\rx}}{(4 \pi D_1 t)^{3/2}},\, \lambda(\tau) = \frac{1}{(4 \pi D_2 \tau)^{3/2}},\,
	\alpha = \frac{1}{4 D_1 t},\,\beta(\tau) = \frac{1}{4 D_2 \tau},\,\,\text{and} \nonumber \\
	\theta(\tau_1,\tau_2) = \left( \alpha + \beta \left( \tau_1 \right) \right) \left( \alpha + \beta \left( \tau_2 - \tau_1 \right) \right) + \alpha \beta \left( \tau_2 - \tau_1 \right).
\end{IEEEeqnarray}
}{%
\begin{IEEEeqnarray}{C}
	\label{Eq.Notations}
	\phi = \frac{V_{\rx}}{(4 \pi D_1 t)^{3/2}},\, 
	\lambda(\tau) = \frac{1}{(4 \pi D_2 \tau)^{3/2}},\,\nonumber \\
	\alpha = \frac{1}{4 D_1 t},\,\beta(\tau) = \frac{1}{4 D_2 \tau},\,\,\text{and} \nonumber \\
	\theta(\tau_1,\tau_2) = \left( \alpha + \beta \left( \tau_1 \right) \right) \left( \alpha + \beta \left( \tau_2 - \tau_1 \right) \right) + \alpha \beta \left( \tau_2 - \tau_1 \right).\nonumber
\end{IEEEeqnarray}
}

In order to quantify the time variations of the end-to-end \gls{MC} channel, we define the coherence time, $T_{\mathrm{c}}$, as the minimum time $\Delta \tau$  for which $\rho_{\tau}(\tau_1,\tau_1+\Delta \tau)$ falls below a certain threshold value $0 < \zeta_{\tau} < 1 $, i.e., \cite{Arman_TimeVariant}
\begin{IEEEeqnarray}{C}
	\label{Eq.CoherenceTime_def}
	T_{\text{c}} = \operatorname*{arg\,min} \limits_{\forall \Delta \tau > 0} \left( \rho_{\tau}(\tau_1,\tau_1+\Delta \tau) < \zeta_{\tau} \right).
\end{IEEEeqnarray}
The coherence time of the channel is a metric which determines the time over which the channel does not change substantially. As such, the particular choice of $\zeta_{\tau}$ depends on the application of interest. Future applications of synthetic \gls{MC} systems that are more robust to CIR variations can assume smaller values of $\zeta_{\tau}$, whereas applications that are more sensitive to \gls{CIR} variations may require larger values of $\zeta_{\tau}$. For example, typical values of
$\zeta_{\tau}$ reported in the conventional wireless communications literature span the range from $0.5$ to $1$ \cite{lopez2009opportunistic,Lett_CSI,ma2009error}. Smaller values of $\zeta_{\tau}$ are often employed for resource allocation problems, while larger values of $\zeta_{\tau}$ are used for channel estimation.
In  Fig.~\ref{Fig:Rho_tau}, we show the absolute correlation $|\rho_{\tau}(\tau_1,\tau_1+\Delta \tau)|$ versus $\Delta \tau$ for different scenarios of transmitter and receiver mobility, i.e., $D_{\mathrm{tx}}=D_{\mathrm{rx}}=\{0.01,0.05,0.1\}\times D$. As can be seen from Fig.~\ref{Fig:Rho_tau}, the channel mean decorrelates as $\Delta \tau$ increases. Moreover, assuming a fixed threshold $\zeta_{\tau}=0.5$, the coherence time decreases as the diffusion coefficients of the transmitter and receiver increase.

\iftoggle{OneColumn}{%
\begin{figure}
	\centering  
	\resizebox{0.7\linewidth}{!}{
		\psfragfig{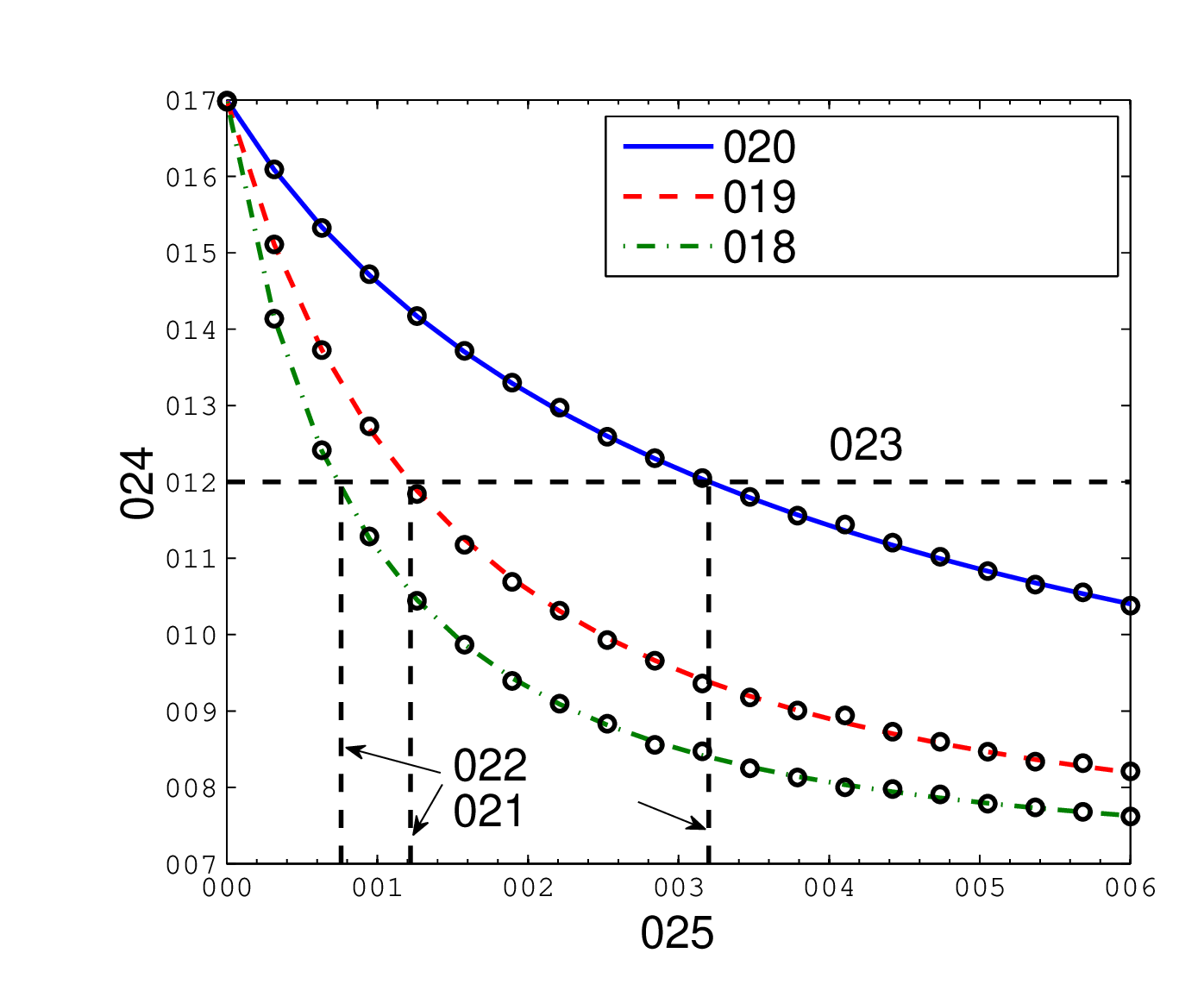}} \vspace{-0.5cm}
	\caption{Absolute correlation $|\rho_{\tau}(\tau_1,\tau_1+\Delta \tau)|$ versus $\Delta \tau$ [ms] for a point transmitter, an unbounded environment, a passive spherical receiver of radius $a_{\rx}=50$~nm, $d_0=200$~nm, $\tau_1=1$~ms, observation time at $t^{\mathrm{p}}=\max_t \bar{r}(t,\tau=0)$, $\Ntx=2000$, $D=10^{-11}\,\,\text{m}^2/\text{s}$, and $D_{\mathrm{tx}}=D_{\mathrm{rx}}=\{0.01,0.05,0.1\}\times D$. Markers denote simulation results and lines denote the analytical results based on (\ref{Eq:Rho_tau}).}
	\label{Fig:Rho_tau}
\end{figure}
}{%
\begin{figure}
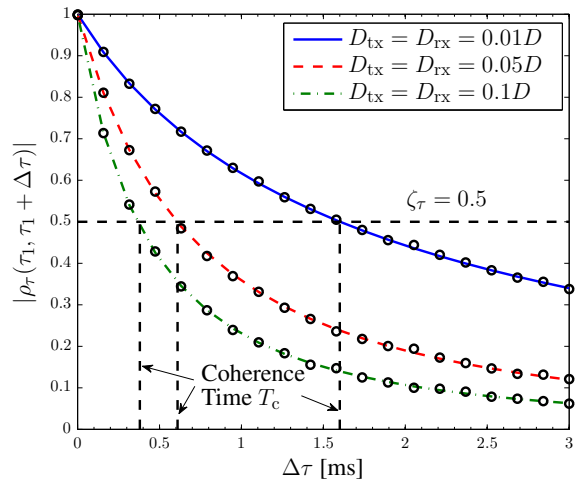

	\centering  
	\resizebox{1\linewidth}{!}{
		\psfragfig{Sections/S4/Fig/Rho_tau_linear/Rho_tau}} \vspace{-0.5cm}
	\caption{Absolute correlation $|\rho_{\tau}(\tau_1,\tau_1+\Delta \tau)|$ versus $\Delta \tau$ [ms] for a point transmitter, an unbounded environment, a passive spherical receiver of radius $a_{\rx}=50$~nm, $d_0=200$~nm, $\tau_1=1$~ms, observation time at $t^{\mathrm{p}}=\max_t \bar{r}(t,\tau=0)$, $\Ntx=2000$, $D=10^{-11}\,\,\text{m}^2/\text{s}$, and $D_{\mathrm{tx}}=D_{\mathrm{rx}}=\{0.01,0.05,0.1\}\times D$. Markers denote simulation results and lines denote the analytical results based on (\ref{Eq:Rho_tau}).}
	\label{Fig:Rho_tau}
\end{figure}
}

\section{Simulation- and Experiment-Driven Models}
\label{Sect:SimExp}


The analytical results presented thus far in this tutorial have focused on tractable solutions based on the underlying physical principles of advection, reaction, and diffusion. In order to arrive at these results, we often had to make assumptions that simplify the physical transmitter, receiver, and channel. However, this approach has limitations. Assumptions are generally constrained by specific channel parameters or the conditions for which they accurately apply. For example, we can assume that an environment's outer boundary is unbounded if it is sufficiently larger than the signaling range; see Fig.~\ref{Fig;BoundedChannels_Analysis} and \cite{Noel2017a}. Similarly, we can model the locally-varying concentration due to a molecule source as uniform if we are observing from a distance that is sufficiently far from the source; see the UCA  in Section~\ref{Sect:TxRxCh} and \cite{NoelPro1}.

Sometimes we are able to relax assumptions and still maintain analytical tractability, cf. Section~\ref{Sect:TxRxCh}. When this occurs, we can define a reliable rule of thumb that dictates explicit conditions under which the assumption can be satisfied to some degree of accuracy. For instance in \cite{NoelPro1}, it was shown that the simplified \gls{CIR} with UCA was within 2\,\% of the ideal \gls{CIR} for most of the time of interest if the radius of a spherical receiver was no more than 15\,\% of the distance from the molecule source to the center of the receiver. However, we generally do not have the option to relax assumptions for analytical tractability while maintaining sufficient accuracy. Furthermore, we might encounter a channel with complex or novel phenomena where we do not yet know what suitable assumptions might be.

In the absence of reliable analytical results, we must rely on data-driven approaches to model a channel. Such approaches can also be used to help verify analytical results. This section reviews simulation and experimental approaches for generating data. Simulations can provide an efficient means for channel modeling, even in the presence of complex and coupled physical phenomena. Reliable experimental data may be preferred, but can be time-consuming and expensive to obtain.


\subsection{Simulation-Driven Models}\label{Sec:Sim}

Simulations of reaction-diffusion systems can be performed over a range of physical scales. As such, there are a range of simulation classes available, which we summarize in Fig.~\ref{Fig:simulation_scales} and also discussed in \cite{Noel2017a}. We refer to these classes as \emph{continuum simulations}, \emph{mesoscopic simulations}, \emph{microscopic simulations}, and \emph{molecular dynamics simulations}. Generally, each class is suitable for a particular scale. Not surprisingly, there is an inherent trade-off between the physical resolution of a simulation and the computational resources (whether measured in time or memory) that are required to simulate it. The continuum approach is most suitable for macroscale systems. Both the microscopic and mesoscopic approaches can be appropriate for microscale systems. The molecular dynamics approach is most suitable for systems at the nanoscale and smaller. While the microscopic approach has been the most common simulation method within the \gls{MC} research community, here we discuss all four approaches, their relevance, and also the potential to combine them in a single simulation. For the microscopic and mesoscopic approaches, we also describe how to implement a simple simulation.

\subsubsection{Continuum Simulations}

\begin{figure}[!t]
	\centering
	\includegraphics[width=0.95\linewidth]{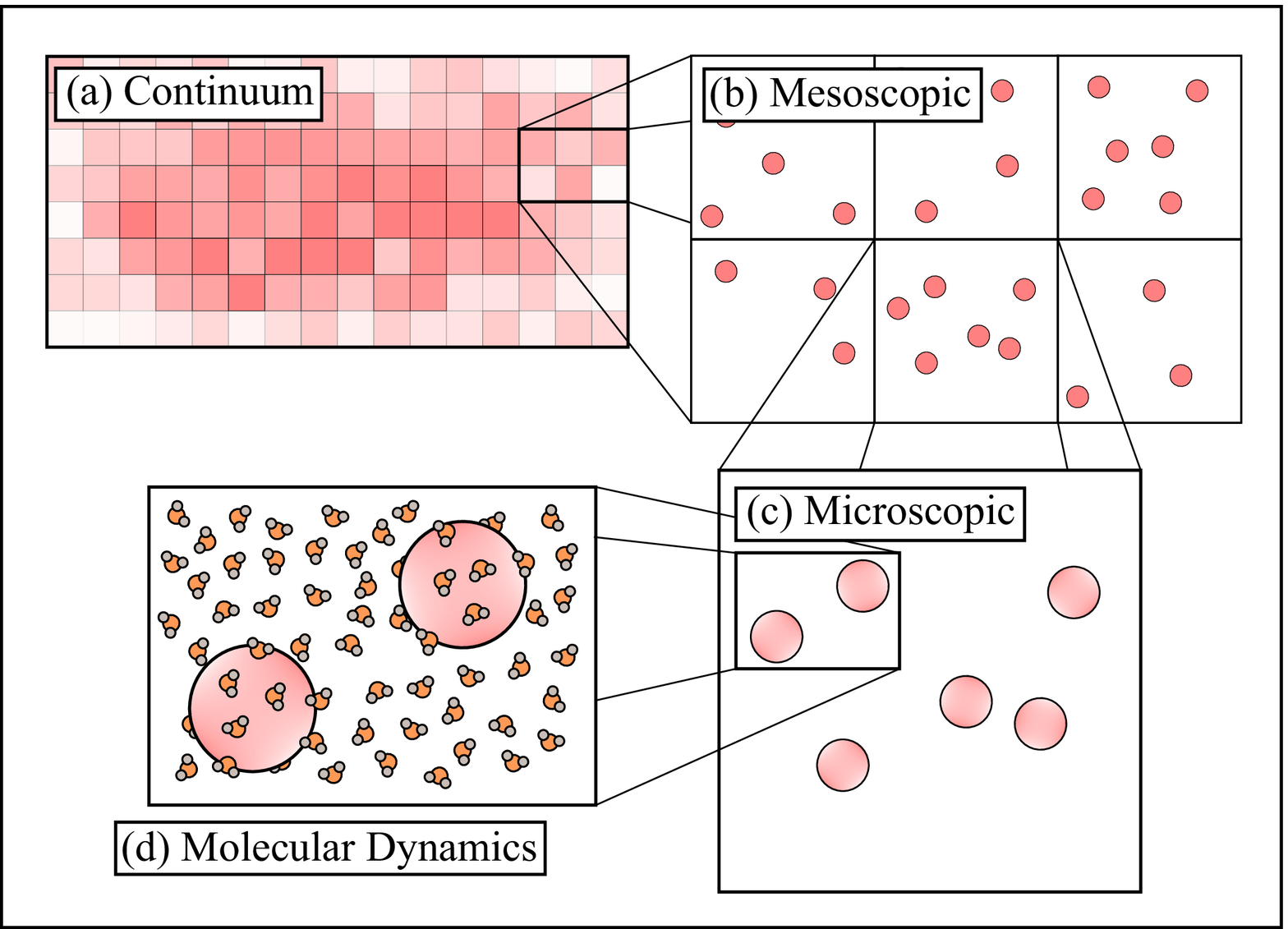}
	\caption{Physical scales of molecular simulation. a) Continuum simulations solve the PDEs that describe the system. Molecular concentrations are non-negative and real-valued. b) Mesoscopic simulations proceed as a sequence of events, where each event is an occurrence of a chemical reaction or a molecule moving between adjacent subvolumes. c) Microscopic simulations individually track each molecule of interest. The solute molecules diffuse within a continuum of solvent molecules. d) Molecular dynamics simulations model \emph{all} individual atoms and molecules, including intermolecular forces and collisions.}
	\label{Fig:simulation_scales}
\end{figure}

When the physical scale of a simulation, including the number of molecules, is sufficiently large, then the evolution of the system can be directly described using the corresponding spatio-temporal \glspl{PDE}, see e.g. \eqref{Eq:Fick_general}. We refer to these as continuum simulations. Specifically, finite element analysis is used to spatially partition the system into a grid (see Fig.~\ref{Fig:simulation_scales}a)), and the system is simulated over a sequence of time steps. The molecule concentrations at each node in the grid are updated in every time step according to the differential equations that describe the phenomena. The updated concentrations are always non-negative real values. Popular commercial solvers that follow this approach include COMSOL Multiphysics \cite{COMSOL} and ANSYS \cite{ANSYS}.

Unless the differential equations are stochastic or explicit noise sources are introduced, then continuum simulation of a system is deterministic. Generally, the accuracy depends on the resolution of the grid and the size of the time step; more accurate simulations can be performed by increasing the grid resolution and decreasing the size of the time step. However, as the nodes in the grid become increasingly close, the number of molecules associated with each node decreases. When the molecule concentrations get sufficiently small, it becomes more appropriate to consider \emph{integer} numbers of molecules instead of continuous-valued concentrations. Thus, we next discuss mesoscopic simulations.

\subsubsection{Mesoscopic Simulations}

Like continuum simulations, mesoscopic simulations also partition the system into a grid. The resulting containers are commonly referred to as \emph{subvolumes} or \emph{voxels}. However, instead of tracking continuous molecule concentrations, mesoscopic modeling counts discrete numbers of molecules in each subvolume; see Fig.~\ref{Fig:simulation_scales}b). Instead of deterministically solving the system's set of \glspl{PDE}, mesoscopic simulations proceed by stochastically generating \emph{event} times, where each event is the occurrence of a chemical reaction or a molecule's transition between two subvolumes. The key assumptions of this scheme are that \textit{i)} molecules within a given subvolume are uniformly distributed, and \textit{ii)} the solvent molecules in a subvolume can be treated as a homogeneous continuum that is in thermal equilibrium. When these assumptions are satisfied, the mesoscopic approach can simulate chemical reactions \emph{exactly} (in a statistical sense, as proven in \cite{Gillespie1992}). Furthermore, if the subvolume sizes are appropriately chosen, advection-reaction-diffusion systems can also be simulated exactly; see \cite{Ramaswamy2011,noel_advection}. An important constraint is that the subvolume size (e.g., cube length) $\ell$ must be much smaller than both $\sqrt{2nDt_\mathrm{r}}$ and $2D/|v|$, where $n$ is the dimension of the subvolume, $t_\mathrm{r}$ is the characteristic time of the fastest reaction in the system, $D$ is the diffusion coefficient of the largest corresponding reactant, and $|v|$ is the magnitude of the flow velocity. If this is not satisfied, then we cannot safely assume that the subvolumes are well-stirred (i.e., that the molecules are uniformly distributed).

\textbf{Simple Implementation:} A basic implementation of a mesoscopic simulation with equal-sized subvolumes (of length $\ell$) is as follows. Let $U_{s,m}$ be the number of molecules of the $m$-th type that are in the $s$-th subvolume. Events are associated with propensities. The propensity $\alpha_{s,q,m}$ of a transition of a molecule of the $m$-th type to diffuse from the $s$-th subvolume to the $q$-th subvolume, where these two subvolumes are adjacent and share a face, is \cite[Eq.~(1.6)]{Flegg2014}
\begin{equation}
\alpha_{s,q,m} = \frac{D_m}{\ell^2}U_{s,m},
\end{equation}
where $D_m$ is the diffusion coefficient of the $m$-th molecule type. The propensity $\beta_{s,p}$ of the $p$-th chemical reaction in the $s$-th subvolume is \cite[Eq.~(6)]{Bernstein2005}
\begin{align}
\beta_{s,p} = &\, \kappa_pV, \\
\beta_{s,p} = &\, \kappa_pU_{s,m}, \\
\beta_{s,p} = &\, \frac{\kappa_pU_{s,m}U_{s,n}}{V},
\end{align}
for zeroth-, first-, and second-order reactions, respectively, where the order corresponds to the number of reactants. $\kappa_p$ is the corresponding reaction rate constant, $V$ is the subvolume volume, and $U_{s,m}$ and $U_{s,n}$ are the corresponding  numbers of reactant molecules. For the entire system, the \emph{total} propensity $\gamma_{\mathrm{tot}}$ is then
\begin{equation}
\gamma_{\mathrm{tot}} = \sum_{s,q,m} \alpha_{s,q,m} + \sum_{s,p} \beta_{s,p},
\end{equation}
and we can simulate the time $t_\mathrm{next}$ of the next event in the system by generating exponential random variable
\begin{equation}
t_\mathrm{next} = -\frac{\log u}{\gamma_{\mathrm{tot}}},
\end{equation}
where $u$ is a random number uniformly distributed between 0 and 1. We can determine which of the possible events occurred by tossing a weighted die, where the likelihood of each event is proportional to its associated propensity. Once the event is determined, we update the molecule counts, update the corresponding propensities, and repeat the process to find the next event.

\begin{remk}
We note that there are mathematically equivalent but more computationally efficient implementations, particularly when updating propensities. These include Gibson and Bruck's Next Reaction Method; see \cite{Gibson2000}. Furthermore, different accuracy-efficiency trade-offs can be introduced to provide more flexible scalability. For example, \emph{tau-leaping} can be used to execute multiple events in a constant time step, where ``tau'' refers to the time step size; see \cite{Gillespie2001}. Tau-leaping enables a transition between continuum and mesoscopic simulations; if the number of events during one ``leap'' is sufficiently large, then it can be treated as a deterministic value. As long as the propensities do not significantly change between time steps, then tau-leaping's computational efficiency gains can be made with minimal losses in accuracy. \QEDwhite
\end{remk}

\subsubsection{Microscopic Simulations}

In some sense, microscopic simulations are the dual of the mesoscopic approach. Whereas the (non-leaping) mesoscopic approach is continuous over time and discrete over space, the common microscopic approach implementation is discrete over time and continuous over space; see \cite{Andrews2004}. Instead of relying on well-stirred subvolumes, microscopic simulations track every molecule individually (i.e., particle-based simulation); see Fig.~\ref{Fig:simulation_scales}c). Nevertheless, they still assume that the solvent is a continuum of molecules, which means that the diffusion of the molecules of interest is still governed by a diffusion coefficient.

\textbf{Simple Implementation:} A basic implementation of a microscopic simulation with flow and first-order reactions in the propagation environment is as follows. In each time step $\Delta t$, every molecule is tested for every possible first-order reaction. If there is only one potential reaction, and the associated reaction rate is $\kappa$, then the corresponding reaction probability $P_\mathrm{rxn}$ is \cite[Eq.~(14)]{Andrews2004}
\begin{equation}
P_{\mathrm{rxn}} = 1-\exp\left(-\kappa\Delta t\right).
\end{equation}

If a coin flip with this probability is successful, then the molecule is converted to the corresponding reaction product. After all of the possible reactions have been tested, the remaining molecules are diffused along every available dimension by adding a displacement of $\sqrt{2D\Delta t}\times\mathcal{N}(0,1)$ towards each dimension of the Cartesian coordinate system, cf. (\ref{Eq:RandomWalk}). The realizations are independent for every molecule and along every dimension. Furthermore, if the environment has a bulk flow with a component $v$ along a particular dimension, then every molecule should have an additional displacement of $v\Delta t$ along that dimension, cf. (\ref{Eq:Flow}). Diffusion should be unimpeded, unless there are boundaries in the environment. For example, if a molecule crosses a solid reflective surface, then the coordinate that is normal to the surface is reverted to its value before diffusion. If a molecule crosses an absorbing surface, then it should be consumed by the absorbing reaction.

Due to their simplicity and their suitability for simulations over a range of nanometers to micrometers, microscopic simulations have been common for cellular systems and also specifically for \gls{MC} systems. Mature tools from the physical chemistry community include Smoldyn (see \cite{Andrews2004,Andrews2009}). Microscopic tools that have been developed specifically for the \gls{MC} community include BiNS2 \cite{Felicetti2013}, N3Sim \cite{Llatser2011}, MUCIN \cite{Yilmaz2014a}, and AcCoRD \cite{Noel2017a}.

\subsubsection{Molecular Dynamics Simulations}

At a more precise scale, solvent molecules and their interactions with solute molecules and with each other can be modeled in detail; see Fig.~\ref{Fig:simulation_scales}d). These are molecular dynamics simulations, and they might account for intermolecular forces (including those imposed by charge potentials) and collision dynamics. One such example is the Large-scale Atomic/Molecular Massively Parallel Simulator (LAMMPS); see \cite{Plimpton1995}. Due to the very large density of molecules to be considered, molecular dynamics simulations are best suited for very small systems, e.g., on a nanoscale. For instance, molecular dynamic simulations can be used to study how the conformation of receptor proteins change after binding to a specific molecule. Thus, they have generally not been applied to study \gls{MC} systems.

\begin{remk}[Hybrid Simulations]
The aforementioned discussion of simulation classes has emphasized their suitability for simulations over different physical scales. However, a particular system might have multiple scales of interest. In order to avoid constraining the entire simulation by the most granular approach needed, hybrid simulation tools have sought to integrate different classes within a single simulation. One approach has been to combine microscopic and mesoscopic models, using hybrid interfaces such as that proposed in \cite{Flegg2014} and later implemented in Smoldyn (see \cite{Robinson2015}) and AcCoRD (see \cite{Noel2017a}). Other examples include the integration of the molecular dynamics solver LAMMPS with a continuum model (see \cite{Wagner2008}), and the integration of the continuum solver Virtual Cell with the microscopic approach in Smoldyn (see \cite{Resasco2012}). \QEDwhite
\end{remk}

\begin{examp}[Example Simulation]
We complete our discussion of simulations with a brief demonstration. We consider an extension of the bounded rectangular-duct channel discussed in Section III-D that has no readily available analytical channel response. Nevertheless, we can simulate the system. The environment is a microfluidic system where two chambers are connected via a long pipe, as shown and described in Fig.~\ref{Fig:simulation_env}. We place $\Ntx=500$ molecules uniformly within one of the chambers (i.e., a cube). These molecules can diffuse out through the rectangular pipe and into the other chamber and no flow is considered. The second chamber has a perfectly-absorbing surface and we count the number of molecules that are absorbed. We assume that the receiver counts the number of molecules absorbed by time $t$. Therefore, the receiver can be classified as nR-AMC, i.e., non-recurrent and accumulative-molecule-counting, with received signal $r(t)=n^{\mathrm{arv}}(t)$, cf. Section~\ref{Sec:UnifSig}. A realization of this system is simulated using a microscopic approach in the AcCoRD simulator with a simulation time step of $\Delta t = 1\,$ms. The number of absorbed molecules $r(t)$ is plotted in Fig.~\ref{Fig:simulation_obs} for different pipe lengths. We see that all released molecules are absorbed within about $850\,$s for the shortest pipe length (i.e., $60\,\mu$m). As the distance between the two chambers increases, fewer molecules get absorbed within the same time. We note that one can obtain the \gls{CIR} of this system, i.e., $h(t)$ defined as the probability of a molecule being absorbed at the receiver in interval  $(0,t]$ after its release by the transmitter at $t=0$, by simulating the system for many realizations and averaging the result, i.e., $h(t)=\Ex{r(t)/\Ntx}$. \QEDwhite
\end{examp}

\iftoggle{OneColumn}{%
\begin{figure}[!t]
	\centering
	\includegraphics[width=0.8\linewidth]{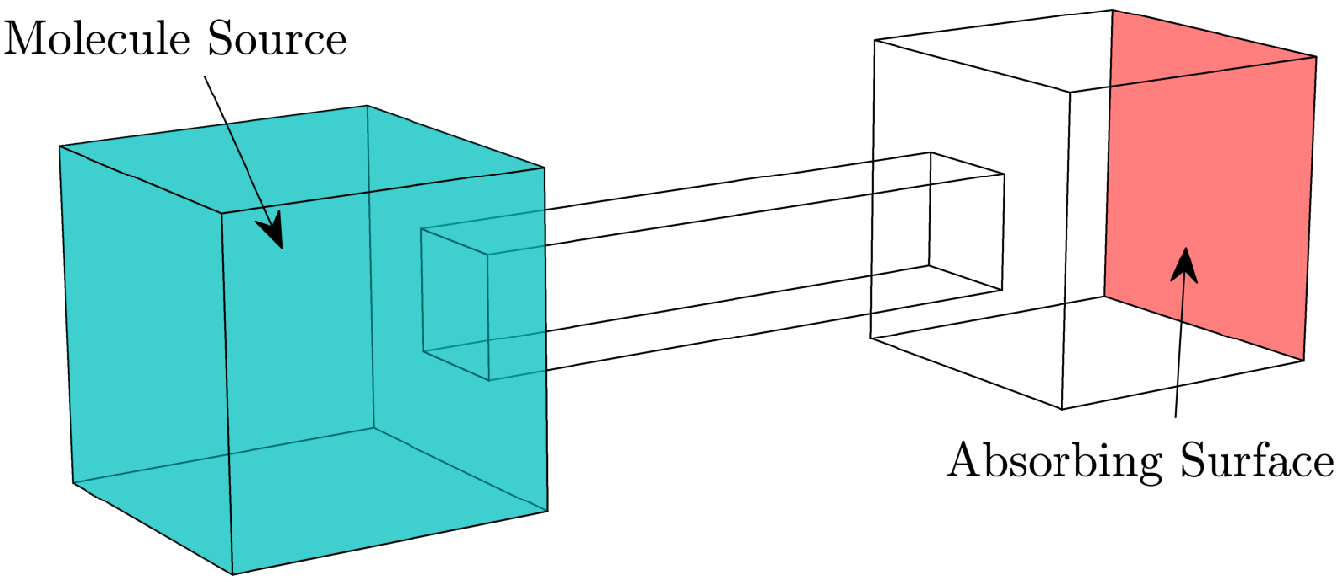}
	\caption{Example environment to simulate. This ``dumbbell''-shaped environment represents two connected microfluidic chambers with a rectangular duct between them. It  is composed of two cubes of length $32\,\mu$m that are connected by a rectangular pipe of size $60\,\mu$m x $12\,\mu$m x $12\,\mu$m. We can also vary the length of the pipe. The left cube has molecules initialized throughout it. The right cube has an absorbing surface on the far side. An analytical channel response for this environment is not readily available.}
	\label{Fig:simulation_env}
\end{figure}	
}{%
\begin{figure}[!t]
	\centering
	\includegraphics[width=0.9\linewidth]{Sections/S5/Fig/Simulation_RxTx/fig_0057_env}
	\caption{Example environment to simulate. This ``dumbbell''-shaped environment represents two connected microfluidic chambers with a rectangular duct between them. It  is composed of two cubes of length $32\,\mu$m that are connected by a rectangular pipe of size $60\,\mu$m x $12\,\mu$m x $12\,\mu$m. We can also vary the length of the pipe. The left cube has molecules initialized throughout it. The right cube has an absorbing surface on the far side. An analytical channel response for this environment is not readily available.}
	\label{Fig:simulation_env}
\end{figure}	
}

\iftoggle{OneColumn}{%
\begin{figure}[!t]
	\centering
	\includegraphics[width=0.7\linewidth]{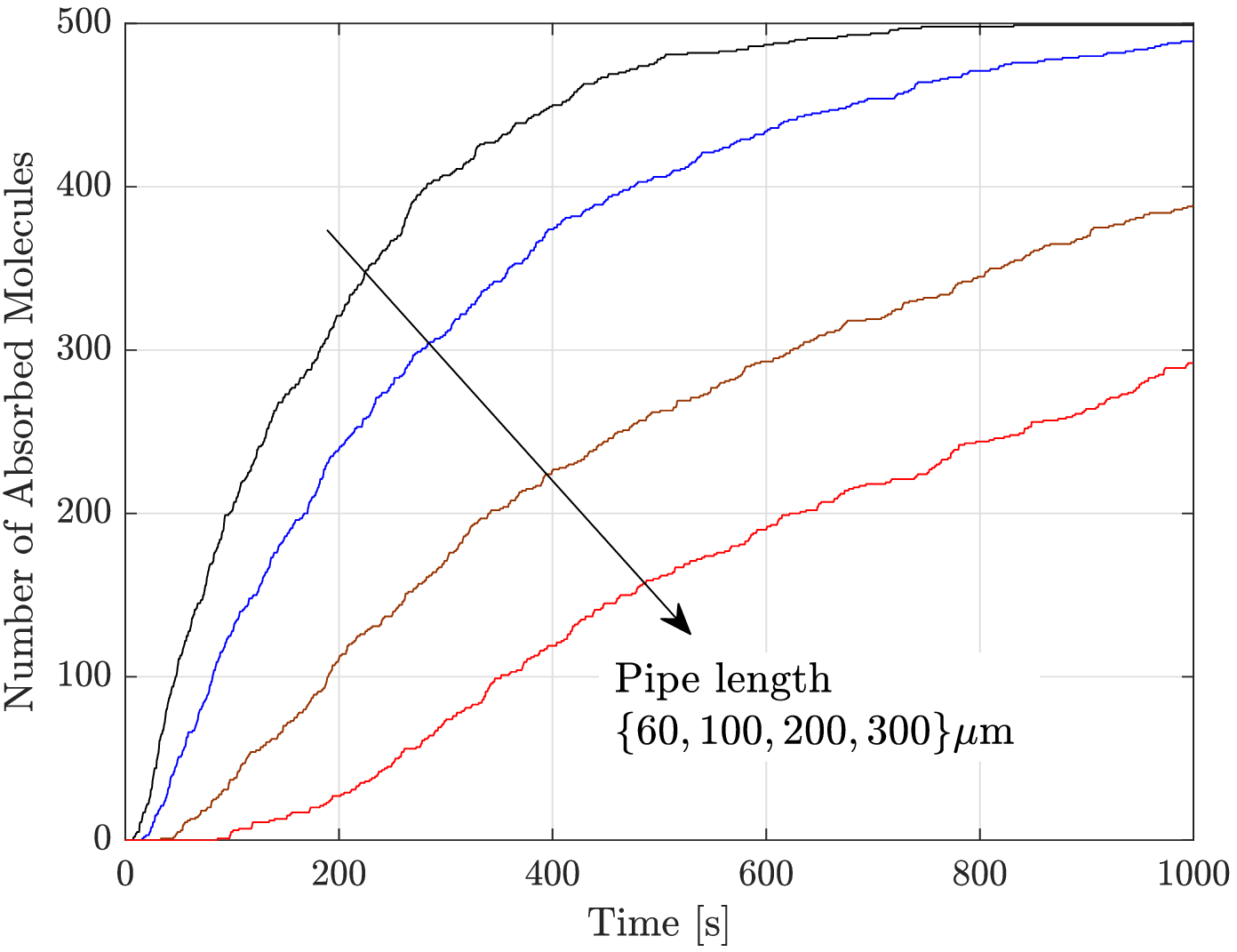}
	\caption{Number of absorbed molecules in the simulation of the system described in Fig.~\ref{Fig:simulation_env}. 500 molecules are instantaneously released throughout the molecule source, which freely diffuse with a diffusion coefficient of $10^{-10}\,\frac{\mathrm{m}^2}{\mathrm{s}}$. A single realization is shown for different pipe lengths using a simulation time step of $1\,$ms.}
	\label{Fig:simulation_obs}
\end{figure}	
}{%
\begin{figure}[!t]
	\centering
	\includegraphics[width=1\linewidth]{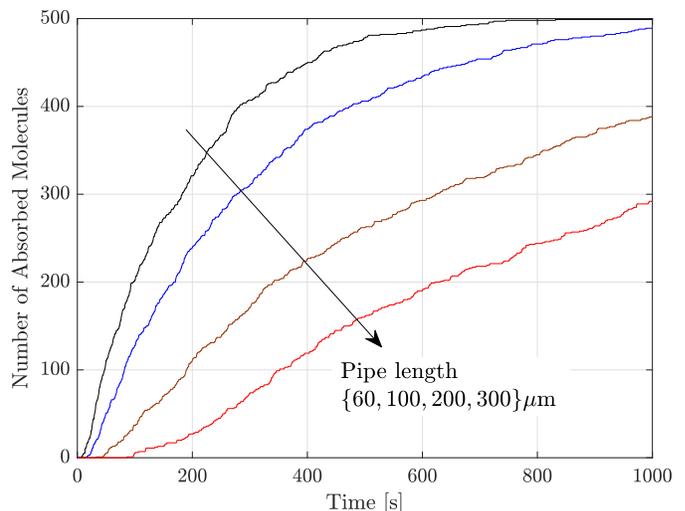}\vspace{-0.5cm}
	\caption{Number of absorbed molecules in the simulation of the system described in Fig.~\ref{Fig:simulation_env}. 500 molecules are instantaneously released throughout the molecule source, which freely diffuse with a diffusion coefficient of $10^{-10}\,\frac{\mathrm{m}^2}{\mathrm{s}}$. A single realization is shown for different pipe lengths using a simulation time step of $1\,$ms.}
	\label{Fig:simulation_obs}
\end{figure}	
}

\subsection{Experimentally-driven Models}
In the previous subsection, we have seen how elaborate simulations can be used for scenarios where it is difficult or even impossible to derive an analytical model based on physical principles. However, for practical systems, we may also face situations where even simulation of certain phenomena is challenging.  In fact, even complex simulation methods typically cannot  account for all characteristics of a real experimental environment. In the following, we first highlight some of the unique characteristics of two existing experimental platforms for \gls{MC} \cite{farsad_tabletop_2013,grebenstein_biological_2018} that cannot be easily modeled or simulated. Subsequently, to cope with the aforementioned challenges, we present a general data-driven modeling approach which is then applied to an example experimental system.

\subsubsection{Challenges in Modeling Existing Experimental Systems}
Several experimental systems exist for demonstrating \gls{MC}.
These testbeds include both  non-biological systems \cite{farsad_tabletop_2013,Nariman_AcidBasePlatform,atthanayake2018experimental,tuccitto_fluorescent_2017,kennedy_spatiotemporal_2018,giannoukos_molecular_2017,deleo_communications_2013,unterweger_experimental_2018} and biological systems \cite{krishnaswamy_timeelapse_2013,felicetti_modeling_2014,Nakano_Microplatform_2008,bicen_efficient_2015,grebenstein_biological_2018}. To show the need for experimentally-driven models, we review the challenges of  channel modeling for two of these testbeds.

\begin{examp}[Non-biological Testbed \cite{farsad_tabletop_2013}]
The first experimental \gls{MC} system was presented in \cite{farsad_tabletop_2013} and is based on spraying and detecting alcohol in open space. In \cite{farsad_channel_2014}, it was shown that a simple model based on diffusion and the flow generated by a fan cannot accurately explain the measurements obtained from the testbed in \cite{farsad_tabletop_2013} due to system nonlinearities whose exact cause is not known. For example, the spray that is used for releasing the chemicals may not produce consistently-sized droplets in the spray stream across different trials, the flow may show turbulent behavior that is difficult to model, and  the receiver sensor is prone to long response and recovery times. \QEDwhite
\end{examp}

\begin{examp}[Biological Testbed \cite{grebenstein_biological_2018}]\label{Exmp:BioTestbed}
The biological \gls{MC} testbed reported in \cite{grebenstein_biological_2018} converts an electrically controlled optical signal into a chemical signal. In particular, for this testbed, \textit{E.~coli} bacteria were genetically modified to incorporate light-driven proton pumps in their cell membranes. Upon a light stimulus, the modified bacteria then pump protons out into the environment which increases the proton concentration outside the bacteria. The resulting proton concentration was measured by a pH sensor playing the role of the receiver. Although complex models were developed in the biology literature for describing the proton release rate of proton pumps as a function of a given induced optical intensity \cite{zifarelli2008buffered,Choi_Cyanobacterial_2014,Lanyi_Bacteriorhodopsin_2004},  they typically do not account for all of the dynamics inherent to living cells. In fact, the growth, dying, and varying living conditions of the bacteria due to constant exposure to light may impact the channel model of the \gls{MC} system in \cite{grebenstein_biological_2018} and cannot be easily captured analytically or via simulation.	 \QEDwhite
\end{examp}

We note that similar inherent randomness and nonlinearities as discussed for the two examples above also exist for other experimental testbeds \cite{Nariman_AcidBasePlatform,atthanayake2018experimental,tuccitto_fluorescent_2017,kennedy_spatiotemporal_2018,giannoukos_molecular_2017,deleo_communications_2013,unterweger_experimental_2018,krishnaswamy_timeelapse_2013,felicetti_modeling_2014,Nakano_Microplatform_2008,bicen_efficient_2015} and are challenging to model
analytically or even simulate since their exact cause is unknown.

\iftoggle{OneColumn}{%
	\begin{figure}[t]
		\centering
		\begin{minipage}{0.49\linewidth}
			\includegraphics[scale=0.5]{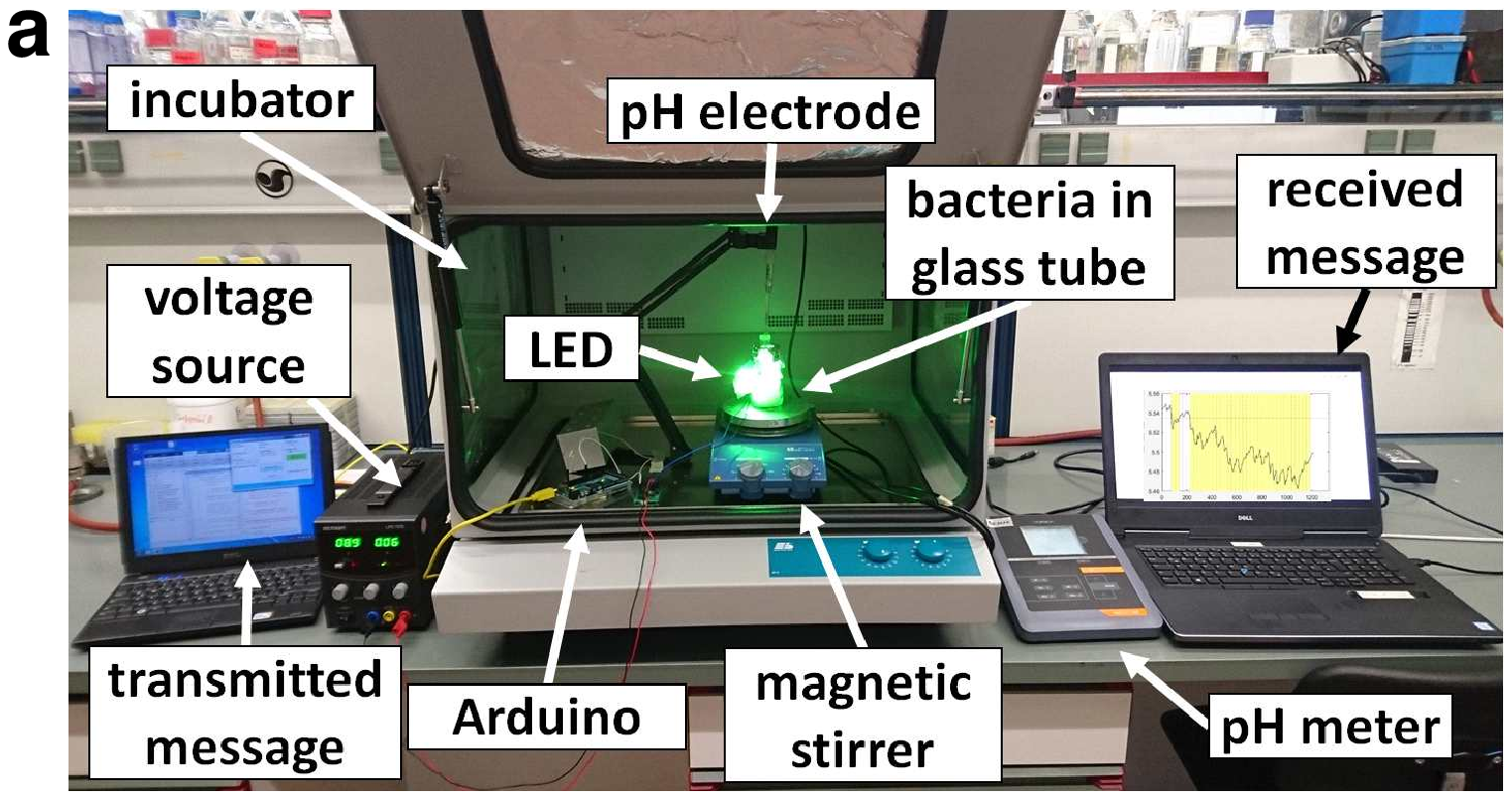}
		\end{minipage}
		\begin{minipage}{0.49\linewidth}
			\includegraphics[scale=0.48]{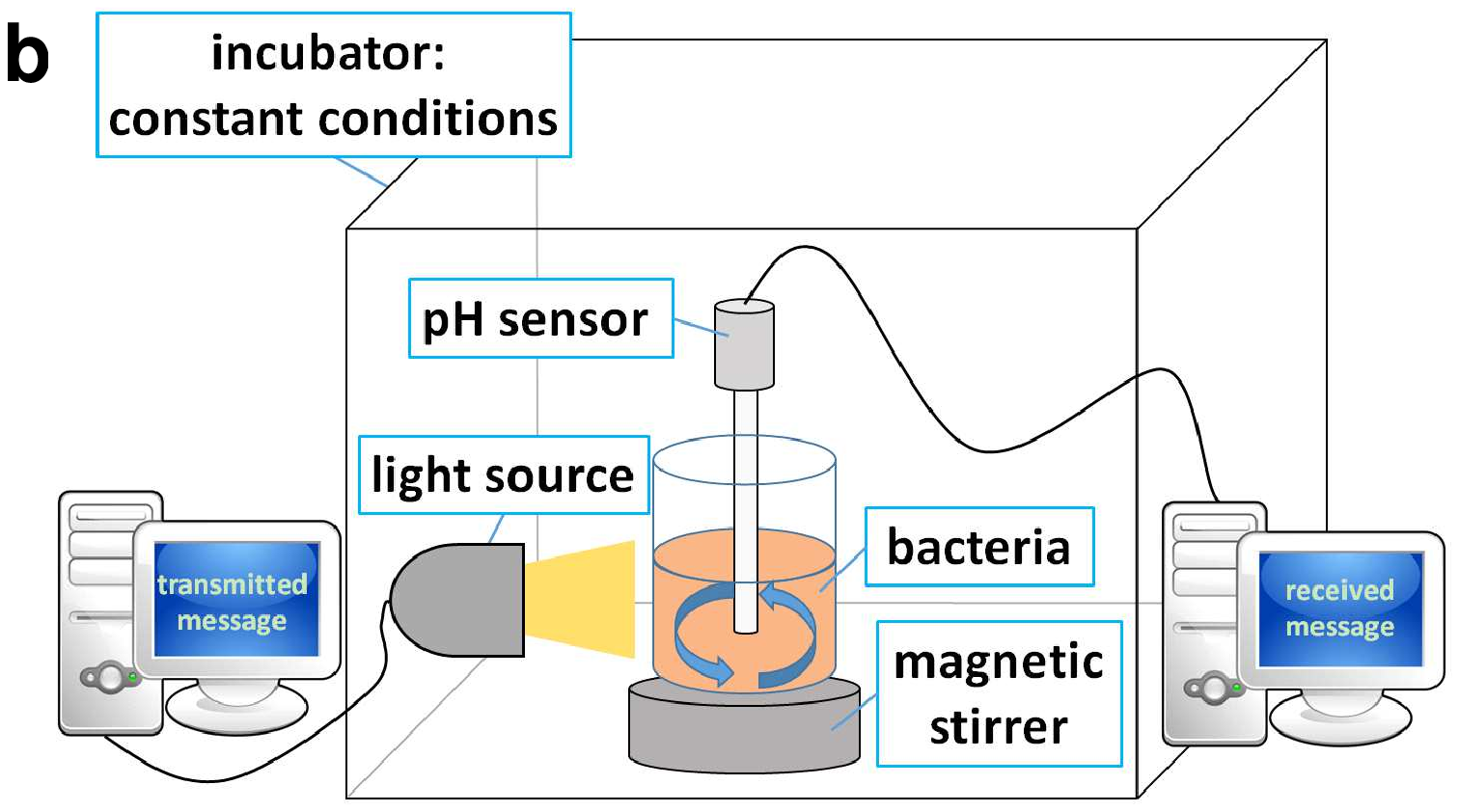}
		\end{minipage}	
		\caption{%
			Biological testbed.
			(a) Benchtop experimental setup;
			(b) Schematic illustration.
			Taken from \cite{grebenstein_biological_2018}.%
		}
		\label{Fig:BacteriaTestbed}
	\end{figure}	
}{%
	\begin{figure*}[t]
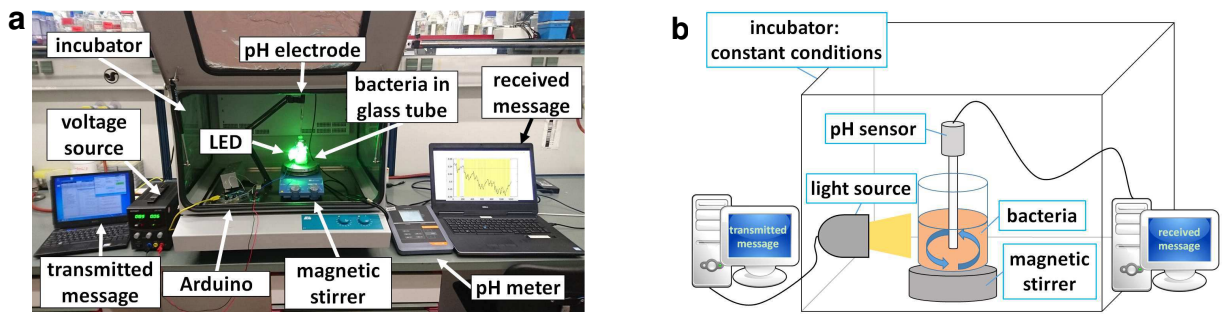

		\begin{center}
			\begin{minipage}{0.04\linewidth}
			\end{minipage}
			\begin{minipage}{0.48\linewidth}
				\includegraphics[scale=0.5]{Sections/S5/Fig/Expriment_TxRx/TxRx1}
			\end{minipage}
			\begin{minipage}{0.48\linewidth}
				\includegraphics[scale=0.48]{Sections/S5/Fig/Expriment_TxRx/TxRx2}
			\end{minipage}
		\end{center}	
		\caption{%
			Biological testbed.
			(a) Benchtop experimental setup;
			(b) Schematic illustration.
			Taken from \cite{grebenstein_biological_2018}.%
		}
		\label{Fig:BacteriaTestbed}
	\end{figure*}	
}

\subsubsection{Data-Driven Model}

To address the aforementioned shortcomings of analytical and simulation models, we propose to employ data-driven models to account for the unpredictable randomness and nonlinearities of real \gls{MC} systems. The basic idea behind these models is to select an appropriate parametric model and choose its corresponding parameters to fit the measurement data. In the following, we describe two different approaches for selecting a suitable parametric model.

\textbf{Physically-Motivated Parametric Models:} Here, the model is chosen based on physics' first principles. For instance, in \cite{farsad_channel_2014}, a mathematical model is developed for the testbed in \cite{farsad_tabletop_2013} which is based on the solution to the advection-diffusion equation with uniform flow, cf.  \eqref{Eq:Solution_DiffAdvec}. Nevertheless, the parameters of the original analytical model were modified to fit the model to the experimental data. As another example, we consider the system in \cite{unterweger_experimental_2018}. The model uses magnetic nanoparticles in duct flow that are detected upon moving through a coil enclosing the duct. Here, the parametric model is based on laminar flow, cf. \eqref{Eq:Poiseuille}, and depends on the initial distribution of the particles released across the cross-section of the duct. The adopted parametric model was then shown to accurately model the complex advection-diffusion process in the duct after fitting its parameters to the measurement data. In general, after choosing the parametric model, standard curve fitting toolboxes can be employed to find the model parameters. One common approach is to use the parameter set that minimizes the mean square error between the model and the measurement data \cite{farsad_channel_2014,unterweger_experimental_2018,grebenstein_biological_2018}.

\textbf{Blind Models based on Neural Networks:} In the absence of an appropriate physically-motivated model, an alternative option is to employ blind models based on  neural networks to jointly learn the model and its parameters \cite{goodfellow2016deep,lee2017machine}. One suitable network architecture for this purpose is the generative adversarial network (GAN) which is able to generate a model that creates artificial data very similar to the measurement data \cite{goodfellow2014generative}. The advantage of such blind parametric models is that they can be universally applied to general \gls{MC} systems, whereas physically-motivated models have to be carefully chosen  according to the \gls{MC} system under consideration. On the other hand, the parameters of a physically-motivated model have physical meaning, which is not the case for the parameters of a trained neural network. The other challenge of channel modeling based on neural networks is that they typically require much more experimental data than parametric models to construct the model. This is not surprising since without domain knowledge, the number of parameters to be learned for a neural network is much larger than that for a parametric model.


\subsubsection{Example of an Experimentally-driven Model}

In order to further familiarize the reader with the main steps of developing an experimentally-driven channel model and to highlight some peculiarities that may arise, we present the modeling methodology for the biological testbed in \cite{grebenstein_biological_2018} in some detail, cf. Example~\ref{Exmp:BioTestbed} and Fig.~\ref{Fig:BacteriaTestbed}.

\textbf{Simple Physically-Motivated Parametric Model:} In order to arrive at an analytical model, the following assumptions are made in \cite{grebenstein_biological_2018}.
It is assumed that the bacteria (i.e., the transmitter) are uniformly distributed in their container and that all bacteria are subject to the same light stimulus at the same intensity because the bacteria suspension is continuously stirred. It is further assumed that the bacteria begin and stop  pumping protons (i.e., signaling molecules) instantly when the light is activated and deactivated, respectively. Furthermore, it is assumed that the bacteria take up protons in a passive manner, i.e., protons are consumed by the bacteria which lowers the measured proton concentration.
Finally, it is assumed that the pH measuring device (i.e., the receiver) is passive and that its presence does not change the proton concentration or the behavior of the bacteria. These assumptions do not strictly hold but are reasonable in consideration of the size of the setup, the pumping speed of a proton pump, and the characteristics of the  bacteria~\cite{grebenstein_biological_2018}. We note that counting the individual molecules observed at the receiver, $r(t)$, might be a reasonable assumption for  nanomachines; however, for experimental testbeds such as \cite{grebenstein_biological_2018}, computing $r(t)$ is not feasible. Hence, in \cite{grebenstein_biological_2018}, the proton concentration obtained from the measured pH was considered as the received signal and was modeled as 
\begin{equation}
\label{Eq:Signal_expriment}
r_c(t)=\frac{r(t)}{V_{\rx}}=\cb(t)+w(t), 
\end{equation}
where $\cb(t)$ is the expected proton concentration and $w(t)$ is a random additive noise. It was shown that $w(t)$ follows a Gaussian distribution and this was justified using the \gls{CLT} since $w(t)$ consists of different types of noises including diffusion (counting) noise, pH sensor circuitry noise,
and the noise inherent to the biological machinery of the bacteria. We assume that the light stimulates the bacteria over a time interval (i.e., a rectangular input pulse) since an impulsive stimulus (i.e., a delta input) does not effectively stimulate the bacteria to release a sufficient number of protons into the \gls{MC} channel, see Remark~\ref{Remk:Pulse}.  Under the aforementioned assumptions,  the expected proton concentration $\cb(t)$, depending on whether the light is on ($i=1$) or off ($i=0$), can be obtained as
\begin{equation}
    \label{Eq:BaselineConcentration}
    \cb(t) = \cb(t_0) + (\cbinf-\cb(t_0)) \left(1 - \exp\left(-\frac{t-t_0}{\tau_i}\right)\right),
\end{equation}
where $\cb(t_0)$ is the initial concentration at starting time $t_0$, $\cbinf$ is the saturation concentration, and $\tau_i$ is a time constant.
The parameters of this model are $\cb(t_0)$, $\cbinf$, $t_0$, and $\tau_i$, which are found using nonlinear least square error minimization to fit the measurement data. For example, in Fig.~\ref{Fig:SingleShot}a), we consider constant illumination for $54$ minutes followed by darkness. The corresponding measurement signal and the fitted model using (\ref{Eq:BaselineConcentration}) are shown in Fig.~\ref{Fig:SingleShot}b) using blue and black lines, respectively. As expected, the concentration increases upon illumination and decreases quickly in darkness. Nevertheless, the model in (\ref{Eq:BaselineConcentration}) fails to accurately follow the measurement data. In fact, there exists an additional persistent decreasing bias in the measurement signal which is not anticipated by the saturation model in \eqref{Eq:BaselineConcentration}.

\textbf{Enhanced Parametric Model:} The assumptions made to arrive at (\ref{Eq:BaselineConcentration}) do not account for the dynamics inherent to living cells.
As such, cells can be growing in number or dying, or their fidelity can change. Motivated by the observations from the measurement data in Fig.~\ref{Fig:SingleShot}b), it was suggested in \cite{grebenstein_biological_2018} to enhance the model (\ref{Eq:BaselineConcentration}) with a simple additive linear offset as follows
\begin{equation}
    \label{Eq:BacteriaDrift}
    \cd(t) = \md\cdot(t-t_0),
\end{equation}
where $\md$ is a parameter controlling the slope of the bias.
The extended model is then given by $r_c(t)=c(t)+w(t)$ where $c(t)=\cb(t)+\cd(t)$. From Fig.~\ref{Fig:SingleShot}b), we can observe that the enhanced model, shown in red, fits the measurement data well. This example shows that further modification of a model that was obtained solely based on physical principles may be needed to arrive at an appropriate parametric model for an experimental system.

\iftoggle{OneColumn}{%
\begin{figure}[t]
	\centering
	\includegraphics{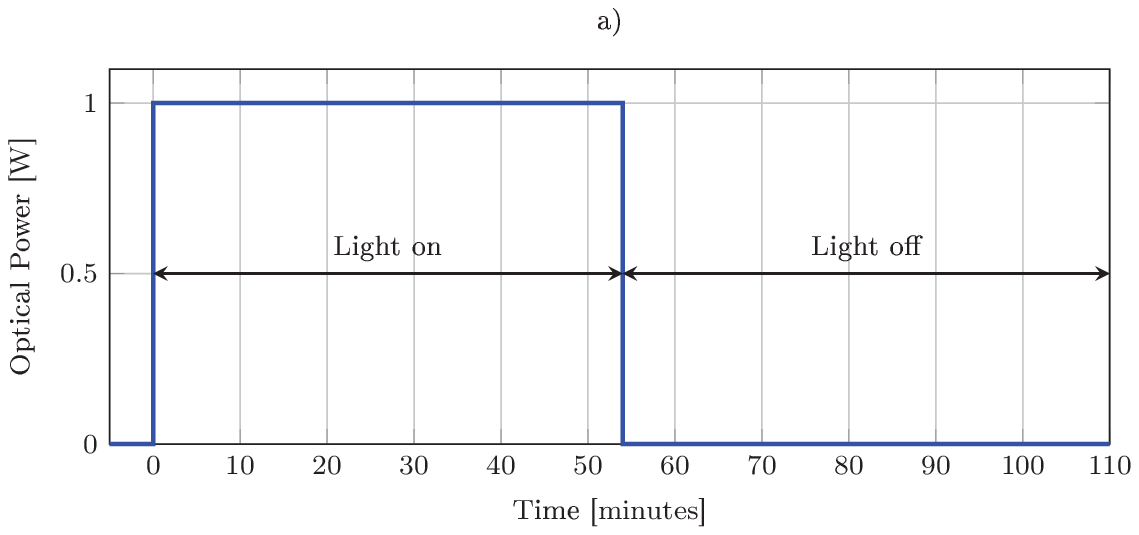}
	\vspace*{3mm}
	
	\includegraphics{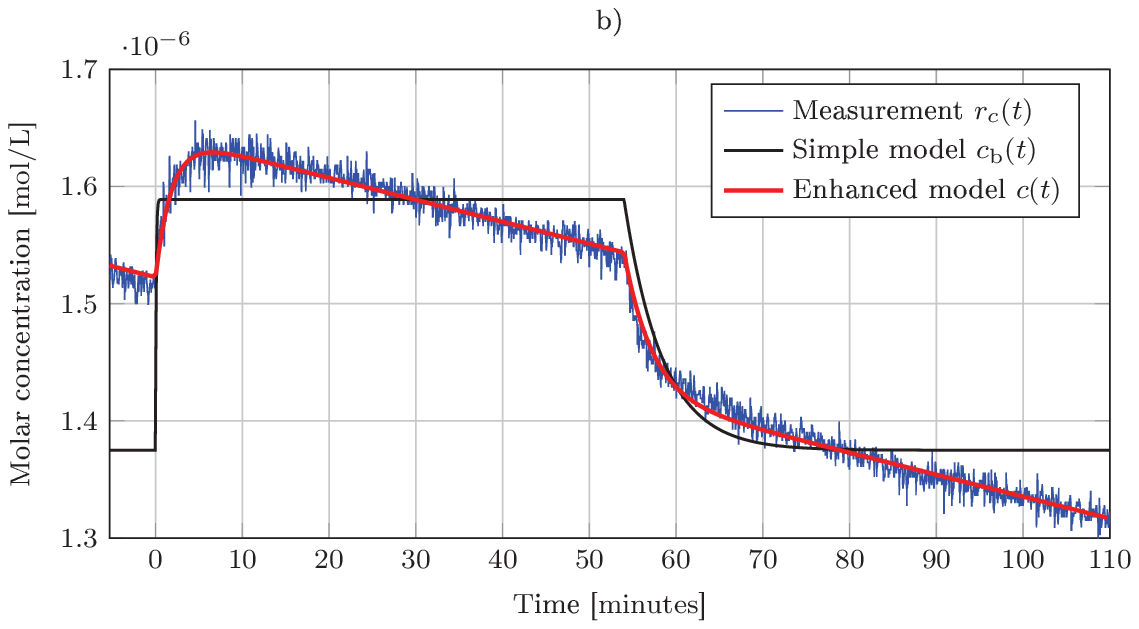}
	\caption{Experimental data. a) Optical signal versus time. b) The measured proton concentration as well as the simple and enhanced models versus time. Data taken from \cite{grebenstein_biological_2018}.}
	\label{Fig:SingleShot}
\end{figure}	
}{%
\begin{figure}[t]
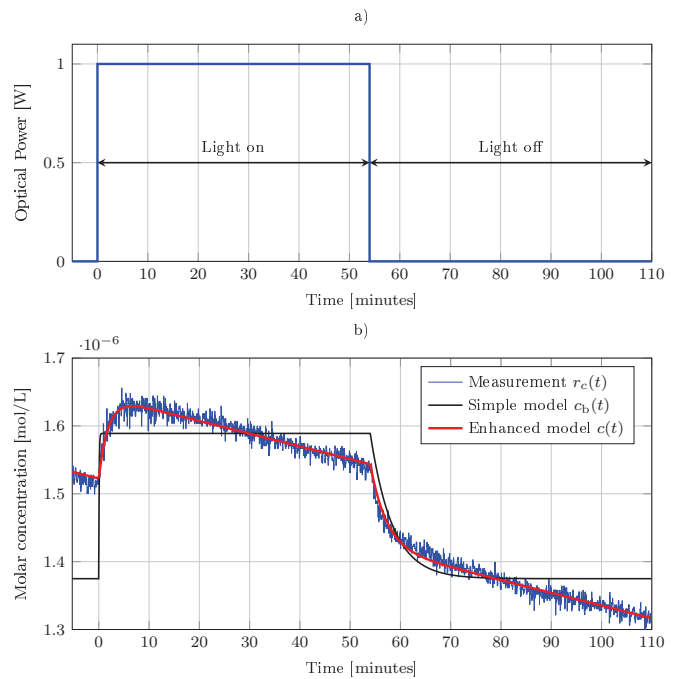

	\centering
	\includegraphics[width=1\linewidth]{Sections/S5/Fig/Expriment_model/single_shot_onoff}
	
	\includegraphics[width=1\linewidth]{Sections/S5/Fig/Expriment_model/single_shot} \vspace{-0.5cm}
	\caption{Experimental data. a) Optical signal versus time. b) The measured proton concentration as well as the simple and enhanced models versus time. Data taken from \cite{grebenstein_biological_2018}.}
	\label{Fig:SingleShot}
\end{figure}	
}

\section{Challenges and Directions for Future Work}
\label{Sect:ChalFuture}
\gls{MC} is still in its early stages of development and our understanding of \gls{MC} channels is still quite limited. In the following, we  review some potential challenges and open research problems which have to be addressed for successful deployment of \gls{MC} systems.

\textbf{Particle Generation and Signaling Pathways:} 
Although the impact of CRNs in the \emph{physical channel} (e.g. degradation reactions) on the \gls{CIR} of the \gls{MC} systems has been studied, see Section~\ref{Sub;ChaMod} and \cite{Adam_Enzyme,Nariman_AcidBasePlatform,PhY_MC,NoelPro3,ArmanJ2,Deng2015,ICC_Reactive}, less attention has been dedicated to the analysis of the influence of the CRNs at the transmitter and receiver. Such CRNs include particle generation reaction networks at the transmitter and signaling pathways at the receiver. In the \gls{MC} literature, there are some preliminary works that have studied the impact of particle generation reaction networks and also simplified signaling pathways; e.g., see  \cite{Chou2015, Chou_Rx}. However, the corresponding models are derived for a mesoscopic modeling approach. Analytical \gls{CIR} models that take the impact of these CRNs into account are crucial for system design and hence constitute an interesting research challenge.  

\textbf{Turbulent Flow:} 
The majority of the \gls{CIR} models in the \gls{MC} literature for advection channels have been developed based on the assumption of a uniform or laminar flow velocity field, cf. Section~\ref{Sub;ChaMod}. However, for several \gls{MC} environments, such as large arteries (e.g. the aorta) and macroscale environments (e.g. oil pipe lines), flow may exhibit turbulent behavior \cite{Arendt_Flow}. In particular, turbulence can occur when the \gls{MC} channel is non-homogeneous, e.g., due to the presence of obstacles in the physical channel. Therefore, studying and analyzing advection channels with a \emph{turbulent} velocity field is an important open research problem. First results towards analyzing turbulent flow for \gls{MC} systems were reported in \cite{atthanayake2018experimental}.

\textbf{Sample Correlation:} 
Multiple-sample detectors are used in the \gls{MC} literature to improve the detection performance \cite{Equ_MC,ConsCIR,Akyl_Receiver_MC,Adam_Enzyme,NoelPro3,Adam_OptReciever,Adam_Channel,TNBC_Sync,Yilmaz2017}. It is typically assumed that different samples are statistically independent from each other. However, this assumption holds only when the sampling interval is chosen large enough such that the independence of consecutive samples is ensured. In Section~\ref{Sect:SampCorrelation}, we have numerically evaluated the correlation among consecutive samples, and in \cite{Adam_OptReciever}, the mutual information between consecutive samples is numerically evaluated. We note that sample correlation significantly depends on the type of receiver, e.g., a recurrent,  non-recurrent, AMC, or IMC receiver, and its physical and chemical properties, e.g., the size, the number of receptors, and the reaction rate constants of the binding and unbinding reactions for a reactive receiver. Therefore, a careful study of sample correlation and the minimum sampling interval needed to ensure sample independence for different receiver types is essential for the applicability and performance analysis of the multiple-sample detectors proposed in the literature. 

\textbf{Complex Networks:} 
In this tutorial, we focused on a single one-way communication link from a transmitter to a receiver. This is the simplest communication architecture and hence the basis for more complex network typologies. We note that although multi-node networks can often be decomposed into a superposition of individual links, there are certain scenarios where such a decomposition is invalid. For instance, if multiple reactive receivers are in the environment, the presence of each receiver will impact the signal received at \emph{any} other receiver, see e.g. \cite{Lu2016} and \cite{Arifler2017}. Moreover, in Section~\ref{Sect:TxRxCh}, we considered \gls{MC} environments with simple boundary and initial conditions in order to derive analytical channel models. However, some important \gls{MC} environments, such as the cardiovascular system, are quite complex and cannot be fully modeled based on physics' first principals. One approach is to develop simulation environments for such complex networks, see  \cite{Felicetti2013} and Section~\ref{Sec:Sim}. Nevertheless, for system design, it is desirable to have simple yet sufficiently accurate analytical models for complex multi-node networks. Developing such analytical models constitutes an important future research topic, see \cite{touchMC2015,chahibi2015molecular,felicetti_modeling_2014,Deng2017} for some related works. 

\textbf{Microscale and Macroscale Models:} \gls{MC} systems have numerous potential applications which range from targeted drug delivery and health monitoring for microscale systems to communication in oil pipelines or chemical reactors and environmental monitoring for macroscale systems. Nevertheless, most of the current literature has targeted microscale applications and the available models are typically developed for microscale \gls{MC} environments.  However, macroscale and microscale \gls{MC} systems may require quite different considerations. For instance, the number of molecules needed for communication at macroscale is typically much larger than that needed for communication at microscale. Moreover, while at microscale, molecules can be counted at the receivers (e.g. via ligand-receptors), at macroscale,  receivers usually measure a quantity that is a function of the molecule concentration (e.g. a pH sensor was used in  \cite{Nariman_AcidBasePlatform,unterweger_experimental_2018} and mass spectroscopy was used in  \cite{McGuiness2018macroscale,giannoukos_molecular_2017}). In summary, the development of channel models for macroscale \gls{MC} systems is an important and interesting topic for future research. We refer the interested reader  to  \cite{PARCERISAGINE20092753,2018arXiv180104627C,McGuiness2018macroscale,giannoukos_molecular_2017,unterweger_experimental_2018,farsad_channel_2014} for preliminary results on channel modeling for macroscale \gls{MC} systems. 

\textbf{Generally-Accepted and Experimentally-Verified Models:} Over the past years, several non-biological experimental testbeds \cite{farsad_tabletop_2013,Nariman_AcidBasePlatform,atthanayake2018experimental,tuccitto_fluorescent_2017,kennedy_spatiotemporal_2018,giannoukos_molecular_2017,deleo_communications_2013,unterweger_experimental_2018} and biological experimental testbeds \cite{krishnaswamy_timeelapse_2013,felicetti_modeling_2014,Nakano_Microplatform_2008,bicen_efficient_2015,grebenstein_biological_2018} have been developed to demonstrate \gls{MC}. Most of these experimental testbeds were developed as proofs-of-concept for human-designed \gls{MC} and have not led to the development of general mathematical \gls{MC} channel models. However, for the advancement of \gls{MC} research, it will be crucial to specify generally-accepted test channels with corresponding experimentally-verified mathematical channel models. Then, researchers in the \gls{MC} community can use these established models for the design and performance analysis of newly-developed communication schemes.

\section{Conclusions}
\label{Sect:Concl}
This paper provided a comprehensive tutorial review of the diffusive \gls{MC} channel models available in the literature. To this end, we first presented the underlying fundamental laws that govern diffusion, advection, and chemical reactions in \gls{MC} channels and constitute the essential mathematical tools from biology, chemistry, and physics required for the development of \gls{MC} channel models. Subsequently, we reviewed the main end-to-end channel models reported in the diffusive \gls{MC} literature and showed how they were developed from basic physical principles. The reviewed end-to-end channel models included the joint effects of release mechanisms, the physical channel, and reception mechanisms. Moreover, we provided a unified definition for the received signal that included the representation obtained by both timing and counting receivers as special cases. Furthermore, for counting receivers, we derived signal models relevant for different time scales. We generalized these models to account for interfering noise molecules and \gls{ISI} and studied the correlation among the received signals observed at different time scales. In addition, simulation-driven and experimentally-driven channel models were investigated for complex scenarios where simple \gls{MC} channel models cannot be obtained from basic physical principles. Finally, we provided a discussion of challenges, open research problems, and future directions for channel modeling of diffusive \gls{MC} systems. 


\bibliographystyle{IEEEtran}
\bibliography{./Sections/S0/References}

\begin{thebibliography}{100}
\providecommand{\url}[1]{#1}
\csname url@samestyle\endcsname
\providecommand{\newblock}{\relax}
\providecommand{\bibinfo}[2]{#2}
\providecommand{\BIBentrySTDinterwordspacing}{\spaceskip=0pt\relax}
\providecommand{\BIBentryALTinterwordstretchfactor}{4}
\providecommand{\BIBentryALTinterwordspacing}{\spaceskip=\fontdimen2\font plus
\BIBentryALTinterwordstretchfactor\fontdimen3\font minus
  \fontdimen4\font\relax}
\providecommand{\BIBforeignlanguage}[2]{{%
\expandafter\ifx\csname l@#1\endcsname\relax
\typeout{** WARNING: IEEEtran.bst: No hyphenation pattern has been}%
\typeout{** loaded for the language `#1'. Using the pattern for}%
\typeout{** the default language instead.}%
\else
\language=\csname l@#1\endcsname
\fi
#2}}
\providecommand{\BIBdecl}{\relax}
\BIBdecl

\bibitem{Hiyama2005}
S.~Hiyama, Y.~Moritani, T.~Suda, R.~Egashira, A.~Enomoto, M.~J. Moore, and
  T.~Nakano, ``{Molecular Communication},'' in \emph{Proc. NSTI Nanotech}, May
  2005, pp. 391--394.

\bibitem{Survey_Mol_Nono}
I.~Akyildiz, F.~Brunetti, and C.~Blazquez, ``{Nanonetworks: A New Communication
  Paradigm},'' \emph{Comput. Net.}, vol.~52, pp. 2260--2279, Apr. 2008.

\bibitem{Nariman_Survey}
N.~Farsad, H.~Yilmaz, A.~Eckford, C.~Chae, and W.~Guo, ``{A Comprehensive
  Survey of Recent Advancements in Molecular Communication},'' \emph{IEEE
  Commun. Surveys Tutorials}, vol.~18, no.~3, pp. 1887--1919, third quarter
  2016.

\bibitem{PARCERISAGINE20092753}
L.~P. Giné and I.~F. Akyildiz, ``{Molecular Communication Options for Long
  Range Nanonetworks},'' \emph{Computer Netw.}, vol.~53, no.~16, pp. 2753 --
  2766, 2009.

\bibitem{PierobonJ3}
M.~Pierobon and I.~Akyildiz, ``{Noise Analysis in Ligand-Binding Reception for
  Molecular Communication in Nanonetworks},'' \emph{{IEEE} Trans. Sig.
  Process.}, vol.~59, no.~9, pp. 4168--4182, Sep. 2011.

\bibitem{EckfordDrift}
S.~Kadloor, R.~Adve, and A.~Eckford, ``{Molecular Communication Using Brownian
  Motion With Drift},'' \emph{IEEE Trans. NanoBiosci.}, vol.~11, no.~2, pp.
  89--99, Jun. 2012.

\bibitem{NakanoGapJunc}
T.~Nakano, T.~Suda, T.~Koujin, T.~Haraguchi, and Y.~Hiraoka, \emph{{Molecular
  Communication Through Gap Junction Channels}}.\hskip 1em plus 0.5em minus
  0.4em\relax Berlin, Heidelberg: Springer Berlin Heidelberg, 2008, pp. 81--99.

\bibitem{Moore_Mmotor}
M.~Moore, A.~Enomoto, T.~Nakano, R.~Egashira, T.~Suda, A.~Kayasuga, H.~Kojima,
  H.~Sakakibara, and K.~Oiwa, ``{A Design of a Molecular Communication System
  for Nanomachines Using Molecular Motors},'' in \emph{Proc. Pervasive Comput.
  Commun. Workshops}, Mar. 2006, pp. 6 pp.--559.

\bibitem{Akyl_Bmototr}
M.~Gregori and I.~F. Akyildiz, ``{A New Nanonetwork Architecture Using
  Flagellated Bacteria and Catalytic Nanomotors},'' \emph{IEEE J. Select. Areas
  in Commun.}, vol.~28, no.~4, pp. 612--619, May 2010.

\bibitem{AlbertsBook}
B.~Alberts, D.~Bray, K.~Hopkin, A.~D. Johnson, A.~Johnson, J.~Lewis, M.~Raff,
  K.~Roberts, and P.~Walter, \emph{Essential Cell Biology}, 3rd~ed.\hskip 1em
  plus 0.5em minus 0.4em\relax Garland Science, 2009.

\bibitem{mallik2004molecular}
R.~Mallik and S.~P. Gross, ``{Molecular Motors: Strategies to Get Along},''
  \emph{Current Biology}, vol.~14, no.~22, pp. R971--R982, 2004.

\bibitem{Crank1979}
J.~Crank, \emph{{The Mathematics of Diffusion}}, 2nd~ed.\hskip 1em plus 0.5em
  minus 0.4em\relax Oxford University Press, 1979.

\bibitem{Carslaw}
H.~S. Carslaw and J.~C. Jaeger, \emph{Conduction of Heat in Solids (Oxford
  Science Publications)}.\hskip 1em plus 0.5em minus 0.4em\relax Oxford
  University Press, 1986.

\bibitem{Survey_Mol_Net}
T.~Nakano, M.~Moore, F.~Wei, A.~Vasilakos, and J.~Shuai, ``{Molecular
  Communication and Networking: Opportunities and Challenges},'' \emph{IEEE
  Trans. NanoBiosci.}, vol.~11, no.~2, pp. 135--148, Jun. 2012.

\bibitem{MC_Book}
T.~Nakano, A.~Eckford, , and T.~Haraguchi, \emph{Molecular
  Communication}.\hskip 1em plus 0.5em minus 0.4em\relax Cambridge, U.K.:
  Cambridge Univ. Press, 2013.

\bibitem{Atakan2010}
B.~Atakan and O.~B. Akan, ``{Deterministic Capacity of Information Flow in
  Molecular Nanonetworks},'' \emph{Nano Commun. Net.}, vol.~1, no.~1, pp.
  31--42, Mar. 2010.

\bibitem{Einolghozati2013a}
A.~Einolghozati, M.~Sardari, and F.~Fekri, ``{Design and Analysis of Wireless
  Communication Systems Using Diffusion-based Molecular Communication Among
  Bacteria},'' \emph{IEEE Trans. Wireless Commun.}, vol.~12, no.~12, pp.
  6096--6105, Dec. 2013.

\bibitem{Nakano2013a}
T.~Nakano, Y.~Okaie, and A.~V. Vasilakos, ``{Transmission Rate Control for
  Molecular Communication Among Biological Nanomachines},'' \emph{IEEE J.
  Select. Areas Commun.}, vol.~31, no.~12, pp. 835--846, 2013.

\bibitem{Ahmadzadeh2015a}
A.~Ahmadzadeh, A.~Noel, and R.~Schober, ``{Analysis and Design of Multi-Hop
  Diffusion-Based Molecular Communication Networks},'' \emph{IEEE Trans.
  Molecular, Biol., and Multi-Scale Commun.}, vol.~1, no.~2, pp. 144--157, Jun.
  2015.

\bibitem{Fang2017a}
Y.~Fang, A.~Noel, N.~Yang, A.~W. Eckford, and R.~A. Kennedy, ``{Convex
  Optimization of Distributed Cooperative Detection in Multi-Receiver Molecular
  Communication},'' \emph{IEEE Trans. Molecular, Biol., and Multi-Scale
  Commun.}, vol.~3, no.~3, pp. 166--182, Sep. 2017.

\bibitem{Lu2016}
Y.~Lu, M.~D. Higgins, A.~Noel, M.~S. Leeson, and Y.~Chen, ``{The Effect of Two
  Receivers on Broadcast Molecular Communication Systems},'' \emph{IEEE Trans.
  NanoBiosci.}, vol.~15, no.~8, pp. 891--900, Dec. 2016.

\bibitem{Arifler2017}
D.~Arifler and D.~Arifler, ``{Monte Carlo Analysis of Molecule Absorption
  Probabilities in Diffusion-Based Nanoscale Communication Systems with
  Multiple Receivers},'' \emph{IEEE Trans. NanoBiosci.}, vol.~16, no.~3, pp.
  157--165, Apr. 2017.

\bibitem{Berg}
H.~C. Berg, \emph{{Random Walks in Biology}}.\hskip 1em plus 0.5em minus
  0.4em\relax Princeton University Press, 1993.

\bibitem{BioPhysic}
P.~Nelson, \emph{{Biological Physics: Energy, Information, Life}}.\hskip 1em
  plus 0.5em minus 0.4em\relax Freeman, 1st ed., 2008.

\bibitem{Subdiffusion}
M.~Mahfuz, D.~Makrakis, and H.~T. Mouftah, ``{Concentration-Encoded
  Subdiffusive Molecular Communication: Theory, Channel Characteristics, and
  Optimum Signal Detection},'' \emph{IEEE Trans. NanoBiosci.}, vol.~15, no.~6,
  pp. 533--548, 2016.

\bibitem{desposito2011active}
M.~A. Desp{\'o}sito, C.~Pallavicini, V.~Levi, and L.~Bruno, ``{Active Transport
  in Complex Media: Relationship Between Persistence and Superdiffusion},''
  \emph{Physica A: Statistical Mechanics and its Applications}, vol. 390,
  no.~6, pp. 1026--1032, 2011.

\bibitem{Diffusion_book}
E.~L. Cussler, \emph{{Diffusion: Mass Transfer in Fluid Systems}}.\hskip 1em
  plus 0.5em minus 0.4em\relax Cambridge University Press, 2009.

\bibitem{Adam_Thesis}
A.~Noel, ``{Modeling and Analysis of Diffusive Molecular Communication
  Systems},'' Ph.D. dissertation, University of British Columbia, 2015.

\bibitem{zoppou1999analytical}
C.~Zoppou and J.~Knight, ``{Analytical Solution of a Spatially Variable
  Coefficient Advection--Diffusion Equation in up to Three Dimensions},''
  \emph{Applied Mathematical Modelling}, vol.~23, no.~9, pp. 667--685, 1999.

\bibitem{kornyshev2003kinetics}
A.~A. Kornyshev, A.~M. Kuznetsov, E.~Spohr, and J.~Ulstrup, ``{Kinetics of
  Proton Transport in Water},'' \emph{J. Physical Chemistry B}, vol. 107,
  no.~15, pp. 3351--3366, 2003.

\bibitem{Akyl_Receiver_MC}
L.~S. Meng, P.~C. Yeh, K.~C. Chen, and I.~F. Akyildiz, ``{On Receiver Design
  for Diffusion-Based Molecular Communication},'' \emph{IEEE Trans. Sig.
  Process.}, vol.~62, no.~22, pp. 6032--6044, Nov. 2014.

\bibitem{ConsCIR}
M.~Mahfuz, D.~Makrakis, and H.~Mouftah, ``{A Comprehensive Study of
  Sampling-Based Optimum Signal Detection in Concentration-Encoded Molecular
  Communication},'' \emph{IEEE Trans. NanoBiosci.}, vol.~13, no.~3, pp.
  208--222, Sep. 2014.

\bibitem{NoelPro1}
A.~Noel, K.~C. Cheung, and R.~Schober, ``{Using Dimensional Analysis to Assess
  Scalability and Accuracy in Molecular Communication},'' in \emph{Proc. IEEE
  Int. Conf. Commun. (ICC)}, Jun. 2013, pp. 818--823.

\bibitem{Akyildiz_MC_E2E}
M.~Pierobon and I.~Akyildiz, ``{A Physical End-to-End Model for Molecular
  Communication in Nanonetworks},'' \emph{IEEE J. Sel. Areas Commun.}, vol.~28,
  no.~4, pp. 602--611, May 2010.

\bibitem{Equ_MC}
D.~Kilinc and O.~B. Akan, ``{Receiver Design for Molecular Communication},''
  \emph{IEEE J. Select. Areas in Commun.}, vol.~31, no.~12, pp. 705--714, Dec.
  2013.

\bibitem{PhY_MC}
W.~Guo, T.~Asyhari, N.~Farsad, H.~B. Yilmaz, B.~Li, A.~Eckford, and C.~B. Chae,
  ``{Molecular Communications: Channel Model and Physical Layer Techniques},''
  \emph{IEEE Wireless Commun.}, vol.~23, no.~4, pp. 120--127, Aug. 2016.

\bibitem{NoelPro3}
A.~Noel, K.~C. Cheung, and R.~Schober, ``{Diffusive Molecular Communication
  with Disruptive Flows},'' in \emph{Proc. IEEE Int. Conf. Commun. (ICC)}, Jun.
  2014, pp. 3600--3606.

\bibitem{TCOM_MC_CSI}
V.~Jamali, A.~Ahmadzadeh, C.~Jardin, C.~Sticht, and R.~Schober, ``{Channel
  Estimation for Diffusive Molecular Communications},'' \emph{IEEE Trans.
  Commun.}, vol.~64, no.~10, pp. 4238--4252, Oct. 2016.

\bibitem{DistanceEstLett}
X.~Wang, M.~Higgins, and M.~Leeson, ``{Distance Estimation Schemes for
  Diffusion Based Molecular Communication Systems},'' \emph{IEEE Commun.
  Lett.}, vol.~19, no.~3, pp. 399--402, Mar. 2015.

\bibitem{Robert_MCnote}
R.~Schober, \emph{{Lecture Notes on Molecular Communications}}.\hskip 1em plus
  0.5em minus 0.4em\relax Friedrich-Alexander University Erlangen-N\"urnberg,
  2017.

\bibitem{FELICETTI201627}
L.~Felicetti, M.~Femminella, G.~Reali, and P.~Liò, ``{Applications of
  Molecular Communications to Medicine: A Survey},'' \emph{Nano Commun. Netw.},
  vol.~7, pp. 27 -- 45, 2016.

\bibitem{chahibi2015molecular}
Y.~Chahibi, I.~F. Akyildiz, S.~Balasubramaniam, and Y.~Koucheryavy,
  ``{Molecular Communication Modeling of Antibody-Mediated Drug Delivery
  Systems},'' \emph{IEEE Trans. Biomed. Eng.}, vol.~62, no.~7, pp. 1683--1695,
  2015.

\bibitem{femminella2015molecular}
M.~Femminella, G.~Reali, and A.~V. Vasilakos, ``{A Molecular Communications
  Model for Drug Delivery},'' \emph{IEEE Trans. NanoBiosci.}, vol.~14, no.~8,
  pp. 935--945, 2015.

\bibitem{chude2017molecular}
U.~A. Chude-Okonkwo, R.~Malekian, B.~T. Maharaj, and A.~V. Vasilakos,
  ``{Molecular Communication and Nanonetwork for Targeted Drug Delivery: A
  Survey},'' \emph{IEEE Commun. Surveys Tutorials}, vol.~19, no.~4, pp.
  3046--3096, 2017.

\bibitem{mosayebi2018early}
\BIBentryALTinterwordspacing
R.~Mosayebi, A.~Ahmadzadeh, W.~Wicke, V.~Jamali, R.~Schober, and
  M.~Nasiri-Kenari, ``{Early Cancer Detection in Blood Vessels Using Mobile
  Nanosensors},'' \emph{submitted to IEEE Trans. NanoBiosci.}, 2018. [Online].
  Available: \url{arXiv:1805.08777}
\BIBentrySTDinterwordspacing

\bibitem{Adam_OptReciever}
A.~Noel, K.~Cheung, and R.~Schober, ``{Optimal Receiver Design for Diffusive
  Molecular Communication with Flow and Additive Noise},'' \emph{IEEE Trans.
  NanoBiosci.}, vol.~13, no.~3, pp. 350--362, Sep. 2014.

\bibitem{WayanPro2}
W.~Wicke, A.~Ahmadzadeh, V.~Jamali, R.~Schober, H.~Unterweger, and C.~Alexiou,
  ``{Molecular Communication Using Magnetic Nanoparticles},'' in \emph{Proc.
  IEEE Wireless Commun. Netw. Conf. (WCNC)}, Apr. 2018, pp. 1--6.

\bibitem{tehrani2015novel}
M.~D. Tehrani, J.-H. Yoon, M.~O. Kim, and J.~Yoon, ``{A Novel Scheme for
  Nanoparticle Steering in Blood Vessels Using a Functionalized Magnetic
  Field},'' \emph{IEEE Trans. Biomed. Eng.}, vol.~62, no.~1, pp. 303--313, Jan.
  2015.

\bibitem{bruus_theoretical_2007}
H.~Bruus, \emph{\BIBforeignlanguage{English}{Theoretical {{Microfluidics}}}},
  1st~ed.\hskip 1em plus 0.5em minus 0.4em\relax Oxford ; New York: {Oxford
  University Press}, Nov. 2007.

\bibitem{probstein_physicochemical_2005}
R.~F. Probstein, \emph{\BIBforeignlanguage{en}{Physicochemical
  {{Hydrodynamics}}: {{An Introduction}}}}.\hskip 1em plus 0.5em minus
  0.4em\relax {John Wiley \& Sons}, Feb. 2005.

\bibitem{white_fluid_2016}
F.~M. White and R.~Y. Chul, \emph{\BIBforeignlanguage{en}{Fluid
  {{Mechanics}}}}.\hskip 1em plus 0.5em minus 0.4em\relax {McGraw-Hill
  Education}, 2016.

\bibitem{back1986measurement}
L.~Back, J.~Radbill, Y.~Cho, and D.~Crawford, ``{Measurement and Prediction of
  Flow Through a Replica Segment of a Mildly Atherosclerotic Coronary Artery of
  Man},'' \emph{J. Biomechanics}, vol.~19, no.~1, pp. 1--17, 1986.

\bibitem{Arendt_Flow}
J.~A. Jensen, \emph{{Lectures Notes on Medical Imaging Systems -- Lecture 5:
  Blood Flow in the Human Body}}.\hskip 1em plus 0.5em minus 0.4em\relax
  Technical University of Denmark, 2018.

\bibitem{WayanPro1}
W.~Wicke, T.~Schwering, A.~Ahmadzadeh, V.~Jamali, A.~Noel, and R.~Schober,
  ``{Modeling Duct Flow for Molecular Communication},'' in \emph{Proc. IEEE
  Global Commun. Conf. (Globecom)}, Dec. 2018, pp. 1--6.

\bibitem{koike2017molecular}
T.~Koike-Akino, J.~Suzuki, and P.~V. Orlik, ``{Molecular Signaling Design
  Exploiting Cyclostationary Drift-Diffusion Fluid},'' in \emph{Proc. IEEE
  Global Commun. Conf. (Globecom)}, 2017, pp. 1--7.

\bibitem{chung_computational_2010}
T.~J. Chung, \emph{\BIBforeignlanguage{en}{Computational {{Fluid
  Dynamics}}}}.\hskip 1em plus 0.5em minus 0.4em\relax {Cambridge University
  Press}, Sep. 2010.

\bibitem{farsad_tabletop_2013}
N.~Farsad, W.~Guo, and A.~W. Eckford, ``{Tabletop {{Molecular Communication}}:
  {{Text Messages}} Through {{Chemical Signals}}},'' \emph{PLOS ONE}, vol.~8,
  no.~12, p. e82935, Dec. 2013.

\bibitem{Adam_Enzyme}
A.~Noel, K.~Cheung, and R.~Schober, ``{Improving Receiver Performance of
  Diffusive Molecular Communication with Enzymes},'' \emph{IEEE Trans.
  NanoBiosci.}, vol.~13, no.~1, pp. 31--43, Mar. 2014.

\bibitem{Nariman_AcidBasePlatform}
N.~Farsad, D.~Pan, and A.~Goldsmith, ``{A Novel Experimental Platform for
  In-Vessel Multi-Chemical Molecular Communications},'' in \emph{Proc. IEEE
  Global Commun. Conf. (Globecom)}, Dec. 2017, pp. 1--6.

\bibitem{ArmanJ2}
A.~Ahmadzadeh, H.~Arjmandi, A.~Burkovski, and R.~Schober, ``{Comprehensive
  Reactive Receiver Modeling for Diffusive Molecular Communication Systems:
  Reversible Binding, Molecule Degradation, and Finite Number of Receptors},''
  \emph{IEEE Trans. NanoBiosci.}, vol.~15, no.~7, pp. 713 -- 727, Oct. 2016.

\bibitem{Deng2015}
Y.~Deng, A.~Noel, M.~Elkashlan, A.~Nallanathan, and K.~C. Cheung, ``{Modeling
  and Simulation of Molecular Communication Systems With a Reversible
  Adsorption Receiver},'' \emph{{IEEE} Trans. Mol. Biol. Multi-Scale Commun.},
  vol.~1, no.~4, pp. 347--362, Dec. 2015.

\bibitem{CoxNatureCommun}
D.~Schnoerr, R.~Grima, and G.~Sanguinetti, ``{Cox Process Representation and
  Inference for Stochastic Reaction-Diffusion Processes},'' \emph{Nature
  Commun.}, vol.~7, 2016.

\bibitem{PhysicChemistry}
R.~Chang, \emph{{Physical Chemistry for the Biosciences}}.\hskip 1em plus 0.5em
  minus 0.4em\relax University Science Books, 2005.

\bibitem{Heren}
A.~C. Heren, H.~B. Yilmaz, C.-B. Chae, and T.~Tugcu, ``{Effect of Degradation
  in Molecular Communication: Impairment or Enhancement?}'' \emph{IEEE Trans.
  Molecular, Biol., and Multi-Scale Commun.}, vol.~1, no.~2, pp. 217--229, Jun.
  2015.

\bibitem{TCOM_NonCoherent}
V.~Jamali, N.~Farsad, R.~Schober, and A.~Goldsmith, ``{Non-Coherent Detection
  for Diffusion-Based Molecular Communications},'' \emph{IEEE Trans. Commun.},
  vol.~66, no.~6, pp. 2515--2531, Jun. 2018.

\bibitem{cho2017effective}
Y.~J. Cho, H.~B. Yilmaz, W.~Guo, and C.-B. Chae, ``{Effective Enzyme Deployment
  for Degradation of Interference Molecules in Molecular Communication},'' in
  \emph{Proc. IEEE Wireless Commun. Netw. Conf. (WCNC)}.\hskip 1em plus 0.5em
  minus 0.4em\relax IEEE, 2017, pp. 1--6.

\bibitem{ICC_Reactive}
V.~Jamali, N.~Farsad, R.~Schober, and A.~Goldsmith, ``{Diffusive Molecular
  Communications with Reactive Signaling},'' in \emph{Proc. IEEE Int. Conf.
  Commun. (ICC)}, May 2018.

\bibitem{Tro2015Chemistry}
N.~J. Tro, \emph{{Principles of Chemistry: A Molecular Approach}}.\hskip 1em
  plus 0.5em minus 0.4em\relax Pearson Higher Ed, 2015.

\bibitem{Andrews2004}
S.~S. Andrews and D.~Bray, ``{Stochastic Simulation of Chemical Reactions with
  Spatial Resolution and Single Molecule Detail},'' \emph{Physical Biology},
  vol.~1, no. 3-4, pp. 137--151, Sep. 2004.

\bibitem{NonlinearPDE_Debnath}
L.~Debnath, \emph{{Nonlinear Partial Differential Equations for Scientists and
  Engineers}}.\hskip 1em plus 0.5em minus 0.4em\relax Springer Science \&
  Business Media, 2011.

\bibitem{Adam_Universal_Noise}
A.~Noel, K.~Cheung, and R.~Schober, ``{A Unifying Model for External Noise
  Sources and ISI in Diffusive Molecular Communication},'' \emph{IEEE J. Sel.
  Areas Commun.}, vol.~32, no.~12, pp. 2330--2343, Dec. 2014.

\bibitem{unterweger_experimental_2018}
H.~Unterweger, J.~Kirchner, W.~Wicket, A.~Ahmadzadeh, D.~Ahmed, V.~Jamali,
  C.~Alexiou, G.~Fischer, and R.~Schober, ``Experimental {{Molecular
  Communication Testbed Based}} on {{Magnetic Nanoparticles}} in {{Duct
  Flow}},'' in \emph{IEEE Int. Workshop Sig. Process. Advances in Wireless
  Commun. (SPAWC)}, Jun. 2018, pp. 1--5.

\bibitem{YilmazL1}
H.~Yilmaz, A.~Heren, T.~Tugcu, and C.-B. Chae, ``{Three-Dimensional Channel
  Characteristics for Molecular Communications With an Absorbing Receiver},''
  \emph{{IEEE} Commun. Lett.}, vol.~18, no.~6, pp. 929--932, Jun. 2014.

\bibitem{Akkaya}
A.~Akkaya, H.~B. Yilmaz, C.~B. Chae, and T.~Tugcu, ``{Effect of Receptor
  Density and Size on Signal Reception in Molecular Communication via Diffusion
  With an Absorbing Receiver},'' \emph{{IEEE} Commun. Lett.}, vol.~19, no.~2,
  pp. 155--158, Feb. 2015.

\bibitem{Deng2017}
Y.~Deng, A.~Noel, W.~Guo, A.~Nallanathan, and M.~Elkashlan, ``{Analyzing
  Large-Scale Multiuser Molecular Communication via 3-D Stochastic Geometry},''
  \emph{{IEEE} Trans. Mol. Biol. Multi-Scale Commun.}, vol.~3, no.~2, pp.
  118--133, Jun. 2017.

\bibitem{Zabini2018}
F.~Zabini, ``{Spatially Distributed Molecular Communications: An Asynchronous
  Stochastic Model},'' \emph{{IEEE} Commun. Lett.}, vol.~22, no.~7, pp.
  1326--1329, Jul. 2018.

\bibitem{Dinc2017}
E.~Dinc and O.~B. Akan, ``{Theoretical Limits on Multiuser Molecular
  Communication in Internet of Nano-Bio Things},'' \emph{IEEE Trans.
  NanoBiosci.}, vol.~16, no.~4, pp. 266--270, Jun. 2017.

\bibitem{SchultenL1}
K.~Schulten and I.~Kosztin, \emph{Lectures in Theoretical Biophysics}.\hskip
  1em plus 0.5em minus 0.4em\relax Champaign, IL, USA: University of Illinois
  at Urbana-Champaign, 2000.

\bibitem{Eckford2012}
S.~Kadloor, R.~S. Adve, and A.~W. Eckford, ``Molecular communication using
  brownian motion with drift,'' \emph{IEEE Trans. NanoBiosci.}, vol.~11, no.~2,
  pp. 89--99, Jun. 2012.

\bibitem{ShahMohammadianProc1}
H.~ShahMohammadian, G.~Messier, and S.~Magierowski, ``{Modelling the Reception
  Process in Diffusion-based Molecular Communication Channels},'' in
  \emph{Proc. IEEE Int. Conf. Commun. (ICC)}, Jun. 2013, pp. 782--786.

\bibitem{Chou_Rx}
C.~T. Chou, ``{Impact of Receiver Reaction Mechanisms on the Performance of
  Molecular Communication Networks},'' \emph{IEEE Trans. Nanotechnol.},
  vol.~14, no.~2, pp. 304--317, Mar. 2015.

\bibitem{Noel_CT}
\BIBentryALTinterwordspacing
A.~Noel, D.~Makrakis, and A.~Hafid, ``{Channel Impulse Responses in Diffusive
  Molecular Communication with Spherical Transmitters},'' in \emph{Biennial
  Symp. Commun.}, 2016. [Online]. Available:
  \url{http://arxiv.org/abs/1604.04684}
\BIBentrySTDinterwordspacing

\bibitem{Yilmaz2017}
H.~B. Yilmaz, G.~Suk, and C.~Chae, ``{Chemical Propagation Pattern for
  Molecular Communications},'' \emph{{IEEE} Wireless Commun. Lett.}, vol.~6,
  no.~2, pp. 226--229, Apr. 2017.

\bibitem{Arjmandi2016}
H.~Arjmandi, A.~Ahmadzadeh, R.~Schober, and M.~N. Kenari, ``{Ion Channel Based
  Bio-Synthetic Modulator for Diffusive Molecular Communication},''
  \emph{{IEEE} Trans. Nanobiosci.}, vol.~15, no.~5, pp. 418--432, Jul. 2016.

\bibitem{Chou2015}
C.~T. Chou, ``{A Markovian Approach to the Optimal Demodulation of
  Diffusion-Based Molecular Communication Networks},'' \emph{{IEEE} Trans.
  Commun.}, vol.~63, no.~10, pp. 3728--3743, Oct. 2015.

\bibitem{Awan2017a}
H.~Awan and C.~T. Chou, ``{Generalized Solution for the Demodulation of
  Reaction Shift Keying Signals in Molecular Communication Networks},''
  \emph{{IEEE} Trans. Commun.}, vol.~65, no.~2, pp. 715--727, Feb. 2017.

\bibitem{Awan2017b}
------, ``{Molecular Circuit-Based Transmitters and Receivers for Molecular
  Communication Networks},'' in \emph{IEEE Int. Workshop Sig. Process. Advances
  in Wireless Commun. (SPAWC)}, Jul. 2017, pp. 1--5.

\bibitem{Chahibi2013}
Y.~Chahibi, M.~Pierobon, S.~O. Song, and I.~F. Akyildiz, ``{A Molecular
  Communication System Model for Particulate Drug Delivery Systems},''
  \emph{{IEEE} Trans. Biomed. Eng.}, vol.~60, no.~12, pp. 3468--3483, Dec.
  2013.

\bibitem{Iwasaki2018}
S.~Iwasaki and T.~Nakano, ``{Graph-Based Modeling of Mobile Molecular
  Communication Systems},'' \emph{{IEEE} Commun. Lett.}, vol.~22, no.~2, pp.
  376--379, Feb. 2018.

\bibitem{Adam_Channel}
A.~Noel, K.~C. Cheung, and R.~Schober, ``{Joint Channel Parameter Estimation
  via Diffusive Molecular Communication},'' \emph{IEEE Trans. Molecular, Biol.,
  and Multi-Scale Commun.}, vol.~1, no.~1, pp. 4--17, Mar. 2015.

\bibitem{Yilmaz_Poiss}
H.~B. Yilmaz and C.~B. Chae, ``{Arrival Modelling for Molecular Communication
  via Diffusion},'' \emph{Electron. Lett.}, vol.~50, no.~23, pp. 1667--1669,
  Nov. 2014.

\bibitem{Yilmaz2014a}
H.~B. Yilmaz and C.-B. Chae, ``{Simulation Study of Molecular Communication
  Systems with an Absorbing Receiver: Modulation and {ISI} Mitigation
  Techniques},'' \emph{Simulation Modelling Practice and Theory}, vol.~49, pp.
  136--150, Dec. 2014.

\bibitem{damrath2017equivalent}
M.~Damrath, S.~Korte, and P.~A. Hoeher, ``{Equivalent Discrete-Time Channel
  Modeling for Molecular Communication with Emphasize on an Absorbing
  Receiver},'' \emph{IEEE Trans. NanoBiosci.}, vol.~16, no.~1, pp. 60--68,
  2017.

\bibitem{cao2018diffusive}
T.~N. Cao, A.~Ahmadzadeh, V.~Jamali, W.~Wicke, P.~L. Yeoh, J.~Evans, and
  R.~Schober, ``{Diffusive Mobile MC for Controlled-Release Drug Delivery with
  Absorbing Receiver},'' \emph{arXiv preprint arXiv:1811.00417}, 2018.

\bibitem{InvGaussian}
K.~V. Srinivas, A.~W. Eckford, and R.~S. Adve, ``{Molecular Communication in
  Fluid Media: The Additive Inverse Gaussian Noise Channel},'' \emph{IEEE
  Trans. Inf. Theory}, vol.~58, no.~7, pp. 4678--4692, Jul. 2012.

\bibitem{Yonathan_Timimg}
Y.~Murin, N.~Farsad, M.~Chowdhury, and A.~Goldsmith, ``{Communication over
  Diffusion-Based Molecular Timing Channels},'' in \emph{Proc. IEEE Global
  Commun. Conf. (Globecom)}, Dec. 2016, pp. 1--6.

\bibitem{Nariman_Timing}
N.~Farsad, Y.~Murin, A.~Eckford, and A.~Goldsmith, ``{On the Capacity of
  Diffusion-Based Molecular Timing Channels},'' in \emph{Proc. IEEE Int. Symp.
  Inf. Theory (ISIT)}, Jul. 2016, pp. 1023--1027.

\bibitem{Rose_T1}
C.~Rose and I.~S. Mian, ``{Inscribed Matter Communication: Part I},''
  \emph{IEEE Trans. Molecular, Biol., and Multi-Scale Commun.}, vol.~2, no.~2,
  pp. 209--227, Dec. 2016.

\bibitem{Rose_T2}
------, ``{Inscribed Matter Communication: Part II},'' \emph{IEEE Trans.
  Molecular, Biol., and Multi-Scale Commun.}, vol.~2, no.~2, pp. 228--239, Dec.
  2016.

\bibitem{li2014capacity}
H.~Li, S.~M. Moser, and D.~Guo, ``{Capacity of the Memoryless Additive Inverse
  Gaussian Noise Channel},'' \emph{IEEE J. Select. Areas Commun.}, vol.~32,
  no.~12, pp. 2315--2329, 2014.

\bibitem{TNBC_Sync}
V.~Jamali, A.~Ahmadzadeh, and R.~Schober, ``{Symbol Synchronization for
  Diffusion-Based Molecular Communications},'' \emph{IEEE Trans. NanoBiosci.},
  vol.~16, no.~8, pp. 873--887, Dec. 2017.

\bibitem{Arman_TimeVariant}
A.~Ahmadzadeh, V.~Jamali, and R.~Schober, ``{Stochastic Channel Modeling for
  Diffusive Mobile Molecular Communication Systems},'' \emph{{IEEE} Trans.
  Commun.}, 2018.

\bibitem{Qiu2017}
S.~Qiu, T.~Asyhari, W.~Guo, S.~Wang, B.~Li, C.~Zhao, and M.~Leeson,
  ``{Molecular Channel Fading Due to Diffusivity Fluctuations},'' \emph{{IEEE}
  Commun. Lett.}, vol.~21, no.~3, pp. 676--679, Mar. 2017.

\bibitem{rappaport2002wireless}
T.~S. Rappaport, ``{Wireless Communications: Principles and Practice},''
  \emph{Microwave J.}, vol.~45, no.~12, pp. 128--129, 2002.

\bibitem{TableIntegSerie}
I.~S. Gradshteyn and I.~M. Ryzhik, \emph{Table of Integrals, Series, and
  Products}.\hskip 1em plus 0.5em minus 0.4em\relax 7th ed. Academic, 2007.

\bibitem{marcone2017gaussian}
A.~Marcone, M.~Pierobon, and M.~Magarini, ``{The Gaussian Approximation in Soft
  Detection for Molecular Communication via Biological Circuits},'' in
  \emph{IEEE Int. Workshop Sig. Process. Advances in Wireless Commun.
  (SPAWC)}.\hskip 1em plus 0.5em minus 0.4em\relax IEEE, 2017, pp. 1--6.

\bibitem{alt1980biased}
W.~Alt, ``{Biased Rrandom Walk Models for Chemotaxis and Related Diffusion
  Approximations},'' \emph{J. Mathematical Biology}, vol.~9, no.~2, pp.
  147--177, 1980.

\bibitem{hong2007chemotaxis}
Y.~Hong, N.~M. Blackman, N.~D. Kopp, A.~Sen, and D.~Velegol, ``{Chemotaxis of
  Nonbiological Colloidal Rods},'' \emph{Physical Review Lett.}, vol.~99,
  no.~17, p. 178103, 2007.

\bibitem{ArmanMobileMC}
A.~Ahmadzadeh, V.~Jamali, A.~Noel, and R.~Schober, ``{Diffusive Mobile
  Molecular Communications Over Time-Variant Channels},'' \emph{IEEE Commun.
  Lett.}, vol.~21, no.~6, pp. 1265--1268, Jun. 2017.

\bibitem{Falk2010LRE}
M.~Falk, J.~H{\"u}sler, and R.-D. Reiss, \emph{{Laws of Small Numbers: Extremes
  and Rare Events}}.\hskip 1em plus 0.5em minus 0.4em\relax Springer Science \&
  Business Media, 2010.

\bibitem{Bialek2012biophysics}
W.~Bialek, \emph{{Biophysics: Searching for Principles}}.\hskip 1em plus 0.5em
  minus 0.4em\relax Princeton Univ. Press, 2012.

\bibitem{Poisson_Channel}
D.~Guo, S.~Shamai, and S.~Verdu, ``{Mutual Information and Conditional Mean
  Estimation in Poisson Channels},'' \emph{IEEE Trans. Inf. Theory}, vol.~54,
  no.~5, pp. 1837--1849, May 2008.

\bibitem{CC_TCOM}
V.~Jamali, A.~Ahmadzadeh, N.~Farsad, and R.~Schober, ``{Constant-Composition
  Codes for Maximum Likelihood Detection without CSI in Diffusive Molecular
  Communications},'' \emph{IEEE Trans. Commun.}, vol.~66, no.~5, pp.
  1981--1995, May 2018.

\bibitem{UysalFSOsurvey}
M.~Khalighi and M.~Uysal, ``{Survey on Free Space Optical Communication: A
  Communication Theory Perspective},'' \emph{IEEE Commun. Surveys Tutorials},
  vol.~16, no.~4, pp. 2231--2258, 2014.

\bibitem{benesty2009pearson}
J.~Benesty, J.~Chen, Y.~Huang, and I.~Cohen, ``{Pearson Correlation
  Coefficient},'' in \emph{Noise Reduction in Speech Processing}.\hskip 1em
  plus 0.5em minus 0.4em\relax Springer, 2009, pp. 1--4.

\bibitem{lopez2009opportunistic}
J.~Lopez, A.~Bel, J.~A. Lopez-Salcedo, and G.~Seco-Granados, ``{Opportunistic
  Relay Selection with Outdated CSI: Outage Probability and Diversity
  Analysis},'' \emph{IEEE Trans. Wireless Commun.}, vol.~8, no.~6, pp.
  2872--2876, Jun. 2009.

\bibitem{Lett_CSI}
V.~Jamali, N.~Waly, N.~Zlatanov, and R.~Schober, ``{Optimal Buffer-Aided
  Relaying With Imperfect CSI},'' \emph{IEEE Commun. Lett.}, vol.~20, no.~7,
  pp. 1309--1312, Jul. 2016.

\bibitem{ma2009error}
Y.~Ma, D.~Zhang, A.~Leith, and Z.~Wang, ``{Error Performance of Transmit
  Beamforming with Delayed and Limited Feedback},'' \emph{IEEE Trans. Wireless
  Commun.}, vol.~8, no.~3, pp. 1164--1170, Mar. 2009.

\bibitem{Noel2017a}
A.~Noel, K.~C. Cheung, R.~Schober, D.~Makrakis, and A.~Hafid, ``{Simulating
  with {AcCoRD}: Actor-based Communication via Reaction--Diffusion},''
  \emph{Nano Commun. Net.}, vol.~11, pp. 44--75, Mar. 2017.

\bibitem{COMSOL}
\BIBentryALTinterwordspacing
{COMSOL Inc.}, ``{COMSOL Multiphysics}.'' [Online]. Available:
  \url{http://www.comsol.com}
\BIBentrySTDinterwordspacing

\bibitem{ANSYS}
\BIBentryALTinterwordspacing
{ANSYS Inc.}, ``{ANSYS}.'' [Online]. Available: \url{http://www.ansys.com}
\BIBentrySTDinterwordspacing

\bibitem{Gillespie1992}
D.~T. Gillespie, ``{A Rigorous Derivation of the Chemical Master Equation},''
  \emph{Physica A: Statistical Mechanics and its Applications}, vol. 188, no.
  1-3, pp. 404--425, Sep. 1992.

\bibitem{Ramaswamy2011}
R.~Ramaswamy and I.~F. Sbalzarini, ``{Exact On-Lattice Stochastic
  Reaction-Diffusion Simulations Using Partial-Propensity Methods},'' \emph{J.
  Chemical Physics}, vol. 135, no.~24, p. 244103, 2011.

\bibitem{noel_advection}
A.~Noel and D.~Makrakis, ``{Algorithm for Mesoscopic Advection-Diffusion},''
  \emph{IEEE Trans. NanoBiosci.}, vol.~17, no.~4, pp. 543--554, Oct. 2018.

\bibitem{Flegg2014}
M.~B. Flegg, S.~J. Chapman, L.~Zheng, and R.~Erban, ``{Analysis of the
  Two-Regime Method on Square Meshes},'' \emph{SIAM J. Sci. Comput.}, vol.~36,
  no.~3, pp. 561--588, Jun. 2014.

\bibitem{Bernstein2005}
D.~Bernstein, ``{Simulating Mesoscopic Reaction-Diffusion Systems Using the
  {Gillespie} Algorithm},'' \emph{Physical Review E - Statistical, Nonlinear,
  and Soft Matter Physics}, vol.~71, no.~4, pp. 1--13, Apr. 2005.

\bibitem{Gibson2000}
M.~A. Gibson and J.~Bruck, ``{Efficient Exact Stochastic Simulation of Chemical
  Systems with Many Species and Many Channels},'' \emph{J. Physical Chemistry
  A}, vol. 104, no.~9, pp. 1876--1889, 2000.

\bibitem{Gillespie2001}
D.~T. Gillespie, ``{Approximate Accelerated Stochastic Simulation of Chemically
  Reacting Systems},'' \emph{J. Chemistry Physics}, vol. 115, no.~4, pp. 1716
  -- 1733, 2001.

\bibitem{Andrews2009}
S.~S. Andrews, ``{Accurate Particle-based Simulation of Adsorption, Desorption
  and Partial Transmission},'' \emph{Physical Biology}, vol.~6, no.~4, p.
  046015, 2009.

\bibitem{Felicetti2013}
L.~Felicetti, M.~Femminella, and G.~Reali, ``{Simulation of Molecular Signaling
  in Blood Vessels: Software Design and Application to Atherogenesis},''
  \emph{Nano Commun. Net.}, vol.~4, no.~3, pp. 98--119, 2013.

\bibitem{Llatser2011}
I.~Llatser, I.~Pascual, N.~Garralda, A.~Cabellos-Aparicio, M.~Pierobon,
  E.~Alarc{\'{o}}n, and J.~{Sol{\'{e}} Pareta}, ``{Exploring the Physical
  Channel of Diffusion-based Molecular Communication by Simulation},'' in
  \emph{Proc. IEEE Global Commun. Conf. (Globecom)}, 2011, pp. 1--5.

\bibitem{Plimpton1995}
S.~Plimpton, ``{Fast Parallel Algorithms for Short-Range Molecular Dynamics},''
  \emph{J. Computational Physics}, vol. 117, no.~1, pp. 1--19, Mar. 1995.

\bibitem{Robinson2015}
M.~Robinson, S.~S. Andrews, and R.~Erban, ``{Multiscale Reaction-Diffusion
  Simulations with {Smoldyn}},'' \emph{Bioinformatics}, vol.~31, no.~14, pp.
  2406--2408, Jul. 2015.

\bibitem{Wagner2008}
G.~Wagner, R.~Jones, J.~Templeton, and M.~Parks, ``{An Atomistic-to-Continuum
  Coupling Method for Heat Transfer in Solids},'' \emph{Computer Methods in
  Applied Mechanics and Engineering}, vol. 197, no.~41, pp. 3351--3365, 2008.

\bibitem{Resasco2012}
D.~C. Resasco, F.~Gao, F.~Morgan, I.~L. Novak, J.~C. Schaff, and B.~M.
  Slepchenko, ``{Virtual Cell}: Computational tools for modeling in cell
  biology,'' \emph{Wiley Interdisciplinary Reviews: Systems Biology and
  Medicine}, vol.~4, no.~2, pp. 129--140, Mar. 2012.

\bibitem{grebenstein_biological_2018}
L.~Grebenstein, J.~Kirchner, R.~S. Peixoto, W.~Zimmermann, F.~Irnstorfer,
  W.~Wicke, A.~Ahmadzadeh, V.~Jamali, G.~Fischer, R.~Weigel, A.~Burkovski, and
  R.~Schober, ``{Biological {{Optical}}-to-{{Chemical Signal Conversion
  Interface}}: {{A Small}}-Scale {{Modulator}} for {{Molecular
  Communications}}},'' \emph{IEEE Trans. NanoBiosci.}, 2018.

\bibitem{atthanayake2018experimental}
I.~Atthanayake, S.~Esfahani, P.~Denissenko, I.~Guymer, P.~J. Thomas, and
  W.~Guo, ``{Experimental Molecular Communications in Obstacle Rich Fluids},''
  in \emph{Proc. ACM Nanocom}, 2018, pp. 32:1--32:2.

\bibitem{tuccitto_fluorescent_2017}
N.~Tuccitto, G.~{Li-Destri}, G.~M.~L. Messina, and G.~Marletta, ``{Fluorescent
  {{Quantum Dots Make Feasible Long}}-{{Range Transmission}} of {{Molecular
  Bits}}},'' \emph{J. Physical Chemistry Lett.}, vol.~8, no.~16, pp.
  3861--3866, Aug. 2017.

\bibitem{kennedy_spatiotemporal_2018}
E.~Kennedy, P.~Shakya, M.~Ozmen, C.~Rose, and J.~K. Rosenstein,
  ``{Spatiotemporal Information Preservation in Turbulent Vapor Plumes},''
  \emph{Applied Physics Lett.}, vol. 112, no.~26, p. 264103, Jun. 2018.

\bibitem{giannoukos_molecular_2017}
S.~Giannoukos, A.~Marshall, S.~Taylor, and J.~Smith,
  ``\BIBforeignlanguage{en}{{Molecular {{Communication}} over {{Gas Stream
  Channels}} Using {{Portable Mass Spectrometry}}}},''
  \emph{\BIBforeignlanguage{en}{J. American Society for Mass Spectrometry}},
  vol.~28, no.~11, pp. 2371--2383, Nov. 2017.

\bibitem{deleo_communications_2013}
E.~De~Leo, L.~Donvito, L.~Galluccio, A.~Lombardo, G.~Morabito, and L.~M.
  Zanoli, ``Communications and {{Switching}} in {{Microfluidic Systems}}:
  {{Pure Hydrodynamic Control}} for {{Networking Labs}}-on-a-{{Chip}},''
  \emph{IEEE Trans. Commun.}, vol.~61, no.~11, pp. 4663--4677, Nov. 2013.

\bibitem{krishnaswamy_timeelapse_2013}
B.~Krishnaswamy, C.~M. Austin, J.~P. Bardill, D.~Russakow, G.~L. Holst, B.~K.
  Hammer, C.~R. Forest, and R.~Sivakumar, ``Time-{{Elapse Communication}}:
  {{Bacterial Communication}} on a {{Microfluidic Chip}},'' \emph{IEEE Trans.
  Commun.}, vol.~61, no.~12, pp. 5139--5151, Dec. 2013.

\bibitem{felicetti_modeling_2014}
L.~Felicetti, M.~Femminella, G.~Reali, P.~Gresele, M.~Malvestiti, and J.~N.
  Daigle, ``{Modeling {CD40}-Based Molecular Communications in Blood
  Vessels},'' \emph{IEEE Trans. NanoBiosci.}, vol.~13, no.~3, pp. 230--243,
  Sep. 2014.

\bibitem{Nakano_Microplatform_2008}
T.~Nakano, Y.-H. Hsu, W.~C. Tang, T.~Suda, D.~Lin, T.~Koujin, T.~Haraguchi, and
  Y.~Hiraoka, ``{Microplatform for Intercellular Communication},'' in
  \emph{Proc. IEEE Int. Conf. Nano/Micro Eng. Molecular Syst. (NEMS)}, Jan.
  2008, pp. 476--479.

\bibitem{bicen_efficient_2015}
A.~O. Bicen, C.~M. Austin, I.~F. Akyildiz, and C.~R. Forest, ``Efficient
  {{Sampling}} of {{Bacterial Signal Transduction}} for {{Detection}} of
  {{Pulse}}-{{Amplitude Modulated Molecular Signals}},'' \emph{IEEE Trans.
  Biomed. Circuits Syst.}, vol.~9, no.~4, pp. 505--517, Aug. 2015.

\bibitem{farsad_channel_2014}
N.~Farsad, N.-R. Kim, A.~W. Eckford, and C.-B. Chae, ``Channel and {{Noise
  Models}} for {{Nonlinear Molecular Communication Systems}},'' \emph{IEEE J.
  Select. Areas Commun.}, vol.~32, no.~12, pp. 2392--2401, Dec. 2014.

\bibitem{zifarelli2008buffered}
G.~Zifarelli, P.~Soliani, and M.~Pusch, ``{Buffered Diffusion Around a
  Spherical Proton Pumping Cell: A Theoretical Analysis},'' \emph{Biophys. J.},
  vol.~94, no.~1, pp. 53--62, Sep. 2008.

\bibitem{Choi_Cyanobacterial_2014}
A.~R. Choi, L.~Shi, L.~S. Brown, and K.-H. Jung, ``{Cyanobacterial Light-driven
  Proton Pump, Gloeobacter Rhodopsin: Complementarity Between Rhodopsin-based
  Energy Production and Photosynthesis},'' \emph{PLoS One}, vol.~9, no.~10, p.
  e110643, 2014.

\bibitem{Lanyi_Bacteriorhodopsin_2004}
J.~K. Lanyi, ``Bacteriorhodopsin,'' \emph{Annu. Rev. Physiol.}, vol.~66, pp.
  665--688, 2004.

\bibitem{goodfellow2016deep}
I.~Goodfellow, Y.~Bengio, A.~Courville, and Y.~Bengio, \emph{{Deep
  Learning}}.\hskip 1em plus 0.5em minus 0.4em\relax MIT Press Cambridge, 2016,
  vol.~1.

\bibitem{lee2017machine}
C.~Lee, H.~B. Yilmaz, C.-B. Chae, N.~Farsad, and A.~Goldsmith, ``{Machine
  Learning Based Channel Modeling for Molecular MIMO Communications},'' in
  \emph{IEEE Int. Workshop Sig. Process. Advances in Wireless Commun. (SPAWC)},
  2017, pp. 1--5.

\bibitem{goodfellow2014generative}
I.~Goodfellow, J.~Pouget-Abadie, M.~Mirza, B.~Xu, D.~Warde-Farley, S.~Ozair,
  A.~Courville, and Y.~Bengio, ``{Generative Adversarial Nets},'' in
  \emph{Advances in Neural Information Processing Systems}, 2014, pp.
  2672--2680.

\bibitem{touchMC2015}
Y.~Chen, P.~Kosmas, P.~S. Anwar, and L.~Huang, ``{A Touch-Communication
  Framework for Drug Delivery Based on a Transient Microbot System},''
  \emph{IEEE Trans. NanoBiosci.}, vol.~14, no.~4, pp. 397--408, Jun. 2015.

\bibitem{McGuiness2018macroscale}
D.~T. McGuiness, S.~Giannoukos, A.~Marshall, and S.~Taylor, ``{Parameter
  Analysis in Macro-Scale Molecular Communications Using
  Advection-Diffusion},'' \emph{IEEE Access}, vol.~6, pp. 46\,706--46\,717,
  2018.

\bibitem{2018arXiv180104627C}
Y.~{Chen}, Y.~{Zhou}, R.~{Murch}, and T.~{Nakano}, ``{Molecular Communications
  at the Macroscale: A Novel Framework for Modeling Epidemic Spreading and
  Mitigation},'' \emph{ArXiv e-prints}, Jan. 2018.

\end{thebibliography}

\end{document}